\documentclass{aa}  

\usepackage{graphicx}
\usepackage{lscape}
\usepackage{natbib} 
\usepackage{siunitx}
\usepackage{url}

\usepackage[varg]{txfonts}

\bibpunct{(}{)}{;}{a}{}{,} 
\defcitealias{HK2000}{HK2000}
\defcitealias{Hamann2006}{HGL06}
\defcitealias{BAT99}{BAT99}

\begin{document}

\title{The Wolf-Rayet stars in the Large Magellanic
Cloud\thanks{Partly based on observations made with the NASA/ESA Hubble
Space Telescope, and obtained from the Hubble Legacy Archive, which is a
collaboration between the Space Telescope Science Institute
(STScI/NASA), the Space Telescope European Coordinating Facility
(ST-ECF/ESA), and the Canadian Astronomy Data Centre (CADC/NRC/CSA).}}

   \subtitle{A comprehensive analysis of the WN class}

   \author{R. Hainich\inst{1}
          \and U. R\"uhling\inst{1}
          \and H. Todt\inst{1}
          \and L. M. Oskinova\inst{1}
          \and A. Liermann\inst{2}
          \and G. Gr{\"a}fener\inst{3}
          \and C. Foellmi\inst{4}
          \and O.~Schnurr\inst{2}
          \and W.-R. Hamann\inst{1}
          }

   \institute{Institut f\"ur Physik und Astronomie,
              Universit\"at Potsdam,
              Karl-Liebknecht-Str. 24/25, D-14476 Potsdam, Germany \\
              \email{rhainich@astro.physik.uni-potsdam.de}
              \and 
              Leibniz-Institut f\"ur Astrophysik Potsdam,
              An der Sternwarte 16, D-14482 Potsdam, Germany
              \and 
              Armagh Observatory, College Hill,
              Armagh BT6\,9D, UK
              \and
              12 rue Servan, 38000 Grenoble, France
              }
   \date{Received <date> / Accepted <date>}


\abstract
%
%
{Massive stars, although being important building blocks of galaxies, 
are still not fully understood. This especially holds true for 
Wolf-Rayet (WR) stars with their strong mass loss, whose
spectral analysis requires adequate model atmospheres.
} 
%
%
{Following our comprehensive studies of the WR stars in the Milky Way,
we now present spectroscopic analyses of almost all known WN stars in
the LMC. 
}
%
%
{For the quantitative analysis of the wind-dominated emission-line spectra, 
we employ the Potsdam Wolf-Rayet (PoWR) model atmosphere code. 
By fitting synthetic spectra to the observed spectral energy distribution
and the available spectra (ultraviolet and optical), we obtain the physical 
properties of 107 stars.
} 
%
%
{We present the fundamental stellar and wind parameters for an almost
complete sample of WN stars in the LMC. Among those stars that are 
putatively single, two different groups can be clearly distinguished. While 
12\,\% of our sample are more luminous than $10^6\,L_\odot$ and 
contain a significant amount of hydrogen, 88\,\% of the WN stars, with 
little or no hydrogen, populate the luminosity range between 
$\log\,(L/L_\odot) = 5.3\,...\,5.8$.
}
%
%
{While the few extremely luminous stars ($\log\,(L/L_\odot)\,>\,6$), if indeed
single stars, descended directly from the main sequence at very high initial 
masses, the bulk of WN stars have gone through the red-supergiant phase. 
According to their luminosities in the range of $\log\,(L/L_\odot) = 
5.3\,...\,5.8$, these stars originate from initial masses
between 20 and 40\,$M_\odot$. This mass range is similar to the one found
in the Galaxy, i.e.\ the expected metallicity dependence of the
evolution is not seen.  
Current stellar evolution tracks, even when accounting for rotationally induced
mixing, still partly fail to reproduce the observed
ranges of luminosities and initial masses. Moreover, 
stellar radii are generally larger and effective temperatures
correspondingly lower than predicted from stellar evolution models, probably
due to subphotospheric inflation.   
}


\maketitle

\section{Introduction}
\label{sect:intro}

The Large Magellanic Cloud (LMC) is one of the closest galaxies to the
Milky Way (MW), allowing detailed spectroscopy of its brighter stars. 
Its distance modulus of only $DM = 18.5$\,mag is well constrained
\citep{Madore1998,Pietrzynski2013}. Another advantage in analyzing
stars of the LMC is the marginal reddening along the line of sight 
\citep{Subramaniam2005,Haschke2011}, which is in general below 
$E_{b-v}=0.25\,\mathrm{mag}$ \citep{Larsen2000}.

Compared to our Galaxy, the LMC is much smaller and has a deviating
structure that is intermediate between a dwarf spiral and an irregular
type. The LMC exhibits a very different history of star formation than the MW.
The metallicity observed in LMC stars is, in general, subsolar 
\citep[$Z / Z_\odot \sim 0.4$,][]{Dufour1982}, but with a
strong age-dependence \citep[e.g.,][]{Piatti2013}. For young massive
stars, it may reach nearly solar values. In the stellar evolution
calculations that we will discuss below, \citet{Meynet2005} adopted $Z =
0.008$, which is about 60\,\% of the solar value \citep{Asplund2009}. 

The metallicity is expected to have significant influence on the 
evolution of massive stars as it has impact on the mass loss due to 
stellar winds. As far as these winds are driven by radiation 
pressure on spectral lines of metals like iron, the mass-loss rate 
is expected to scale with $Z^{m}$ with $m \approx 0.5$ \citep[e.g.,][]{Kudritzki1989}.
Based on their theoretical models, \citet{Vink2005} derived an exponent 
of $ m = 0.86$ for late-type Wolf-Rayet (WR) stars. The fact that the distribution of the
WR stars on the subclasses  (i.e., the nitrogen sequence: WN
and the carbon sequence: WC) strongly differs between the LMC and the
MW is generally attributed to this metallicity effect. \citet{Eldridge2006} 
found that the mass-loss rates from \citet{Vink2005} can account for the 
observed WC/WN ratio as a function of the metallicity. The metallicity 
dependence of the WN mass-loss was affirmed by the hydrodynamic 
stellar wind models presented by \citet{graefener+hamann-2008}. 

In previous papers, we have concentrated on analyzing the WR
population of the MW. In \citet[][hereafter HGL06]{Hamann2006}, we 
presented a comprehensive analysis of the Galactic WN stars, 
while the WC subtypes were studied by \citet{Sander2012}. For both
classes we found discrepancies between the parameters of the observed 
WR population and the predictions of the available stellar evolution 
calculations.   

The current paper focuses on the WR stars in the LMC. With more
than 100 objects, our sample comprise nearly all WN-type stars known 
in the LMC. In contrast, earlier analyses of WN stars in the LMC 
were limited to a sample size below 20 objects 
\citep{Crowther1997,Crowther1998,HK2000} and were often confined to 
specific subclasses \citep{Crowther1995b,Pasquali1997}.

At the time of these studies, stellar atmosphere models commonly 
did not yet account for iron-line blanketing 
\citep{Hillier1999,Graefener2002} and wind inhomogeneities. The 
inclusion of these two effects, the latter by means of the 
microclumping approach \citep[cf.][]{HK98}, significantly improved stellar 
atmosphere models and entailed a pervasive revision of the derived 
stellar parameters \citep[e.g.,][]{HK2000,Crowther2002,Crowther2010,Sander2012}. 
Similar profound improvements were achieved in the field of stellar 
evolution by the inclusion of physical processes such as stellar 
rotation \citep{Meynet2003,Meynet2005} and, more recently, magnetic fields 
\citep{Maeder2005,Yoon2012}.

In the last decade, high signal to noise spectra in the optical spectral range 
of almost all WN stars in the LMC were obtained in extensive spectroscopic 
studies realized by \citet{Foellmi2003b} and \citet{Schnurr2008}. For the 
first time, the spectra obtained by these two studies make it possible to 
analyze a comprehensive sample of LMC WN stars. By this means, we  
obtain a general overview of a nearly complete WN-star population,
which we employ to test state-of-the-art evolution models.

This paper is organized as follows: In the next Section, we
introduce our sample of stars and the observational data employed. 
In Sect.\,\ref{sect:models}, we briefly characterize the Potsdam
Wolf-Rayet (PoWR) model atmospheres. The method of our analyses 
is described in Sect.\,\ref{sect:method}. The results are compiled 
in Sect.\,\ref{sect:results}. In Sect.\,\ref{sect:evolution}, we 
discuss our results with respect to the stellar evolution theory. 
A summary and conclusions are presented in Sect.\,\ref{sect:conclusions}.

The {\em Online Material} gives details about the
observational data (Appendix\,\ref{sec:addtables}) and comments on
the individual stars (Appendix\,\ref{sec:comments}). Finally, we provide 
spectral fits for all sample stars (Appendix\,\ref{sec:specfits}).

\section{The sample}
\label{sect:sample}

\subsection{Sample selection}
\label{subsect:sample}

Our sample is based on the fourth catalog of WR stars in the
LMC \citep[][hereafter BAT99]{BAT99}. Throughout this paper, we
identify the stars by their running number in that list. In this
catalog, a spectral type of the WN sequence was assigned to each
of the 109 objects. 

In a few cases, the spectral classification had to be revised. The stars
BAT99\,45 and BAT99\,83 are actually luminous blue variables (LBVs)
\citep[see][]{Humphreys1994,Schnurr2008} and are, therefore, excluded 
from our sample. 

Five of the stars listed with a WN classification in the BAT99 catalog 
have been reclassified as Of-types: 
BAT99\,107 has been identified as a massive spectroscopic binary system
comprising two Of-type stars \citep{Taylor2011}. \citet{Niemela2001} 
found BAT99\,6 to be an O-type binary system as well.
\citet{Crowther2011} have reclassified BAT99\,105 and BAT99\,110 as O2If*
stars. The spectral type O3If* has been assigned to BAT99\,93 by
\citet{Evans2011}. Despite their reclassifications, we keep these
O-type objects in our sample. Thus, the number of proper WN stars from 
the original BAT99 catalog is reduced to 102. 

Since the publication of the BAT99 catalog, only a few additional WN 
stars have been identified in the LMC. A list of the seven newly discovered 
WR stars, six of them WN-type stars, can be found in Table\,3 of 
\citet{Neugent2012}. Thus, the number of known WN stars in the LMC
amounts to 108, although two of these new detections are precarious.
\citet{Massey2000} identified Sk\,-69$^\circ$\,194 as
B0\,Ia\,+\,WN. However, \citet{Foellmi2003b} could not confirm this 
detection. \citet{Neugent2012} also list \mbox{LH\,90$\beta$-6} as a 
new WN star. However, according to \citet{Massey2000}, this is an 
alias of TSWR\,1, which was resolved into multiple components by 
\citet{Walborn1999} and incorporated in the BAT99 catalog with the 
number BAT99\,78. Therefore, the basis of this new detection is not clear.
Another six new WR stars in the LMC are 
reported by \citet{Reid2012}, but without giving coordinates or closer 
classifications. We do not include any of these newly discovered 
WN stars in our analyses.  

\begin{figure*}
\centering
\includegraphics[width=0.7\textwidth]{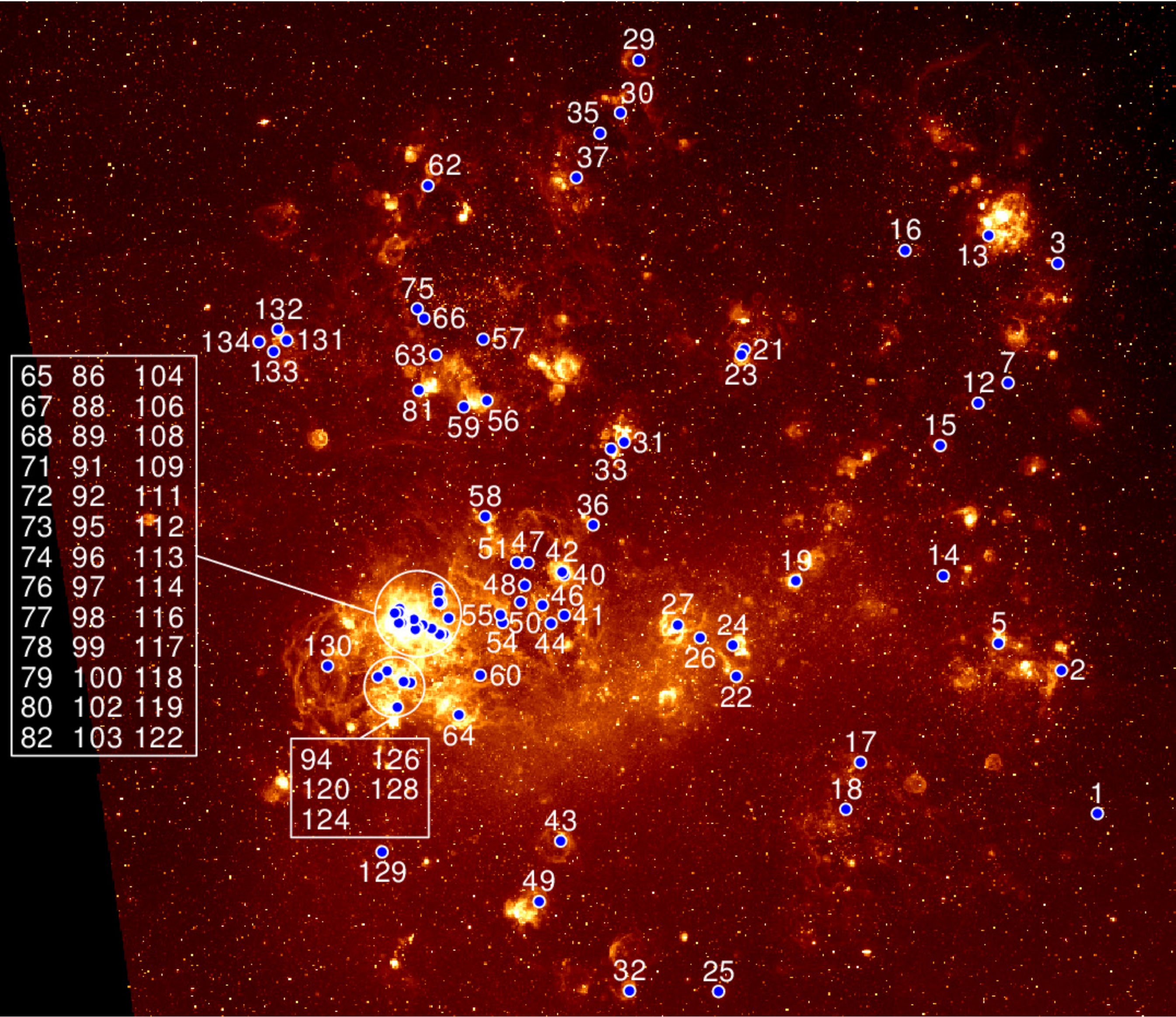}
\caption{The WN stars of our sample, identified by their
number in the BAT99 catalog. The two boxes refer to the 
very crowded region of 30\,Doradus. The H$\alpha$ image 
in the background is from the Magellanic
Cloud Emission-Line Survey \citep[MCELS, ][]{Smith2005}.
} 
\label{fig:Halpha_wn}
\end{figure*}

With 102 out of 108 known WN stars in the LMC, our sample covers this
class nearly completely with all subtypes present. The spatial 
distribution of our program stars is illustrated in 
Fig.\,\ref{fig:Halpha_wn}, and the complete list of analyzed objects 
(including the five Of stars) is compiled in Table\,\ref{table:parameters}. 
For the majority of our sample, the spectral types  have been 
determined by \citet{Foellmi2003b} and \citet{Schnurr2008}, 
respectively, based on the classification scheme elaborated by 
\citet{Smith1996}. For the handful of stars missing in their samples, 
we adopt the spectral type from the BAT99 catalog. A couple of our
stars have been reclassified by various authors since the publication
of these catalogs. The present classification of each star is quoted
in Table \ref{table:parameters}. The subtypes WN2 to WN5 are sometimes
referred to as WNE (``early''), while WN6 to WN11 are referred to 
as WNL (``late''). 

The total census of WR stars in the LMC, as far as they are
assigned to their subclass, amounts to 134. In addition to the 108 WN
stars, only 24 WC stars plus two WR stars with prominent oxygen lines 
(WO stars) were discovered \citep{Barlow1982,Neugent2012}. The 
composition of the WR population is thus very different from our Galaxy,
where the ratio of WN to WC stars is close to unity.

\subsection{Binaries}
\label{subsect:binaries}

Among the objects in our sample some may be binary (or multiple) systems.
We, therefore, carefully consider the binary status of each object. 
All stars for which \citet{Foellmi2003b}, \citet{Schnurr2008}, or BAT99 
list periodic radial-velocity variations are considered as 
confirmed binaries (cf.\ Table\,\ref{table:parameters}). 

For some of our targets there are less conclusive radial-velocity measurements or
binary classifications based on spectral peculiarities. Such cases are
considered binary suspects, as indicated by a question mark in
Table\,\ref{table:parameters} with the corresponding references.  

Another method to identify WR stars as binaries is to evaluate their
X-ray luminosity. According to studies of Galactic WR-stars, single
WC-type stars are not X-ray sources at all \citep{Oskinova2003}, while
single WN-type stars in general are relatively X-ray faint. Some of
them remain undetected in X-rays despite quite sensitive observations,
setting strict upper limits on the X-ray luminosity. For example, 
\citet{Gosset2005} obtained $L_{\mathrm X} < 2 \times 10^{30}$\,erg\,s$^{-1}$ for
WR\,40 (WN8). The X-ray luminosities of those WN
stars that were detected are relatively small, not exceeding a
few times $10^{32}$\,erg\,s$^{-1}$ \citep{Ignace2000, Ignace2003,
Skinner2012}. The mechanism of X-ray production in single WN stars is
not fully understood, but is thought to be related to the presence of
clumps and large scale structures in their winds
\citep{Chene2011, Oskinova2012}.

In contrast, binary WR stars with colliding winds are significantly
more X-ray luminous than single stars.  In such systems, the production
of X-rays is explained by the heating of gas in a strong shock that
results when two stellar winds collide \citep{Stevens1992}. Therefore, a
higher than usual X-ray luminosity can serve as a good indicator for a
colliding wind binary. As an example, the high X-ray luminosity of
the Galactic star WR\,25 ($L_{\mathrm X} = 1.3 \times
10^{34}$\,erg\,s$^{-1}$) provided strong indications that this star is
a colliding-wind binary \citep{Raassen2003}, as confirmed later from
radial velocity measurements \citep{Gamen2006}.

Binary WR-stars with a compact companion, i.e., with a neutron star
or black hole, are expected to have even higher X-ray luminosities 
exceeding $10^{35}$\,erg\,s$^{-1}$. The X-ray luminosities in these
systems are powered by the wind accretion onto the companion. An 
intriguing example is Cyg\,X-3, a Galactic high-mass X-ray binary 
with a WN-type primary, which has an X-ray luminosity of 
$10^{38}$\,erg\,s$^{-1}$ \citep[e.g.,][]{Lommen2005}.

The X-ray properties of WR stars in the Magellanic Clouds were studied
systematically by \citet{Guerrero2008I, Guerrero2008II}, using
observations with the X-ray observatories  {\em Rosat} and {\em
Chandra}. The sensitivity of these surveys was limited to X-ray
luminosities of a few times $ 10^{32}$\,erg\,s$^{-1}$. They detected
X-rays from 27 of the WR stars in the LMC, with X-ray luminosities
being similar to those of Galactic colliding-wind binaries. Since there
is no reason to assume that single WR-stars in the LMC are
intrinsically more bright in X-rays than in the Galaxy, we suspect all WR
stars detected by Guerrero \& Chu to be colliding wind binaries and mark
them accordingly in Table\,\ref{table:parameters}.

Altogether, our sample (without the Of stars) includes 17 confirmed 
binaries plus 22 binary suspects. From 108 known WN stars in the LMC, 
this corresponds to a binary frequency of only $16-36\,\%$. Although 
this binary fraction seems to be a bit low, it is in line with expectations 
from binary population studies 
\citep[e.g.,][]{Foellmi2003a,Chini2012,Sana2012,Sana2013b}. Moreover, 
there are most likely more binaries in our sample that are not yet recognized.

\subsection{Observational data}
\label{subsect:data}

This study was facilitated by optical spectra obtained by \citet{Foellmi2003b}
who observed 61 WNE stars with various instruments between 1998 and 
2002. These data are publicly
available\footnote{\url{http://wikimbad.obs.ujf-grenoble.fr/Category_Wolf-Rayet_Star.html}}.
The completeness of this study was only possible due to spectroscopic 
observations of 42 late-type WN stars carried out by \citet{Schnurr2008}. For 
details on the instrumentation and data reduction, we refer to 
\citet{Foellmi2003b} and \citet{Schnurr2008}. These two sets of data 
were primarily designed to search for radial velocity variations, the 
results being published in \citet{Foellmi2003b} and \citet{Schnurr2008}. 
These spectra were not flux-calibrated and have been normalized by the 
respective authors. 

From the VizieR archive we retrieved flux-calibrated, low-resolution
optical spectra for most of our targets, recorded by
\citet{Torres-Dodgen1988} on a SIT-vidicon detector at the Cassegrain
spectrograph of the 1.5\,m telescope of the Cerro Tololo Inter-American
Observatory (CTIO). Furthermore, we reused
19 observations dating back to 1989 \citep[cf.][]{Koesterke1991},
obtained with the the ESO Faint Object Spectrograph and Camera (EFOSC)
at the 3.6\,m telescope. Unreduced spectra of WN and Of stars 
observed with the Anglo-Australian Telescope (AAT) 
were obtained from the study by \citet{Crowther1997}. We
performed the wavelength calibration with given arc lamp data and
normalized them "by eye" if no other optical spectra were at hand. 

Ultraviolet spectra secured with the International Ultraviolet Explorer
(IUE) are available from the archives for almost all of the stars of our
sample, except for those located in the very crowded 30\,Dor region. 
Especially for some of the latter, UV and optical spectra were recorded with
spectrographs aboard the {\em Hubble Space Telescope} (HST). 
A subset of 19 stars have been observed with the Far Ultraviolet Spectroscopic
Explorer (FUSE), but these data were not used in the current study. 
The FUSE spectra will be the subject of a detailed abundance analysis
of LMC WN stars in a subsequent paper. 

Before fitting the observed spectra, we corrected the wavelengths  for
the radial velocities of the individual stars, mostly taken from 
\citet{Foellmi2003b} and \citet{Schnurr2008}. The details about the 
origin of all spectra employed in this paper are compiled in
Table\,\ref{table:data} in the {\em Online Material}.

We used narrowband optical photometry ($u, b, v$) obtained by 
\citet{Crowther2006b} whenever available. Otherwise, 
we used the older measurements from \citet{Torres-Dodgen1988}, 
and finally complemented the data with values 
from BAT99. Near-infrared magnitudes ($J, H, K_S$) were
retrieved from the 2MASS catalog \citep{2MASS}, except for those stars 
located in the crowded field of 30\,Dor.
Photometry from the InfraRed Array Camera (IRAC, $3.6, 4.5, 5.8$, and 
$8.0\,\mu$m) of the {\em Spitzer Space Telescope} is available for most
stars from the catalog by \citet{Bonanos2009}.

\section{The models}
\label{sect:models}

Our spectral analyses are based on non-local thermodynamic equilibrium 
(non-LTE) model atmospheres calculated with the PoWR code. 
Its basic assumptions are spherical symmetry and stationarity of the 
flow. The radiative transfer equation is solved in the comoving frame, 
iteratively with the equations of statistical equilibrium  and radiative 
equilibrium. For more details of the PoWR code, see \citet{Hamann2004}.

The main parameters of a model atmosphere are the luminosity $L$ and the
``stellar temperature''  $T_*$. The latter is the effective temperature
related to the stellar radius $R_*$ via the Stefan-Boltzmann law 
\begin{equation}
\label{eq:sblaw}
L = 4 \pi \sigma R_*^2 T_*^4~.
\end{equation}
The stellar radius $R_*$ is per definition located at a radial Rosseland 
optical depth of 20, which represents the lower boundary of the model atmosphere. 

Additional parameters, which describe the stellar wind, can 
be combined in the so-called transformed radius $R_{\mathrm t}$. This
quantity was introduced by \citet{Schmutz1989}; we define it as
\begin{equation}
\label{eq:rt}
R_{\mathrm{t}} = 
R_* \left(\frac{\varv_\infty}{2500 \, \mathrm{km}\,\mathrm{s^{-1}}} 
\left/
\frac{\dot M \sqrt{D}}{10^{-4} \, M_\odot \, \mathrm{
    yr^{-1}}}\right)^{2/3}
\right. 
\end{equation}
with $\varv_\infty$ denoting the terminal wind velocity, $\dot M$ the
mass-loss rate, and $D$ the clumping contrast (see below). 
\citet{Schmutz1989} noticed that  model spectra with equal $R_{\mathrm
t}$ exhibit approximately the same emission line strengths, independent
of the specific combination of the particular wind parameters as long
as $T_*$ and the chemical composition are the same. Even the line
profile is conserved under the additional condition that 
$v_\infty$ is also kept constant. One can understand this invariance when
realizing that $R_{\mathrm t}$  is related to the ratio between the
volume emission measure and the stellar surface area. 

According to this scaling invariance, a model can be scaled to a different 
luminosity as long as $R_{\mathrm t}$ and $T_*$ are unchanged. 
Equation\,(\ref{eq:rt}) implies that the mass-loss rate then must be scaled 
proportional to $L^{3/4}$ in order to preserve the normalized line spectrum.

Allowing for wind inhomogeneities, the ``density contrast''  $D$ is the 
factor by which the density in the clumps is enhanced compared  to a
homogeneous wind of the same $\dot{M}$. We account for wind clumping in the
approximation of optically thin structures \citep{Hillier1991,HK98}. 
From the analysis of the electron-scattering line wings in 
Galactic WN stars, \citet{HK98} found that a density contrast of
$D=4$ is adequate. To the contrary, \citet{Crowther2010} and \cite{Doran2013} inferred 
$D=10$ in their analyses of WN stars in the 30\,Doradus region. For the 
current study, we uniformly adopt a density contrast of $D=10$, because we noticed in a detailed 
investigation of a subsample that with $D=4$, the line-scattering wings 
in the models are stronger than observed. Note that the empirical 
mass-loss rates derived in this work scale with $D^{-1/2}$ 
(cf.\ Eq.\,\ref{eq:rt}). 

For the Doppler velocity $\varv_{\mathrm D}$, describing the line
broadening due to microturbulence and thermal motion, we adopt
a value of $100\,\mathrm{km\,s^{-1}}$, which provides a good fit to 
the data and is approved in previous studies
(e.g.,\,\citealt{HK2000}, hereafter HK2000; \citetalias{Hamann2006}).

For the velocity law $\varv(r)$ in the supersonic part of the wind, we
adopt the so-called $\beta$-law. For the exponent $\beta$, the 
radiation-driven wind theory predicts about 0.8 in agreement with
observations \citep[e.g.,][]{Pauldrach1986}. In WN stars, the law is more shallow because of 
multiple-scattering effects. We adopt $\beta=1$, which better resembles
the hydrodynamic prediction \citep{GH2007},  and yields consistent spectral
fits.  In the subsonic part,  the velocity field is implied by the
hydrostatic density stratification according to the continuity
equation. 

The models are calculated using complex atomic data of H, He, C, and N.
Iron group elements are considered in the ``superlevel approach'' that
encompasses $\sim 10^7$ line transitions between $\sim 10^5$ levels
within 72 superlevels \citep{Graefener2002}. 


\section{Method}
\label{sect:method}

To facilitate the analysis of a large number of WN stars, we first
establish grids of models. As explained above, the main parameters are
the stellar temperature $T_\ast$ and the transformed radius 
$R_{\mathrm t}$. 

\subsection{Abundances}
\label{sect:abundances}

We calculated three grids of models for different hydrogen abundances:
one hydrogen-free ``WNE'' grid and two ``WNL'' grids with hydrogen mass
fractions of 0.2 and 0.4, respectively. From the trace elements, 
we account for carbon, nitrogen, and a generic model atom 
re\-presenting the iron-group elements in relative solar mixture 
\citep{Graefener2002}.  

The material in the WN atmosphere has undergone at least partial CNO
burning due to mixing processes in the stellar interior, such as 
rotational induced mixing \citep[e.g.,][]{Heger2000}. Accordingly, 
most of the oxygen and carbon was transformed into nitrogen. Assuming 
equilibrium, the remaining mass fractions of oxygen and carbon relative 
to that of nitrogen should be only 1/60 \citep{Schaerer1993}. Hence, 
the nitrogen abundance should roughly equal the sum of the C, N, and O
abundances of the initial material from which the star was formed.
Note that \citetalias{HK2000} inferred a nitrogen abundance for the 
Galactic WN stars nearly twice the sum of solar CNO 
(cf.\ Table\,\ref{table:abundances}). 

As a reference for LMC abundances, we use spectral analyses of B-type stars. 
The results from two such studies \citep{Hunter2007, Korn2005} are
listed in Table\,\ref{table:abundances}. Moreover, the table gives
abundances found in \ion{H}{ii} regions of the LMC
\citep{Dufour1998}. Compared to solar abundances, the sum of C, N, and
O in these LMC objects is roughly half the solar value. We, therefore,
adopt a nitrogen mass fraction of 0.004 for the models throughout
this paper. The carbon abundance is set to 1/60 of this value. We
neglect oxygen in our WN star models, since no prominent O lines are
present in the optical wavelength range, nor do we expect it to influence 
the atmospheric stratification. 

The iron abundance in B-type stars from several clusters in the LMC has 
been studied by \citet{Trundle2007} (see 
Table\,\ref{table:abundances}). On average, this value is 0.0007 (mass fraction),
i.e.,\ again roughly half the solar iron abundance. We adopted this value
as  the iron-group abundance for our LMC models. 

A detailed abundance analysis is beyond the scope of the present paper.
However, the spectral fits presented below reveal that the models 
cannot reproduce the observed nitrogen lines for a subset 
of our sample. Thus, it seems that the N abundance in these LMC WN stars
is slightly higher than our adopted value.

\begin{table}
\caption{Chemical composition (mass fractions in percent)} 
\label{table:abundances}      
\centering
\small
\tabcolsep 0.4ex
\begin{tabular}{lSSSSSS}
\hline\hline 
 & \multicolumn{1}{r}{Sun\tablefootmark{a}} 
 & \multicolumn{1}{r}{Gal.} 
 & \multicolumn{4}{c}{\rule[1ex]{16mm}{0.5pt} LMC \rule[1ex]{16mm}{0.3pt} }
 \rule[0mm]{0mm}{2.2ex}\\
&& \multicolumn{1}{r}{WN\tablefootmark{b}} 
 & \multicolumn{1}{r}{B\,stars\tablefootmark{c}} 
 & \multicolumn{1}{r}{B\,stars\tablefootmark{d}} 
 & \multicolumn{1}{r}{\ion{H}{ii}\tablefootmark{e}}
 & \multicolumn{1}{r}{WN\tablefootmark{f}} \\
\hline 
C        & 0.237  & 0.01  & 0.054 & 0.086 & 0.058  &
                                            0.0067\rule[0mm]{0mm}{2.2ex} \\
N        & 0.069 & 1.5   & 0.0083 & 0.011 & 0.0087 & 0.40 \\
O        & 0.573  & \multicolumn{1}{c}{-}    
                         & 0.27  & 0.30  & 0.281  & \multicolumn{1}{c}{-} \\
$\Sigma$CNO & 0.88  & 1.5   & 0.33  & 0.40  & 0.35   & 0.41\\
Fe       & 0.129    & 0.14\tablefootmark{g}  & 0.07\tablefootmark{h}    
                    & \multicolumn{1}{c}{-}  &  \multicolumn{1}{c}{-}  
                    & .07\tablefootmark{g} \\
\hline 
\end{tabular}
\tablefoot{
\tablefoottext{a}{\citet{Asplund2009}}
\tablefoottext{b}{as used in \citetalias{Hamann2006}}
\tablefoottext{c}{\citet{Hunter2007}}
\tablefoottext{d}{\citet{Korn2005}}
\tablefoottext{e}{\ion{H}{ii} regions \citep{Dufour1998}}
\tablefoottext{f}{as adopted in this work}
\tablefoottext{g}{including the whole iron group}
\tablefoottext{h}{mean value from \citet{Trundle2007}}
}
\end{table}

\subsection{The model grids}
\label{subsec:modelgrid}

Three large grids of WN models were computed. The parameter domain
of each grid is spanned by $T_*$ and $R_{\mathrm t}$. The grid spacing
is 0.05\,dex in $\log\,(T_*/\mathrm{kK})$ and 0.1\,dex in 
$\log\,(R_{\mathrm t}/R_{\odot})$ \citep[see][for details]{Hamann2004}. 
The luminosity is fixed at $\log\,(L/L_\odot) = 5.3$. Thanks to the scaling 
invariance described in Sect.\,\ref{sect:models}, the normalized line spectra 
apply in good approximation to different luminosities, while the absolute 
fluxes scale with $L$. 

\begin{figure}
\centering
\includegraphics[angle=-90,width=\hsize]{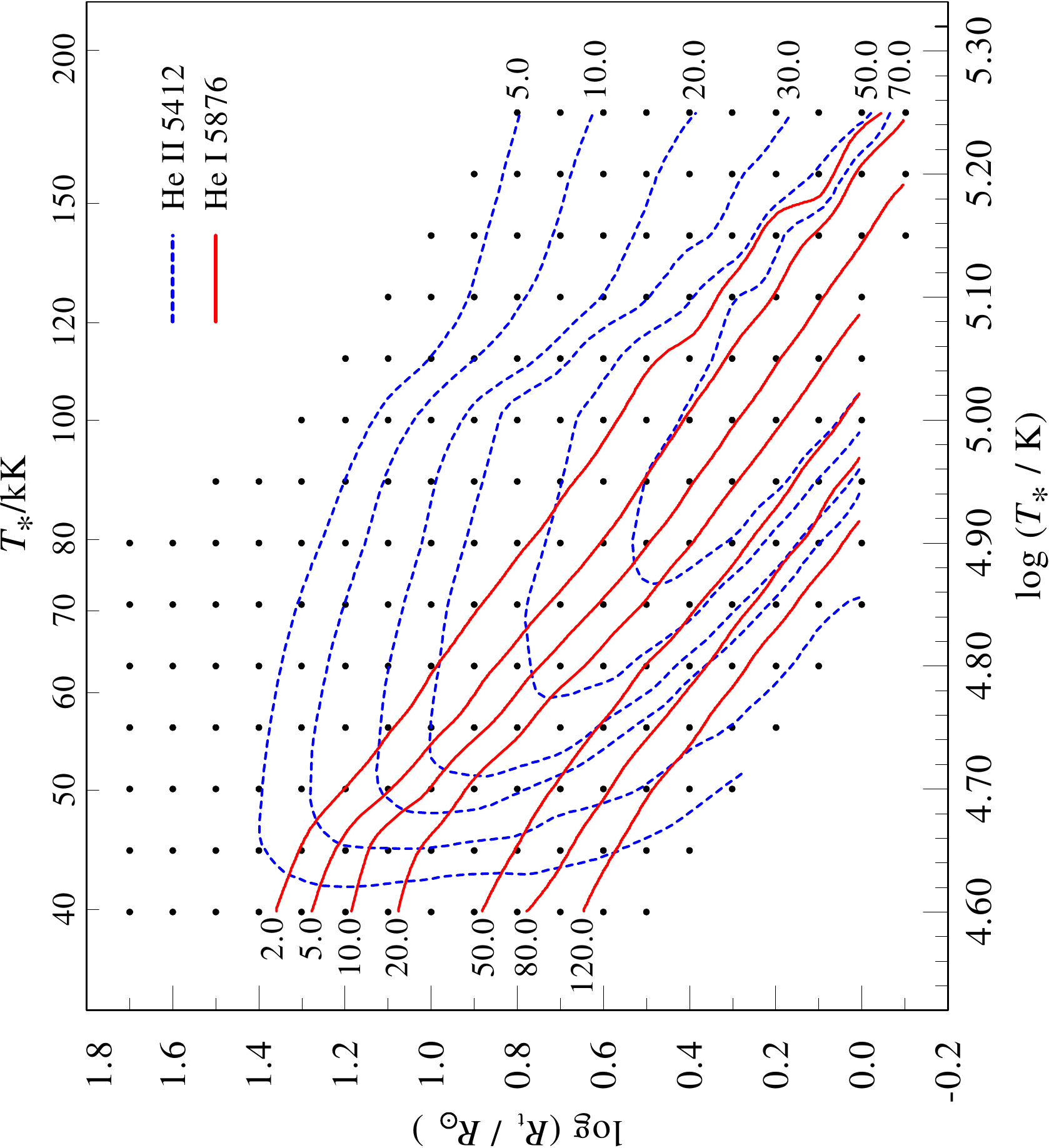}
\caption{Grid of models for hydrogen-free WN stars in the LMC: 
contours of constant line emission, labeled with the
equivalent width in \AA; thick (red) contours: \ion{He}{i} line at 5876\,\AA,
thin-dashed (blue) contours: \ion{He}{ii} at 5412\,\AA. Tiny dots
indicate calculated grid models.}
\label{fig:he-contours}
\end{figure}

\begin{figure*}
  \centering
  \includegraphics[width=0.8\textwidth]{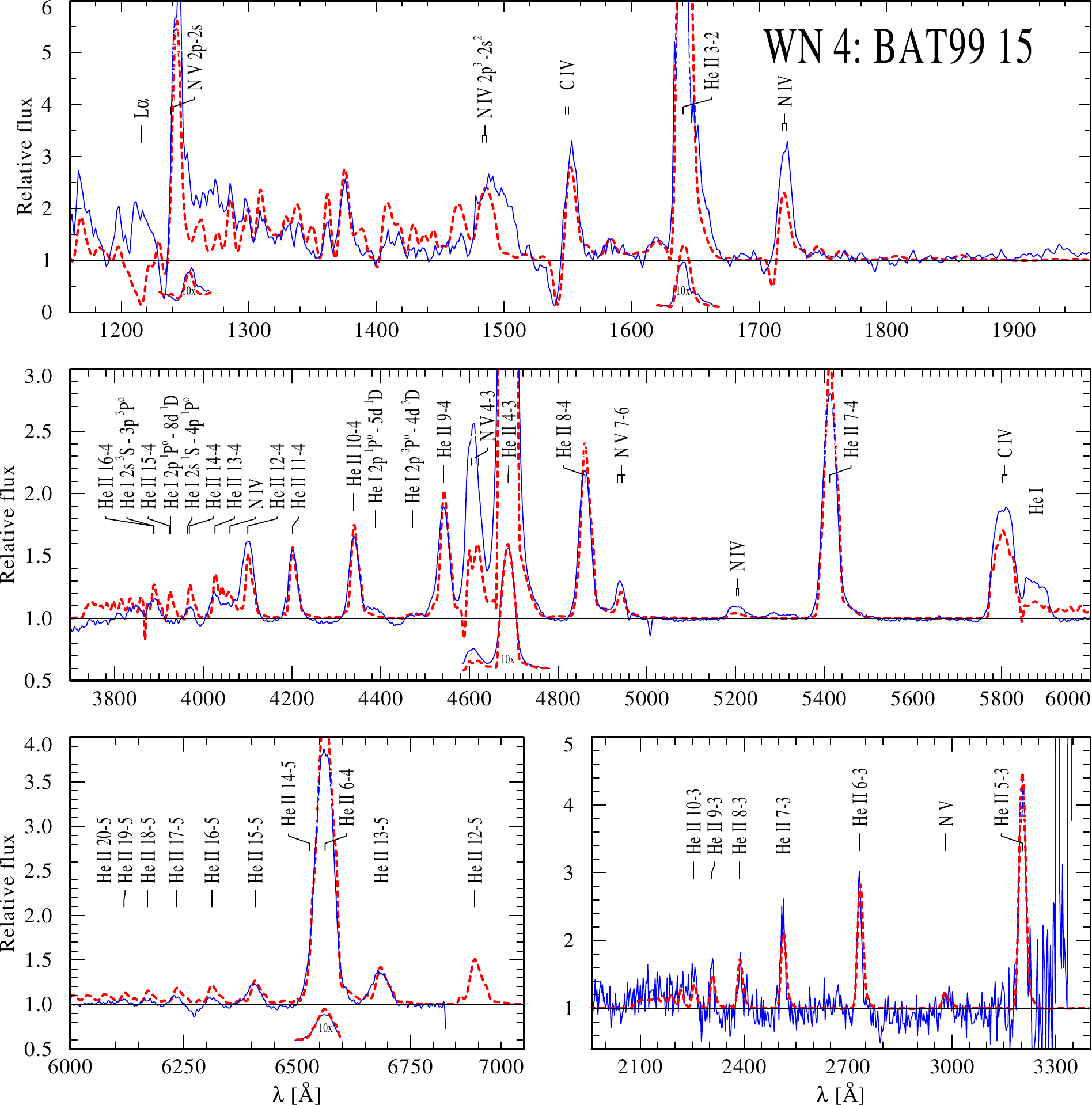}
\caption{Normalized spectrum of the WNE star BAT99\,015. The thin
(blue) solid line depicts the observation, whereas the thick (red)
dashed line represents the synthetic spectrum of the best fitting
PoWR model.}
\label{fig:bat015_linefit}
\end{figure*}

The three grids differ not only in the hydrogen mass fraction (0, 0.2,
and 0.4, respectively), but also in the terminal wind velocity.  
One grid has been computed for WNE stars with $X_{\element{H}}=0$ and
$\varv_\infty=1600\,\mathrm{km\,s^{-1}}$. The other two grids were
established for WNL stars ($X_{\element{H}}=0.2$ and
$X_{\element{H}}=0.4$), respectively, with
$\varv_\infty=1000\,\mathrm{km\,s^{-1}}$ in both cases. The trace
element abundances are set to the values described in
Sect.\,\ref{sect:abundances}. These model grids are publicly available on 
our website\footnote{\url{http://www.astro.physik.uni-potsdam.de/PoWR.html}}. 

Analyzing a star thus means to identify the specific model which 
gives the best fit to the observations. A first orientation can be obtained 
from contour plots like the one shown in Fig.\,\ref{fig:he-contours}. If,
for example, the \ion{He}{ii} emission line at 5412\,\AA\ is observed
with an equivalent width of 30\,\AA, the temperature is 
restricted to values above 50\,kK. Combining this with the measured 
equivalent width of the \ion{He}{i} line at 5876\,\AA, preliminary
model parameters can  already be estimated from the intersection point
of the corresponding contours. 

This method works, of course, only if the \ion{He}{i}
and  the \ion{He}{ii} lines are both present in the spectrum of the
considered star. For those stars where this is not the case, such as very 
hot stars, other ions or elements must be employed. These contour plots 
are provided on the PoWR homepage for several transitions of the 
ions \ion{He}{i}, \ion{He}{ii}, \ion{N}{iii}, \ion{N}{iv}, and \ion{N}{v}.
As is evident from Fig.\,\ref{fig:he-contours}, the method may also 
fail in the lowest part of the diagram, i.e.,\ for the densest winds, 
because the contours do not intersect in this parameter regime. 
At these parameters, the winds are so thick that the whole spectrum, 
including the continuum, is formed in the rapidly moving part of the 
wind. For a fixed luminosity, such models have only the mass-loss 
rate as  significant parameter, while the stellar radius and the 
related effective temperature become meaningless. Due to the spacing 
chosen for our grid, models of the same mass-loss rate lie on a 
diagonal of the grid cells, like the parallel contours. 
Along these diagonals, the models exhibit fairly 
similar line spectra. Thus, the stellar parameters derived for stars 
in the regime of parameter degeneracy solely depend on small differences 
of the synthetic spectra. 
There are indeed some stars in our sample that fall into this 
regime of parameter degeneracy (see Appendix\,\ref{sec:comments}).

For 27 stars, we preselect suitable grid models by
means of a $\chi_{\nu}^2$-fitting technique. This method is based on a reduced 
$\chi_{\nu}^2$-statistic, which is calculated for the rectified observations 
with respect to the model spectra in our grids. For details of this fitting 
technique, we refer to \citet{Todt2013}.

\subsection{Spectral fitting}
\label{subsec:spefit}
\label{sect:rotation}
 
\begin{figure*}
  \centering
  \includegraphics[width=17cm]{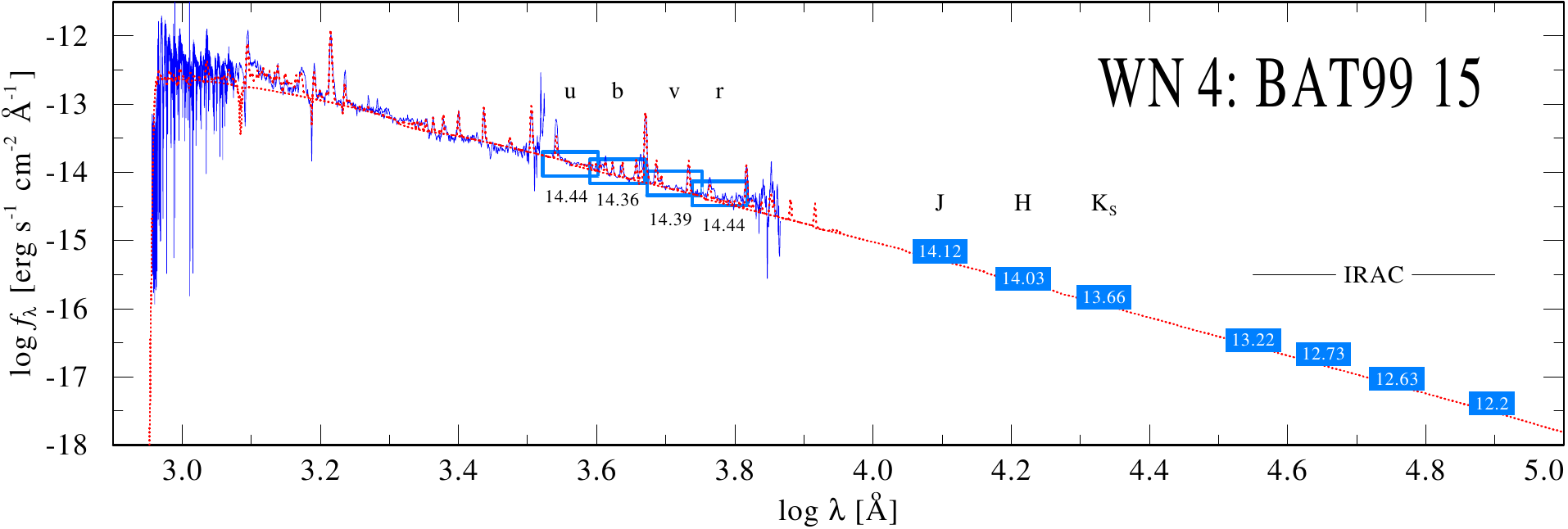}
\caption{Spectral energy distribution of the WNE star BAT99\,015.
Flux-calibrated FUSE, IUE, and CTIO spectra (blue noisy line) and
multiband photometry (blue boxes, labeled with the magnitudes) 
are compared to the model flux (red dotted lines),
accounting for interstellar extinction.}
\label{fig:bat015_sed}
\end{figure*}

\begin{figure*}
\centering
\includegraphics[angle=-90,width=0.8\textwidth]{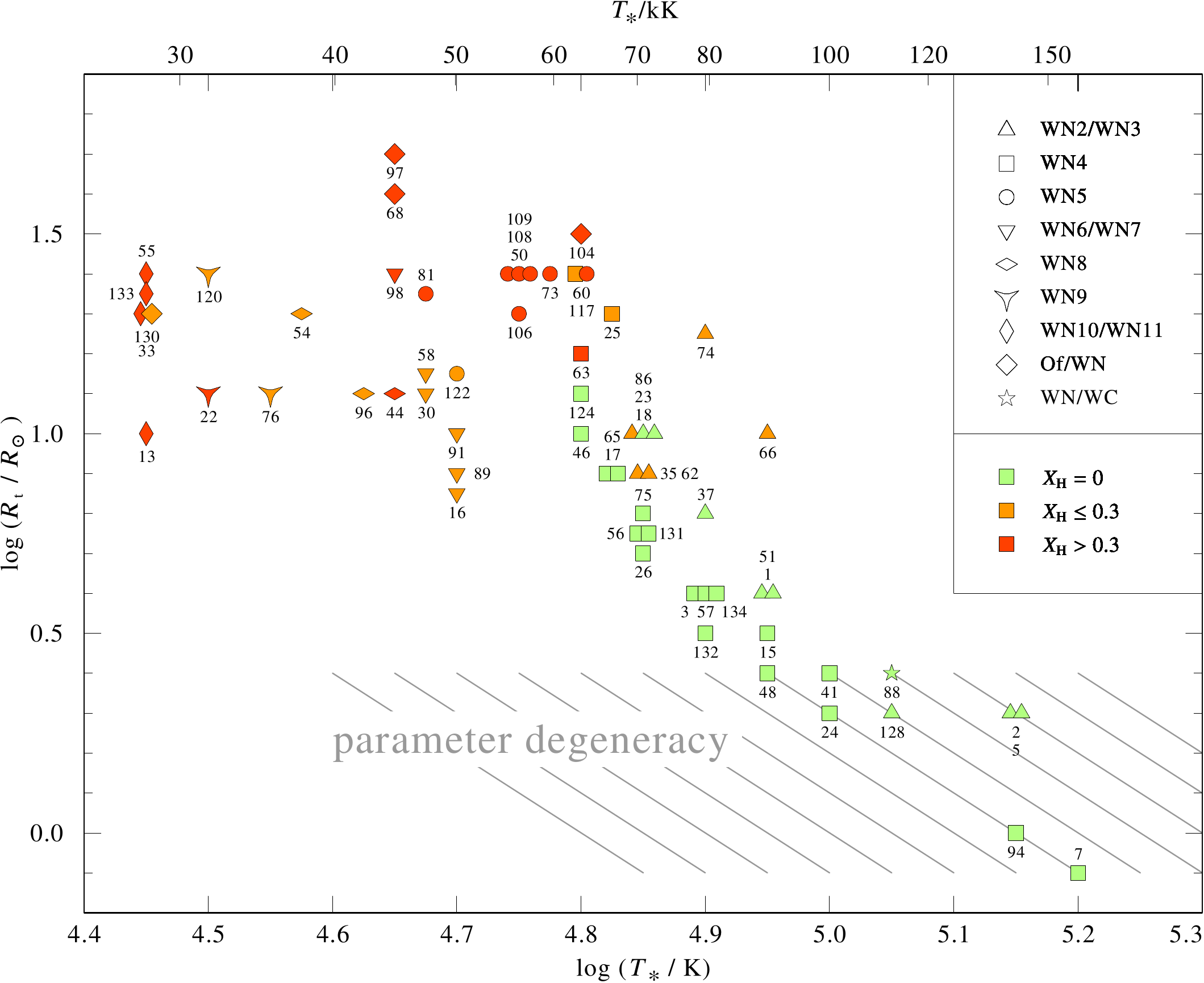}
\caption{The positions of the analyzed WN stars in the $\log T_*$-$\log
R_\mathrm{t}$-plane. The labels refer to the BAT99 catalog. Different
WN subtypes are distinguished by the shape of the symbols, as shown in 
the inlet. The hydrogen abundance is color-coded in three steps
(undetectable, about 0.2, and about 0.4 mass fraction). 
The lower hatched part roughly indicates the region where the 
parameter space becomes degenerate because of large
optical thickness of the wind. In this part, the stars can be shifted 
parallel to the gray lines without significant changes in the 
synthetic normalized emission line spectrum.
}
\label{fig:wnstats}
\end{figure*}

After preliminary parameters have been estimated either by the 
$\chi_{\nu}^2$-fit or by the contour plots, we compare 
observations and models in detail for each star. A typical fit of the
normalized line spectrum is shown in  Fig.\,\ref{fig:bat015_linefit},
while analogous plots for each star of the sample can be found in the
{\em Online Material}. 

While most of the observed spectra are given in normalized form, some of
the spectra \citep[from IUE, HST, and][]{Torres-Dodgen1988} are
flux-calibrated. These data are normalized consistently through division 
by the reddened model continuum. In this respect, spectral fitting is an
iterative process, coupled with the fitting of the spectral energy
distribution described below. 

With the starting estimates for $T_\ast$ and $R_{\mathrm t}$, 
we carefully compare the observed line spectrum with 
models of neighboring parameters, and finally determine the 
best fitting values. The uncertainty is usually smaller than one grid
cell, i.e.,\ the error margins are smaller than  $\pm0.05$\,dex in
$T_\ast$ and $\pm0.1$\,dex in $R_{\mathrm t}$. The latter translates to
an uncertainty of $\pm0.15$\,dex for the mass-loss rate (cf.\
Eq.\,\ref{eq:rt}). This, of course, does not account for systematic
errors, because, for instance the model assumptions are not exactly 
fulfilled.  

The terminal wind velocity, $v_\infty$, mainly influences the width of
the line profiles. For 62 of our program stars, \citet{Niedzielski2002}
and  \citet{Niedzielski2004} measured the wind velocities from
P-Cygni profiles in the UV. Depending on the considered line, they
obtain quite different values for the same star. Two possible reasons
are: (a) While the $\beta$-law for the velocity field quickly
approaches the terminal velocity, the winds are in fact further
accelerated even at large distances from the star. Therefore, the
strongest lines give the highest wind velocity. (b) The velocity field
in the wind has some nonuniform, stochastic component that is not
perfectly described by our assumption of a constant and isotropic
microturbulence. In any case, the largest of the velocities
given by \citet{Niedzielski2004} often yield optical emission line profiles 
that are considerably broader than observed.

Therefore, we prefer to perform our own estimates of $v_\infty$ from the
width of the optical emission lines. First, we inspect whether the standard
$v_\infty$ of the respective grid is sufficient to reproduce the observed
line width. If not, we recalculate the model for the considered star 
with a more appropriate estimate, aiming at an accuracy of about 
$\pm 200\,\mathrm{km\,s^{-1}}$. The values of $v_\infty$ used for the final 
fits and the subsequent discussion are compiled in 
Table\,\ref{table:parameters}. 

Among the studied sample, we found a couple of stars (especially the
putatively single  WNE stars BAT99\,7, 51, 88, and 94) to exhibit very
unique spectra. Their emission lines have a round shape that is distinctly 
different from those of all other stars, but similar to the shape of WR\,2 
in the Milky Way \citepalias{Hamann2006}. Such profiles can be reproduced
by convolving the model spectrum with a rotation profile of very high
$\varv \sin i$. A more adequate treatment of rotational broadening in WR
winds presently confirmed that rotation might in principle account for
these spectra \citep{Shenar2014}.  

A further important model parameter is the hydrogen abundance.  Its
determination is one of the major aims of this paper.  For this purpose, we
calculated three extended model grids for hydrogen mass fractions of 0,
0.2, and 0.4, respectively, plus  a couple of models for 0.6. By
comparison and tentative interpolation between these grids, we can
estimate the hydrogen mass fraction with $\pm0.1$ accuracy. 

After the appropriate model has been selected from the line fit, 
it must be scaled to the correct luminosity by fitting the spectral energy
distribution (SED) to the photometric and flux-calibrated observations 
(for example, see Fig. \ref{fig:bat015_sed}). The scaling corresponds to 
a simple vertical shift in this logarithmic plot, while the normalized emission
line spectrum  does not change between models with same transformed
radius (cf.\ Sect.\,\ref{sect:models}). The model flux is geometrically
diluted according to the LMC distance modulus of 18.5\,mag
\citep{Madore1998,Pietrzynski2013}, corresponding to a distance of 50\,kpc.

The color excess $E_{b-v}$ must be adjusted simultaneously. The
reddening encompasses contributions from both the internal LMC
reddening and Galactic foreground reddening, assuming the Seaton
reddening law \citep{seaton} with $E_{b-v} = 0.03$\,mag for the latter. The
remaining LMC excess is determined by adjusting the free $E_{b-v}$
parameter of the LMC reddening law determined by \citet{Howarth}. 

Since the stellar flux in the optical and IR depends roughly linearly on the
stellar temperature (Rayleigh-Jeans domain), the error in $T_\ast$ 
($\pm0.05$\,dex, see above) influences the luminosity estimate directly.  
Additional uncertainties are inferred from the reddening correction, which is 
relatively small for our LMC stars, the imperfect SED fit, and the 
error margins of the photometry. These errors combine to a final 
accuracy of about $\pm0.1$\,dex in $\log\,(L/L_\odot)$ for those stars where 
photometry and flux-calibrated spectra are available. If only photometry is 
accessible, the accuracy is reduced to $\pm0.2$\,dex, due to a larger 
uncertainty in the SED fit.

\begin{longtab}
\small
\setlength{\tabcolsep}{4.6pt}
\begin{longtable}{ccccScccScccccll}
\caption{Parameters of LMC WN stars }
\label{table:parameters}\\
\hline \hline
    BAT99 &
    Subtype &
    Ref. &
    $T_*$ &
    $\mathrm{log}\, R_\mathrm{t}$ &
    $\varv_{\infty}$ &
    $E_{b-v}$ &
    $M_{v}$ &
    $R_*$ &
    $\log \dot M$ &
    $\log L$ &
    $\eta$ &
    $M$\tablefootmark{a} &
    $X_{\element{H}}$ &
    Bin.\tablefootmark{b} &
    Ref. \rule[0mm]{0mm}{3.2mm}\\
    &
    &
    &
    $[\mathrm{kK}]$ &
    $[R_{\odot}]$ &
    $[\mathrm{km/s}]$ &
    $[\mathrm{mag}]$ &
    $[\mathrm{mag}]$ &
    $[R_{\odot}]$ &
    $[M_{\odot}/\mathrm{yr}]$ &
    $[L_{\odot}]$ &
    &
    $[M_{\odot}]$ &
    &
    & \\
\hline 
\endfirsthead
\caption{continued.}\\
\hline\hline
    BAT99 &
    Subtype &
    Ref. &
    $T_*$ &
    $\mathrm{log}\, R_\mathrm{t}$ &
    $\varv_{\infty}$ &
    $E_{b-v}$ &
    $M_{v}$ &
    $R_*$ &
    $\log \dot M$ &
    $\log L$ &
    $\eta$ &
    $M$\tablefootmark{a} &
    $X_{\element{H}}$ &
    Bin.\tablefootmark{b} &
    Ref. \rule[0mm]{0mm}{3.2mm}\\
    &
    &
    &
    $[\mathrm{kK}]$ &
    $[R_{\odot}]$ &
    $[\mathrm{km/s}]$ &
    $[\mathrm{mag}]$ &
    $[\mathrm{mag}]$ &
    $[R_{\odot}]$ &
    $[M_{\odot}/\mathrm{yr}]$ &
    $[L_{\odot}]$ &
    &
    $[M_{\odot}]$ &
    &
    & \\
\hline 
\endhead
\hline 
\endfoot
001  & WN3b \rule[0mm]{0mm}{3.5mm} & 1   	& 89 	& 0.60 & 1600 	& 0.14 & -3.32 & 1.9 	& -5.18 & 5.30 & 2.6 & 12 & 0.0 &  &  \\ 
002  & WN2b(h) 	& 2   	& 141 	& 0.30 & 1600 	& 0.13 & -2.36 & 0.8 	& -5.28 & 5.37 & 1.8 & 13 & 0.0 &  &  \\ 
003  & WN4b 	& 1   	& 79 	& 0.60 & 1600 	& 0.12 & -4.36 & 3.0 	& -4.88 & 5.51 & 3.2 & 16 & 0.0 &  &  \\ 
005  & WN2b 	& 2   	& 141 	& 0.30 & 1600 	& 0.27 & -2.94 & 0.9 	& -5.22 & 5.45 & 1.7 & 15 & 0.0 &  &  \\ 
006  & O3\,f*+O 	& 3   	& 56 	& 1.80 & 1600 	& 0.08 & -6.59 & 17.7 	& -5.52 & 6.45 & 0.1 & 94 & 0.2 & x & 4,3,5 \\ 
007  & WN4b 	& 1   	& 158 	& -0.10 & 1600 	& 0.08 & -5.02 & 1.1 	& -4.48 & 5.84 & 3.8 & 25 & 0.0 &  &  \\ 
012  & O2\,If*/WN5 	& 6   	& 50 	& 1.70 & 2400 	& 0.10 & -5.19 & 10.6 	& -5.53 & 5.80 & 0.5 & 53 & 0.5 & x & 7 \\ 
013  & WN10 	& 1   	& 28 	& 1.00 & 400 	& 0.20 & -6.34 & 25.3 	& -4.69 & 5.56 & 1.1 & 35 & 0.4 &  &  \\ 
014  & WN4o(+OB) 	& 2   	& 67 	& 1.15 & 1600 	& 0.09 & -5.17 & 6.4 	& -5.21 & 5.86 & 0.7 & 26 & 0.0 & ? & 2,8 \\ 
015  & WN4b 	& 1   	& 89 	& 0.50 & 1600 	& 0.08 & -4.44 & 2.6 	& -4.83 & 5.57 & 3.1 & 17 & 0.0 &  &  \\ 
016  & WN7h 	& 7   	& 50 	& 0.85 & 1000 	& 0.09 & -6.12 & 10.6 	& -4.64 & 5.80 & 1.8 & 42 & 0.3 &  &  \\ 
017  & WN4o 	& 1   	& 67 	& 0.90 & 1600 	& 0.11 & -4.79 & 5.2 	& -4.97 & 5.69 & 1.7 & 20 & 0.0 &  &  \\ 
018  & WN3(h) 	& 2   	& 71 	& 1.00 & 1600 	& 0.10 & -4.31 & 4.4 	& -5.24 & 5.63 & 1.1 & 29 & 0.2 &  &  \\ 
019  & WN4b+O5: 	& 2   	& 79 	& 0.75 & 1600 	& 0.16 & -5.33 & 6.2 	& -4.63 & 6.14 & 1.3 & 39 & 0.0 & x\,\tablefootmark{c} & 2 \\ 
021  & WN4o(+OB) 	& 2   	& 67 	& 1.30 & 1600 	& 0.09 & -5.76 & 10.6 	& -5.11 & 6.30 & 0.3 & 51 & 0.0 & ? & 2,8 \\ 
022  & WN9h 	& 1   	& 32 	& 1.10 & 400 	& 0.13 & -7.00 & 25.1 	& -4.85 & 5.75 & 0.5 & 44 & 0.4 &  &  \\ 
023  & WN3(h) 	& 2   	& 71 	& 1.00 & 1600 	& 0.60 & -3.98 & 4.0 	& -5.30 & 5.55 & 1.1 & 17 & 0.0 &  &  \\ 
024  & WN4b 	& 1   	& 100 	& 0.30 & 2400 	& 0.10 & -4.39 & 2.0 	& -4.53 & 5.54 & 10.1 & 17 & 0.0 &  &  \\ 
025  & WN4ha 	& 2   	& 67 	& 1.30 & 1600 	& 0.15 & -4.01 & 4.5 	& -5.67 & 5.55 & 0.5 & 26 & 0.2 &  &  \\ 
026  & WN4b 	& 1   	& 71 	& 0.70 & 1600 	& 0.14 & -4.38 & 4.3 	& -4.79 & 5.62 & 3.0 & 18 & 0.0 &  &  \\ 
027  & WN5b(+B1\,Ia) 	& 2   	& 71 	& 1.40 & 1000 	& 0.23 & -8.22 & 29.8 	& -4.79 & 7.30 & 0.0 & 587 & 0.2 & ? & 9 \\ 
029  & WN4b+OB 	& 2   	& 71 	& 0.80 & 1600 	& 0.12 & -4.37 & 3.7 	& -5.03 & 5.50 & 2.3 & 16 & 0.0 & x & 2 \\ 
030  & WN6h 	& 1   	& 47 	& 1.10 & 1000 	& 0.07 & -5.48 & 10.0 	& -5.05 & 5.65 & 1.0 & 34 & 0.3 &  &  \\ 
031  & WN4b 	& 2   	& 75 	& 0.70 & 1600 	& 0.17 & -3.81 & 2.7 	& -5.09 & 5.33 & 3.0 & 12 & 0.0 & ? & 2 \\ 
032  & WN6(h) 	& 1   	& 47 	& 1.10 & 1600 	& 0.08 & -6.14 & 13.9 	& -4.63 & 5.94 & 2.1 & 44 & 0.2 & x & 7,10 \\ 
033  & O\,fpe/WN9? 	& 7   	& 28 	& 1.30 & 400 	& 0.37 & -8.48 & 74.8 	& -4.43 & 6.50 & 0.2 & 103 & 0.2 &  &  \\ 
035  & WN3(h) 	& 2   	& 71 	& 0.90 & 1600 	& 0.11 & -4.11 & 4.2 	& -5.11 & 5.60 & 1.5 & 24 & 0.1 &  &  \\ 
036  & WN4b/WCE+OB 	& 2   	& 79 	& 0.70 & 1600 	& 0.13 & -4.33 & 3.8 	& -4.88 & 5.71 & 2.0 & 21 & 0.0 & ? & 2,11 \\ 
037  & WN3o 	& 2   	& 79 	& 0.80 & 1600 	& 0.50 & -4.12 & 3.5 	& -5.07 & 5.65 & 1.5 & 19 & 0.0 &  &  \\ 
040  & WN4(h)a 	& 2   	& 63 	& 1.20 & 1600 	& 0.15 & -4.41 & 5.4 	& -5.39 & 5.62 & 0.8 & 29 & 0.2 & ?\,\tablefootmark{c} &  \\ 
041  & WN4b 	& 1   	& 100 	& 0.40 & 1300 	& 0.12 & -4.11 & 2.1 	& -4.90 & 5.60 & 2.0 & 18 & 0.0 &  &  \\ 
042  & WN5b(h)(+B3\,I) 	& 2   	& 71 	& 1.70 & 1000 	& 0.30 & -9.88 & 66.6 	& -4.71 & 8.00 & 0.0 & - & 0.4 & ?\,\tablefootmark{c} & 2,9,12 \\ 
043  & WN4o+OB 	& 2   	& 67 	& 1.10 & 1600 	& 0.13 & -4.84 & 6.3 	& -5.15 & 5.85 & 0.8 & 25 & 0.0 & x & 2 \\ 
044  & WN8ha 	& 7   	& 45 	& 1.10 & 700 	& 0.12 & -5.59 & 11.3 	& -5.12 & 5.66 & 0.6 & 40 & 0.4 &  &  \\ 
046  & WN4o 	& 1   	& 63 	& 1.00 & 1600 	& 0.21 & -4.09 & 4.4 	& -5.23 & 5.44 & 1.7 & 14 & 0.0 &  &  \\ 
047  & WN3b 	& 2   	& 89 	& 0.60 & 1300 	& 0.20 & -3.97 & 2.6 	& -5.06 & 5.59 & 1.4 & 18 & 0.0 & ?\,\tablefootmark{c} &  \\ 
048  & WN4b 	& 1   	& 89 	& 0.40 & 1600 	& 0.10 & -4.22 & 2.1 	& -4.81 & 5.40 & 4.9 & 14 & 0.0 &  &  \\ 
049  & WN4:b+O8V 	& 2   	& 71 	& 1.80 & 2400 	& 0.15 & -5.49 & 9.9 	& -5.73 & 6.34 & 0.1 & 122 & 0.6 & x & 2,13 \\ 
050  & WN5h 	& 14   	& 56 	& 1.40 & 1600 	& 0.18 & -4.75 & 7.1 	& -5.52 & 5.65 & 0.5 & 39 & 0.4 &  &  \\ 
051  & WN3b 	& 1   	& 89 	& 0.60 & 1600 	& 0.02 & -3.39 & 1.9 	& -5.18 & 5.30 & 2.6 & 12 & 0.0 &  &  \\ 
054  & WN8ha 	& 7   	& 38 	& 1.30 & 1000 	& 0.50 & -6.23 & 17.7 	& -4.97 & 5.75 & 0.9 & 34 & 0.2 &  &  \\ 
055  & WN11h 	& 1   	& 28 	& 1.40 & 400 	& 0.13 & -7.04 & 32.3 	& -5.13 & 5.77 & 0.2 & 45 & 0.4 &  &  \\ 
056  & WN4b 	& 1   	& 71 	& 0.75 & 1600 	& 0.12 & -4.46 & 4.0 	& -4.91 & 5.56 & 2.6 & 17 & 0.0 &  &  \\ 
057  & WN4b 	& 1   	& 79 	& 0.60 & 1600 	& 0.10 & -4.04 & 2.7 	& -4.96 & 5.40 & 3.4 & 14 & 0.0 &  &  \\ 
058  & WN7h 	& 7   	& 47 	& 1.15 & 1000 	& 0.50 & -5.35 & 9.9 	& -5.13 & 5.64 & 0.8 & 34 & 0.3 &  &  \\ 
059  & WN4b+O8: 	& 2   	& 71 	& 1.30 & 1600 	& 0.16 & -6.01 & 11.2 	& -5.07 & 6.45 & 0.2 & 66 & 0.0 & ? & 2 \\ 
060  & WN4(h)a 	& 2   	& 63 	& 1.40 & 2400 	& 0.15 & -4.82 & 6.5 	& -5.40 & 5.78 & 0.8 & 35 & 0.2 &  &  \\ 
062  & WN3(h) 	& 2   	& 71 	& 0.90 & 1600 	& 0.12 & -3.85 & 3.4 	& -5.25 & 5.41 & 1.7 & 19 & 0.1 &  &  \\ 
063  & WN4ha: 	& 2   	& 63 	& 1.20 & 1600 	& 0.10 & -4.33 & 5.2 	& -5.42 & 5.58 & 0.8 & 36 & 0.4 &  &  \\ 
064  & WN4o+O9: 	& 2   	& 71 	& 1.10 & 1600 	& 0.26 & -5.18 & 7.1 	& -5.07 & 6.05 & 0.6 & 34 & 0.0 & x & 2 \\ 
065  & WN4o 	& 2   	& 67 	& 0.90 & 1600 	& 0.45 & -4.84 & 5.6 	& -4.92 & 5.75 & 1.7 & 22 & 0.0 &  &  \\ 
066  & WN3(h) 	& 2   	& 89 	& 1.00 & 1600 	& 0.13 & -3.73 & 3.3 	& -5.42 & 5.78 & 0.5 & 35 & 0.2 &  &  \\ 
067  & WN5ha 	& 2   	& 47 	& 1.30 & 1600 	& 0.33 & -6.11 & 14.3 	& -4.91 & 5.96 & 1.1 & 51 & 0.3 & ?\,\tablefootmark{c} &  \\ 
068  & O3.5\,If*/WN7 	& 6   	& 45 	& 1.60 & 1000 	& 0.52 & -6.22 & 16.7 	& -5.46 & 6.00 & 0.2 & 76 & 0.6 &  &  \\ 
071  & WN4+O8: 	& 2   	& 63 	& 1.30 & 1600 	& 0.38 & -5.16 & 8.2 	& -5.27 & 5.98 & 0.4 & 31 & 0.0 & x & 2 \\ 
072  & WN4h+O3: 	& 2   	& 71 	& 1.40 & 1600 	& 0.40 & -4.31 & 5.3 	& -5.71 & 5.80 & 0.2 & 47 & 0.4 & ? & 2 \\ 
073  & WN5ha 	& 14   	& 60 	& 1.40 & 1600 	& 0.20 & -4.67 & 6.8 	& -5.54 & 5.72 & 0.4 & 43 & 0.4 &  &  \\ 
074  & WN3(h)a 	& 2   	& 79 	& 1.25 & 2000 	& 0.20 & -3.82 & 3.7 	& -5.62 & 5.69 & 0.5 & 32 & 0.2 &  &  \\ 
075  & WN4o 	& 2   	& 71 	& 0.80 & 1600 	& 0.07 & -4.32 & 4.0 	& -4.99 & 5.56 & 2.2 & 17 & 0.0 &  &  \\ 
076  & WN9ha 	& 7   	& 35 	& 1.10 & 400 	& 0.26 & -6.31 & 17.9 	& -5.07 & 5.66 & 0.4 & 30 & 0.2 &  &  \\ 
077  & WN7ha 	& 7   	& 45 	& 1.60 & 1000 	& 0.27 & -5.18 & 41.6 	& -4.87 & 6.79 & 0.1 & 305 & 0.7 & x\,\tablefootmark{c} & 7,10 \\ 
078  & WN6(+O8\,V) 	& 2   	& 71 	& 0.85 & 1600 	& 0.20 & -4.48 & 4.7 	& -4.96 & 5.70 & 1.7 & 32 & 0.2 & ?\,\tablefootmark{c} &  \\ 
079  & WN7ha+OB 	& 7   	& 42 	& 1.20 & 1600 	& 0.50 & -7.03 & 22.8 	& -4.46 & 6.17 & 1.9 & 61 & 0.2 & ?\,\tablefootmark{c} &  \\ 
080  & WN5h:a 	& 7   	& 45 	& 1.70 & 2400 	& 0.50 & -7.31 & 26.5 	& -4.93 & 6.40 & 0.5 & 87 & 0.2 & ?\,\tablefootmark{c} &  \\ 
081  & WN5h 	& 2   	& 47 	& 1.35 & 1000 	& 0.33 & -4.47 & 8.2 	& -5.55 & 5.48 & 0.5 & 32 & 0.4 &  &  \\ 
082  & WN3b \rule[0mm]{0mm}{3.5mm} & 1   	& 100 	& 0.60 & 1600 	& 0.27 & -3.68 & 1.9 	& -5.16 & 5.53 & 1.6 & 16 & 0.0 & ?\,\tablefootmark{c} &  \\ 
086  & WN3(h) 	& 15   	& 71 	& 1.00 & 1600 	& 0.36 & -3.37 & 3.1 	& -5.46 & 5.33 & 1.3 & 12 & 0.0 &  &  \\ 
088  & WN4b/WCE 	& 2   	& 112 	& 0.40 & 1600 	& 0.84 & -4.19 & 2.1 	& -4.81 & 5.80 & 1.9 & 24 & 0.0 &  &  \\ 
089  & WN7h 	& 1   	& 50 	& 0.90 & 1000 	& 0.28 & -5.37 & 10.3 	& -4.73 & 5.78 & 1.5 & 35 & 0.2 &  &  \\ 
091  & WN6(h) 	& 16   	& 50 	& 1.00 & 1000 	& 0.33 & -5.87 & 6.8 	& -5.15 & 5.42 & 1.3 & 23 & 0.2 &  &  \\ 
092  & WN3:b(+O)+B1\,Ia 	& 7   	& 45 	& 1.50 & 1000 	& 0.39 & -8.69 & 50.0 	& -4.60 & 6.95 & 0.1 & 240 & 0.2 & x\,\tablefootmark{c} & 7,10 \\ 
093  & O3\,If* 	& 6,16   	& 45 	& 1.80 & 1600 	& 0.24 & -5.65 & 14.9 	& -5.63 & 5.90 & 0.2 & 67 & 0.6 & ?\,\tablefootmark{c} &  \\ 
094  & WN4b 	& 1   	& 141 	& 0.00 & 1600 	& 0.29 & -4.80 & 1.3 	& -4.51 & 5.80 & 3.9 & 24 & 0.0 &  &  \\ 
095  & WN7h+OB 	& 16   	& 50 	& 0.80 & 1600 	& 0.25 & -6.36 & 13.3 	& -4.21 & 6.00 & 4.9 & 48 & 0.2 & x & 7 \\ 
096  & WN8 	& 7   	& 42 	& 1.10 & 1000 	& 0.70 & -7.55 & 28.1 	& -4.37 & 6.35 & 0.9 & 80 & 0.2 &  &  \\ 
097  & O3.5\,If*/WN7 	& 6,16   	& 45 	& 1.70 & 1600 	& 0.60 & -7.19 & 23.7 	& -5.18 & 6.30 & 0.3 & 115 & 0.6 &  &  \\ 
098  & WN6 	& 7   	& 45 	& 1.40 & 1600 	& 0.80 & -8.11 & 37.5 	& -4.43 & 6.70 & 0.6 & 226 & 0.6 &  &  \\ 
099  & O2.5\,If*/WN6 	& 6,16   	& 45 	& 1.80 & 1600 	& 0.30 & -6.77 & 14.9 	& -5.63 & 5.90 & 0.2 & 42 & 0.2 & x\,\tablefootmark{c} & 7 \\ 
100  & WN7 	& 7   	& 47 	& 1.00 & 1000 	& 0.28 & -6.80 & 17.7 	& -4.52 & 6.15 & 1.0 & 59 & 0.2 & ?\,\tablefootmark{c} &  \\ 
102  & WN6 	& 7   	& 45 	& 1.30 & 1600 	& 0.70 & -8.38 & 42.1 	& -4.21 & 6.80 & 0.8 & 221 & 0.4 & ?\,\tablefootmark{c} & 7 \\ 
103  & WN5(h)+O 	& 16   	& 47 	& 1.30 & 1600 	& 0.40 & -7.13 & 19.9 	& -4.70 & 6.25 & 0.9 & 87 & 0.4 & x\,\tablefootmark{c} & 7,10 \\ 
104  & O2\,If*/WN5 	& 6   	& 63 	& 1.50 & 2400 	& 0.38 & -5.48 & 9.0 	& -5.34 & 6.06 & 0.5 & 66 & 0.4 &  &  \\ 
105  & O2\,If* 	& 6   	& 50 	& 1.80 & 1600 	& 0.30 & -6.93 & 21.1 	& -5.41 & 6.40 & 0.1 & 134 & 0.6 & ?\,\tablefootmark{c} &  \\ 
106  & WN5h 	& 1   	& 56 	& 1.30 & 2400 	& 0.35 & -6.86 & 19.0 	& -4.55 & 6.51 & 1.0 & 130 & 0.4 &  &  \\ 
107  & O6.5\,Iafc+O6\,Iaf 	& 17   	& 35 	& 1.50 & 1000 	& 0.26 & -7.45 & 37.9 	& -4.78 & 6.31 & 0.4 & 95 & 0.4 & x\,\tablefootmark{c} & 10,17 \\ 
108  & WN5h 	& 1   	& 56 	& 1.40 & 2400 	& 0.37 & -7.10 & 28.8 	& -4.43 & 6.87 & 0.6 & 256 & 0.4 &  &  \\ 
109  & WN5h 	& 1   	& 56 	& 1.40 & 2400 	& 0.39 & -6.50 & 23.4 	& -4.56 & 6.69 & 0.7 & 179 & 0.4 &  &  \\ 
110  & O2\,If* 	& 6   	& 50 	& 1.70 & 2400 	& 0.41 & -6.36 & 17.1 	& -5.22 & 6.22 & 0.4 & 113 & 0.7 &  &  \\ 
111  & WN9ha 	& 1   	& 45 	& 1.70 & 1000 	& 0.43 & -7.00 & 22.3 	& -5.42 & 6.25 & 0.1 & 118 & 0.7 & ?\,\tablefootmark{c} &  \\ 
112  & WN5h 	& 1   	& 56 	& 1.30 & 2400 	& 0.44 & -7.20 & 18.4 	& -4.57 & 6.48 & 1.0 & 99 & 0.2 & ?\,\tablefootmark{c} & 18 \\ 
113  & O2\,If*/WN5 	& 6,16   	& 50 	& 1.70 & 1600 	& 0.28 & -6.08 & 14.8 	& -5.49 & 6.09 & 0.2 & 54 & 0.2 & x\,\tablefootmark{c} & 7 \\ 
114  & O2\,If*/WN5 	& 6,16   	& 63 	& 1.70 & 2400 	& 0.31 & -6.18 & 13.9 	& -5.35 & 6.44 & 0.2 & 116 & 0.4 & ?\,\tablefootmark{c} &  \\ 
116  & WN5h:a 	& 7   	& 63 	& 1.30 & 2400 	& 0.75 & -7.93 & 28.1 	& -4.29 & 7.05 & 0.5 & 390 & 0.4 & ?\,\tablefootmark{c} & 7 \\ 
117  & WN5ha 	& 2   	& 63 	& 1.40 & 2400 	& 0.19 & -6.33 & 13.3 	& -4.93 & 6.40 & 0.5 & 109 & 0.4 &  &  \\ 
118  & WN6h 	& 1   	& 47 	& 1.10 & 1600 	& 0.16 & -7.96 & 31.9 	& -4.09 & 6.66 & 1.4 & 136 & 0.2 & x\,\tablefootmark{c} & 7,19 \\ 
119  & WN6h+? 	& 1   	& 47 	& 1.20 & 1600 	& 0.29 & -7.64 & 28.8 	& -4.31 & 6.57 & 1.0 & 116 & 0.2 & x\,\tablefootmark{c} & 7,10 \\ 
120  & WN9h 	& 1   	& 32 	& 1.40 & 500 	& 0.15 & -6.53 & 20.6 	& -5.33 & 5.58 & 0.3 & 32 & 0.3 &  &  \\ 
122  & WN5h 	& 2   	& 50 	& 1.15 & 1600 	& 0.28 & -6.90 & 17.3 	& -4.56 & 6.23 & 1.3 & 67 & 0.2 &  &  \\ 
124  & WN4 	& 2   	& 63 	& 1.10 & 1600 	& 0.30 & -4.32 & 4.5 	& -5.37 & 5.45 & 1.2 & 15 & 0.0 &  &  \\ 
126  & WN4b+O8: 	& 2   	& 71 	& 1.10 & 1600 	& 0.22 & -6.05 & 11.1 	& -4.78 & 6.44 & 0.5 & 65 & 0.0 & ?\,\tablefootmark{c} & 2 \\ 
128  & WN3b 	& 2   	& 112 	& 0.30 & 1600 	& 0.17 & -3.74 & 1.4 	& -4.93 & 5.44 & 3.4 & 14 & 0.0 &  &  \\ 
129  & WN3(h)a+O5\,V 	& 20   	& 79 	& 1.25 & 2000 	& 0.35 & -5.01 & 6.7 	& -5.24 & 6.20 & 0.4 & 64 & 0.2 & x & 7 \\ 
130  & WN11h 	& 1   	& 28 	& 1.30 & 200 	& 0.25 & -6.70 & 29.1 	& -5.35 & 5.68 & 0.1 & 41 & 0.4 &  &  \\ 
131  & WN4b 	& 2   	& 71 	& 0.75 & 1600 	& 0.13 & -4.76 & 4.6 	& -4.83 & 5.67 & 2.5 & 20 & 0.0 &  &  \\ 
132  & WN4b(h) 	& 2   	& 79 	& 0.50 & 1600 	& 0.23 & -4.82 & 3.3 	& -4.67 & 5.58 & 4.4 & 17 & 0.0 &  &  \\ 
133  & WN11h 	& 1   	& 28 	& 1.35 & 200 	& 0.11 & -6.85 & 29.4 	& -5.42 & 5.69 & 0.1 & 41 & 0.4 &  &  \\ 
134  & WN4b 	& 1   	& 79 	& 0.60 & 1600 	& 0.06 & -4.24 & 3.0 	& -4.88 & 5.51 & 3.2 & 16 & 0.0 &  &  \\ 

\hline
\end{longtable}
\tablebib{
(1)~\citetalias{BAT99};
(2) \citet{Foellmi2003b};
(3) \citet{Niemela2001};
(4) \citet{Niemela1995};
(5) \citet{Koenigsberger2003};
(6) \citet{Crowther2011};
(7) \citet{Schnurr2008};
(8) \citet{Breysacher1981};
(9) \cite[][and references therein]{Smith1996};
(10) \citet{Moffat1989};
(11) \citet{Crowther1995};
(12) \citet{Seggewiss1991};
(13) \citet{Niemela1991};
(14) \citet{Crowther2006b};
(15) \citet{Doran2013};
(16) \citet{Evans2011};
(17) \citet{Taylor2011};
(18) \citet{Schnurr2009};
(19) \citet{Sana2013};
(20) \citet{Foellmi2006}.
}
\tablefoot{
\tablefoottext{a}{Masses calculated from the luminosity, using the mass–luminosity relation derived by \citet{Graefener2011}}
\tablefoottext{b}{x = detected, ? = questionable}
\tablefoottext{c}{high X-ray emission}
}
\end{longtab}

\section{Results}
\label{sect:results}

\subsection{Stellar parameters}
\label{sect:parameters}

The analysis of the line spectrum yields the stellar temperature and
the ``transformed radius'' as an immediate result from the PoWR models (cf.\
Sect.\,\ref{sect:method}). The obtained values are compiled in
Table\,\ref{table:parameters} for all stars. Note that the parameters 
obtained for the binaries and binary candidates in our sample are 
biased, since we analyzed the spectra as if they were from a single
star. A detailed analysis that accounts for the composite nature of
these spectra is planned for a forthcoming paper. 

The location of the WN stars in the $\log T_*$-$\log
R_\mathrm{t}$-plane is shown in Fig.\,\ref{fig:wnstats}, omitting
the binaries (even the questionable ones). The two parameters ($\log
T_*$ and $\log R_\mathrm{t}$) appear well correlated, although there is
some scatter that exceeds their error margins. Compared to the
corresponding diagram for the Galactic WN stars \citepalias[cf.\ Fig.\,2
in][]{Hamann2006}, the correlation is similar, while the LMC stars
cover wider range of spectral subtypes at the cool end (WN10-11). 

The WN atmospheres are dominated by helium, while hydrogen is generally 
depleted. The precise determination of the hydrogen abundance is
delicate, because all \element{H} lines are blended with \ion{He}{ii} 
lines due to the wind broadening. Among the assumably 63 single WN stars of
our sample, we find 27 stars where hydrogen is below detectability
(see Table\,\ref{table:parameters}). The detection limit depends on the
individual circumstances such as stellar parameters, quality of the
observation, and consistency of the fit. Based on our experience, we
estimate that hydrogen abundances higher than 0.05 (mass fraction) do not escape
detection.  

Line contributions from hydrogen can be found in the spectra of 36 
putatively single WN stars, including the four Of/WN stars. There are about 
equal numbers of stars that fall into the 0.2 and the 0.4 category, 
respectively. 

Overall, Fig.\,\ref{fig:wnstats} shows a clear dichotomy regarding the
hydrogen abundance. Hydrogen is typically undetectable in the hotter
stars (early subtypes, WNE), while in the cooler stars (late subtypes,
WNL) hydrogen is clearly present, albeit depleted. This pattern is
similar to the one found for the Galactic WN stars \citepalias{Hamann2006}.
Even the rough equality between the numbers of WN stars with and without 
hydrogen is similar to the Galactic sample. 

Based on the obtained luminosity, we estimate the current stellar mass 
using the mass-luminosity relations for chemically homogeneous stars
from \citet{Graefener2011}. For stars with hydrogen at their surface, 
we use their Eq.\,9 for core \element{H}-burning stars, and for stars 
without hydrogen we use Eq.\,10 for core \element{He}-burning stars.
Among the putatively single stars, eight stars (BAT99\,33, 97, 98, 106, 
108, 109, 110, 117) exhibit current stellar masses in excess of 
$100\,M_\odot$ and thus belong to the category of very massive 
stars \citep{Vink2013}. 

In recent years, evidence is growing that canonical upper mass limit 
($150\,M_\odot$) is exceeded \citep[e.g.,\ ][]{Crowther2010,Vink2013}. 
In their study of the stellar population in the core of R136, 
\citet{Crowther2010} argued in favor of an upper mass limit that 
is roughly two times the canonical value. Although we achieve 
slightly lower values in our new analysis, we can confirm the range 
of stellar masses derived by these authors. 

In addition to the stellar parameters listed in Table\,\ref{table:parameters},
we have compiled in Table\,\ref{table:zansT} the number of 
hydrogen and helium ionizing photons for each star as well as the corresponding 
Zanstra temperatures. These values have been derived from the ionizing 
flux of the best fitting model. In some cases however, the stellar wind is 
so opaque that only an insignificant number of ionizing photons can escape. 

\subsection{Mass-loss rates} 
\label{subsect:mdot}

Mass loss of massive stars, especially WR stars, is of key importance
for understanding their evolution and their influence on their
environment. The huge injection of nuclear-enriched material is one of
the main drivers of the chemical evolution of their host clusters and
galaxies. For the evolution of WR stars, mass loss  can be more
important than nuclear fuel consumption. 

The mass-loss rates obtained from our analyses are given in
Table\,\ref{table:parameters}. One must keep in mind that the empirical $\dot{M}$
scales with the square-root of the clumping contrast, $\sqrt{D}$. The
value $D=10$ is not accurately constrained, and may in fact vary from
star to star or as function of radius \citep{Nugis1998,Puls2006,Liermann2008}.  

\begin{figure}
\centering
\includegraphics[width=\hsize]{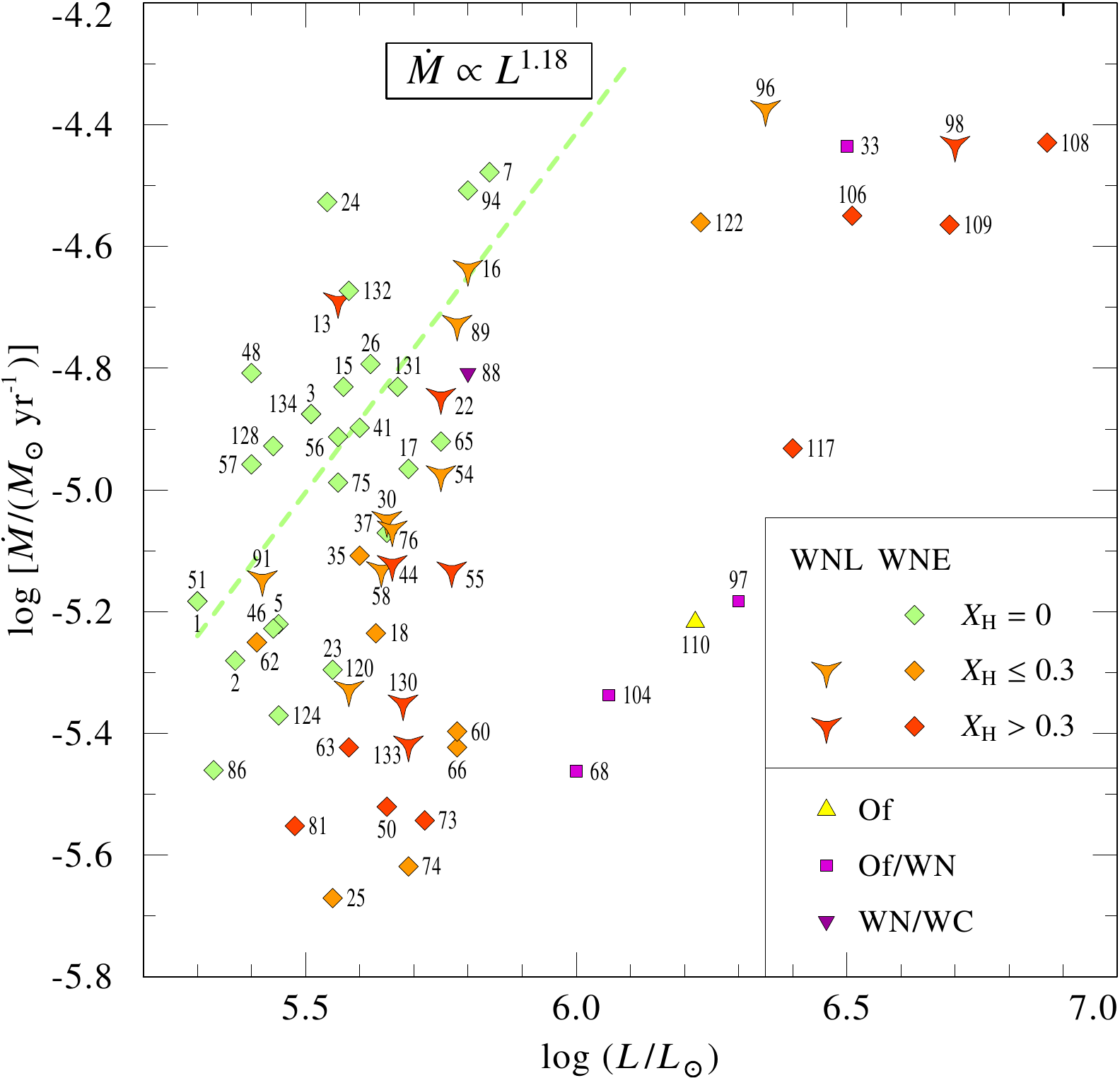}
\caption{Mass-loss rates versus luminosity for the putatively
single WN stars. The symbol shapes refer to the WNE and WNL subclass,
respectively. The atmospheric hydrogen mass fraction is color-coded, as 
indicated in the inlet. Also shown is a fit (green dashed line) to the hydrogen-free WNE stars.
}
\label{fig:mdot_l}
\end{figure}

In Fig. \ref{fig:mdot_l}, we plot the inferred mass-loss rates versus the
luminosities of our sample stars. There is no simple correlation, but there is a 
characteristic pattern. The very luminous stars ($\log\,(L/L_\odot) > 6.0$), 
which all show atmospheric hydrogen, have the highest mass-loss rates 
($\log\,(\dot M / ({\mathrm M_\odot}/\mathrm{yr})) \approx -4.5$). These stars may 
be very massive stars directly evolving off the main sequence, possibly 
still burning hydrogen. However, there are also three Of/WN-type 
stars with luminosities $\ge 6.0$\,dex that have weaker winds.

\begin{figure*}
\centering
\includegraphics[width=0.7\textwidth]{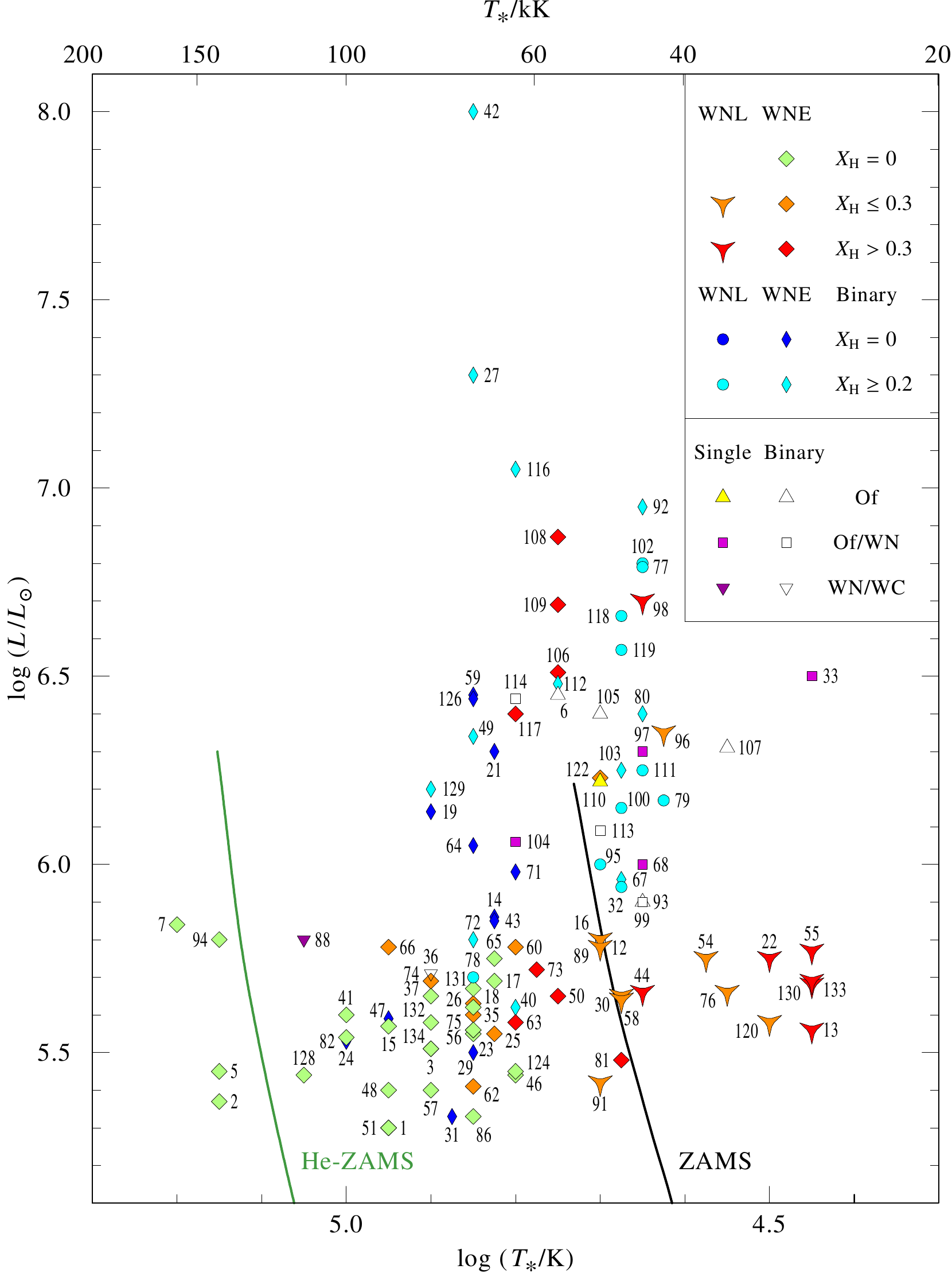}
\caption{The HRD of our sample of LMC stars. As explained in the inlet,
different symbols represent WNL and WNE stars, respectively. Binaries
are also included, but distinguished by different symbols; their spectra
have been analyzed as if they were single stars. The five O-type stars in
our sample are plotted with their own symbols. Among the WN
stars, the color codes the hydrogen mass fraction as defined in
the inlet. The zero-age main sequences (ZAMS) for hydrogen-rich and
pure-helium stars  are shown for orientation.
}
\label{fig:hrd}
\end{figure*}

The bulk of ``proper'' WN stars populate the luminosity range from $\log\,
(L/L_\odot) = 5.3$ to $5.8$. Their mass-loss rates scatter over more than
one order of magnitude (from $-5.7$\,dex to $-4.5$\,dex), but in clear
correlation with the hydrogen abundance. The hydrogen-free stars, 
which are obviously helium burners, exhibit the strongest
winds. Their $\log \dot{M}$ can be fitted to a linear relation with
$\log L$, giving 
\begin{equation}
\dot{M} = 
\left( \frac{L}{10^6\ L_\odot} \right)^{1.18}\ 
10^{-4.42}\ M_\odot/\mathrm{yr}\ .
\end{equation}

In the luminosity range below $5.9$\,dex, the mass-loss rates of
those WN stars, which show a detectable amount of atmospheric hydrogen,
scatter a significantly. Nevertheless, it seems that the ``proper'' WN stars increase
their mass-loss rates while evolving toward the Eddington limit, in line
with hydrodynamical models calculated by \citet{graefener+hamann-2008}. 

Table\,\ref{table:parameters} also gives the wind efficiency
$\eta$, defined as the ratio between the rates of the wind momentum,
$\dot{M} v_\infty$, and of the momentum of the radiation field, $L/c$:
\begin{equation}
 \eta := \frac{\dot{M} v_\infty c}{L}\ .
\end{equation}

Wind efficiencies exceeding unity (the ``single scattering limit'')
imply that an average photon undergoes multiple scatterings in the
wind.  Only specific hydrodynamic wind models can account for this
effect. \citet{Graefener2005} obtained $\eta = 2.5$ for a model of
the Galactic WC star WR\,111. More adequate for our sample,
\citet{graefener+hamann-2008} calculated WN models for different
metallicities and found that, under LMC conditions,  the wind
efficiency hardly exceeds unity. The empirical wind efficiencies
obtained from our analysis are also moderate; the average values (only
for the single stars) are 0.8 for the WNL and 2.1 for the WNE
subtypes, respectively. 

The mass-loss rates of the WN stars in the LMC obtained in the present 
study can be compared with those of the Galactic WN stars from 
\citet{Hamann2006}. Note that different values for the clumping contrast 
have been adopted in these studies (LMC: $D=10$, MW: $D=4$). Since the 
empirical $\dot{M}$ depends on the degree of clumping, one must assume 
for such a comparison that the clumping properties do not differ between 
the LMC and MW, and scale the mass-loss rates according to Eq.\,(\ref{eq:rt}). 

The comparison reveals that the WN stars in the LMC have on average 
lower mass-loss rates by roughly a factor of two compared to their 
Galactic counterparts. This is consistent with a dependence of $\dot{M}$ 
with metallicity $Z$ to the power $0.9 \pm 0.3$, depending on the metallicity 
assumed for the LMC and the MW. This agrees well with the results obtained 
by \citet{Crowther2006} as well as the exponent 0.86 theoretically derived 
by \cite{Vink2005} for late-type WR stars. A detailed investigation of the 
mass-loss rate as a function of the metallicity will be the subject of a 
forthcoming paper, where we will incorporate the results from our study of 
the WN stars in the SMC.

\subsection{The Hertzsprung-Russell diagram}
\label{sect:hrd}

The Hertzsprung-Russell diagram (HRD, Fig.\,\ref{fig:hrd}) shows 
all stars analyzed in this paper. Notably, the diagram includes
those stars that are already known to be binaries,
but were analyzed here as if they were single stars. The highest
luminosities in the HRD refer to such multiple stars.

\begin{figure}
\centering
\includegraphics[width=\hsize]{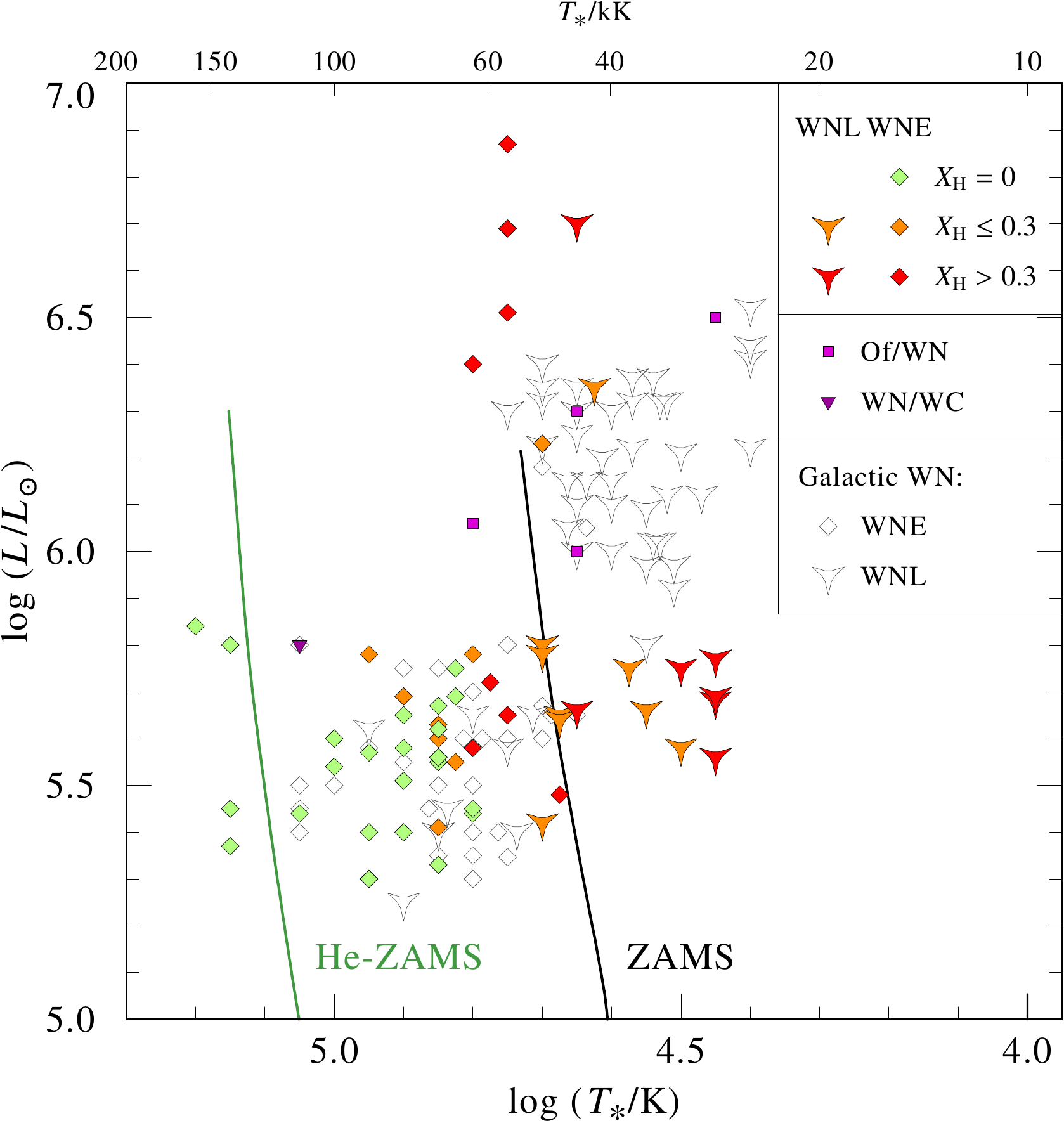}
\caption{The HRD of the single WN stars. The color-filled symbols refer to
the LMC stars analyzed in the present paper. The open symbols in
the background represent the Galactic WN stars for comparison (cf.
inlet).} 
\label{fig:hrd_galWN}
\end{figure}

In a second version of the HRD (Fig.\,\ref{fig:hrd_galWN}), we restrict 
the sample to the WN single stars (including the Of/WN types). While
the LMC stars are represented by color-filled symbols, the open
symbols in the background are the Galactic WN stars analyzed by
\citetalias{Hamann2006}, \citet{Martins2008},  \citet{Liermann2010}, and
\citet{Barniske2008}. 

One of the striking features in the HRD is the occurrence of a few
extremely luminous stars. All these stars show atmospheric hydrogen. 
In the Galactic sample, there is also a detached group of very luminous
stars, but the most luminous WN star encountered in the Galaxy -- the 
``Peony star'' WR\,102ka \citep{Barniske2008}, reaches only about 
6.5\,dex solar luminosities. 

The very luminous WN stars in the LMC are mostly of early subtypes
(WNE) or Of/WN, while the Galactic ones are WNL types. Moreover, the
number of WN stars in the high-luminosity domain seems to be much
larger in the Galactic sample. This might actually indicate a problem
with the Galactic analyses that arises from the uncertainty of the
stellar distances.  Many of these Galactic WNL stars were ``brightness
calibrated'' by means of those few WNL representatives that belong to 
clusters or associations. However, these young WNL stars may be
exceptionally luminous. By employing them for the brightness
calibration, the luminosities of other Galactic WNL stars might have
been overestimated. Due to the known distance, the LMC results are
free from such uncertainties. 

\begin{figure}
\centering
\includegraphics[width=\hsize]{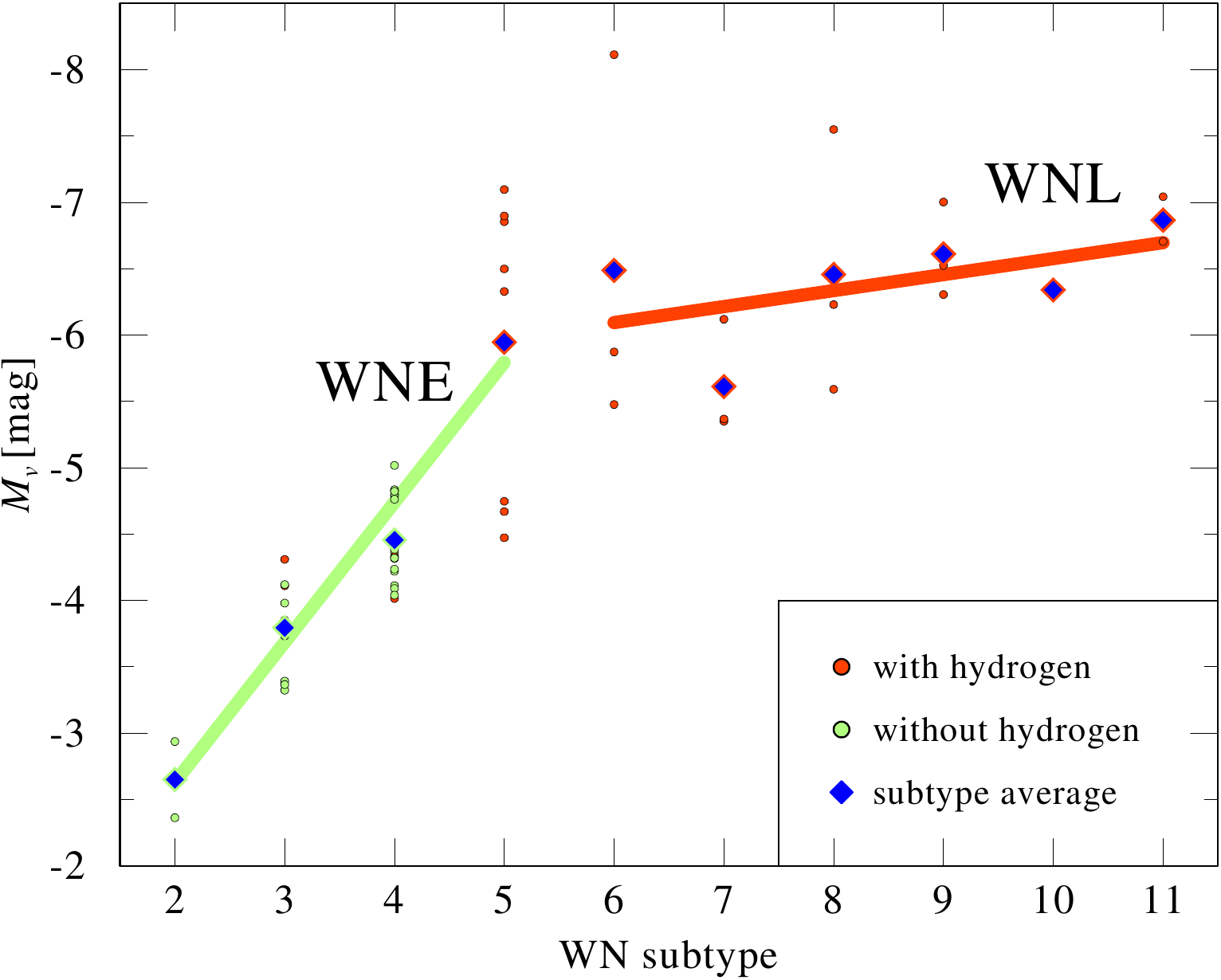}
\caption{Absolute visual magnitudes (narrowband color as defined by \citealt{Smith1968}) versus spectral subtype number for 
the putatively single stars in our sample. Colors code the absence
(green/light) or presence (red/dark) of hydrogen. While small
symbols represent individual stars, the thick symbols mark the average
$M_v$ of each subtype. The thick lines indicate linear fits to the
early (WN2--5) and late (WN6--11) subtypes, respectively.
}
\label{fig:mv}
\end{figure}

Based on their known, uniform distance, we can now check for our LMC
stars if such relation between absolute brightness and spectral subtype
really exists. As Fig.\,\ref{fig:mv} reveals, there is indeed some
correlation, but the scatter within each subtype is large 
($\rm$1\,mag, typically). The relation obtained by linear regression 
(thick shaded lines in Fig.\,\ref{fig:mv}) for the WNE stars is similar 
to the one adopted in \cite{Hamann2006}, while the hydrogen-containing WNL 
stars in the LMC are on average less bright than $M_v = -7.22$\,mag as 
used for the Galactic calibration. We note that the highest $M_v$ is 
associated with BAT99\,98, which we treat as a single star, although 
the moderate fit quality might indicate a hidden companion 
(cf.\ Appendix\,\ref{sec:comments}). The average $M_v$ value
of each WN subtype is compiled in Table\,\ref{table:magnitudes}. 

\begin{table}
\caption{Averaged absolute visual magnitudes of each WN subtype} 
\label{table:magnitudes}      
\centering
\begin{tabular}{lSS}
\hline\hline
Subtype \rule[0mm]{0mm}{4.0mm} & $M_v$ & $\sigma_{M_v}$ \\
\hline
WN\,2 \rule[0mm]{0mm}{4.0mm}  & -2.65 & 0.29 \\
WN\,3  & -3.8   & 0.31 \\
WN\,4  & -4.46  & 0.3  \\
WN\,5  & -5.95  & 1.05 \\
WN\,6  & -6.49  & 1.16 \\
WN\,7  & -5.61  & 0.36 \\
WN\,8  & -6.46  & 0.82 \\
WN\,9  & -6.61  & 0.29 \\
WN\,10 & -6.34  & \multicolumn{1}{c}{-}  \\
WN\,11 & -6.87  & 0.14 \\
\hline
\end{tabular}
\end{table}

The bulk of ``proper'' WN stars populate the luminosity range from $\log\,
(L/L_\odot) = 5.3$ to $5.8$.  The hydrogen containing stars are mainly
found on the cool side of the zero-age main-sequence (ZAMS), where all
WN stars are of late subtypes (WNL).  The hydrogen-free stars, all of
early subtypes (WNE), gather at the hot side of the ZAMS and near 
the theoretical zero-age main-sequence for helium stars (He-ZAMS). 

In the Galactic sample, the group of WNL stars with hydrogen was not
encountered in this luminosity range, possibly because of the erroneous 
brightness calibration applied to those stars as discussed above.

\section{Discussion}
\label{sect:evolution}

\begin{figure}
\centering
\includegraphics[width=\hsize]{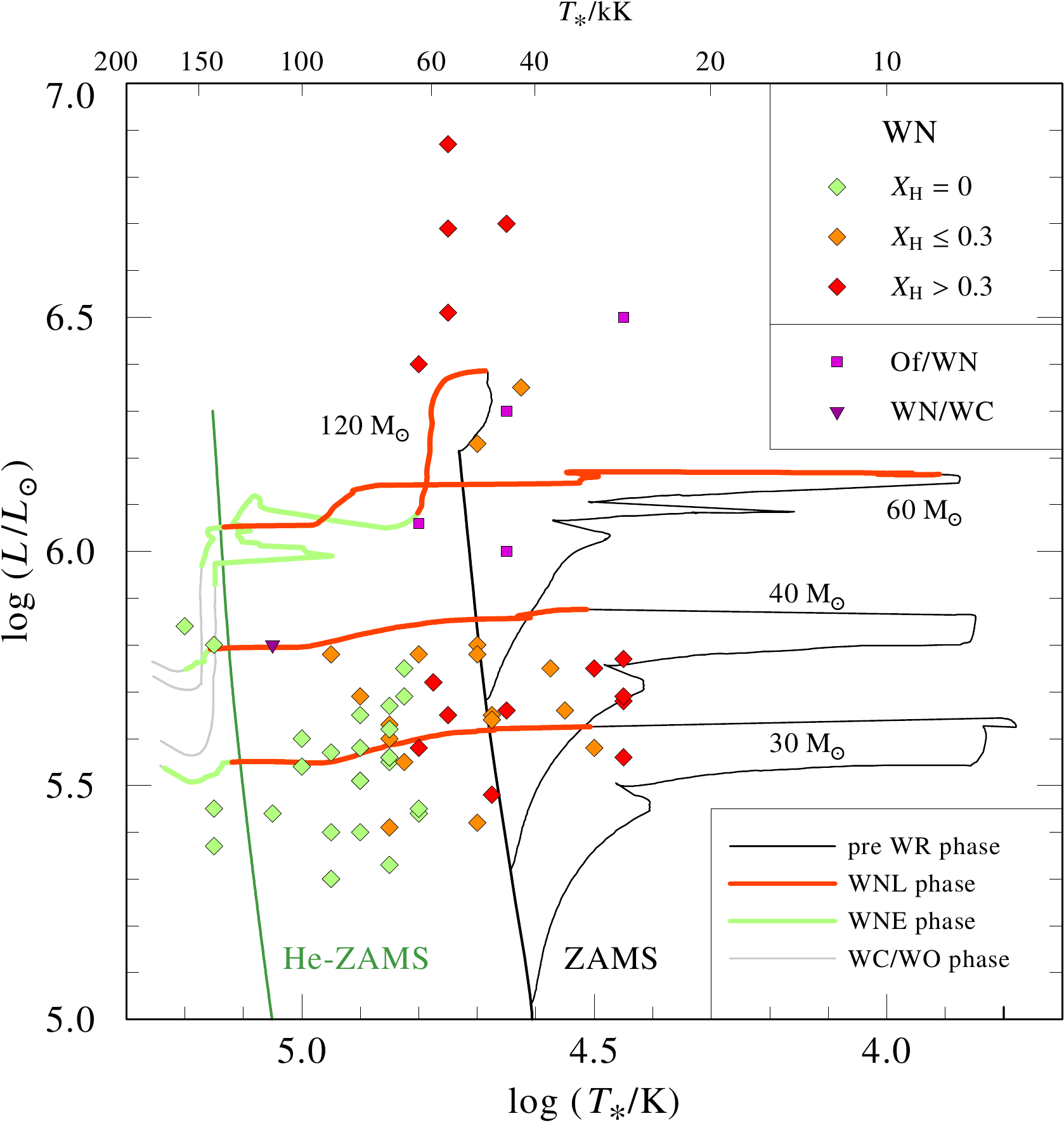}
\caption{The HRD with the single WN stars (discrete symbols) and stellar
evolution tracks from \citet{Meynet2005}, which account for the
effects of rotation. The labels indicate the initial mass. 
The color coding of the tracks during the WR
phases (thick lines) corresponds to the filling color of the symbols,
which reflects the observed atmospheric composition (see inlet).
}
\label{fig:hrd_geneva}
\end{figure}

Our large sample offers an excellent possibility to compare the
almost complete WN population of the LMC with the predictions of the stellar
evolution theory. Figure\,\ref{fig:hrd_geneva} shows the HRD of our
program stars in comparison to the stellar evolution tracks calculated
by the Geneva group for LMC metallicity. In the version shown here,
stellar evolution models account for the effects of rotation, but neglect 
the metallicity scaling in the WR phase \citep{Meynet2005}. 

The individual evolution phases are distinguished by different drawing
styles, according to the chemical composition at the stellar surface.
At hydrogen surface abundances $X_{\element{H}} > 0.4$ (mass fraction), the
star is considered to be in a pre-WR phase represented  by a thin black
line. The WNL stage, which per definition initiates when the hydrogen abundance
drops below 0.4 (mass fraction) in the atmosphere, is highlighted by thick red lines.
Hydrogen abundances below 0.05 (mass fraction) are considered to correspond to the WNE
stage, and the track is plotted as a thick green line. Finally, for
carbon abundances above 0.2 (mass fraction), the  star reaches the WC and WO phase and
the track is drawn as a gray line. 

Until now, we classified stars as WNE and WNL according
to their spectroscopic subtype. With regard to stellar evolution, the
terms ``WNE'' and ``WNL''  are defined differently and refer only to
the atmospheric composition, i.e.,\ the absence or presence of hydrogen.  
Note also that the temperature axis in Fig.\,\ref{fig:hrd_geneva}
refers to the stellar temperature, i.e.,\ to the effective
temperature related to the hydrostatic core, but not to the photosphere
(cf.\ Sect.\,\ref{sect:models}). 

Let us first look at the very luminous stars, separated from the others 
by a gap at about $\log\,(L/L_\odot) = 5.9$. These stars can be explained
by tracks similar to the one for $120\,M_\odot$ shown in Fig.\,\ref{fig:hrd_geneva}. 
This track stays close to the ZAMS, increases in luminosity until it enters the
WNL stage, then drops, and finally evolves toward the helium main
sequence after having lost all hydrogen. While the highest initial mass
for which tracks are provided by \citet{Meynet2005} is $120\,M_\odot$, the
most luminous star of this group (BAT99\,108), if indeed a single star, 
requires about $300\,M_\odot$, as already pointed out by
\citet{Crowther2010}. 

Similar to the WN stars in the Galaxy, only objects with a substantial 
amount of hydrogen are found at luminosities of $\log\,(L/L_\odot) > 5.9$, 
although, according to stellar evolution models \citep{Meynet2005,Yusof2013}, 
hydrogen-free stars are also expected in this parameter regime. This 
mismatch might be partly explained by very luminous type IIn 
supernovae (SN). \citet{Smith2008} unveiled that the progenitors of these rare 
SN are probably high-mass stars ($M_{\mathrm{init}} > 50\,M_\odot$) that 
explode with a significant amount of hydrogen left in their stellar atmospheres. 
The observational evidence discussed by this author argues for episodic 
mass loss prior to the type IIn SNe in excess of typical WR mass-loss rates, 
suggesting that progenitor candidates are classical LBVs. Since the WN stars 
observed in this mass range seem to be rather normal WN stars, apart from 
their high luminosities, the origin and fate of the very massive stars still 
remains puzzling.

With the exception of the $120\,M_\odot$ track, all stellar evolution tracks (for 
LMC metallicities) presented by \citet{Meynet2005} evolve toward the red supergiant
(RSG) stage, but barely reach the corresponding temperature range before 
evolving back to the hot part of the HRD. Nevertheless, in this paper, we denote all 
evolution stages in the cool part of the HRD as RSG even if the tracks only 
pass the blue supergiant and LBV domain of the HRD.
As can be seen in Fig.\,\ref{fig:hrd_geneva}, luminosities of about 6.2\,dex
are reached by the post RSG evolution of stars with initially $60\,M_\odot$ 
\citep{Meynet2005}. However, no ``cool'' WNL stars are found in this 
range, which might indicate that the RSG stage is in fact not reached 
in this mass range, possibly because of the LBV instability barrier.

The bulk of WN stars are found in the luminosity range $\log\,(L/L_\odot) =
5.3\,...\,5.8$. The upper end of this range corresponds quite well to the
post-RSG track with initially $40\,M_\odot$. Interestingly, this upper
mass limit agrees well with analyses of WC stars in the LMC by
\citet{Grafener1998} and \citet{Crowther2002}. 

The lowest initial mass for which a post-RSG track is provided by 
\citet{Meynet2005} for LMC metallicities is $30\,M_\odot$. The
luminosities of the analyzed WN stars show a lower cutoff at about
$10^{5.3}\,L_\odot$, which rather corresponds to an initial mass of
about $20\,M_\odot$. \citet{Meynet2005}, based on interpolation of
their tracks, estimated that the minimum initial mass for reaching WR
phases is about $25\,M_\odot$ for LMC metallicity. This limit seems to
be too high, compared to our results. The older Geneva tracks, which did 
not account for rotation \citep{Schaerer1993}, are even more contradictory; 
these tracks fail to reach post-RSG stages for initial masses below $60 M_\odot$. 
We also compared our empirical HRD positions with non rotating models from 
\citet{Eldridge2006}. These models predict WR stars for initial masses 
exceeding $33 M_\odot$.

As stated in Sect.\,\ref{sect:intro}, it is theoretically expected 
that the evolution of massive stars depends on their initial metallicity. 
Stellar evolution models predict a higher minimum for the initial 
mass of WR stars with decreasing metallicity due to the reduced 
mass loss by stellar winds anticipated at lower metallicities. Indeed, 
our study shows that the mass-loss rate of an average WN star in the 
LMC is lower than in the Galaxy (see Sect.\,\ref{subsect:mdot} for details). 
However, Fig.\,\ref{fig:hrd_galWN} illustrates that the initial masses 
for the LMC WN stars are comparable to their Galactic counterparts. 
Thus, the conventional expectation that WN stars in the LMC originate 
from a higher mass range compared to their Galactic twins is not 
supported from our sample. 

In the mass range from $20\,M_\odot$ to $40\,M_\odot$, the stellar 
evolution in the Galaxy and the LMC results in comparable HRD 
positions for the WN stars, although the mass-loss rates and the 
metallicity are lower in the LMC. This may be attributed to the 
relatively small differences between the metallicity of these two 
galaxies. Another implication might be that the WN stars and their 
progenitors rotate faster in the LMC than in the Galaxy, because their 
lower mass-loss rates imply a reduced loss of angular momentum. A 
faster rotation increases the WR lifetime and decreases the minimum 
initial mass for the WR phase, mainly due to a more efficient 
internal mixing \citep{Maeder2005}.               

Do the stellar evolution calculations successfully predict the observed 
number ratio between WNE and WNL stars? We used the four tracks from
\citet{Meynet2005} for computing synthetic populations, assuming a
Salpeter initial mass function and a constant star formation rate
\citepalias[see][for similar simulations of the Galactic WR stars]{Hamann2006}. 
From our simulations, we expect roughly 20\,\% of all WN stars in the 
WNE phase, i.e.,\ much less than the 40\,\% observed. Even worse, almost 
half of these WNE stars originate from the tracks for 60 and 
120\,$M_\odot$ initial mass and thus lie at luminosities where no WNE 
stars at all are found in our sample. 

The synthetic population yields about the same 
number of WC stars as WNE stars. In fact, the WNE stars residing in the 
LMC are twice as frequent as the LMC-WC stars. Moreover,
half of these WC stars are predicted to evolve from $120\,M_\odot$ initial
mass, which does not match the low WC luminosities actually observed  
\citep{Grafener1998, Crowther2002}. Based on stellar evolution models, 
\citet{Eldridge2006} show that a better agreement with the observed 
WC/WN ration can be achieved by means of a metallicity scaling of the 
mass-loss rate during the WR phase. However, like the Geneva models, 
these models also fail to reproduce the observed luminosity ranges for 
the different WR subtypes.

The color-coded evolution tracks in Fig.\,\ref{fig:hrd_geneva} reveal a
discrepancy in stellar temperature between these tracks and
the HRD position of the WNE stars. Theoretically, it is expected that
the hydrogen-free WNE stars are located on the He-ZAMS. Instead, a
clear gap is seen between most of these stars and the He-ZAMS. This
discrepancy in the effective temperature is attributed to the long known
"radius problem" of hydrogen-free WR stars. The observed WR radii are up
to an order of magnitude larger than the radii predicted by stellar evolution
models. We note that the effective temperature of the stellar evolution
models refer to the hydrostatic core radius whereas the effective
temperature of our atmosphere models is defined at the inner boundary
radius $R_*$, where the Rosseland optical depth reaches 20 (cf.\ Sect.\,
\ref{sect:models}). However, this cannot be the decisive reason because
$R_*$ is close to the hydrostatic core radius as long as $R_*$ is located
in the hydrostatic part of the wind, which is the case for most of
our final-fit models. A solution for this radius problem has been
studied by \citet{Graefener2012}. These authors show that an inflated
subphotospheric layer, which incorporates the effect of clumping, can
bring the observed WR radii in agreement with the theory. 

The binaries in our sample have been thoroughly identified, but some may 
have escaped detection, as discussed in Sect.\,\ref{subsect:binaries}. 
Moreover, a single star could be the product of binary 
evolution, e.g., a merger or a binary system where the companion already 
exploded \citep{Sana2012}. The consequences of binary evolution with 
respect to the population of WR stars have been discussed by various authors 
\citep[e.g.,][]{Paczynski1967,Vanbeveren2007,Eldridge2008,Eldridge2013}. 
According to these authors, the minimum initial mass for the WR phases 
is considerably decreased in binary systems due to a significant pre-WR 
mass-loss through Roche lobe overflow. If the least luminous stars of 
our sample had evolved through the binary channel, this would 
explain the discrepancy with the minimum initial mass of WR stars as 
predicted from single star evolution.

Alternatively, single star evolution may produce WR stars from lower 
initial masses when higher initial rotation velocities (more than the 
$300\,\mathrm{km\,s^{-1}}$ as assumed by \citealt{Meynet2005}) are 
adopted. More plausibly, mass-loss rates that are enhanced over those 
usually adopted during the RSG stage promote the evolution of massive stars 
toward the blue part of the HRD \citep[e.g.,][]{Vanbeveren1998,Vanbeveren2007,Georgy2012}.  

Binary evolution was shown to significantly affect the WC/WN ratio 
\citep[e.g.,][]{Vanbeveren2007,Eldridge2008}. \citet{Eldridge2008} 
demonstrate that, while their single star models reproduce the observed 
WC/WN ratio, the binary population models predict too many WN stars. 
The bias from close binary evolution on the WN population should be 
subject to future investigations.

The WNE and WC stars are thought to be the progenitors of type Ib and 
type Ic supernovae \citep[e.g.,][]{Gaskell1986,Begelman1986}, if they do
not directly collapse to a black hole without a bright SN. The first 
tentative identification of a WR star as a type Ib SN progenitor has 
been reported recently for SN\,iPTF13bvn \citep{Cao2013}. Alternatively, 
\citet{Eldridge2013} show that the bulk of the SNe of type Ibc can 
originate from relatively low-mass pure helium stars whose hydrogen-rich 
atmospheres have been stripped by close binary interactions. Our 
empirical HDR positions (Fig.\,\ref{fig:hrd_geneva}) suggest that the 
LMC-WNE stars and their successors are restricted to initial masses 
below $40\,M_\odot$ similar to the situation in our Galaxy 
\citep[cf.][]{Sander2012}.

Summarizing the discussion on stellar evolution, we found some 
general agreement between our WN analyses and the Geneva tracks for
LMC metallicity that account for stellar rotation \citep{Meynet2005}. 
However, in a quantitative sense, the stellar evolution tracks are not 
consistent with our empirical results. The discrepancies refer
especially to the range of initial masses required for reaching the respective
WR phases, and to the number ratios and luminosities of the different WR
subtypes. Interestingly, similar conclusions have been drawn from our
study of the Galactic WN stars and the comparison with the
corresponding Geneva tracks \citepalias{Hamann2006}. Moreover,
\citet{Sander2012} analyzed the Galactic WC stars and found that their
relatively low luminosities are not reproduced by the Geneva tracks.
Calculations by \citet{Vanbeveren1998}, who adopted higher mass-loss
rates during the RSG stage, yielded a better agreement. We may also
mention here the work of \citet{Hunter2008}, who determined the chemical
composition of 135 early B-type stars in the LMC to test the prediction
of rotationally induced mixing. They demonstrated that about 40\,\% of
their sample do not agree with the predicted correlation between
rotation and nitrogen enrichment. Hence it seems that massive-star
evolution is still not fully understood.  

\section{Summary and Conclusions}
\label{sect:conclusions}

\begin{enumerate}

\item 
The spectra of 107 stars in the LMC have been
analyzed by means of state-of-the-art model atmospheres.

\item 
The sample contains 102 WR stars of the nitrogen sequence (WN,
including five Of/WN), and thus comprises nearly the complete population of WN stars
known in the LMC. 

\item 
Of these WN stars, 63 do not show any indications of multiplicity, and
thus are putatively single stars. 

\item
Two groups of WN stars can be distinguished from their luminosity: a small
group (12\,\%) of very luminous stars $\log\,(L/L_\odot) > 5.9$, and a large 
group (88\,\%) populating the range of moderate luminosities between 
$\log\,(L/L_\odot) = 5.3\,...\,5.8$.  

\item
Of the 63 single, 27 WN stars (i.e.,\ 43\,\%) do not show hydrogen in
their wind. These hydrogen-free stars are only found in the group with
moderate luminosities.

\item
Stellar radii are generally larger, and effective temperatures
correspondingly lower, than predicted from stellar evolution models. This
may indicate a subphotospheric inflation, as discussed in the recent
literature.   

\item
The bulk of WN stars (with moderate luminosities)
seem to be in a post-RSG evolution phase. 

\item
According to their luminosities in the range $\log\,(L/L_\odot) = 
5.3\,...\,5.8$, these WN stars originate from initial stellar masses between 
$20\,M_\odot$ and $40\,M_\odot$. This mass range is similar to the range found
for the Galaxy. Hence, the expected metallicity dependence of the
evolution is not seen.  

\item
Stellar evolution tracks, when accounting for rotationally induced
mixing, in principle, can explain the extremely luminous WN stars as well as
the bulk of WN stars with moderate luminosities. The former evolve
directly from the ZAMS, while the latter go through the RSG stage.
However, the stellar evolution models still fail to correctly reproduce the observed
ranges of luminosities and initial masses. 

\end{enumerate}

\begin{acknowledgements}

We would like to thank the referee, John Eldridge, for his constructive comments
that have significantly improved the present work.
This research made use of the SIMBAD database, operated at CDS,
Strasbourg, France, and of data products from the Two Micron All Sky
Survey, which is a joint project of the University of Massachusetts and
the Infrared Processing and Analysis Center/California Institute of
Technology, funded by the National Aeronautics and Space Administration
and the National Science Foundation (NASA). 
This work is partly based on INES data from the IUE satellite, and on  observations
with the Spitzer Space Telescope, which is operated by the Jet
Propulsion Laboratory, California Institute of Technology under a
contract with NASA. 
This research made also use of NASA's Astrophysics Data System
and of the VizieR catalog access tool, CDS, Strasbourg, France.
Some of the data presented in this paper were retrieved from the
Mikulski Archive for Space Telescopes (MAST). STScI is operated by the
Association of Universities for Research in Astronomy, Inc., under NASA
contract NAS5-26555. Support for MAST for non-HST data is provided by
the NASA Office of Space Science via grant NNX09AF08G. The work also based on data 
made available through Wikimbad, hosted in the LAOG, France (http://wikimbad.org). 
LMO acknowledges the funding by DLR grant 50 OR 1302.

\end{acknowledgements}

\bibliographystyle{aa}
\bibliography{paper}


\Online
\label{onlinematerial}

\begin{appendix} 
\section{Additional tables}
\label{sec:addtables}

\begin{table}[hb]
\caption{Available spectra for WN stars in the LMC} 
\label{table:data}
\centering  
\begin{tabular}{lclllclc}
\hline\hline \rule[0mm]{0mm}{4.0mm} 
   No. & 
   $\Delta\lambda$\,/\,{\AA} &
   Instrument &  
   Flux & 
   Archive & 
   HST Proposal ID & 
   PI &
   No. of stars \\
\hline   
   1  & 1090-2330 & HST-FOS   & abs.  & MAST    & 4260 & C. Leitherer \rule[0mm]{0mm}{4.0mm} & 5  \\
   2  & 1090-2330 & HST-FOS   & abs.  & MAST    & 5702 & C. Leitherer  & 1  \\ 
   3  & 1120-1720 & HST-STIS  & abs.  & MAST    & 9412 & P. Massey     & 2  \\
      & 6300-6860 &           &       &         &                      &    \\
   4  & 1140-1730 & HST-STIS  & abs.  & MAST    & 9434 & J. Lauroesch  & 1  \\
   5  & 1150-3200 & IUE       & abs.  & INES    & -    & diverse       & 63 \\
   6  & 1160-1750 & HST-GHRS  & abs.  & MAST    & 5297 & S. Heap       & 2  \\
   7  & 1160-1750 & HST-FOS   & abs.  & MAST    & 6018 & S. Heap       & 1  \\
   8  & 1200-1760 & HST-GHRS  & abs.  & MAST    & 5157 & D. Ebbets     & 2  \\
   9  & 2020-2030 & HST-GHRS  & abs.  & MAST    & 5887 & K. Roth       & 2  \\
      & 2050-2070 &           &       &         &                      &    \\
   10 & 3020-3300 & HST-STIS  & abs.  & MAST    & 9412 & P. Massey     & 1  \\ 
   11 & 3240-4780 & HST-FOS   & abs.  & MAST    & 6508 & N. Walborn    & 1  \\
   12 & 3240-4780 & HST-FOS   & abs.  & MAST    & 6417 & P. Massey     & 6  \\
   13 & 3240-4780 & HST-FOS   & abs.  & MAST    & 6032 & N. Walborn    & 1  \\
   14 & 3240-6820 & HST-FOS   & abs.  & MAST    & 6018 & S. Heap       & 5  \\
   15 & 3240-6820 & HST-FOS   & abs.  & MAST    & 6110 & C. Leitherer  & 1  \\
   16 & 3400-7300 & CTIO      & abs.  & VizieR  & -    & A.~V. Torres-Dodgen  & 53 \\
   17 & 3670-6000 & AAT-RGO   & norm. & AAT     & -    & Smith, P. Crowther   & 28 \\
   18 & 3700-6800 & diverse   & norm. & private & -    & C.~Foellmi    & 62 \\
   19 & 3770-9055 & ESO-EFOSC & norm. & private & -    & U. Wessolowski       & 19 \\
   20 & 4000-5500 & diverse   & norm. & private & -    & O.~Schnurr    & 42 \\
   21 & 4310-4590 & HST-STIS  & abs.  & MAST    & 7739 & P. Massey     & 1  \\
      & 6300-6870 &           &       &         &                      &    \\
\hline 

\end{tabular}
\end{table}

\begin{table*}[p]
\caption{LMC WN stars analyzed in this work}
\label{table:wnsample}
\centering  
\begin{tabular}{cccl | cccl}
\hline\hline 
   BAT99 &
   Brey &
   Figure &
   Observations\tablefootmark{a} &
   BAT99 &
   Brey &
   Figure &
   Observations\tablefootmark{a}   \rule[0mm]{0mm}{4.0mm}\\
\hline  
001  &  01  & \ref{fig:bat001}   &  5,16,18,19     &  071  &  60  & \ref{fig:bat071}   &  5,16,18   \rule[0mm]{0mm}{4.0mm}       \\
002  &  02  & \ref{fig:bat002}   &  5,18           &  072  &  61  & \ref{fig:bat072}   &  18               \\
003  &  03  & \ref{fig:bat003}   &  5,16,18        &  073  &  63  & \ref{fig:bat073}   &  18               \\
005  &  04  & \ref{fig:bat005}   &  5,16,18,19     &  074  &  63a & \ref{fig:bat074}   &  18               \\
006  &  05  & \ref{fig:bat006}   &  5              &  075  &  59  & \ref{fig:bat075}   &  5,18             \\
007  &  06  & \ref{fig:bat007}   &  5,16,17,18,19  &  076  &  64  & \ref{fig:bat076}   &  2,5,16,17,19,20  \\
012  &  10a & \ref{fig:bat012}   &  5,10,17,20     &  077  &  65  & \ref{fig:bat077}   &  5,16,20          \\
013  &  -   & \ref{fig:bat013}   &  1,5,17,20      &  078  &  65b & \ref{fig:bat078}   &  11,18            \\
014  &  11  & \ref{fig:bat014}   &  5,18           &  079  &  57  & \ref{fig:bat079}   &  5,16,17,20       \\
015  &  12  & \ref{fig:bat015}   &  5,16,17,18,19  &  080  &  65c & \ref{fig:bat080}   &  20               \\
016  &  13  & \ref{fig:bat016}   &  5,16,17,19     &  081  &  65a & \ref{fig:bat081}   &  5,18             \\
017  &  14  & \ref{fig:bat017}   &  5,16,18        &  082  &  66  & \ref{fig:bat082}   &  5,16,18          \\
018  &  15  & \ref{fig:bat018}   &  5,16,18        &  086  &  69  & \ref{fig:bat086}   &  16,18            \\
019  &  16  & \ref{fig:bat019}   &  5,16,17,18,19  &  088  &  70a & \ref{fig:bat088}   &  18               \\
021  &  17  & \ref{fig:bat021}   &  18             &  089  &  71  & \ref{fig:bat089}   &  5,17,20          \\
022  &  18  & \ref{fig:bat022}   &  5,16,17,20     &  091  &  73  & \ref{fig:bat091}   &  13,18,20         \\
023  &  -   & \ref{fig:bat023}   &  18,17          &  092  &  72  & \ref{fig:bat092}   &  5,16,20          \\
024  &  19  & \ref{fig:bat024}   &  5,16,18        &  093  &  74a & \ref{fig:bat093}   &  5,20             \\
025  &  19a & \ref{fig:bat025}   &  18             &  094  &  85  & \ref{fig:bat094}   &  16,18,19         \\
026  &  20  & \ref{fig:bat026}   &  5,16,18        &  095  &  80  & \ref{fig:bat095}   &  5,16,17,20       \\
027  &  21  & \ref{fig:bat027}   &  5,16,18        &  096  &  81  & \ref{fig:bat096}   &  20               \\
029  &  23  & \ref{fig:bat029}   &  5,16,18        &  097  &  -   & \ref{fig:bat097}   &  20               \\
030  &  24  & \ref{fig:bat030}   &  5,16,17,19,20  &  098  &  79  & \ref{fig:bat098}   &  20               \\
031  &  25  & \ref{fig:bat031}   &  5,16,18        &  099  &  78  & \ref{fig:bat099}   &  3,12,20          \\
032  &  26  & \ref{fig:bat032}   &  5,16,17,19,20  &  100  &  75  & \ref{fig:bat100}   &  12,20            \\
033  &  -   & \ref{fig:bat033}   &  1,5,9,17,20    &  102  &  87  & \ref{fig:bat102}   &  20               \\
035  &  27  & \ref{fig:bat035}   &  5,16,18,19     &  103  &  87  & \ref{fig:bat103}   &  20               \\
036  &  29  & \ref{fig:bat036}   &  5,16,17,18,19  &  104  &  76  & \ref{fig:bat104}   &  12,20,21         \\
037  &  30  & \ref{fig:bat037}   &  16,18          &  105  &  77  & \ref{fig:bat105}   &  5,12,20          \\
040  &  33  & \ref{fig:bat040}   &  16,17,18       &  106  &  82  & \ref{fig:bat106}   &  8,14             \\
041  &  35  & \ref{fig:bat041}   &  5,16,18        &  107  &  86  & \ref{fig:bat107}   &  5,20             \\
042  &  34  & \ref{fig:bat042}   &  5,16,18        &  108  &  82  & \ref{fig:bat108}   &  8,14             \\
043  &  37  & \ref{fig:bat043}   &  5,16,18,19     &  109  &  82  & \ref{fig:bat109}   &  6,14             \\
044  &  36  & \ref{fig:bat044}   &  5,16,17,20     &  110  &  82  & \ref{fig:bat110}   &  6,14             \\
046  &  38  & \ref{fig:bat046}   &  5,16,18        &  111  &  82  & \ref{fig:bat111}   &  7,14             \\
047  &  39  & \ref{fig:bat047}   &  18             &  112  &  82  & \ref{fig:bat112}   &  12               \\
048  &  40  & \ref{fig:bat048}   &  5,16,18,19     &  113  &  -   & \ref{fig:bat113}   &  12,20            \\
049  &  40a & \ref{fig:bat049}   &  5,18           &  114  &  -   & \ref{fig:bat114}   &  12,20            \\
050  &  41  & \ref{fig:bat050}   &  18             &  116  &  84  & \ref{fig:bat116}   &  20               \\
051  &  42  & \ref{fig:bat051}   &  5,16,18        &  117  &  88  & \ref{fig:bat117}   &  5,16,18,19       \\
054  &  44a & \ref{fig:bat054}   &  17,20          &  118  &  89  & \ref{fig:bat118}   &  4,5,16,20        \\
055  &  -   & \ref{fig:bat055}   &  1,5,9,15,17,20 &  119  &  90  & \ref{fig:bat119}   &  5,16,17,20       \\
056  &  46  & \ref{fig:bat056}   &  5,16,18        &  120  &  91  & \ref{fig:bat120}   &  1,5,17,20        \\
057  &  45  & \ref{fig:bat057}   &  5,16,18        &  122  &  92  & \ref{fig:bat122}   &  5,16,17,18       \\
058  &  47  & \ref{fig:bat058}   &  5,16,17,19,20  &  124  &  93a & \ref{fig:bat124}   &  18               \\
059  &  48  & \ref{fig:bat059}   &  5,16,18,19     &  126  &  95  & \ref{fig:bat126}   &  18               \\
060  &  49  & \ref{fig:bat060}   &  18,16          &  128  &  96  & \ref{fig:bat128}   &  16,18            \\
062  &  51  & \ref{fig:bat062}   &  18,16          &  129  &  97  & \ref{fig:bat129}   &  18               \\
063  &  52  & \ref{fig:bat063}   &  5,16,17,18,19  &  130  &  -   & \ref{fig:bat130}   &  5,17,20          \\
064  &  53  & \ref{fig:bat064}   &  18             &  131  &  98  & \ref{fig:bat131}   &  5,18             \\
065  &  55  & \ref{fig:bat065}   &  17,18          &  132  &  99  & \ref{fig:bat132}   &  5,16,18          \\
066  &  54  & \ref{fig:bat066}   &  16,18          &  133  &  -   & \ref{fig:bat133}   &  1,5,17,20        \\
067  &  56  & \ref{fig:bat067}   &  16,18,19       &  134  &  160 & \ref{fig:bat134}   &  5,16,18          \\
068  &  58  & \ref{fig:bat068}   &  3,5,16,20      &       &      &    &                   \\
\hline 
\end{tabular}
\tablefoot{
\tablefoottext{a}{Sources of observations; the numbers refer to the
entries in Table\,\ref{table:data}} 
}
\end{table*}

\begin{table*}[p]
\caption{Number of ionizing photons and Zanstra temperatures for WN stars in the LMC} 
\label{table:zansT}
\centering  
\begin{tabular}{cccccc | cccccc}
\hline\hline  \rule[0mm]{0mm}{4.0mm} 
   BAT99 &
   \multicolumn{2}{c}{\ion{H}{i}}  & 
   \ion{He}{i} &
   \multicolumn{2}{c}{\ion{He}{ii}} &
   BAT99 &
   \multicolumn{2}{c}{\ion{H}{i}} &
   \ion{He}{i} &
   \multicolumn{2}{c}{\ion{He}{ii}}
   \\
    &  
    $\log Q$ &
    $T_\mathrm{Zanstra}$ &
    $\log Q$ &
    $\log Q$ &
    $T_\mathrm{Zanstra}$ &
    &
    $\log Q$ &
    $T_\mathrm{Zanstra}$ &
    $\log Q$ &
    $\log Q$ &
    $T_\mathrm{Zanstra}$
    \\
    &  
    $[\mathrm{s^{-1}}]$ &
    $[\mathrm{K}]$ &
    $[\mathrm{s^{-1}}]$ &
    $[\mathrm{s^{-1}}]$ &
    $[\mathrm{K}]$ &
    &
    $[\mathrm{s^{-1}}]$ &
    $[\mathrm{K}]$ &
    $[\mathrm{s^{-1}}]$ &
    $[\mathrm{s^{-1}}]$ &
    $[\mathrm{K}]$
    \\
\hline 
001 & 49.16 & 67981 & 48.92 & 38.01 & 20714 & 071 & 49.86 & 66841 & 49.47 & - & -\rule[0mm]{0mm}{4.0mm} \\ 
002 & 49.22 & 84379 & 49.04 & 47.52 & 80957 & 072 & 49.68 & 75882 & 49.37 & 46.16 & 51455 \\ 
003 & 49.37 & 59612 & 49.08 & - & - & 073 & 49.59 & 62324 & 49.17 & - & - \\ 
005 & 49.30 & 84379 & 49.12 & 47.60 & 80957 & 074 & 49.57 & 84145 & 49.31 & 46.49 & 57497 \\ 
006 & 50.31 & 60738 & 49.82 & 39.80 & 21522 & 075 & 49.44 & 62180 & 49.11 & - & - \\ 
007 & 49.67 & 57958 & 49.45 & - & - & 076 & 48.97 & 29996 & - & - & - \\ 
012 & 49.62 & 52361 & 49.00 & 39.26 & 21416 & 077 & 50.55 & 45662 & 49.79 & 39.77 & 20389 \\ 
013 & - & - & - & 29.98 & - & 078 & 49.58 & 57920 & 49.25 & - & - \\ 
014 & 49.75 & 69099 & 49.40 & 39.08 & 21545 & 079 & 49.92 & 41794 & 48.91 & - & - \\ 
015 & 49.43 & 61599 & 49.16 & 38.59 & 21001 & 080 & 50.17 & 46145 & 49.38 & 38.36 & 18931 \\ 
016 & 49.61 & 43162 & 48.84 & - & - & 081 & 49.30 & 48116 & 48.58 & 37.72 & 19339 \\ 
017 & 49.57 & 62210 & 49.22 & - & - & 082 & 49.40 & 77611 & 49.17 & 39.27 & 22783 \\ 
018 & 49.51 & 64794 & 49.20 & - & - & 086 & 49.21 & 69922 & 48.90 & - & - \\ 
019 & 50.01 & 68147 & 49.74 & - & - & 088 & 49.66 & 70897 & 49.43 & 38.82 & 21288 \\ 
021 & 50.19 & 71312 & 49.84 & - & - & 089 & 49.60 & 44234 & 48.90 & - & - \\ 
022 & 48.50 & 22617 & 40.85 & - & - & 091 & 49.25 & 45949 & 48.58 & - & - \\ 
023 & 49.43 & 69922 & 49.12 & - & - & 092 & 50.74 & 46069 & 49.95 & 38.59 & 18501 \\ 
024 & 49.39 & 55730 & 49.10 & 38.17 & 20193 & 093 & 49.65 & 45639 & 48.89 & 38.73 & 20177 \\ 
025 & 49.43 & 69718 & 49.09 & - & - & 094 & 49.64 & 57867 & 49.40 & 38.75 & 20782 \\ 
026 & 49.49 & 57392 & 49.15 & - & - & 095 & 49.78 & 41802 & 48.85 & - & - \\ 
027 & 51.18 & 76456 & 50.87 & 47.26 & 47867 & 096 & 50.09 & 41175 & 48.96 & - & - \\ 
029 & 49.38 & 62180 & 49.05 & - & - & 097 & 50.06 & 45563 & 49.29 & 38.98 & 19936 \\ 
030 & 49.47 & 45563 & 48.69 & 37.58 & 18810 & 098 & 50.47 & 45042 & 49.67 & 38.96 & 19286 \\ 
031 & 49.20 & 61206 & 48.89 & - & - & 099 & 49.67 & 46213 & 48.89 & 38.12 & 19294 \\ 
032 & 49.76 & 45314 & 48.98 & 37.30 & 18063 & 100 & 49.97 & 44385 & 49.17 & 37.72 & 18289 \\ 
033 & 49.16 & 21529 & 41.08 & - & - & 102 & 50.58 & 44687 & 49.76 & 38.54 & 18554 \\ 
035 & 49.48 & 60383 & 49.16 & - & - & 103 & 50.07 & 47527 & 49.34 & - & - \\ 
036 & 49.58 & 65035 & 49.30 & - & - & 104 & 49.93 & 67173 & 49.55 & 40.23 & 23274 \\ 
037 & 49.52 & 70919 & 49.25 & - & - & 105 & 50.22 & 52407 & 49.60 & - & - \\ 
040 & 49.50 & 63102 & 49.12 & - & - & 106 & 50.37 & 57074 & 49.89 & - & - \\ 
041 & 49.46 & 63646 & 49.20 & 38.79 & 21348 & 107 & 49.78 & 33161 & 47.89 & - & - \\ 
042 & 51.88 & 77528 & 51.57 & 48.76 & 55866 & 108 & 50.73 & 58502 & 50.25 & 40.78 & 22436 \\ 
043 & 49.74 & 68087 & 49.39 & - & - & 109 & 50.55 & 58502 & 50.07 & 40.60 & 22436 \\ 
044 & 49.44 & 43479 & 48.56 & 37.36 & 18455 & 110 & 50.04 & 51984 & 49.41 & 40.10 & 22156 \\ 
046 & 49.32 & 61500 & 48.93 & - & - & 111 & 50.01 & 45707 & 49.25 & 39.34 & 20578 \\ 
047 & 49.46 & 68540 & 49.21 & 39.06 & 22005 & 112 & 50.35 & 57278 & 49.85 & 39.74 & 21220 \\ 
048 & 49.25 & 55443 & 48.96 & 38.67 & 21190 & 113 & 49.92 & 53011 & 49.30 & - & - \\ 
049 & 50.21 & 76811 & 49.90 & 47.21 & 57157 & 114 & 50.31 & 68041 & 49.92 & 46.59 & 48366 \\ 
050 & 49.51 & 58615 & 49.03 & - & - & 116 & 50.93 & 64922 & 50.54 & 41.50 & 23758 \\ 
051 & 49.17 & 67981 & 48.91 & 38.69 & 21847 & 117 & 50.28 & 66229 & 49.89 & 40.56 & 23206 \\ 
054 & 49.35 & 36243 & 47.70 & - & - & 118 & 50.48 & 45254 & 49.71 & 37.82 & 17821 \\ 
055 & 48.39 & 21258 & 41.51 & - & - & 119 & 50.39 & 46621 & 49.65 & - & - \\ 
056 & 49.43 & 59756 & 49.10 & - & - & 120 & 48.68 & 26530 & - & - & - \\ 
057 & 49.26 & 59612 & 48.97 & - & - & 122 & 50.07 & 48441 & 49.42 & - & - \\ 
058 & 49.46 & 46183 & 48.72 & 37.37 & 18569 & 124 & 49.33 & 64016 & 48.95 & - & - \\ 
059 & 50.34 & 76184 & 50.02 & - & - & 126 & 50.33 & 72906 & 50.01 & - & - \\ 
060 & 49.66 & 66531 & 49.27 & 39.38 & 22141 & 128 & 49.30 & 64076 & 49.05 & - & - \\ 
062 & 49.29 & 60383 & 48.97 & - & - & 129 & 50.08 & 84145 & 49.82 & 47.00 & 57497 \\ 
063 & 49.46 & 62875 & 49.08 & - & - & 130 & 48.22 & 20638 & 39.86 & - & - \\ 
064 & 49.94 & 72906 & 49.62 & - & - & 131 & 49.54 & 59756 & 49.21 & - & - \\ 
065 & 49.63 & 62210 & 49.28 & - & - & 132 & 49.43 & 54053 & 49.12 & - & - \\ 
066 & 49.65 & 85134 & 49.43 & 46.51 & 56916 & 133 & 48.29 & 21099 & - & - & - \\ 
067 & 49.78 & 47709 & 49.06 & - & - & 134 & 49.37 & 59612 & 49.08 & - & - \\ 
068 & 49.76 & 45699 & 48.99 & 38.66 & 19906 & \\ 

\hline 
\end{tabular}
\tablefoot{
In these cases, where only an insignificant number of ionizing photons
can escape the stellar atmosphere, a hyphen is used for
$T_\mathrm{Zanstra}$ and $\log Q$.  
}
\end{table*}

\clearpage

\section{Comments on individual stars}
\label{sec:comments}

{\em Preliminary remark on mass-loss rates.} In our models, clumping is 
parameterized by means of the ``density contrast'' $D$ (see
Sect.\,\ref{sect:models}). We adopt $D=10$ throughout our analyses.
Other authors used different assumptions for the degree of clumping, or
neglected the wind inhomogeneities. As a consequence of the scaling
invariance described by Eq.\,(\ref{eq:rt}), the impact of the clumping
contrast on the empirical mass-loss rates that are derived from
recombination lines is simply $\dot{M} \propto D^{-1/2}$. Therefore,
when comparing mass-loss rates with the results from other authors in
the following text, we scale their $\dot{M}$ values to our assumptions
of $D=10$.

\paragraph{BAT99\,1} is classified as WN3b in the BAT99 catalog. It
shows no periodic radial velocity variations, but seems to be a runaway
star, according to \citet{Foellmi2003b}. We find no hints for binarity in
the spectra and treat them as single stars. \citet{Bonanos2009} derived a
stellar temperature of $T_* = 85\, \mathrm{kK}$ on the basis of CMFGEN
models from \citet{Smith2002}, which is slightly lower than $T_* =
89\, \mathrm{kK}$, as derived in our analysis. The same value has been obtained
by \citetalias{HK2000} with a previous version of our code. \citet{Willis2004} have
used their FUSE spectra to derive a terminal velocity of $v_\infty =
2745\,\mathrm{km\,s^{-1}}$ from the black edge of the P Cygni profiles
present in the far UV (FUV), whereas \citet{Niedzielski2004} estimated
terminal velocities of $v_\infty = 1265 - 2506\,\mathrm{km\,s^{-1}}$
from the P Cygni lines in the IUE-range with the same method. We prefer
$v_\infty = 1600\,\mathrm{km\,s^{-1}}$ from the
width of the optical emission lines in accordance with \citetalias{HK2000}. A
terminal velocity of $v_\infty = 2754\,\mathrm{km\,s^{-1}}$ would result
in emission lines which are considerably broader than observed (see Sect.\,\ref{subsec:spefit}
for details).

\paragraph{BAT99\,2} is one of the two WN2 stars (the other one is 
BAT99 49) in the LMC, and clearly among the hottest stars of the sample. 
Actually, BAT99\,2 is one of only two WR stars in the LMC able to
ionize a \ion{He}{ii}-region \citep{Naze2003b}. These authors conclude
from the nebular \ion{He}{ii}-flux that the exciting star 
delivers $4 \cdot 10^{47}$\,\ion{He}{ii} ionizing photons per second. 
This agrees with our final model, which 
produces $3.3 \cdot 10^{47}$\,\ion{He}{ii} ionizing photons per second.

A model with a stellar temperature of $90-100$\,kK \citep{Naze2003}
cannot reproduce the observed spectra. The synthetic spectra below $T_*
= 110\,\mathrm{kK}$ show \ion{C}{iv}\,$\lambda\, 1548$ and
\ion{C}{iv}\,$\lambda\, 5801$ lines, which are not observed. All 
appropriate models with stellar temperatures above $T_* =
110\,\mathrm{kK}$ fall into the regime of parameter degeneracy of the model
grid (see Sect.\,\ref{subsec:modelgrid}). In this case, the determination
of the temperature was based on a slightly better fit for the
\ion{N}{v}\,$\lambda\, 1242$ and \ion{N}{v}\,$\lambda\, 4603$ lines at
$T_* = 141\,\mathrm{kK}$. Between fits with temperatures of 130\,kK and
160\,kK, the slope of the optical/UV continuum (and thus the inferred
reddening parameter) hardly changes, but the luminosity
increases from $\log\,(L/L_\odot) = 5.35$ to $5.58$ due to the bolometric
correction.

\citet{Foellmi2003b} reported the detection of hydrogen emission
in the spectrum of BAT99\,2. We cannot exclude a hydrogen mass-fraction of
$X_{\element{H}} \sim 0.1$, but the observation is perfectly
consistent with zero hydrogen. The presence of hydrogen would be 
unexpected for a WN star of such high effective temperature and, thus, 
advanced evolution stage. 

\citet{Foellmi2003b} find that the radial velocity of BAT99\,2 
differs from the mean $v_\mathrm{rad}$ of their WN sample by about
$-120\,\mathrm{km\,s}^{-1}$ and suggest that this star might be a runaway
object. 

\paragraph{BAT99\,3} is listed as spectral type WN4b in the BAT99
catalog. For the first time, stellar and wind parameters for
this object are presented here. 

\paragraph{BAT99\,5} is the second of the two WN2b stars
\citep{Foellmi2003b} in the LMC. BAT99\,2 and BAT99\,5 exhibit very similar
spectra that can be reproduced by the same grid model. The temperature
of BAT99\,5 was previously determined to be only $71\,\mathrm{kK}$ by
HK2000, but with today's line-blanketed models much higher temperatures
are adequate, as discussed above for BAT99\,2. However, we note that this star is 
located within the regime of parameter degeneracy 
(see Sect.\,\ref{subsec:modelgrid}).
\citet{Massey2000} suggest an initial mass of $M_{\mathrm{init}} >
40\,M_\odot$, while the Geneva tracks  in Fig.\,\ref{fig:hrd_geneva}
indicate slightly less than $30\,M_\odot$.

BAT99\,5 is suspected to have an OB-companion by \citet{Smith1996},
in reference to absorption features visible in the spectrum from
\citet{Torres-Dodgen1988}, but the authors mention that these features
could also be artifacts from the subtraction of nebular lines. 
According to \citet{Foellmi2003b}, the spectrum shows no radial velocity
variations. Like BAT99\,2, BAT99\,5 was not detected in
X-rays by \citet{Guerrero2008II}. As we also find no indications of a
companion in the spectrum, we consider this star to be single. If the
spectroscopic twin of BAT99\,5, namely BAT99\,2, is indeed a runaway
star as suggested by \citet{Foellmi2003b}, the single star evolution
of BAT99\,5 and the binary evolution of BAT99\,2 have led to almost
identical products.

\paragraph{BAT99\,6} is one of the WN binary systems (period of 2\,d) listed in \citetalias{BAT99}, but has been demoted since the publication of this catalog. \citet{Niemela2001} reclassified it as O3\,f*+O on the basis of its optical spectrum, suggesting that the system contains four stars (two close pairs). On the other hand, \citet{Koenigsberger2003} conclude that the system does not comprise more than two luminous stars. Unfortunately, we do not have optical spectra of this star, so that the stellar parameters are derived from the UV spectrum and the photometry alone. We achieved a reasonable fit, although this binary is fitted as a single star. According to \citet{Niemela2001} the total mass of the system is probably over $80\,M_\odot$. 

\paragraph{BAT99\,7,} classified as WN4b \citepalias{BAT99}, shows strong emission lines with a round line shape. These line shapes can only be reproduced assuming a high rotational velocity of $v_{\mathrm{rot}} = 2200\,\mathrm{km\,s^{-1}}$. Thus, this star is a prototype for the so-called ``round line'' stars, which are characterized by strong and broad emission lines with round line profiles. The correlation of these line shapes with the stellar rotation has recently been investigated by \citet{Shenar2014}. These authors confirm that rotation can account for the spectral characteristics of BAT99\,7, but only in connection with a strong magnetic fields that force the wind to co-rotate.

We derive a stellar temperature of  $T_* = 158\,\mathrm{kK}$, which is the hottest of all WN4 stars. However, we note that this stars falls into the regime of parameter degeneracy (see Sect. \ref{subsec:modelgrid}). The derived stellar temperature is conspicuously higher than values obtained by \citet[][$T_* = 90\,\mathrm{kK}$]{Koesterke1991}, using pure helium models. \citetalias{HK2000} obtained a temperature of $T_* = 100\,\mathrm{kK}$ with unblanketed model atmospheres. 

\paragraph{BAT99\,12} is a transition type O2\,If*/WN5 star, according to \citet{Crowther2011}. \citet{Schnurr2008} argued that the star is most likely a runaway, as already suggested by \citet{Massey2005}. Several spectra are at hand for the analysis of this star. In the UV range an HST spectrum and multiple observations with the IUE satellite are available, although only two IUE short-wavelength spectra are in accordance with the HST observation. All IUE long-wavelength spectra exhibit a substantial offset to the rest of the observed SED. Therefore, we have ignored these IUE data. In the optical spectral range, the AAT spectrum and one spectrum observed by \citet{Foellmi2003b} complement each other in wavelength coverage. All spectra can be fitted with the same model, which gives us confidence in the derived stellar parameters.

We derived a stellar temperature of $T_* = 50\, \mathrm{kK}$, 
confirming the value obtained by \citet{Doran2011} and \citet{Doran2013}.
A lower temperature limit of $T_* > 42\, \mathrm{kK}$ was derived by \citet{Massey2005} with the FASTWIND model atmosphere code \citep{Puls2005}. However, at these cooler temperatures neither the \ion{N}{iv}\,$\lambda\, 4060$ nor the \ion{N}{v}\,$\lambda\lambda\,4604, 4620$ lines can be satisfactorily reproduced. 
In comparison to the study carried out by \citet{Doran2013}, we derived the same luminosity, while the mass-loss rate is 0.1\,dex lower. In contrast, \citet{Massey2005} derived a mass-loss rate that is 0.23\,dex higher. We achieve the best fit with synthetic spectra for a hydrogen mass-fraction of $X_{\element{H}} = 0.5$, which is 0.1\,dex lower than previously derived by \citet{Doran2013}.

\citet{Schnurr2008} reported radial velocity variations of this star with a period of 3.2\,d. For the companion, no spectral features are detected. The SED is well reproduced by a single-star model, thus we expect that the companion does not contribute much to the bolometric luminosity of the binary system. 

\paragraph{BAT99\,13} is the only WN10 star \citepalias{BAT99} in the LMC. It has been analyzed before by \citet{Crowther1995} and \citet{Pasquali1997}. The former derived a stellar temperature of $T_* = 29.7\,\mathrm{kK}$ with unblanketed atmosphere models. The analysis by \citet{Pasquali1997} with line blanketed models obtained a higher temperature of $T_* = 33\,\mathrm{kK}$. Our best fit is obtained with a model of $T_* = 28\,\mathrm{kK}$. At a temperature of $T_* = 32\,\mathrm{kK}$ (one grid step higher) the fit is also reasonable, with the exception of a considerably over-predicted \ion{He}{ii}\,$\lambda\, 4686$ line. Thus, we preferred the model with the lower temperature ($T_* = 28\,\mathrm{kK}$). Since the emission line strength of the \ion{He}{ii}\,$\lambda\, 4686$ line is slightly too low at these temperatures, the ``real'' temperature is probably marginally higher. But this has only a minor impact on the other stellar parameters. The mass-loss rate derived in our analysis is more than a factor of three below the previous values obtained by \citet{Crowther1995} and \citet{Pasquali1997}. 

\begin{figure*}
  \centering
  \includegraphics[width=0.46\hsize]{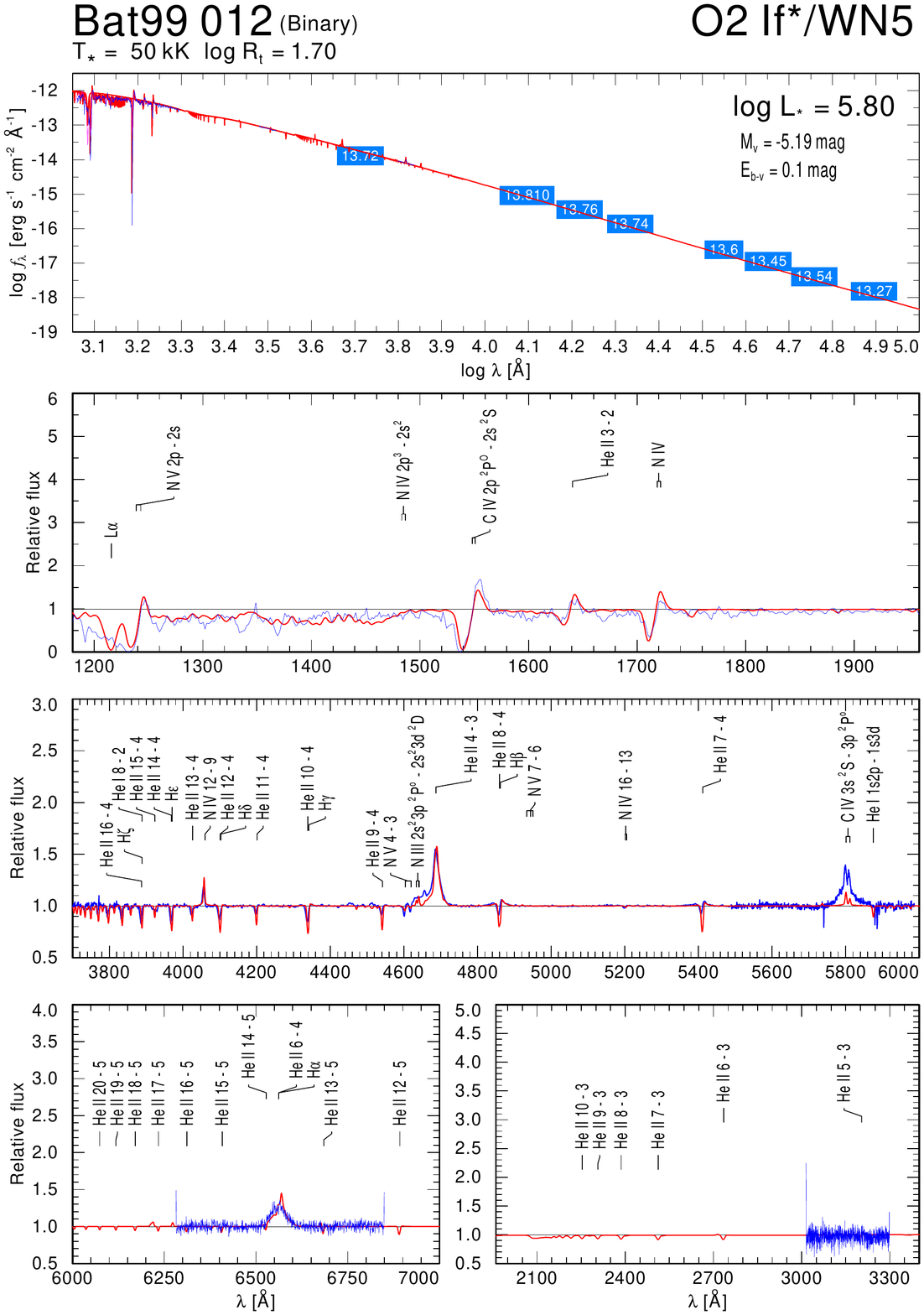}
  \qquad
  \includegraphics[width=0.46\hsize]{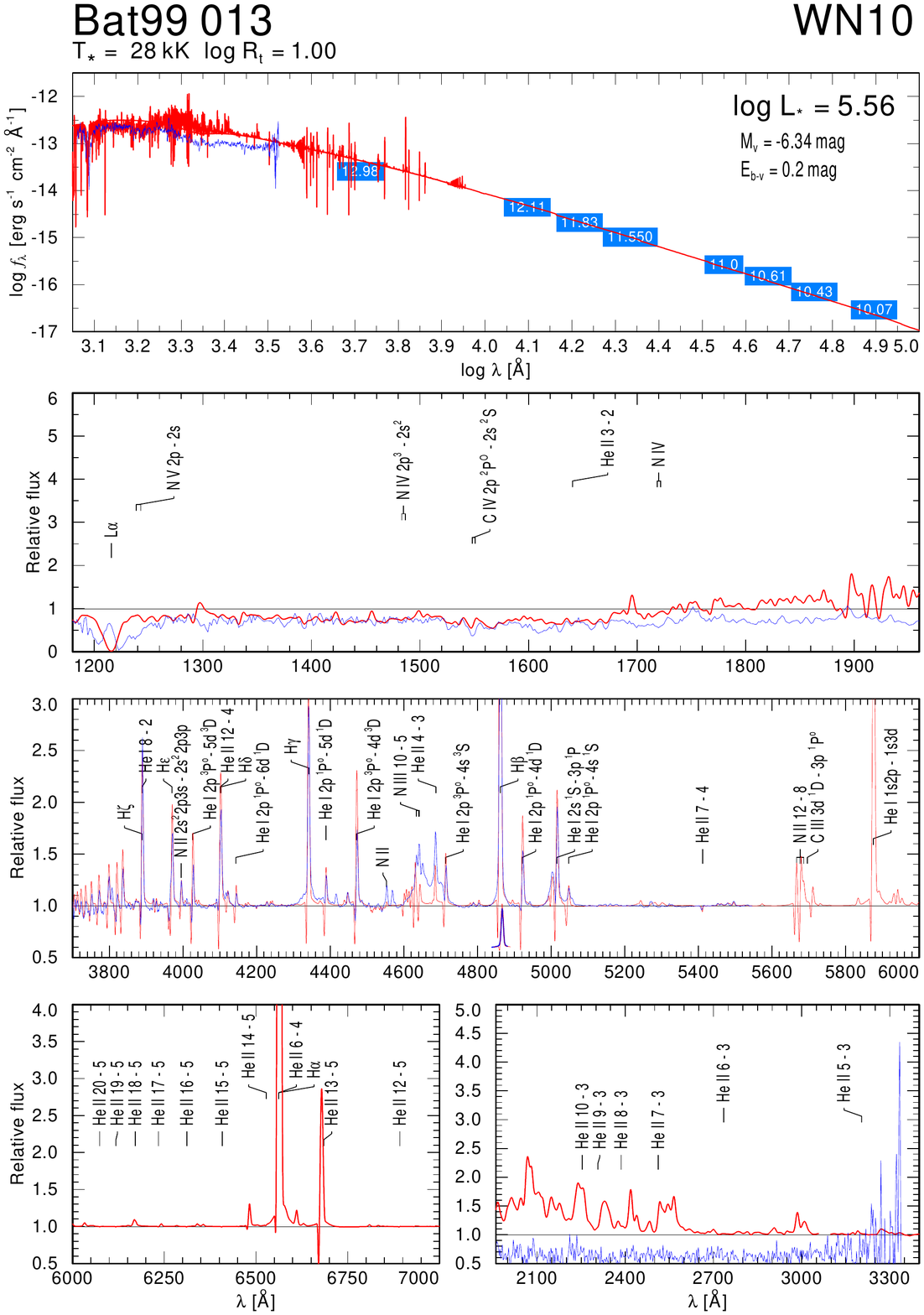}
  \vspace{-0.4cm}
  \caption{Spectral fit for BAT99\,012 and BAT99\,013}
  \label{fig:bat012}
  \label{fig:bat013}
\end{figure*}

\citet{Bonanos2009} compared the observed SED with the continua of a 45\,kk WN model calculated by \citet{Smith2002} with CMFGEN. From this comparison they found an infrared excess, which cannot be confirmed by our analysis (see Fig.\,\ref{fig:bat013}). These different results arise from the high temperature assumed by \citet{Bonanos2009} for this WN10 star. Models calculated with this temperature cannot consistently reproduce the observed spectra. 

\paragraph{BAT99\,14} is listed as WN+OB? binary candidate without a period in the BAT99 catalog. \citet{Foellmi2003b} reclassified it as WN4o(+OB), but they did not find significant periodical variations in their radial velocity data, concluding that this object is probably not a short-period binary. However, they find absorption lines superimposed on the emission lines, which they attribute to a nearby visual companion. Since our analysis is based on the same spectra, we treat this star as a binary suspect until more appropriate data are available. Stellar and wind parameters are derived in this work for the first time.

\paragraph{BAT99\,15} is classified as WN4b \citepalias{BAT99}. It has already been analyzed by \citet{Koesterke1991} and \citetalias{HK2000}. We obtained approximately the same stellar temperature as \citetalias{HK2000}, but a factor of two lower mass-loss rate. A higher mass-loss rate, however, results in emission lines which are substantially stronger than observed. Furthermore, we derived a slightly higher color excess of $E_{b-v}=0.08\,\mathrm{mag}$, which gives rise to the higher luminosity derived in our analysis. 

\paragraph{BAT99\,16} was classified as WN7h by \citet{Schnurr2008}. We achieve the best fit at a stellar temperature of $T_* = 50\,\mathrm{kK}$, whereas $T_* = 33\,\mathrm{kK}$, $T_* = 34.8\,\mathrm{kK}$ and $T_* = 35.5\,\mathrm{kK}$ have been derived by \citet{Koesterke1991}, \citet{Crowther1997}, and \citetalias{HK2000}, respectively. Such a low temperature, however, would completely spoil the fit of the \element{He} and \element{N} lines. For example, the line ratio of the \ion{He}{i}\,$\lambda\, 5877$ to \ion{He}{ii}\,$\lambda\, 5412$ is much higher than observed in the model with a temperature of $35\,\mathrm{kK}$. We attribute these differences to the unblanketed model atmospheres used by the former authors. 

We found a mass-loss rate of $\log\,(\dot M / (\mathrm{ M_\odot}/\mathrm{yr}))=-4.64$. The same value was derived by \citet{Crowther1997}, while \citet{Koesterke1991} obtained a mass-loss rate almost a factor of two lower. \citetalias{HK2000}, on the other hand, derived a mass-loss rate nearly a factor of two higher. We estimate a hydrogen mass-fraction of $X_{\element{H}} = 0.3$ from the best fitting models, slightly higher compared to the value derived by \citetalias{HK2000}.

\paragraph{BAT99\,17} is listed as WN4o in the BAT99 catalog. \citet{Bonanos2009} have analyzed this star, using CMFGEN models by \citet{Crowther2006}. They derived an effective temperature of $T_{\mathrm{eff}} =  52\,\mathrm{kK}$, a luminosity of $\log\,(L/L_\odot) = 5.4$ and a mass-loss rate of $\log\,(\dot M / (\mathrm{ M_\odot}/\mathrm{yr}))=-4.85$. In our analysis, the best overall fit is achieved with a model corresponding to an effective temperature of $T_{\mathrm{eff}} =  65\,\mathrm{kK}$. We note that a slightly better fit of the \ion{He}{ii}\,$\lambda\, 5412$ and the \ion{He}{i}\,$\lambda\, 5877$ lines can be achieved at an effective temperature of $T_{\mathrm{eff}} =  55\,\mathrm{kK}$, although the fit quality of all nitrogen lines and the \ion{He}{ii}\,$\lambda\, 1641, 4201, 4339, 4542$ is reduced compared to the model with $T_{\mathrm{eff}} =  65\,\mathrm{kK}$. The luminosity derived in this work is factor of two higher than the value obtained by \citet{Bonanos2009}, while the mass-loss rate and the terminal velocity are nearly the same. The luminosity increase in comparison to the former study by \citet{Bonanos2009} originates from the higher temperature and thus higher bolometric correction derived in our analysis.

\paragraph{BAT99\,18} is a WN3(h) star \citep{Foellmi2003b}, which had never been analyzed before by means of model atmospheres. We confirm the presence of hydrogen ($X_{\element{H}} = 0.2$), i.e.,\ the corresponding classification.

\paragraph{BAT99\,19} was classified as WN4b+OB? in the BAT99 catalog. \citet{Foellmi2003b} reported a period of 17.99\,d and specified the companion to be an O5: star. In the spectrum from \citet{Foellmi2003b}, small absorption lines are superimposed on the emission lines. Due to the weakness of these lines we expect that the contribution of the companion to the bolometric luminosity is only minor. Actually, we achieved a reasonable fit with our single-star model although a high rotational velocity ($v_\text{rot} \cdot \sin i = 2000\,\mathrm{km\,s^{-1}}$) is necessary to reproduce the round emission lines, as has already been remarked by \citet{Breysacher1981}. Thus, this star belongs to the category of the so-called ``round line'' stars. The correlation of these line shapes with the stellar rotation has recently been investigated by \citet{Shenar2014}.

\citetalias{HK2000} have derived a stellar temperature of $T_{*} = 70.8\,\mathrm{kK}$ one grid step lower compared to our new results, which engender a slightly better fit of the UV and optical line spectra at $T_{*} = 79\,\mathrm{kK}$. We attribute these differences in the models to the line blanketing not incorporated in the models by \citetalias{HK2000}. Both results are considerably above the 50\,kK obtained by \citet{Koesterke1991} with pure helium models. The mass-loss rate presented in this paper is marginally lower compared to \citetalias{HK2000}, if we take the higher terminal velocity $v_\infty = 2500\,\mathrm{km\,s^{-1}}$ derived by these authors into account. The SED fit results in nearly the same luminosity and color excess as obtained by \citetalias{HK2000}. 

\paragraph{BAT99\,21} is listed as WN4+OB binary candidate without a period in the BAT99 catalog. \citet{Foellmi2003b} reclassified this object as WN4o(+OB). According to these authors, the radial velocity variations are only marginal, concluding that this object is probably not a short-period binary. They further mentioned that a visual companion situated $2\arcsec$ away contributes to the observed flux and causes the absorption lines in the spectrum. We analyze the object as a single WN star, although a substantial flux contribution from the companion is expected due to the sizable absorption lines of the companion. More appropriate data is needed to ensure accurate model estimates and to verify the binary status. Until this data is available, we treat this star as a binary suspect. 

\paragraph{BAT99\,22} is an LBV candidate according to \citet{Humphreys1994}, \citet{Crowther1995b}, and \citet{Pasquali1997}, which is listed as WN9h star by \citetalias{BAT99}. So far, similar stellar parameters have been reported by \citet{Schmutz1991}, \citet{Crowther1995b}, and \citet{Pasquali1997}. These authors derived stellar temperatures in the range from $T_{*} = 28.5\,\mathrm{kK}\,...\,35.2\,\mathrm{kK}$ with unblanketed and blanketed model atmosphere codes, respectively. We obtain a stellar temperature of $T_{*} = 32\,\mathrm{kK}$, matching the hitherto known temperature range. However, we can exclude the higher as well as the lower temperature, since the \ion{He}{ii}\,$\lambda\, 4686$ line would be considerably overpredicted at 35\,kK and underpredicted at 28\,kK, respectively. 

This star is one of three LMC WN stars that has been detected at $24\,\mu\mathrm{m}$ with the IRAC instrument aboard the Spitzer space telescope \citep{Bonanos2009}. It is known to show a huge infrared excess \citep{Glass1984,Stahl1984}, which is also visible in our SED fit. According to \citet{Allen1976}, \citet{Cowley1978}, and \citet{Stahl1984}, a M2 supergiant contributes to the near-infrared flux, and thus probably causes the infrared excess. \citet{Schmutz1991} concluded that the M supergiant is not physically bound to BAT99\,22, rather incidentally located along the same line of sight. Observations by \citet{Heydari-Malayeri1997} with ESO NTT SUSI show that the M supergiant is closer than $0.12 \arcsec$, which points to a binary system according to these authors. However, no significant variations have been found in the recent radial-velocity study by \citet{Schnurr2008}, which militates against a short-period binary. 

A comparison of the spectrum obtained by \citet{Schnurr2008} with that shown in \citet{Cowley1978} indicates that at least one faint TiO band at 5167\,{\AA} is visible in the spectrum from \citet{Schnurr2008}. Therefore, our results for BAT99\,22 are also slightly effected by the late-type supergiant, since our analysis is partially based on this spectrum. Due to the weakness of this feature, however, we consider the uncertainty of the WN parameters introduced by the contribution of the M supergiant to the flux in the optical spectral range as small. 

\citet{Vink2007} found intrinsic line depolarization for the \ion{He}{ii}\,$\lambda\, 6560$ line, suggesting an asymmetry in the wind, which is probably either caused by an binary companion or rapid rotation. \citet{Weis2003} found evidence for a nebula associate to BAT99\,22, which is not spatially resolved by the available observations. A circumstellar shell has already been proposed by \citet{Stahl1984}. These authors concluded that the excess in the L band is too large to be explained by a late supergiant only. 

The UV and optical spectra (flux-calibrated) are consistent with a luminosity of $\log\,(L/L_\odot) = 5.75$. Within the uncertainties, this is equal to the findings of \citet{Schmutz1991} and \citet{Crowther1995b}, whereas $\log\,(L/L_\odot) = 5.9$ was obtained by \citet{Pasquali1997}. The best fitting model requires a mass-loss rate of $\log\,(\dot M / (\mathrm{ M_\odot}/\mathrm{yr}))=-4.85$, which is lower but comparable with the previous results \citep{Schmutz1991,Crowther1995b,Pasquali1997}.

\paragraph{BAT99\,23} was classified as WN3(h) by \citet{Foellmi2003b} due to the hydrogen emission detected by these authors. Our best fit is achieved with a hydrogen-free grid-model, suggesting that the hydrogen mass fraction in the atmosphere of this star is below $X_{\element{H}} = 0.1$. As far as we know, no analysis based on stellar atmosphere models has been published for this object yet.

\paragraph{BAT99\,24} is listed as WN4b in the BAT99 catalog. For the first time, we present stellar and wind parameters for this star derived with modern stellar atmospheres. With a stellar temperature of $T_* = 100\,\mathrm{kK}$ and a terminal velocity of $v_\infty = 2400\,\mathrm{km\,s^{-1}}$, it belongs to the hottest stars in our sample with one of the fastest winds. However, we note that this star is located within the regime of parameter degeneracy (see Sect.\,\ref{subsec:modelgrid} for details).

\paragraph{BAT99\,25} was classified as WN4ha by \citet{Foellmi2003b}, rejecting the previous WN3 classification. Since we could not find parameters of this object in the literature for comparison, this object had probably never been analyzed before.

\paragraph{BAT99\,26} is a WN4b star \citepalias{BAT99} with stellar and wind parameters typical for early WN stars. The only exception is the mass-loss rate, which is lower compared to the average of the other WN4b stars in our sample. The parameters derived in our analysis are the first made public for this star. 

\paragraph{BAT99\,27} has a known B-type companion, although \citet{Foellmi2003b} concluded that this star is probably not a short-periodic binary because no radial velocity variations were detected. Since absorption lines of the companion are also visible in the spectra obtained by these authors, they have slightly reclassified the system to WN5b(+B1\,Ia).
High resolution observations with the Wide Field Camera 3 (WFC3) aboard the HST by P. Massey (HST Proposal 12940) could not resolve this object in multiple components. Whether the B-type star is physically bound to the WR star or only located in the line of sight is still unclear. 

The early B-type companion significantly contributes to the flux of the whole system. Therefore, our analysis of BAT99\,27 as a single WN star is considerably affected by the companion. This is the reason that no conclusive fit could be established in this work. The disentanglement of the spectra will be a subject of a forthcoming paper. However, the emission lines have a broad and round line shape, suggesting that the WN component belongs to the so-called ``round line'' stars. To account for the round line shape, the synthetic spectra is convolved with a rotation profile, corresponding to a rotational velocity of $v_\mathrm{rot} = 1000\, \mathrm{km\,s^{-1}}$. The correlation of the line shapes with the stellar rotation has recently been investigated by \citet{Shenar2014}.

\paragraph{BAT99\,29} was revealed as a binary with a period of 2.2\,d by \citet{Foellmi2003b} on the basis of their radial velocity studies. These authors classified this object as WN4b+OB. They note a reduced line intensity of the hydrogen lines due to the presence of a companion. However, a reasonable fit of the Balmer lines can also be established with a single-star model (although the H$\alpha$ line is slightly overpredicted by the model), suggesting that the flux contribution of the OB companion is rather small in this binary system. 

\paragraph{BAT99\,30} is listed as WN6h in the BAT99 catalog. Stellar temperatures in the range $T_* = 33\,...\,39.8\,\mathrm{kK}$ have been derived by \citet{Koesterke1991}, \citet{Crowther1997}, and \citetalias{HK2000} with unblanketed stellar atmosphere models. We achieved the best fit with a stellar temperature of $T_* = 47\,\mathrm{kK}$. At this temperature, our models reproduce the observed \ion{He}{i}/\ion{He}{ii} ratio as well as the \ion{N}{iii}/\ion{N}{iv} ratio (with a trend to slightly higher temperatures), whereas a clear mismatch is obtained with the low temperatures derived in the previous studies. This discrepancy probably originates from the line-blanketing effect included in our models. Similar values for the luminosity and the mass-loss rate have been obtained by \citet{Koesterke1991}, \citet{Crowther1997}, and \citetalias{HK2000}. We confirm the previous estimates of the mass-loss rate, but obtain a higher luminosity of $\log\,(L/L_\odot) = 5.65$, which can be partly attributed to the hotter stellar temperature derived from our line fit. In agreement with the classification, we find a hydrogen mass-fraction of $X_{\element{H}} = 0.3$, which is 0.1\,dex lower than previously derived by \citetalias{HK2000}.

\paragraph{BAT99\,31} was classified as WN4b binary candidate by \citet{Foellmi2003b}. However, these authors could not derive a period from their radial velocity measurements. As we find no indications of a companion in its spectrum, we consider the contribution to the bolometric luminosity of this possible companion as negligible and analyze this object as a single WN star.

\paragraph{BAT99\,32} is long known as a binary system with a short period of $P = 1.91\,\mathrm{d}$ \citep{Moffat1989,Schnurr2008}. It is listed with a WN6(h) spectral type in the BAT99 catalog. We note that emission lines like \ion{He}{ii}\,$\lambda\, 4686$ and \ion{He}{ii}\,$\lambda\, 5412$ exhibit an asymmetric line shape in the spectrum obtained by \citet{Torres-Dodgen1988}, whereas none of these peculiarities can be detected in the spectrum by \citet{Schnurr2008}, which might be an effect of the lower spectral resolution. Since we cannot with certainty estimate the flux contribution of the companion, we consider the derived physical parameters of this object as uncertain. 

Fairly similar stellar parameters were obtained by \citet{Koesterke1991}, \citet{Crowther1997}, and \citetalias{HK2000}, although the derived stellar temperatures exhibit some scatter. Our analysis, on the other hand results in a stellar temperature of $T_* = 47\,\mathrm{kK}$, i.e.,\ 7\,kK higher than the highest temperature obtained in previous studies. Similar to BAT99\,30, the high temperature derived in our analysis probably originates from the line-blanketing effect included in our code. Since the earlier analyses by \citet{Koesterke1991} and \citet{Crowther1997} are based on models that do not account for wind inhomogeneities, the mass-loss rate derived by these authors is certainly overestimated. However, the scaling of these values according to the clumping factor $D = 10$ assumed in this work results in nearly the same mass-loss rate. The same is true for the results obtained by \citetalias{HK2000}, who assumed a lower clumping factor of $D = 4$. Compared to \citet{Koesterke1991}, \citet{Crowther1997}, and \citetalias{HK2000}, we have derived a higher luminosity mainly due to the different stellar temperature obtained in our analysis.

\paragraph{BAT99\,33} belongs to the category of Ofpe/WN9 stars (also designated as ``cool slash-stars''), that were incorporated in the WN-subclass system \citep{Crowther1995b,Crowther1997} as WN9--11 stars. According to \citet{Crowther1997}, this star is exceptional, since its spectral peculiarities prevent a closer classification. A detailed discussion of the spectral morphology can be found in \citet{Crowther1997} and \citet{Pasquali1997}. \citet{Humphreys1994} list this star as a LBV candidate. No clear evidence for a circumstellar nebular was found by \citet{Weis2003}. They attributed the nebulosity previously detected by \citet{Nota1996} to a background \ion{H}{ii} region due to the low densities derived by these authors. In contrast \citet{Gvaramadze2010} found a bow-shock structure to the east of BAT99\,33, using archival $24\,\mu \mathrm{m}$ data obtained with the Multiband Imaging Photometer (MIPS) aboard the {\em Spitzer Space Telescope}. Significant intrinsic line polarization was discovered by \citet{Vink2007}, suggesting the presence of an asymmetric stellar wind, which could explain the peculiarities visible in the spectrum of this star. 

\citet{Pasquali1997} inferred a stellar temperature of $T_* = 35\,\mathrm{kK}$, which is 7\,kK higher than the temperature obtained in our analysis. Our 35\,kK model completely overestimates the strength of the \ion{He}{ii}\,$\lambda\, 4686, 5412$ lines in comparison to the \ion{He}{i} lines, which unambiguously points to lower stellar temperatures. Moreover, the former authors speculated that the \ion{He}{i} and \ion{He}{ii} lines originate from different wind components because the helium line widths do not follow their expectations. Thus, they noted that their stellar parameters may not accurately represent the physical conditions of this star. With the exception of the asymmetric \ion{He}{ii}\,$\lambda\, 4686$ line, the line width of all helium lines can be consistently reproduced in our analysis.

\paragraph{BAT99\,35} was classified as WN3(h) by \citet{Foellmi2003b}. Our analysis results in similar stellar parameters compared to the previous analysis by \citetalias{HK2000}. The only exception is the inferred stellar temperature, which is reduced by one grid step in our present study, leading to a slightly better reproduction of the \ion{He}{i}/\ion{He}{ii} ratio. The best fit is obtained with a hydrogen mass-fraction of $X_{\element{H}} = 0.1$, although a hydrogen-free atmosphere cannot be excluded with certainty.

\paragraph{BAT99\,36} is one of two WNE/WCE transition type stars in the LMC that exhibit a clear carbon enhancement compared to the rest of the LMC WN sample. As noted by \citet{Foellmi2003b}, the top of the Balmer lines are attenuated, which, according to these authors, could be due to a faint absorption-line companion, as already proposed by \citet{Crowther1995}. However, \citet{Foellmi2003b} note that their radial velocities measurements are consistent with a single star. We regard this object as a binary suspect until more appropriate data are available. Considering the small effect on the emission lines of the WN star, we expect that the potential companion does not contribute much to the bolometric luminosity. 

We obtained quite the same stellar parameters as \citetalias{HK2000}. Analyses prior to \citetalias{HK2000} suggested lower values for the stellar temperature and luminosity but nearly the same mass-loss rate \citep{Koesterke1991,Crowther1995}. With a carbon abundance of $X_{\element{C}} = 0.003$ derived in our analysis, we can confirm the previous results obtained by \citet{Crowther1995}.

\paragraph{BAT99\,37} has the spectral type WN3o \citep{Foellmi2003b}. Unfortunately, we do not have an interpretable UV spectrum of this star. The optical spectrum from \citet{Torres-Dodgen1988} is very noisy, so that reddening and luminosity cannot be determined precisely. Fitting this spectrum, the color excess is $E_{b-v}=0.7\,\mathrm{mag}$, while a color excess of $E_{b-v}=0.5\,\mathrm{mag}$ is necessary to obtain a good fit of the photometry measured by \citet{Crowther2006b}; the luminosity does not differ between the fits. 

\paragraph{BAT99\,40} is listed as WN5o+O binary candidate in the BAT99 catalog. However, \citet{Foellmi2003b} did not find a radial velocity period and reclassified this star to WN4(h)a. Furthermore, they attributed the absorption lines visible in their spectra to be intrinsic to the wind of the WN star. These authors, on the other hand, have detected X-ray emission in the archival {\em ROSAT} data, whereas \citet{Guerrero2008II} list this object as undetected by {\em ROSAT} observations. Despite these contradicting results, we treat this star as a binary suspect until its binary status is clarified. The stellar parameters given in this work are the first derived for this object. 

\paragraph{BAT99\,41} is a WN4b star \citepalias{BAT99} that has never been analyzed before. We note that this stars falls into the regime of parameter degeneracy (cf.\ Sect. \ref{subsec:modelgrid}).

\paragraph{BAT99\,42} is the brightest source in our sample, with an extremely high stellar luminosity of $\log\,(L/L_\odot) = 8.0$. This is $1.06\, \mathrm{dex}$ higher than the stellar luminosity derived by \citet{Crowther2010} for BAT99\,108 in the core of R136, and $1.2\, \mathrm{dex}$ higher than the value obtained by \citet{Sana2013} for the binary BAT99\,118 (R144). 

BAT99\,42 is a visual binary classified as WN5b(h) \citep{Foellmi2003b} with a long known B-type supergiant (B3I) companion \citep[][and references therein]{Smith1996}. For the whole system, the detection of X-ray emission was reported by \citet{Guerrero2008II}. \citet{Seggewiss1991} found a spectroscopic period of $P = 30.18\,\mathrm{d}$, though this value is highly uncertain, according to the authors. \citet{Foellmi2003b}, on the other hand, concluded from their radial velocity studies that the WN component is probably not a spectroscopic binary, although some scatter is present in their radial velocity data. 

\begin{figure}
\centering
\includegraphics[width=\hsize]{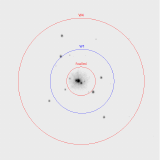}
\caption{An HST-WFC3 F225W image of the region within about $10 \arcsec$ distance of BAT99\,42. The image was requested from the HST archive. The cycles refer to the apertures of the WISE photometer ($6.1 \arcsec$ for the W1 and $12 \arcsec$ for the W4 band, respectively) and the maximum slit width used by \citet{Foellmi2003b}.}
\label{fig:bat42_hst}
\end{figure}

This object is associated with the LH\,58 cluster, which is located about $1.1\degr$ to the northwest of 30 \,Doradus. New UV observations of this cluster were obtained by P. Massey with the WFC3 aboard the HST. These high resolution images dissolve BAT99\,42 in three major components (see Fig.\,\ref{fig:bat42_hst}) that lie within the slit width used by \citet{Foellmi2003b} and \citet{Torres-Dodgen1988} (see Fig.\,\ref{fig:bat42_hst}). 

Figure \ref{fig:bat42_hst} even shows a small cluster around BAT99\,42. This whole cluster is completely covered by the large aperture of the IUE satellite, which was used for the UV observations analyzed in this work. Thus, all available spectra represent at least the three major objects in this small cluster. A disentanglement of the spectra is beyond the scope of this paper and will be the subject of our forthcoming work. The photometry used to construct the observed SED is affected by the same problem since the core of this cluster cannot be resolved by most of the available instruments.  

Apart from the lower signal-to-noise ratio (S/N), the flux-calibrated spectrum obtained by \citet{Torres-Dodgen1988} is almost identical to the spectrum observed by \citet{Foellmi2003b}. The weak round-shaped emission lines are obviously diluted by the contribution of the non-WN components to the overall flux. The round shape of the emission lines requires a convolution of the model spectrum with a rotation profile corresponding to a rotational velocity of $v_\text{rot} \cdot \sin i = 2300\,\mathrm{km\,s^{-1}}$. Thus, the WN component might belong to the so-called ``round line'' stars. 

The narrow absorption lines from the non-WR components of this system are clearly visible in the high S/N spectrum from \citet{Foellmi2003b}. These lines are relatively weak, which can be attributed to dilution effects as well. Therefore, we estimate the contribution of the WN component to the bolometric luminosity to be considerable although the non-WN components probably contribute most to the flux in the optical spectral range. Thus, this cluster probably hosts one of the most luminous WN stars in the LMC. However, without additional data it is hard to constrain the real luminosity of the WR component. 

\paragraph{BAT99\,43} was listed as WN4+OB binary candidate in the BAT99 catalog. \citet{Foellmi2003b} found a 2.8\,d period for this double-line spectroscopic binary (SB2) \citep{Moffat1989}. We expect that the companion does not contribute much to the bolometric luminosity of this binary system \cite[see also][]{Foellmi2003b}, since we do not see any unambiguous features of the companion in our spectra. This object has been previously analyzed by \citet{Koesterke1991} and \citetalias{HK2000}. These authors have derived a stellar temperature of $T_* = 79.4\,\mathrm{kK}$, one grid step above our new results. A higher stellar temperature in our analysis, however, would result in a complete mismatch of the observed \ion{N}{iv}\,/\,\ion{N}{v} ratio. To the contrary, the line strength of the \ion{He}{i}\,$\lambda\, 5877$ line points to an even lower stellar temperature. Compared to \citetalias{HK2000}, the luminosity obtained here is almost a factor of two lower, while the mass-loss rate is approximately the same.

\paragraph{BAT99\,44} was classified as WN8ha by \citet{Schnurr2008}. \citet{Bonanos2009} discovered an infrared excess by comparing the observed SED with line-blanketed atmosphere models \citep{Smith2002}, which exhibit a stellar temperature of $T_* = 45\,\mathrm{kK}$. This temperature is considerably higher than obtained by \citet{Crowther1995}, using unblanketed model atmospheres. Our analysis agrees with the estimates by \citet{Bonanos2009}. However, we cannot find an infrared excess for this object. Compared to \citet{Crowther1995}, the luminosity obtained in this work is a factor of two higher, mainly due to the higher stellar temperature, while the mass-loss rate is fairly the same.      

\paragraph{BAT99\,46} is a WN4o star \citepalias{BAT99}, which was not spectroscopically analyzed by means of model atmospheres before. 

\paragraph{BAT99\,47} is classified as WN3b \citep{Foellmi2003b} and had never been analyzed before. Although \citet{Foellmi2003b} could not find periodic radial velocity variations, we treat this object as a binary suspect because of the X-ray emission reported by \citet{Guerrero2008II}. Unfortunately, we do not have flux-calibrated spectra for BAT99\,47. However we have photometric data from the UV to the mid-infrared, so that the luminosity and the interstellar reddening can be well determined. We achieved a plausible fit of the SED with our single-star model. Therefore, we expect that the possible companion does not contribute much to the bolometric luminosity.

\paragraph{BAT99\,48} is listed as a WN4b in the BAT99 catalog. This star was previously analyzed by \citet{Koesterke1991} and \citetalias{HK2000}. The former authors have obtained a stellar temperature of $T_* = 57\,\mathrm{kK}$ based on unblanketed model atmospheres, whereas the latter authors derived a stellar temperature of $T_* = 79.4\,\mathrm{kK}$ with blanketed model atmospheres. In comparison to the last work, our new analysis results in a 10\,kK higher stellar temperature, a factor of two higher mass-loss rate but nearly the same luminosity. We note that this star is located within the regime of parameter degeneracy (cf.\ Sect.\,\ref{subsec:modelgrid}).

\paragraph{BAT99\,49} was identified as a SB2 binary with a period of 34\,d \citep{Niemela1991}. A slightly smaller period of 31.7\,d was found by \citet{Foellmi2003b}, who classified the primary as WN4:b and the companion as O8\,V. The stellar parameters that we derived for this object are to be taken with care, since we analyze this object as a single star but the companion may substantially contribute to the overall flux.

BAT99\,49 is one of only two WR stars in the LMC that is able to ionize a \ion{He}{ii} region \citep{Naze2003b}: the other one is BAT99\,2. \citet{Naze2003b} conclude from the nebular \ion{He}{ii}-flux that the exciting star delivers $\sim1 \cdot 10^{47}$\,\ion{He}{ii} ionizing photons per second. This agrees with our final model, which produces $1.6 \cdot 10^{47}$\,\ion{He}{ii} ionizing photons per second. However, we note that our final model probably overestimates the number of ionizing photons, since the proper luminosity of the WN component is certainly lower than the value given in Table\,\ref{table:parameters}, which is derived neglecting its binary nature.

\paragraph{BAT99\,50} was classified as WN5h by \citet{Crowther2006b}. For this object, we have derived the stellar parameters by means of stellar atmosphere models for the first time. Unfortunately, we do not have UV spectra so the stellar parameters are derived from this normalized optical spectrum and photometry alone. Thus, the obtained luminosity is subject to higher uncertainties compared to those luminosities simultaneously derived from flux-calibrated spectra and photometry.

\paragraph{BAT99\,51} is listed with a WN3b classification in the BAT99 catalog. The spectrum of BAT99\,51 is dominated by broad and round emission lines. Therefore, this object belongs to the so-called round-lined stars (cf. comment on BAT99 7). The line shapes can only be reproduced assuming a high rotational velocity of $v_{\mathrm{rot}} = 1000\,\mathrm{km\,s^{-1}}$ as discussed in Sect\,\ref{sect:rotation}.

Unfortunately, we do not have an interpretable UV spectrum for this stars. Its only IUE spectrum (SWP\,04872) is not usable. It is noisy and lacks WR features, and does not show enough flux in relation to the photometric data and flux-calibrated spectrum obtained by \citet{Torres-Dodgen1988}.

\paragraph{BAT99\,54} was classified as WN8ha by \citet{Schnurr2008}. 
Unfortunately, the optical spectra at hand do not cover the H$\alpha$ line. Consequently, the hydrogen abundance is determined from the higher members of the Balmer series alone. We note that the line strength of the H$\beta$ line may point to a slightly higher hydrogen mass-fraction than the $X_{\element{H}} = 0.2$ given in Table\,\ref{table:parameters}. With our line-blanketed models we derived a 7\,kK higher stellar temperature than previously obtained by \citet{Crowther1997} with unblanketed model atmospheres. This higher value results in a roughly 50\,\% higher luminosity, whereas the mass-loss rate is a factor of three higher in our analysis. 

\paragraph{BAT99\,55} is one of only three WN11 stars in the whole sample. According to \citet{Humphreys1994}, this WN star is a LBV candidate. \citet{Schnurr2008} concluded from the radial velocity of this object that it is most likely a runaway star. BAT99\,55 is one of three LMC WN stars detected at $24\,\mu\mathrm{m}$ with the IRAC instrument aboard the {\em Spitzer space telescope}, suggesting the presence of circumstellar dust \citep{Bonanos2009}. \citet{Crowther1997} and \citet{Pasquali1997} derived rather similar stellar parameters for this star. The stellar temperature and the luminosity derived in this work are in good agreement with the properties presented by \citet{Crowther1997}. On the other hand, our study results in a 70\,\% higher mass-loss rate, which can be attributed to the higher terminal wind velocity derived in our analysis. 

\paragraph{BAT99\,56} is a WN4b star \citepalias{BAT99}. This object had never been spectroscopically analyzed by means of model atmospheres before. 

\paragraph{BAT99\,57} is another WN4b star \citepalias{BAT99} with typical stellar parameters, which was not analyzed by means of model atmospheres before.

\paragraph{BAT99\,58} is a WN7h star \citep{Schnurr2008} studied by \citet{Koesterke1991}, \citet{Crowther1997}, \citetalias{HK2000} and \citet{Bonanos2009}. \citet{Koesterke1991} and \citet{Crowther1997} derived nearly equal stellar parameters, including a stellar temperature of about $T_* = 35\,\mathrm{kK}$. \citetalias{HK2000} obtained a stellar temperature of $T_* = 39.8\,\mathrm{kK}$. Our state of the art atmosphere models, however, need a temperature of $T_* = 47\,\mathrm{kK}$ to reproduce the observed \ion{He}{i}/\ion{He}{ii} line ratios. These differences are attributable to the line-blanketing effect included in our models. The higher stellar temperature contributes to an increase in the derived luminosity of about 0.5\,dex compared to \citet{Koesterke1991}, \citet{Crowther1997}, and \citetalias{HK2000}. The derived mass-loss rate, on the other hand, is fairly the same. 

\begin{figure*}
  \centering
  \includegraphics[width=0.46\hsize]{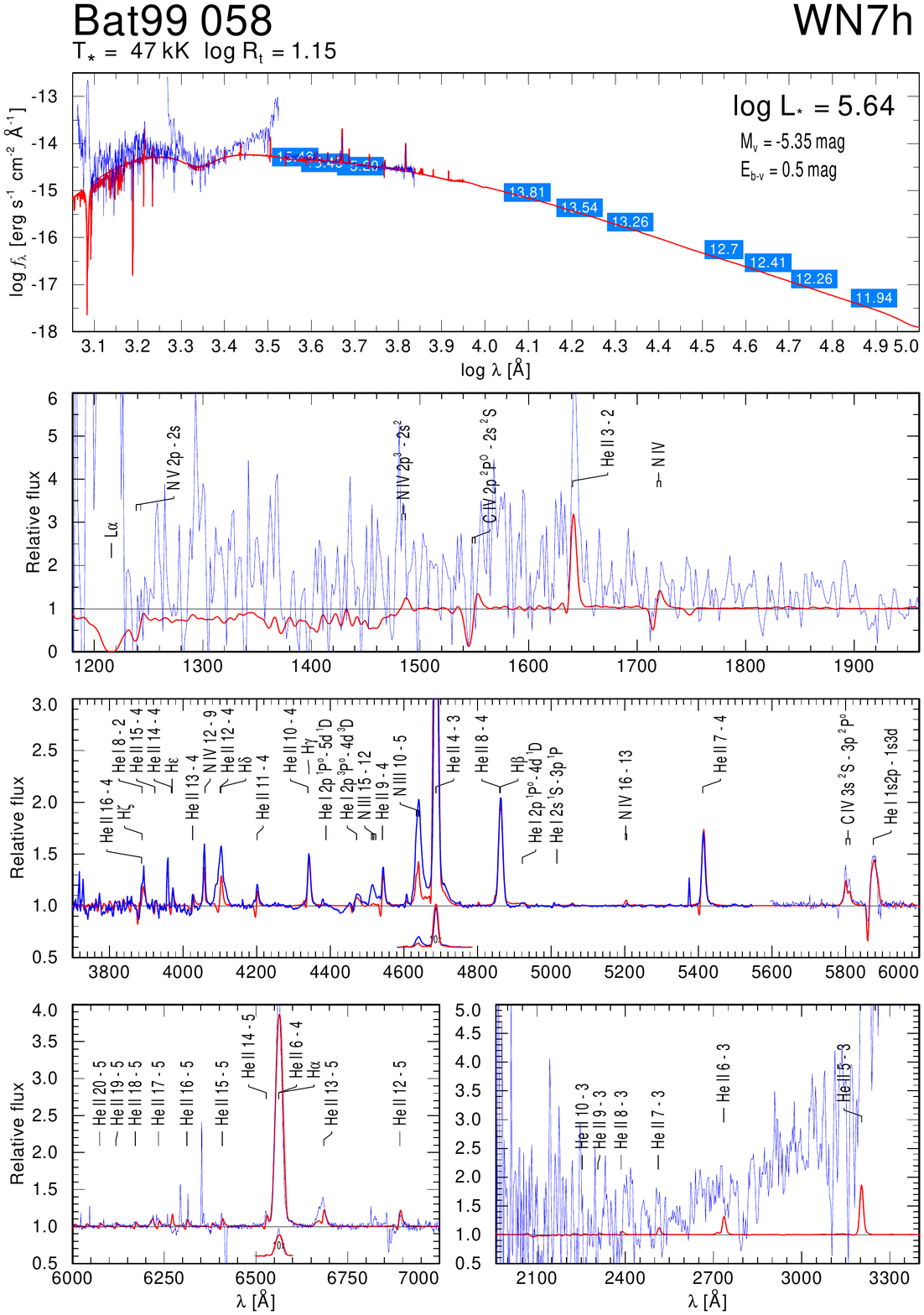}
  \qquad
  \includegraphics[width=0.46\hsize]{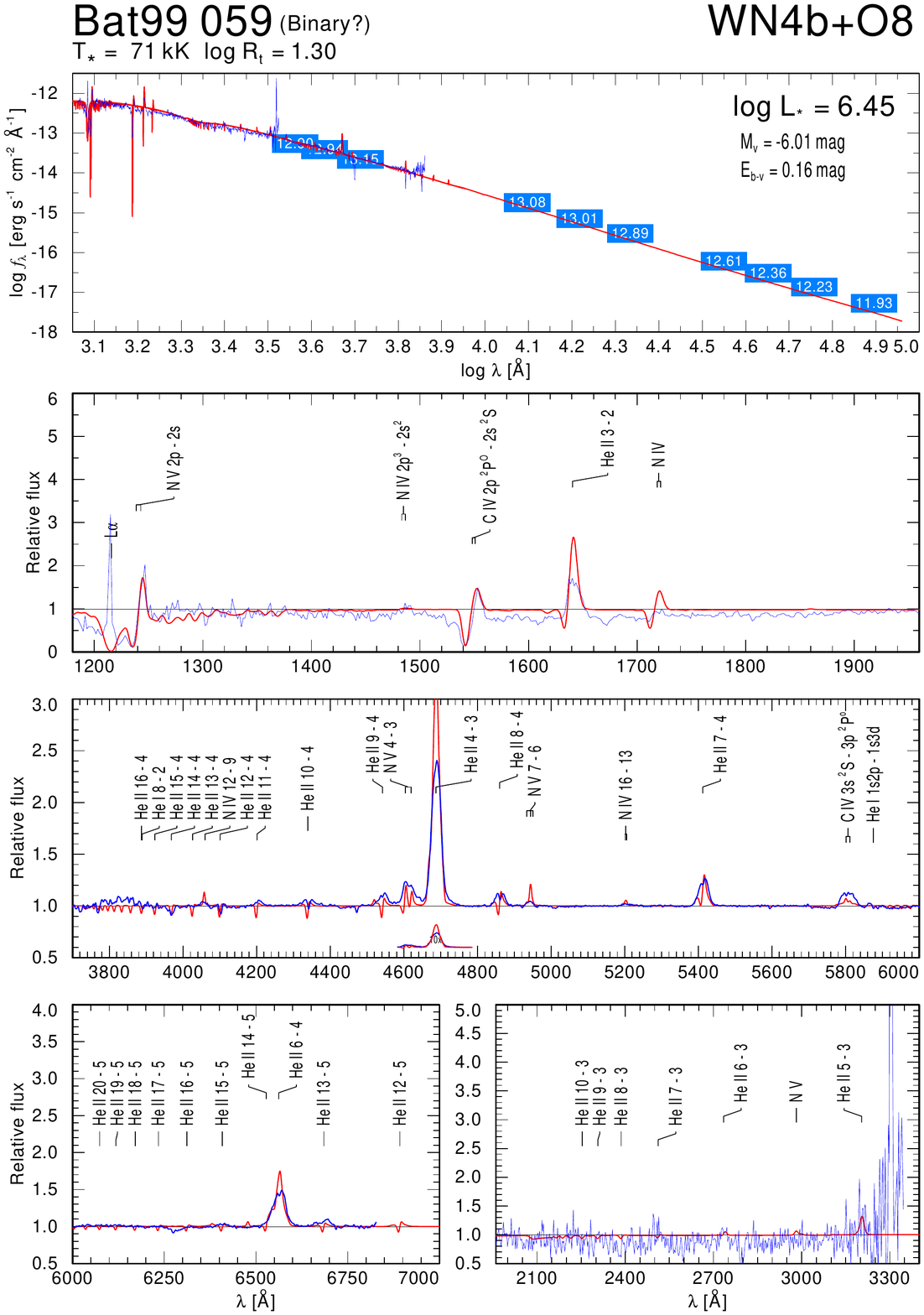}
  \vspace{-0.4cm}
  \caption{Spectral fit for BAT99\,058 and BAT99\,059}
  \label{fig:bat058}
  \label{fig:bat059}
\end{figure*}

\citet{Bonanos2009} find an infrared excess for this object by comparing a $T_* = 45\,\mathrm{kK}$ CMFGEN model from \citet{Smith2002} with the observed SED. However, our fit does not show a clear infrared excess, although a slight mismatch of the infrared photometry can be seen in Fig.\,\ref{fig:bat058}. We note that this discrepancy may be attributed to the IUE spectra, which are of poor quality (bad signal to noise ratio, arbitrary continuum shape of the IUE long-wavelength spectrum). 

\paragraph{BAT99\,59} is listed as WN4o?+B binary candidate in the BAT99 catalog. \citet{Foellmi2003b} found a period of 4.7\,d, reclassified the primary to WN4b and specified the companion to be an O8: star. They still assign a question mark to the binary status because the determined radial velocity amplitude is close to their detection limit. Since distinct absorption lines of the companion are visible in the spectra, the companion should contribute substantially to the total flux. The derived stellar properties (Table\,\ref{table:parameters}) are thus to be considered with caution. \citet{Koesterke1991} used pure helium models to derive lower limits for the stellar temperature and luminosity than our more sophisticated models.

\paragraph{BAT99\,60} is listed as WN3+OB binary candidate in the BAT99 catalog, but \citet{Foellmi2003b} did not find a radial velocity period and reclassified the object to WN4(h)a. Furthermore, these authors identified the absorption lines visible in their spectra to be intrinsic to the wind of the WN star. Therefore, we consider this star to be single for the time being. Unfortunately, we do not have UV spectra of this star, which had never been analyzed with stellar atmosphere models before. We confirm the presence of hydrogen ($X_{\element{H}} = 0.2$) and thus the above classification.

\paragraph{BAT99\,62} was classified as WN3(h) by \citet{Foellmi2003b} due to indications of hydrogen in its spectrum. Our analysis is inconclusive at this point. A reasonable fit can be achieved with a hydrogen-free model as well as with a model of a moderate hydrogen mass-fraction of $X_{\element{H}} = 0.1$. All optical spectra at hand exhibit distinct absorption lines in place of the \ion{O}{iii}-nebular emission lines at 4959\,{\AA} and 5007\,{\AA}, likely caused by an over correction of the diffuse emission. If this is true, the Balmer series will probably be narrowed by the inadequate nebular subtraction, lending credence to the relatively high hydrogen abundance derived in the line fit. In this paper, we present for the first time stellar parameters for this star, derived by spectral analysis. Regrettably, no UV spectra are available for this star.  

\paragraph{BAT99\,63} is listed as a binary candidate in the BAT99 catalog on the basis of absorption lines possibly belonging to a companion star. However, no binary period was found by the radial velocity analysis of \citet{Foellmi2003b}. Moreover, \citet{Cowley1984} and \citet{Foellmi2003b} argue that the absorption components are intrinsic to the WR wind. Thus, the object was classified as WN4ha: by \citet{Foellmi2003b}. We treat this star as single until its binary status is confirmed. \citet{Cowley1984} note that the high radial velocity suggests that the star is a runaway. This conclusion is confirmed by the radial velocity study of \citet{Schnurr2008}. 

A lower limit for the stellar temperature was derived by \citet{Koesterke1991} on the basis of the helium spectrum. More elaborate models are used by \citetalias{HK2000}, deriving a stellar temperature of $T_* = 70,8\,\mathrm{kK}$. In comparison to this work, our best fitting model with $T_* = 63\,\mathrm{kK}$ is one grid step cooler. This lower temperature is justified by a slightly better fit of the nitrogen lines. Apart from that, we obtained similar values for the mass-loss rate, while the luminosity and hydrogen abundance are slightly lower.

The filamentary ring nebula associated with BAT99\,63 was studied by \citet{Naze2003b}. As no \ion{He}{ii} nebular emission is detected by these authors, they obtained an upper limit for the number of \ion{He}{ii} ionizing photons delivered by the exciting star, which amounts to $< 2.5 \cdot 10^{45}$\,\ion{He}{ii} ionizing photons per second. This agrees with our final model, which does not predicts a significant number of \ion{He}{ii} ionizing photons (see Table\,\ref{table:zansT}).

\paragraph{BAT99\,64} was listed as WN4+OB? binary candidate in the BAT99 catalog. A period of 37.6\,d was found by \citet{Foellmi2003b}, who classified the companion as O9 and specified the primary to be a WN4o star. However, we expect that the companion does not contribute much to the bolometric luminosity. 

\paragraph{BAT99\,65} was classified as WN4o by \citet{Foellmi2003b}. For this star we only have a co-added optical spectrum, which is not flux-calibrated so that the observed SED is covered by photometry alone. However, the coverage is complete from UV to mid-infrared. 

\paragraph{BAT99\,66} is a WN3(h) star \citep{Foellmi2003b} for which stellar parameters have been presented for the first time in this paper. Unfortunately, no UV spectra are available for this star.

\paragraph{BAT99\,67} is listed as a binary candidate with a WN5o?+OB classification in the BAT99 catalog, though \citet{Foellmi2003b} concluded from their radial velocity study that this star is probably not a short-period binary. \citet{Foellmi2003b} note that the absorption features visible in their spectra are strongly blue-shifted. They emphasize that if this shift arises from a companion, it would result in distinct radial velocity variations that are not observed. Hence, these authors argue that the absorption lines are intrinsically formed in the wind of the WN star, and accordingly reclassify the object to WN5ha. Nevertheless, we still treat this star as a binary suspect because of the X-ray emission detected by \citet{Guerrero2008I}.

We derived a stellar temperature of $T_* = 47\,\mathrm{kK}$, only slightly higher than the value obtained by \citetalias{HK2000} ($T_* = 44.7\,\mathrm{kK}$). The luminosity and mass-loss rate derived in our analysis are slightly lower compared to the result obtained by \citetalias{HK2000}, while the hydrogen abundance is 0.1\,dex higher. 

\paragraph{BAT99\,68} is a transition type O3.5\,If*/WN7 star \citep{Crowther2011}. The physical parameters were previously estimated with FASTWIND by \citet{Massey2005}, but the authors note that these values are only poorly constrained. Our new analy\-sis results in a stellar temperature of $T_{\mathrm{eff}} = 44\,\mathrm{kK}$, marginally higher than the temperature estimate obtained by \citet[][$T_{\mathrm{eff}} =  40 - 42\,\mathrm{kK}$]{Massey2005}. The mass-loss rate derived in our new analysis is a factor of three lower, even if we account for the impact of different terminal velocities and density contrasts. We note that a mass-loss rate as high as obtained by \citet{Massey2005} results in considerably too strong emission lines in our synthetic spectra.

The IUE spectra associated with this object are noisy and do not show WR features. Therefore, we precluded these spectra from our analysis. We also excluded the optical spectrum measured by \citet{Torres-Dodgen1988}, since the continuum slope of this flux-calibrated spectrum disagrees with the optical photometry \citepalias{BAT99} and the HST spectra. Moreover, the available infrared photometry \citep{Cutri2012,Kato2007} is inconsistent and cannot be fitted in accordance with the UV and optical data. The infrared observations probably suffer from contributions of nearby objects, which are visible on WFPC2 images within $1\arcsec$ around BAT99\,68. Fitting the UV and optical data, we derived a luminosity of $\log\,(L/L_\odot) = 6.0$ which is a factor of two lower than previously obtained by \citet{Massey2005}.

\paragraph{BAT99\,71} was identified as a short-period binary by \citet{Foellmi2003b}. They classified the primary as WN4 and specified the companion to be a O8: star. Its IUE spectrum is not usable: It is noisy and does not have distinct WR features. Therefore, we reject these spectra from our analysis.

\paragraph{BAT99\,72} is probably a medium-period binary according to \citet{Foellmi2003b}, assigning an O3: type to the companion and reclassifying the primary to WN4h. \citet{Foellmi2003b} mention that an absorption-line component is clearly visible in the spectrum. However, we fit the star as a single WN star and achieve a good fit for almost the whole spectrum. For that reason, and due to the well-reproduced SED, we expect that the companion does not contribute much to the bolometric luminosity. We note that the luminosity obtained in our analysis is derived from visual narrowband, 2MASS and IRAC photometry alone since no flux-calibrated spectra were available. 

\paragraph{BAT99\,73} was classified as WN5ha by \citet{Crowther2006b}. Unfortunately, the spectrum from \citet{Foellmi2003b} is the only one available to us. Since this spectrum is not flux-calibrated, the luminosity is derived from photometry alone, but the available photometry covers the whole range from the UV to mid-infrared so that the luminosity is nevertheless well constrained. This WN star was not analyzed by means of stellar atmosphere models before. 

\paragraph{BAT99\,74} is a WN3(h)a star \citep{Foellmi2003b}, which is analyzed in this work for the first time. The spectrum of this star is characterized by small absorption lines and moderate emission lines. \citet{Foellmi2003b} identified the absorption lines to be intrinsic to the wind of the WN star. Both absorption and emission lines can be accordantly reproduced with our single-star model. From the H$\alpha$ and H$\beta$ lines, we estimated a hydrogen mass-fraction of $X_{\element{H}} = 0.2$, which underpins the WN(h) classification. No flux-calibrated spectra are available for BAT99\,74. However, the luminosity is well constrained due to the photometric coverage of the SED from UV to mid-infrared. 

\paragraph{BAT99\,75} was classified as WN4o by \citet{Foellmi2003b}. The prominent emission lines in the spectra of this star are best reproduced with a hydrogen-free model. For the first time, we present here parameters for this star. 

\paragraph{BAT99\,76} is a WN9ha star \citep{Schnurr2008}, which has been analyzed by \citet{Koesterke1991}, \citet{Crowther1995}, \citet{Pasquali1997}, and \citetalias{HK2000}. Two UV spectra and three optical spectra are at hand for the analysis of this star (see Table\,\ref{table:data} and \ref{table:wnsample}). All can be fitted with the same model, which give us confidence in the derived physical parameters. 

The stellar temperature derived in our analysis, $T_* = 35\,\mathrm{kK}$, fits in the range of temperatures $T_* = 30 - 38.7\,\mathrm{kK}$ obtained by the authors mention above. The mass-loss rates presented in these papers are nearly the same with the exception of the results obtained by \citetalias{HK2000}, who inferred a mass-loss rate a factor of two higher. Our analysis confirms the lower mass-loss rates obtained by \citet{Koesterke1991}, \citet{Crowther1995}, and \citet{Pasquali1997}. The luminosity derived in the previous papers range from $\log\,(L/L_\odot) = 5.4$ to $\log\,(L/L_\odot) = 5.8$, which in principle agrees with the value ($\log\,(L/L_\odot) = 5.66$) derived in this work. The hydrogen abundance provided in Table\,\ref{table:parameters} is lower than previously derived by \citet{Crowther1995}, \citet{Pasquali1997}, and \citetalias{HK2000}.

\paragraph{BAT99\,77} was revealed as SB1 binary with a period of 3\,d by \citet{Moffat1989}. This finding was later confirmed by the radial velocity study of \citet{Schnurr2008}. These authors classified the object as WN7ha. Since no spectral features can be unambiguously attributed to the companion, the contribution of the companion to the overall flux is hard to estimate. Thus, the physical parameters listed in Table\,\ref{table:parameters} need to be taken with caution. A tentative detection of X-ray emission has been reported by \citet{Guerrero2008I}.

\paragraph{BAT99\,78} is located in a tight cluster in the western part of LH90 with several additional sources within a radius of less than $1\arcsec$ \citep{Walborn1999}. This star is listed as WN4 in the BAT99 catalog, whereas \citet{Foellmi2003b} assign a WN6 spectral type to it. According to these authors, the spectrum is a superposition of the WN star and nearby objects. This is attributable to the crowded environment and the relatively wide slit width used by these authors. The co-added spectrum observed by \citet{Foellmi2003b} resembles an Of supergiant rather than a WN star since it shows distinct Of features and a H$\beta$ line clearly in absorption \citep{Crowther2011}. A reasonable fit of this spectrum can be achieved with a model of $T_* = 45\,\mathrm{kK}$. 

Beside the spectrum obtained by \citet{Foellmi2003b} we retrieved a relatively short optical spectrum from the HST archive which exhibits a clear WN characteristic. This spectrum can be well fitted with a model corresponding to a stellar temperature of $T_* = 71\,\mathrm{kK}$. Since the HST spectrum is flux-calibrated, it can also be used for the SED fit. This results in an implausibly small luminosity of $\log\,(L/L_\odot) =  4.64$. The flux of the HST spectrum disagrees with the available photometry, which in turn is not uniform. For example, the data of the J-band magnitude varies from $11.6 - 14.68$\,mag. A fit of the 2MASS, Spitzer IRAC, and WISE photometry results in a luminosity of $\log\,(L/L_\odot) = 6.8$. The observational discrepancies might be traced back to the small angular distance of BAT99\,78 to the other cluster members \cite[see, e.g.,][]{Walborn1999}. The smallest WISE aperture covers not only the four closest objects to BAT99\,78, but rather a large part of the whole cluster around it. 

\citet{Walborn1999} were able to chiefly resolve the host cluster of BAT99\,78 with the Wide Field/Planetary Camera 1 (WFPC1) aboard the HST. They obtained visual broadband photometry, which we adjusted to account for the contribution of the emission lines, using the correction factor derived by \citet{Breysacher1986}. This visual magnitude $M_\mathrm{v} = 14.84$ results in a luminosity of $\log\,(L/L_\odot) = 5.7$, assuming that the continuum slope of the HST spectrum and the thereof derived color excess of $E_{b-v}=0.2$ is correct. 

Due to the X-ray emission detected by \citet{Guerrero2008I}, we treat this object as a binary candidate, although \citet{Foellmi2003b} find no significant periodicity in their radial velocity study, which is based on a small number of observations of this object. 

\paragraph{BAT99\,79} is listed as WN7h+OB binary candidate in the BAT99 catalog. However, \citet{Schnurr2008} do not find periodic radial velocity variations. Despite this nondetection, we still regard this object as a binary candidate since a considerable amount of X-ray emission has been detected by \citet{Guerrero2008I}. We note that \citet{Crowther1997} estimated a significant contribution of the companion to the overall flux by means of their own spectra. In contrast, we expect that the companion does not contribute much to the bolometric luminosity, since only very small absorption features can be seen superimposed on the emission lines of the WN star in the spectrum observed by \citet{Schnurr2008}. Therefore, the stellar parameters listed in Table\,\ref{table:parameters} should represent a reliable approximation. 

\paragraph{BAT99\,80} is listed as O4\,If/WN6 transition type star in BAT99 catalog. \citet{Schnurr2008} reclassified the star to WN5h:a, arguing that the spectrum shows a diluted WN star rather than a hot slash star. However, no significant radial velocity variations were detected by \citet{Schnurr2008}. Nevertheless, we treat this object as a binary suspect because of the X-ray emission reported in \citet{Guerrero2008I}.   

The only spectrum available to us is characterized by absorption and moderate emission lines. Although it might show a binary, both absorption and emission lines can be accordingly reproduced with our single-star model.

\paragraph{BAT99\,81} was classified as WN5h by \citet{Foellmi2003b}. The IUE long-wavelength spectrum of this star apparently has a considerably lower flux than the IUE short-wavelength spectrum. The former does not fit to the rest of the observed SED. Therefore, it is not further considered in our analysis of BAT99\,81, which provides stellar and wind parameters of this object for the first time. \citet{Foellmi2003b} highlights the differences between the results of \citet{Cowley1984} and their own radial velocity study, which exhibits no clear radial-velocity deviation from the mean $v_\mathrm{rad}$ of their WN sample, whereas \citet{Cowley1984} argued in favor of a runaway nature.

\paragraph{BAT99\,82} is a WN3b star \citepalias{BAT99}, for which the detection of X-rays has been reported by \citet{Guerrero2008I}. We treat this star as a binary on the basis of the X-ray emission. However, no periodic radial velocity variations were found by \citet{Foellmi2003b}. The IUE spectrum assigned to BAT99\,82 was taken $1.2\,\arcmin$ away from its position and does show some helium emission, but no nitrogen. Furthermore, its flux is not compatible to the SED composed of the optical spectrum by \citet{Torres-Dodgen1988} and the photometry from the 2MASS catalog, so that the IUE spectrum was ignored.

\paragraph{BAT99\,86} has recently been classified as WN3(h) by \citet{Doran2013}. Unfortunately, no UV spectrum is available for this object. The hydrogen abundance is hard to determine because of absorption features superimposed on the Balmer lines. The best fit is obtained with a hydrogen-free model. However, we note that the emission lines without the absorption components might agree with an hydrogen mass-fraction of $X_{\element{H}} = 0.1$. 
On the contrary, \citet{Doran2013} derived a hydrogen abundance of $X_{\element{H}} = 0.2$ on the basis of their stellar atmosphere models. However, a hydrogen abundance this high would result in a considerable overprediction of the Balmer lines by our synthetic spectra. Apart from that, \citet{Doran2013} derived a stellar temperature that is 8\,kK higher and a luminosity that is 0.1\,dex higher than our values. However, the strongest deviation to the results presented by \citet{Doran2013} is obtained in the derived mass-loss rate, which is more than nine times lower in our analysis. Our models with a mass-loss rate as high as derived by \citet{Doran2013} considerably overpredict the equivalent width of all emission lines.

\paragraph{BAT99\,88} is the second WNE/WCE transition type star in the LMC. These stars exhibit a clear carbon enrichment compared to the rest of the LMC WN stars. The best fit is obtained with a carbon mass-fraction of $X_{\element{C}} = 0.005$ in accordance with the estimate by \citet{Doran2013}. 

This is one of the hottest stars in our sample, with a stellar temperature of $T_* = 112\,\mathrm{kK}$. 
This value is about 30\,kK higher than the stellar temperature ($T_* = 80\,\mathrm{kK}$) obtained by \citet{Doran2013}. In contrast, our model with $T_* = 80\,\mathrm{kK}$ underpredicts the equivalent width of the \ion{He}{ii}\,$\lambda\, 4686$ line with respect to the other \ion{He}{ii} lines, while the model with $T_* = 112\,\mathrm{kK}$ is able to simultaneously reproduce the equivalent width of all \ion{He}{ii} lines. Furthermore, the \ion{N}{iv} and \ion{N}{v} lines are slightly better reproduced at this higher temperature. 
However, we note that this star is located within the regime of parameter degeneracy (cf. Sect.\,\ref{subsec:modelgrid}). 

Contrary to the stellar temperature, the luminosity and the mass-loss rate derived in this work are comparable to the values derived by \citet{Doran2013}.
Unfortunately, we neither have UV spectra nor flux-calibrated spectra. Thus, the luminosity is derived from UV, optical, near- and mid-infrared photometry alone. 

The emission lines of this object exhibit round profiles. Thus, this star falls in the category of the so-called round line stars (cf. comment on BAT99 7). A reasonable fit of the observed line profile is achieved by convolving our synthetic spectrum with a rotation profile corresponding to a rotational velocity of $v_\text{rot} \cdot \sin i = 1200\,\mathrm{km\,s^{-1}}$. 

\paragraph{BAT99\,89} is a WN7h star \citepalias{BAT99}, which was previously studied by \citet{Crowther1997} 
and \citet{Doran2013}. The best fit is achieved at a stellar temperature of $T_* = 50\,\mathrm{kK}$, which agrees well with the recent results of \citet{Doran2013}. In contrast, \citet{Crowther1997} obtained a stellar temperature that is 10\,kK lower. However this value is excluded by our analysis, since the synthetic spectra clearly overpredicts the \ion{He}{i}/\ion{He}{ii} ratio at this temperature. This difference in the stellar temperature can be attributed to the line-blanketing effect, which is included in modern atmosphere models, but was not regarded at the time of the study by \citet{Crowther1997}. The luminosity derived in this paper ($\log\,(L/L_\odot) = 5.56$) is only marginally higher than the value obtained by \citet{Doran2013}, while \citet{Crowther1997} derived a luminosity which is a factor of two lower. In terms of the mass-loss rate, we obtain a value of almost 50\,\% lower compared to the recent results by \citet{Doran2013}, whereas the mass-loss rate derived by \citet{Crowther1997} is slightly higher.

\paragraph{BAT99\,91} was resolved into multiple components by \citet{Testor1988} and later by the HST observations of \citet{Walborn1995}. The WN star has been classified as WN6(h) by \citet{Evans2011}, although their ground-based observations were not able to entirely resolve the components of this object. The same seems to hold for the spectrum obtained by \citet{Schnurr2008}, which does not resemble the HST spectrum for BAT99\,91 obtained by \citet{Walborn1995}. For that reason, our spectroscopic analysis relies on the HST spectrum alone, which unfortunately covers only 1500\,{\AA} out of 3300 - 4800\,{\AA}. Since this range comprises neither the H$\alpha$ nor H$\beta$, the determination of the hydrogen abundance is based on the higher members of the Balmer series alone. With a hydrogen mass-fraction of $X_{\element{H}} = 0.2$, we confirm the corresponding classification. In the SED fit, we exclusively used the HST photometry from \citet{Walborn1995} and the HST flux-calibrated spectra from \citet{Walborn1999}, since these observations are distinguished by their particularly high spatial resolution, which seems to be necessary for reliable results in this tight cluster. This star had never been analyzed before. 

\paragraph{BAT99\,92} was classified as WN3:b(+O)+B1\,Ia by \citet{Schnurr2008}. Both the BAT99 catalog and \citet{Schnurr2008} give a binary period of 4.3\,d for this star. We note that the detection of X-ray emission has been reported by \citet{Guerrero2008I}. Although the optical spectra seem to be considerably affected by the companions, they still allow us to assess parameters like the stellar temperature. Furthermore, this object belongs to the round-line stars, since the emission lines exhibit a round shape which can only be reproduced with a rotational velocity of $v_{\mathrm{rot}} = 1500\,\mathrm{km\,s^{-1}}$. Another interesting fact is the significant strength of the \ion{C}{iv}\,$\lambda\,5808$ emission line, suggesting that either the system comprises an additional WC star or that the WN star belongs to the rare WNE/WCE transition type. The SED can be well reproduced with a single-star model. 

\paragraph{BAT99\,93} is one of the stars listed as WN stars in the BAT99 catalog which has been downgraded to O3\,If* by \citet{Evans2011} and \citet{Crowther2011}. Tentative X-ray emission has been detected by \citet{Guerrero2008I}. Thus, we treat this object as a binary suspect, although no radial velocity variations have been detected by \citet{Schnurr2008}. The only optical spectrum at hand lacks a subtraction of the diffuse background and shows only a truncated H$\beta$ line. For these reasons, we are not able to give a precise hydrogen abundances for this star. 

\paragraph{BAT99\,94} is characterized by broad emission lines with a round line shape. Therefore, this WN4b star \citepalias{BAT99} is classified as a round line star (cf. comment on BAT99 7).  The round shape of the emission lines requires a convolution of the model spectrum with a rotation profile corresponding to a rotational velocity of $v_\text{rot} \cdot \sin i = 1600\,\mathrm{ km\,s^{-1}}$. 

Unfortunately, we do not have UV spectra for this star. In this work, we derive a stellar temperature of $T_* = 141\,\mathrm{kK}$, which is significantly higher compared to the previous results by \citetalias{HK2000} ($T_* = 100\,\mathrm{kK}$). In our analysis, the model with the higher temperature results in a sightly better fit of the \ion{He}{ii}\,$\lambda\, 4686$, \ion{C}{iv}\,$\lambda\,5808$ and \ion{N}{iv}\,$\lambda\, 4060$ lines. However, we note that star is located in the regime of parameter degeneracy (cf.\ Sect.\,\ref{subsec:modelgrid}). In comparison with \citetalias{HK2000}, we have obtained a factor of two lower mass-loss rate, while the luminosity is a factor of three higher. The higher luminosity originates from the higher temperature and thus higher bolometric correction.

\paragraph{BAT99\,95} was identified as binary by \citet{Schnurr2008}. These authors find radial velocity variations with a period of 2.1\,d. \citet{Evans2011} classified the object as WN7h+OB. We expect that the companion does not contribute much to the bolometric luminosity, since we do not see spectral features of the companion in any of the available spectra. However, all optical spectra available to us suffer from either an oversubtraction of the diffuse background, or even a missing background subtraction. For this reason we cannot give a reliable value for the hydrogen content in the atmosphere of this star. 

In comparison to the previous analysis by \citet{Crowther1997}, our best fit is achieved at a 14\,kK higher stellar temperature of $T_* = 50\,\mathrm{kK}$. The temperature derived by \citet{Crowther1997}, however, would result in an overprediction of the \ion{He}{i}\,$\lambda\, 5877$ to \ion{He}{ii}\,$\lambda\, 5412$ line ratio. The different temperatures likely arise from the line blanketing, which is incorporated in our stellar atmosphere models, but was not accounted for in the models used by \citet{Crowther1997}. The higher temperature derived in this work results in a luminosity that is almost a factor of three higher compared to the previous results by \citet{Crowther1997}. The mass-loss rate derived in this work, on the other hand, is identical to the value given by \citet{Crowther1997}. 

\paragraph{BAT99\,96} is of subtype WN8 \citep{Schnurr2008} and located in the southern part of 30\,Doradus. We do not have flux-calibrated spectra or intrinsic narrowband photometry for this star, which renders it difficult to obtain precise values for the stellar luminosity and the interstellar reddening. Moreover, the available UBVR photometry is inconsistent since the values derived by various authors differ by up to 2\,mag. The narrowband photometry listed in the BAT99 catalog is derived from visual broadband photometry \cite[][$V = 13.65\,\mathrm{mag}$]{Parker1993} by means of the correction factor found by \citet{Breysacher1986}. The visual magnitude obtained by \citet{Parker1993} is slightly higher than the value ($V = 13.76\,\mathrm{mag}$) observed by \citet{Selman1999}. \citet{Massey2002} and \citet{Zaritsky2004} obtained higher magnitudes of $V = 12.84\,\mathrm{mag}$ and $V = 12.9\,\mathrm{mag}$, respectively. In contrast, \citet{Duflot2010} and \citet{Girard2011} derived lower visual magnitudes of $V = 14.5\,\mathrm{mag}$ and $14.47\,\mathrm{mag}$, respectively.

To construct the SED, we relied on the optical photometry obtained by \citet{Parker1993}, 2MASS and IRAC photometry \citep{Bonanos2009}. Similar to BAT99\,98, the SED fit results in a relatively high value for the color excess ($E_{b-v} = 0.7\,\mathrm{mag}$) and the luminosity ($\log\,(L/L_\odot) =  6.4$). In contrast to our results, \citet{Doran2013} derived a color excess of $E_{B-V} = 0.65\,\mathrm{mag}$, which corresponds to $E_{b-v} = 0.54\,\mathrm{mag}$. We cannot achieve a reasonable SED fit with a color excess as low as derived by these authors. Moreover, we obtain the same color excess and luminosity in our SED fit regardless of whether the optical broadband photometry from \citet{Parker1993} or the optical narrowband magnitudes ($v = 13.82\,\mathrm{mag}$, $b-v = 0.49\,\mathrm{mag}$) given by \citet{Doran2013} are used. 

\begin{figure*}
  \centering
  \includegraphics[width=0.46\hsize]{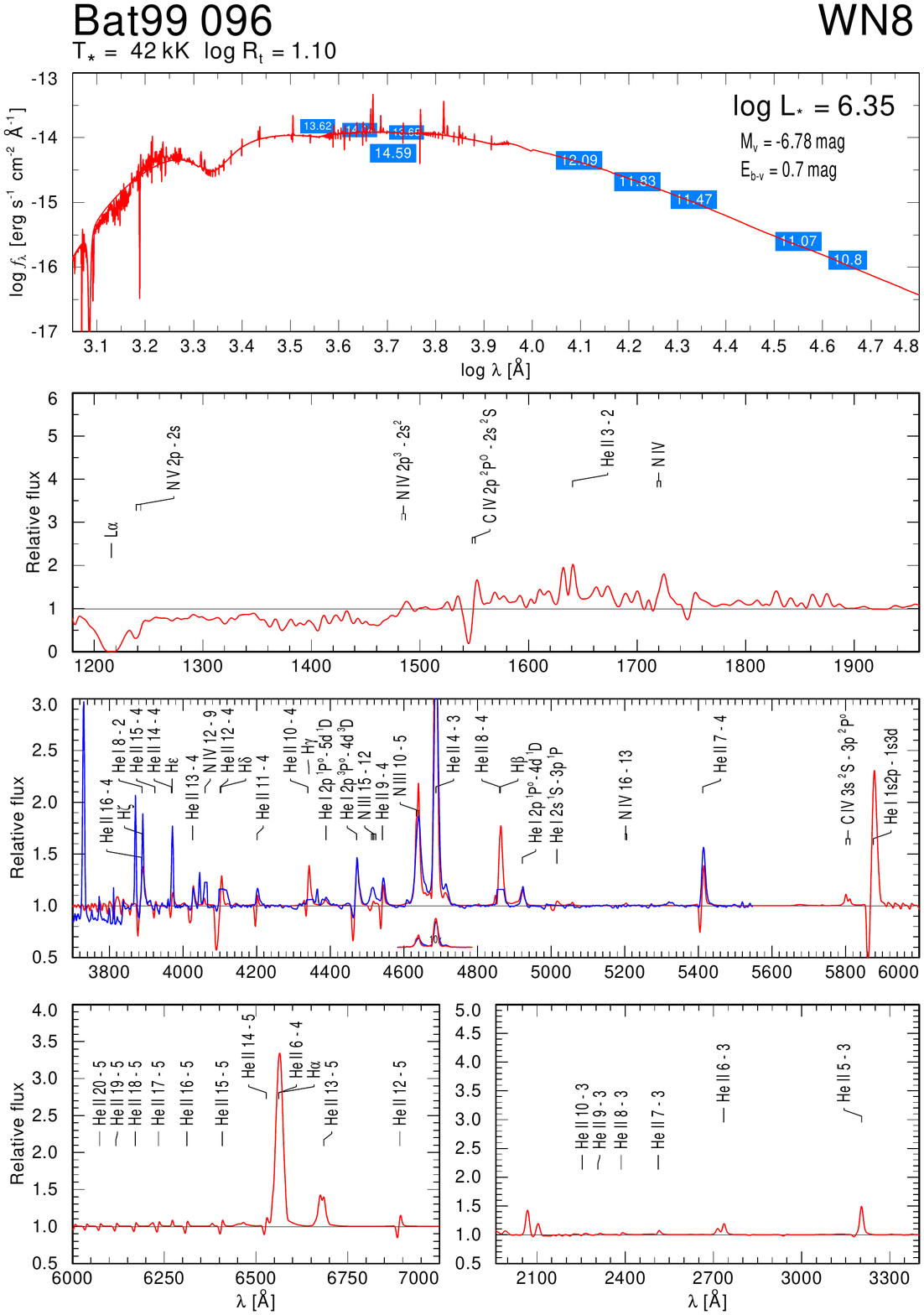}
  \qquad
  \includegraphics[width=0.46\hsize]{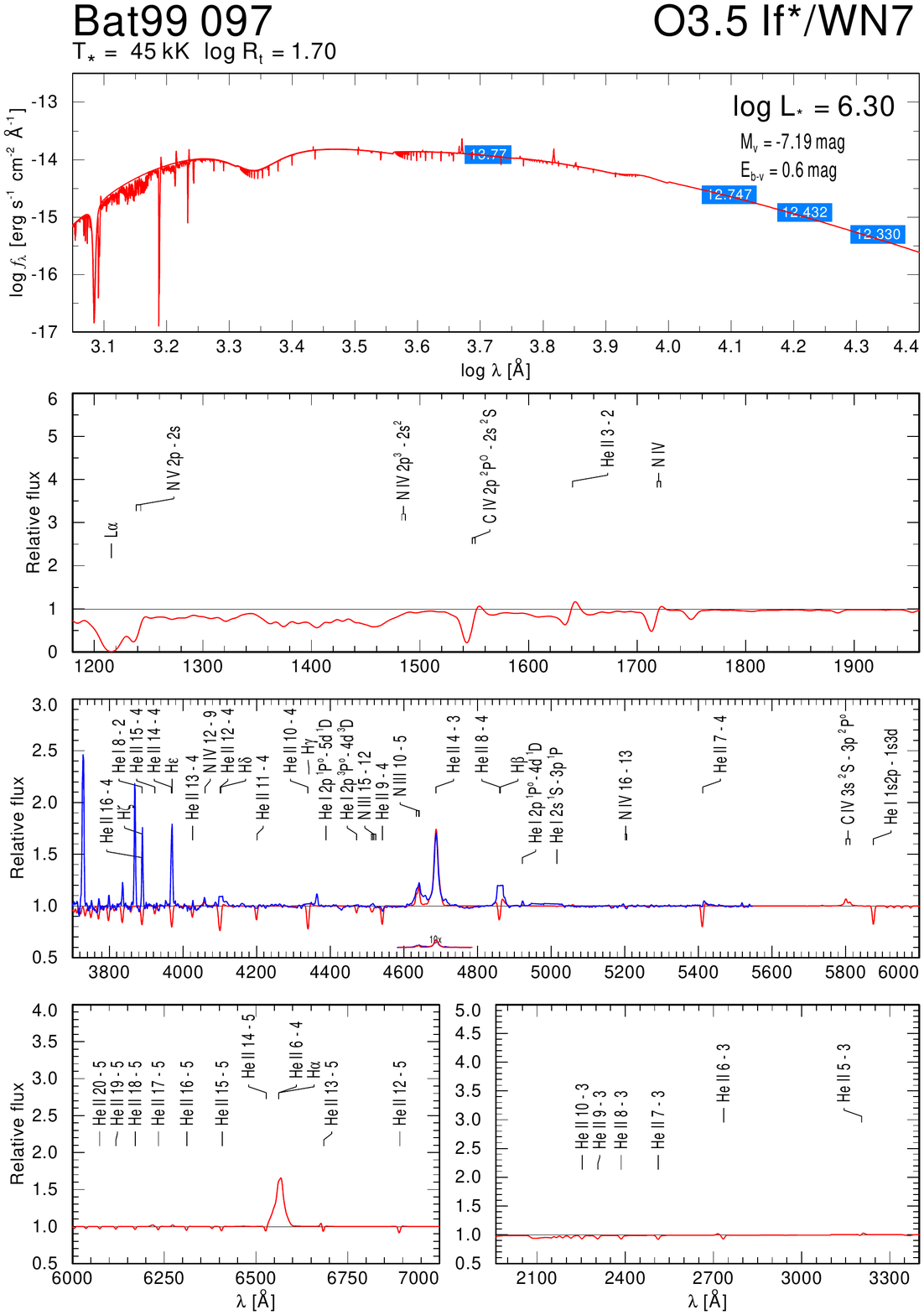}
  \vspace{-0.4cm}
  \caption{Spectral fit for BAT99\,096 and BAT99\,097}
  \label{fig:bat096}
  \label{fig:bat097}
\end{figure*}

The stellar luminosity derived in our analysis is much higher compared to the results derived by \citet{Crowther1997} and \citet{Doran2013}. These authors determined a stellar luminosity of $\log\,(L/L_\odot) =  5.86$ and $\log\,(L/L_\odot) =  6.04$, respectively. In comparison to \citet{Doran2013}, the deviation in the luminosity primarily originates from the different reddening parameters. The additional IRAC photometry incorporated in our analysis enhances the constraints on the shape of the SED (see Fig.\,\ref{fig:bat096}), which gives us confidence in the derived luminosity and color excess. Apart from these two parameters, we achieve a good agreement for the stellar temperature and the mass-loss rate obtained in this work and the values presented by \citet{Doran2013}. 
Due to the missing subtraction of the diffuse background in the spectra obtained by \citet{Schnurr2008}, we note the doubtful hydrogen abundance determined in our analysis.  

We consider this star to be single, since neither periodic radial-velocity variations nor X-ray emission have been detected by \citet{Schnurr2008} and \citet{Guerrero2008I,Guerrero2008II}, respectively. However, \citet{Parker1993} argued in favor of a multiple object on the basis of their ground-based photometry. However, we cannot detect any visual companion either on the images taken with the {\em Visual and Infrared Telescope for Astronomy} (VISTA) for the VISTA survey of the Magellanic Clouds system (VMC) \citep{Emerson2006,Dalton2006} or on the high resolution images with the WFC3 aboard the HST \citep{OConnell2008}. 
From a comparison of the empirical HRD position (see Figs.\,\ref{fig:hrd} and \ref{fig:hrd_geneva}) with the stellar evolution tracks calculated by \citet{Meynet2005}, we estimate an initial mass on the order of $M_{\mathrm{init}} = 100\,M_\odot$. Thus, we consider this object to belong to the category of very massive stars. 

\paragraph{BAT99\,97} is another transition type O3.5\,If*/WN7 star \citep{Crowther2011,Evans2011}. Unfortunately, the only spectrum available to us is affected by nebular emission due to a missing background subtraction. Thus, we are not able to give a precise hydrogen abundances for this star. 
We derive a stellar temperature that is slightly higher compared to the results published by \citet{Doran2013}, while the mass-loss rate and the luminosity are higher by about 0.15\,dex.

\paragraph{BAT99\,98} is a WN6 star \citep{Schnurr2008} located near R136. The derived luminosity, and thus the stellar mass as well, are comparable to those of the very massive stars in the core of R136 analyzed by \citet{Crowther2010}. This star is distinguished by the relatively high extinction of $E_{b-v} = 0.8\, \mathrm{mag}$ derived from the SED fit. Unfortunately, no flux-calibrated spectra and intrinsic narrowband photometry (Smith system) are available. Broadband photometry (e.g.,\ Johnson system), on the other hand, is contaminated by the prominent emission features.

The optical narrowband photometry from the BAT99 catalog is a corrected Johnson $V$ magnitude from \citet{Parker1993}, using the subtype-dependent correction factor derived by \citet{Breysacher1986}. Following this procedure, we obtain a narrowband magnitude of $v=13.70\,\mathrm{mag}$. 
Further broadband photometry is available from \citet{Selman1999} and \citet{Massey2002}, whereas the coordinates quoted by \citet{Massey2002} show the largest deviation from the position stated in {\em Simbad}. The $V$ band magnitudes obtained by these authors, corrected for the contribution of the emission lines, result in $v=13.64\, \mathrm{mag}$, $v=13.61\, \mathrm{mag}$ and $v=13.67\, \mathrm{mag}$, respectively. The $v$ magnitude inferred by \citet{Breysacher1986} is a corrected Str{\"o}mgen $y$ magnitude that amounts to $v=13.65\, \mathrm{mag}$. The optical photometry, together with the near- and mid-infrared photometry, gives rise to the high color excess and a stellar luminosity of $\log\,(L/L_\odot) =  6.7$. 
However, the luminosity has a relatively large uncertainty, since it is derived from photometry alone.

So far, neither periodical radial velocity variations \citep{Schnurr2008} nor X-ray emission \citep{Guerrero2008I, Guerrero2008II} were detected. Thus, we treat this object as a single star, although the moderate fit quality may indicate a line dilution due to a yet undetected companion, as already suggested by \citet{Crowther1997}. The applied model underpredicts the \ion{He}{ii}\,$\lambda\, 5201$, \ion{He}{i}\,$\lambda\, 4471$ lines and overpredicts the \ion{He}{ii}\,$\lambda\, 4686$ line. An adjustment of the temperature in one direction or the other spoils the fit of either the \ion{N}{iv}\,$\lambda\, 4060$ line or the \ion{N}{iii}\,$\lambda\, 4640$ line, used as main diagnostic lines for this object.

Unfortunately, the only spectrum available to us is compromised by a missing subtraction of the diffuse background. This is probably the reason that the spectrum is affected by nebular contamination, a fact that entails an uncertain determination of the hydrogen content. This is aggravated by the fact that H$\beta$ and $H\gamma$ are truncated \citep[see][]{Schnurr2008}.

The luminosity derived in this work corresponds to a current mass of $M_* = 226\, M_\odot$, according to the mass-luminosity relation from \citet{Graefener2011}. However, the error margin of this quantity is large, since it is calculated from the luminosity and the hydrogen abundance, which are in turn affected by considerable uncertainties.
By comparing the empirical HRD position to stellar evolution tracks calculated by \citet{Yusof2013}, we estimated an initial mass of at least $M_{\mathrm{init}} = 250\, M_\odot$. Thus, this star is one of the most massive stars hitherto known in the LMC. 

\paragraph{BAT99\,99} is a transition type O2.5\,If*/WN6 star. Two spectra are used for the analysis of this star, an HST spectrum and a spectrum obtained by \citet{Schnurr2008}. Both spectra are characterized by small absorption lines and relatively weak emission lines. We note that the spectrum taken by \citet{Schnurr2008} shows a substantially weaker \ion{He}{ii}\,$\lambda\, 4686$ line, although the \ion{N}{iv}\,$\lambda\, 4060$ line is of comparable strength in both observations. However, we do not consider the ground-based spectrum observed by \citet{Schnurr2008} in our analysis because it seems to be contaminated with nebular emission and the H$\beta$ line is arbitrary truncated.

\begin{figure*}
  \centering
  \includegraphics[width=0.46\hsize]{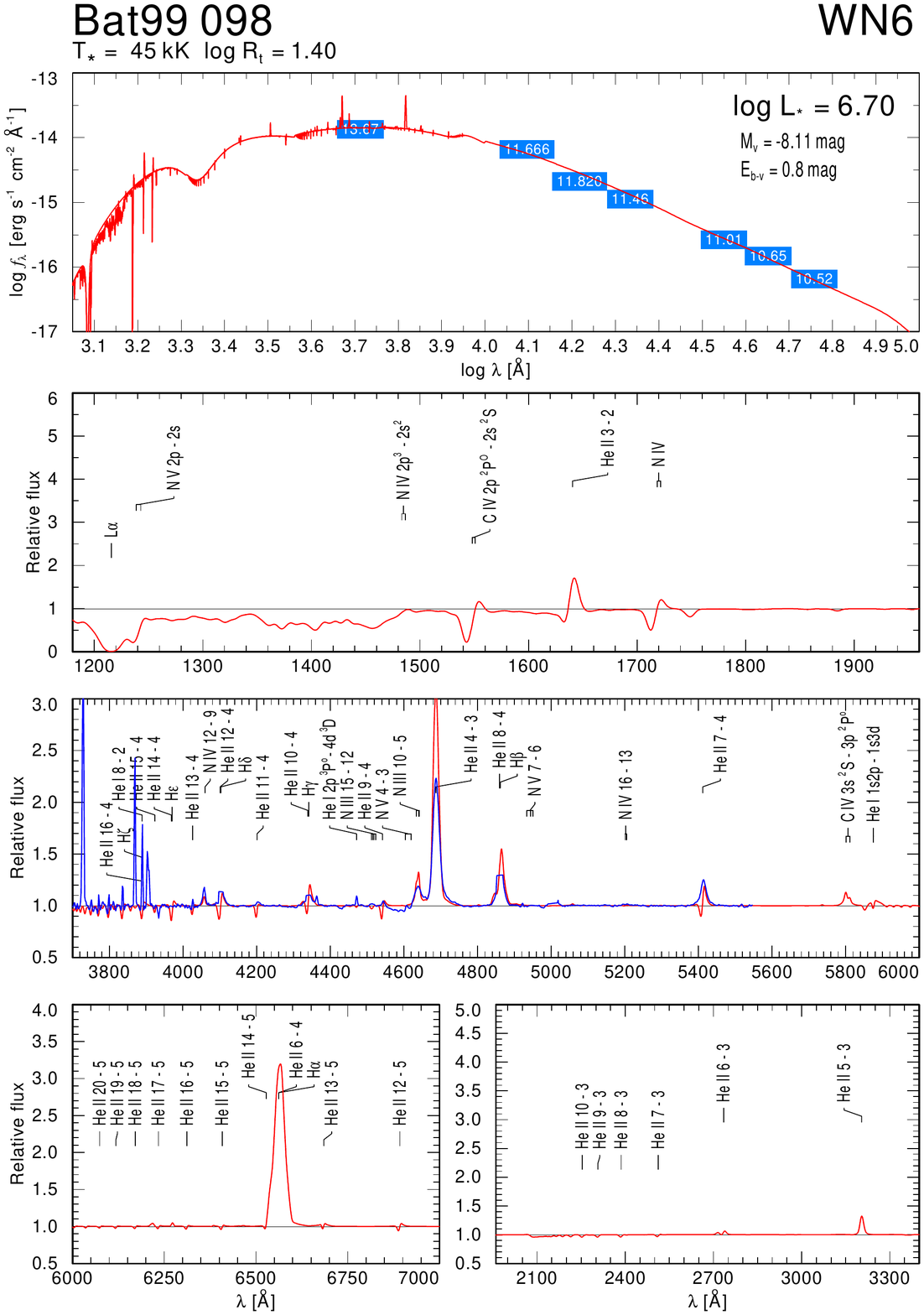}
  \qquad
  \includegraphics[width=0.46\hsize]{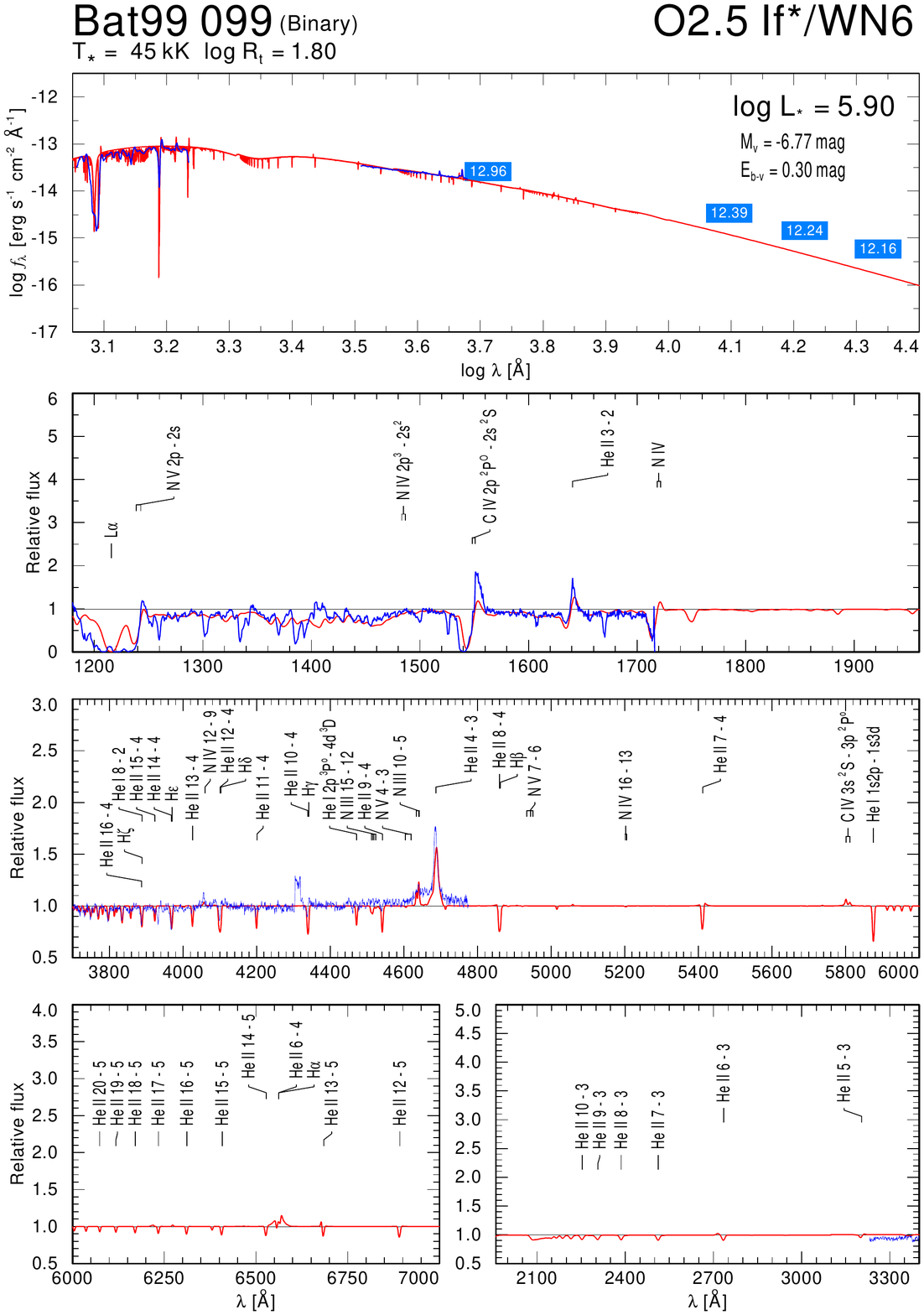}
  \vspace{-0.4cm}
  \caption{Spectral fit for BAT99\,098 and BAT99\,099}
  \label{fig:bat098}
  \label{fig:bat099}
\end{figure*}

The HST spectra exhibit a flux of roughly a factor of two lower in comparison to the photometric data listed in the BAT99 catalog and the infrared photometry by \citet{Kato2007}. Since BAT99\,99 is located in the vicinity of the 30\,Doradus core, this mismatch can be attributable to this crowded environment. Therefore, we have derived the luminosity from the HST spectra alone, which are distinguished by the high spatial resolution of the HST. The infrared excess presented in the SED fit in Fig.\,\ref{fig:bat099} presumably originates from nearby sources visible on high resolution HST images. 

The detection of X-ray emission has been reported by \citet{Guerrero2008I}, which is indirect evidence for the binary nature of this object (see Sect.\,\ref{subsect:binaries}). Direct evidence has been supplied by \citet{Schnurr2008}. These authors have found radial velocity variations, corresponding to a period of 93\,d. We did not detect spectral lines of a companion star in either of the spectra used in our analysis. Therefore, the contribution of an OB companion to the overall flux cannot be properly evaluated. Thus, the parameters for the WN component listed in Table\,\ref{table:parameters} might be affected by this unknown flux contribution.  

\paragraph{BAT99\,100} is a WN7 star located in the crowded environment close to the core of 30\,Doradus. In the optical spectral range, we used two spectra, an archival HST spectrum and a spectrum obtained by \citet{Schnurr2008}. The latter shows strong nebular emission lines, but only small stellar emission lines, whereas the former exhibit much stronger emission lines. Due to the limited spatial resolution of the ground-based observations, our analysis is mainly based on the HST spectrum. 

In a former study, \citet{Crowther1997} analyzed this star with unblanketed stellar atmospheres. In contrast to their work, we achieve the best fit with a model corresponding to a stellar temperature of $T_* = 47\,\mathrm{kK}$, which is 15\,kK higher than the temperature derived by these authors. At stellar temperatures below 47\,kK, our models overpredict the \ion{He}{i}/\ion{He}{ii} line ratio. These differences can be attributed to the line-blanketing effect. The luminosity derived in this work is a factor of two higher compared to the results obtained by \citet{Crowther1997}, which is attributable to the different bolometric correction due to the higher stellar temperature. The mass-loss rate is nearly the same compared to the previous estimate by \citet{Crowther1997}. Since the HST spectrum does not cover the H$\alpha$ and H$\beta$ lines, the hydrogen abundance is derived from the higher Balmer members alone. 

BAT99\,100 was found by \citet{Guerrero2008I} to show X-ray emission. Thus, we consider this object to be a binary suspect, although no radial velocity variations were discovered by \citet{Schnurr2008}. The noteworthy infrared excess of this object might be attributed to a hidden companion. 

\begin{figure*}
  \centering
  \includegraphics[width=0.46\hsize]{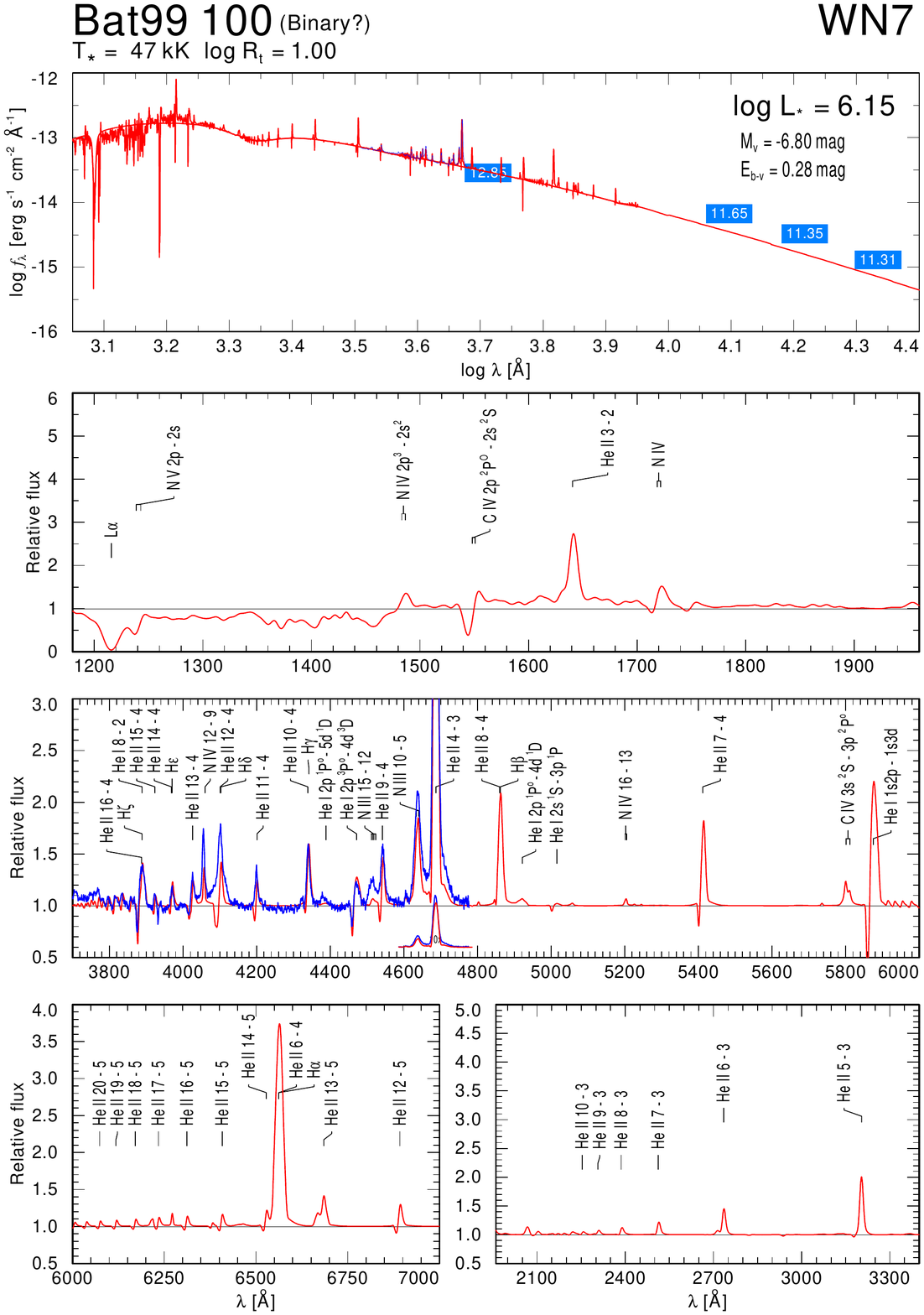}
  \qquad
  \includegraphics[width=0.46\hsize]{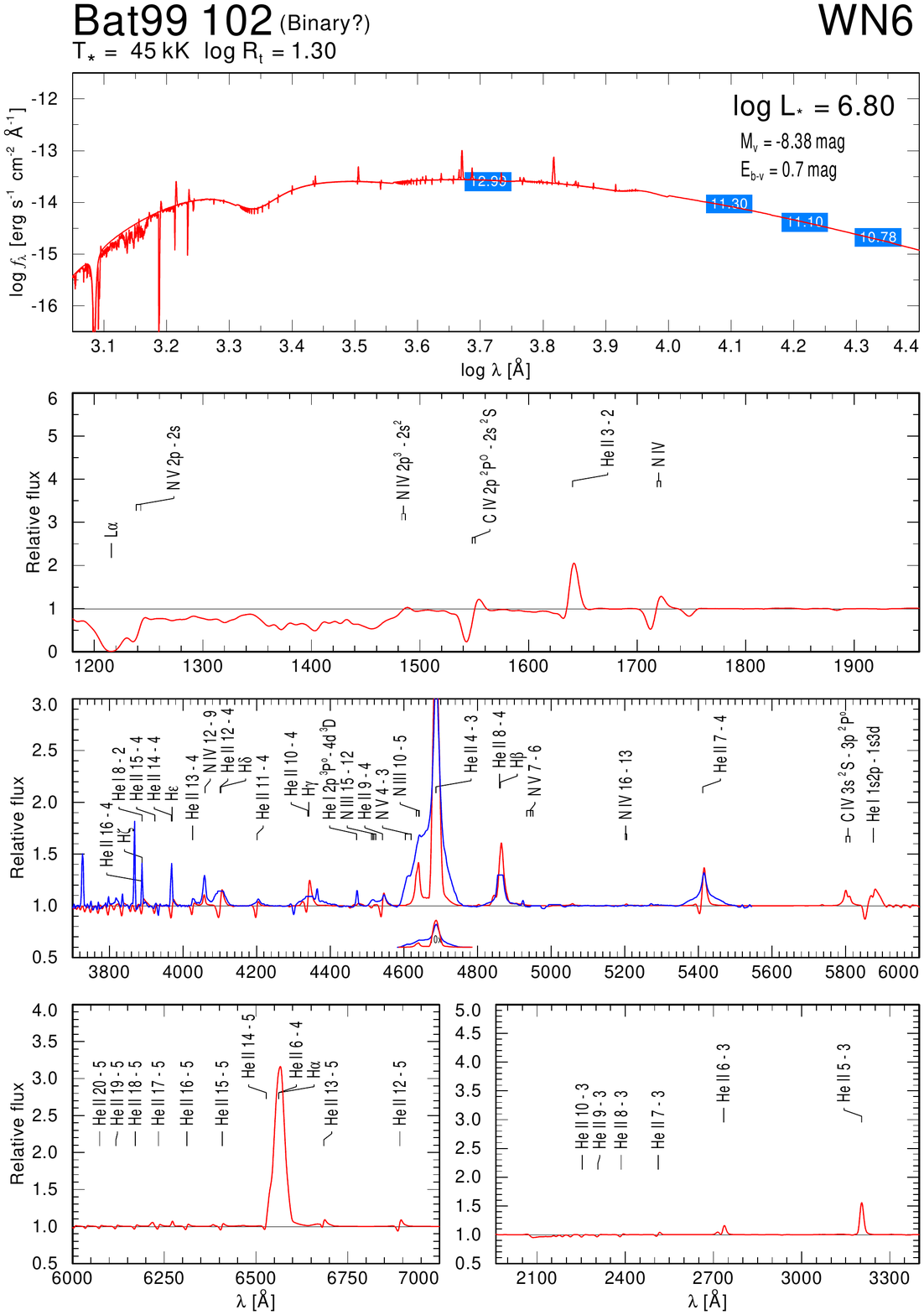}
  \vspace{-0.4cm}
  \caption{Spectral fit for BAT99\,100 and BAT99\,102}
  \label{fig:bat100}
  \label{fig:bat102}
\end{figure*}

\paragraph{BAT99\,102} was classified as WN6 by \citet{Schnurr2008}. According to these authors, the spectra are contaminated by the flux of the WC star BAT99\,101, since it was not possible to resolve these close objects even under the best seeing conditions \citep{Schnurr2008}. Unfortunately, this is the only spectrum available to us. Since this spectrum lacks a sufficient background subtraction, we are not able to derive a reliable hydrogen abundance. Moreover, the flux contribution of BAT99\,101 to the total flux is unknown but not negligible, since the broad line wings of the \ion{He}{ii}\,$\lambda\, 4686$ line and \ion{He}{ii}\,$\lambda\, 5412$ line (Fig.\,\ref{fig:bat102}) probably originate from the WC star. Therefore, the physical parameters listed in Table\,\ref{table:parameters} need to be taken with caution.

According to \citet{Guerrero2008I}, BAT99\,101, together with BAT99\,102 is one of the brightest X-ray source in 30\,Doradus. Unfortunately, the {\em Chandra} ACIS instrument is not able to resolve BAT99\,101 and 102, so that the X-ray emission cannot be attributed to one of these stars alone. The {\em ROSAT} HRI observation analyzed by \citet{Guerrero2008II} results in the same conclusion, even though BAT99\,101--103 could not be resolved into individual objects by {\em ROSAT}. \citet{Moffat1987} found radial velocity variations with a period of 2.76\,d for BAT99\,102, whereas \citet{Schnurr2008} found the same period for the nearby BAT99\,103 instead of for BAT99\,102. Until the binary status is confirmed, we consider this star as a binary suspect. 

\paragraph{BAT99\,103} is a WN5(h)+O binary \citep{Evans2011} located in the direct neighborhood of BAT99\,101 and 102. This star was identified as a binary with a period of 2.76\,d by \citet{Schnurr2008}. Tentative X-ray emission was reported by \citet{Guerrero2008I}. 

\begin{figure*}
  \centering
  \includegraphics[width=0.46\hsize]{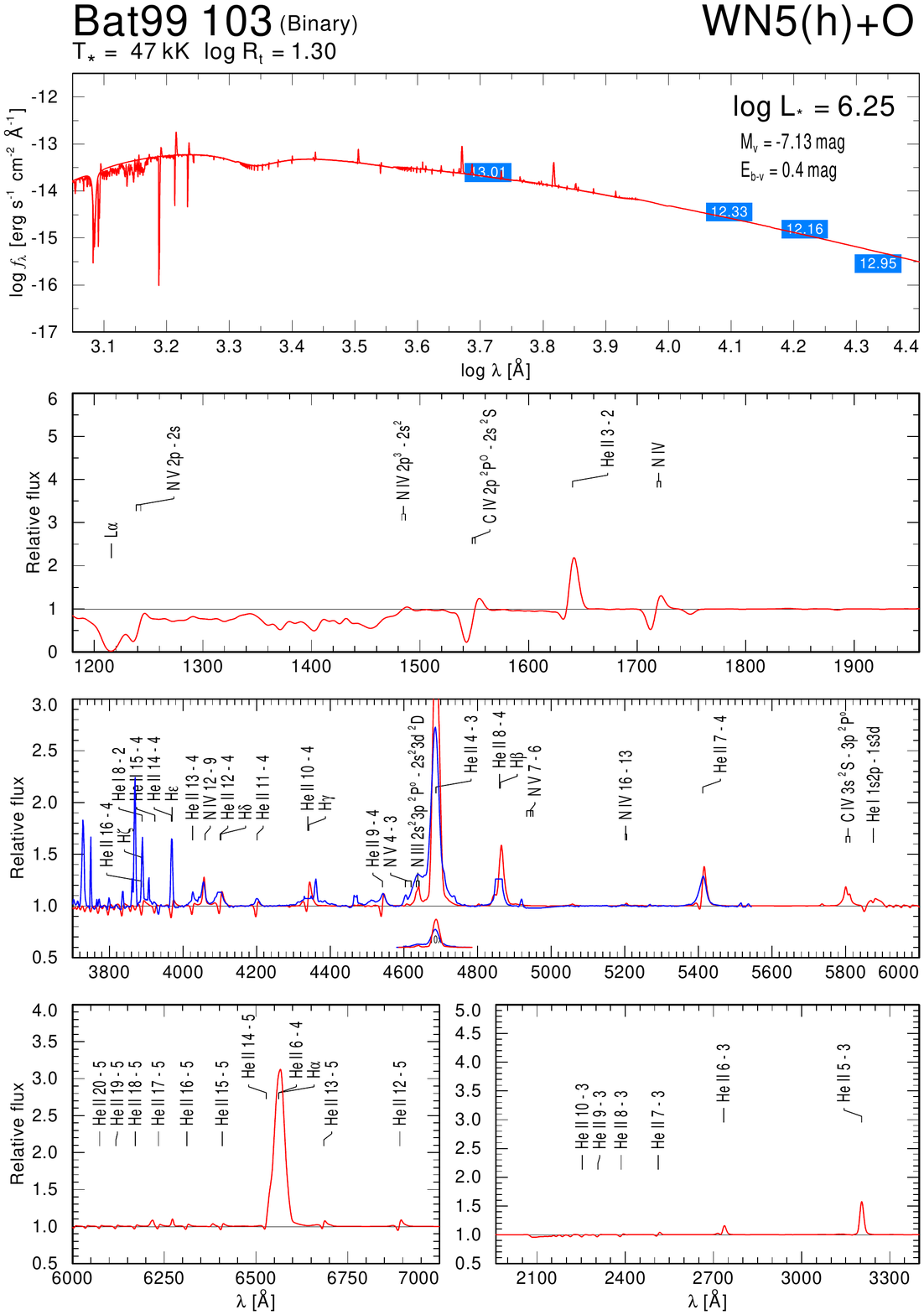}
  \qquad
  \includegraphics[width=0.46\hsize]{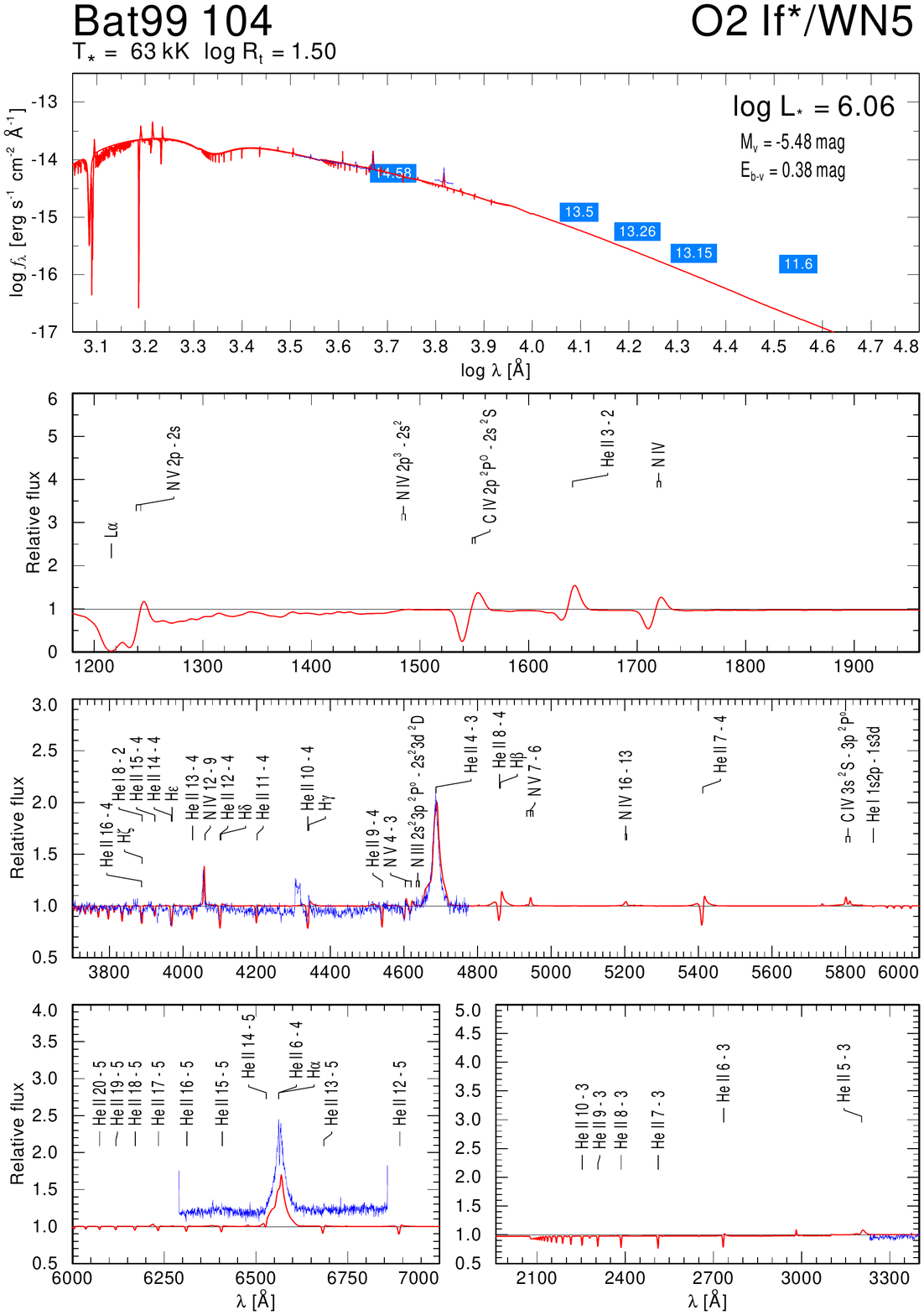}
  \vspace{-0.4cm}
  \caption{Spectral fit for BAT99\,103 and BAT99\,104}
  \label{fig:bat103}
  \label{fig:bat104}
\end{figure*}

Unfortunately, no background subtraction was applied to the spectrum shown in Fig.\,\ref{fig:bat103}. Since this is the only spectrum available to us, we are not able to derive a meaningful hydrogen abundance for this object. 

\paragraph{BAT99\,104} is a O2\,If*/WN5 transition type star located close to the center of 30\,Doradus. We have three optical spectra at hand, two archival HST spectra (H$\alpha$ and $3200-4800$\,{\AA}) and one spectrum obtained by \citet{Schnurr2008}. Considerable differences can be detected between these data. For example, the HST spectrum exhibits significantly higher emission-line strengths of the \ion{He}{ii}\,$\lambda\, 4686$ and the \ion{N}{iv}\,$\lambda\, 4060$ lines. These observational discrepancies might result from a nearby source that contaminated these observations. Due to the higher spatial resolution of the HST compared to the ground-based telescopes, we rely primarily on HST data. 

The optical narrowband magnitude listed in the BAT99 catalog is in excellent agreement with the HST spectrum, which covers the spectral range from 3200\,{\AA} to 4800\,{\AA}. Another HST spectrum covering H$\alpha$, on the other hand, exhibits a flux of approximately 0.1\,dex higher. Thus, the uncertainty in the derived luminosity (Table\,\ref{table:parameters}) is higher than for the other stars in our sample.

For the infrared part of the SED, we used the photometric data obtained by \citet{Kato2007}, instead of the low quality 2MASS data. With the exception of the J-band magnitude, the photometric data measured by \citet{Kato2007} appear to be unaffected by nearby sources. However, in comparison to the optical data, we find an infrared excess that might be caused by nearby sources, a hidden companion, or dust emission. 

\paragraph{BAT99\,105} is listed as a transition-type star in the BAT99 catalog, but has been demoted to O2\,If* by \citet{Crowther2011}. We have UV as well as optical spectra at hand for this star. The optical observations (an archival HST spectrum and a ground-based spectrum observed by \citealt{Schnurr2008}) possibly show two different objects. For example, the HST spectrum exhibits the \ion{He}{ii}\,$\lambda\, 4201, 4542$ lines in absorption, whereas these lines are in emission in the spectrum obtained by \citet{Schnurr2008}. Moreover, the \ion{He}{ii}\,$\lambda\, 4686$ and \ion{N}{iii}\,$\lambda\, 4640$ lines are appreciably stronger in the latter. Since the HST spectrum exhibits the same appearance as the UVES spectrum shown by \citet{Crowther2011}, we choose to rely on the HST data in the optical spectral range. 

BAT99\,105 is suspected to be a binary due to the X-ray emission detected by \citet{Guerrero2008I}, although no significant radial velocity variations were detected by \citet{Schnurr2008}. We note that the optical HST and the IUE short-wavelength spectrum can be consistently reproduced by the same single-star model.

This object was intensively analyzed by \citet{Heap1991}, \citet{Pauldrach1994}, \citet{deKoter1997}, and \citet{Doran2011}. On the basis of modern stellar atmosphere models, \citet{Doran2011} derived a stellar temperature of $T_* = 49.8\,\mathrm{kK}$, which agrees with our own results. In comparison to the latest comprehensive analysis by \citet{deKoter1997}, our fit results in a 5\,kK higher stellar temperature, an identical luminosity, but a considerably lower mass-loss rate (a factor of 3.6 lower). \citet{Pauldrach1994} have derived a mass-loss rate of similar extent as \citet{deKoter1997}. However, a mass-loss rate as high as derived by \citet{Pauldrach1994} and \citet{deKoter1997} results in a considerable overprediction of the emission lines. The studies of \citet{Pauldrach1994} and \citet{deKoter1997} were based on UV spectra alone. 

\paragraph{BAT99\,106} is a WN5h star \citepalias{BAT99} located in the core of R\,136. This star was studied by \citet{deKoter1997}, \citet{Crowther1998}, and \citet{Crowther2010}. \citet{Crowther2010} report it to be one of the most massive stars known so far. Our independent analysis basically confirms the physical parameters derived by \citet{Crowther2010}. 

X-ray emission was detected by \citet{Guerrero2008I} for BAT99\,106, 108, 109, and 110 with the {\em Chandra} satellite. Since these stars in the tight cluster R136 cannot be resolved by this instrument, we treat BAT99\,106 as a single star, although the X-ray emission might be associated with it and indicate colliding winds in a binary system.

We note that in addition to the photometry given in Sect.\,\ref{subsect:data}, we used $b$-band photometry  \citep{Crowther1998} and ${K_S}$-band photometry \citep{Crowther2010} in the SED fit.

\paragraph{BAT99\,107} has been identified by \citet{Taylor2011} as a massive SB2 binary system consisting of two O-type stars. \citet{Moffat1989} found a radial velocity variation with a period of 52.7\,d. However, \citet{Schnurr2008} could not confirm this period. 

\begin{figure*}
  \centering
  \includegraphics[width=0.46\hsize]{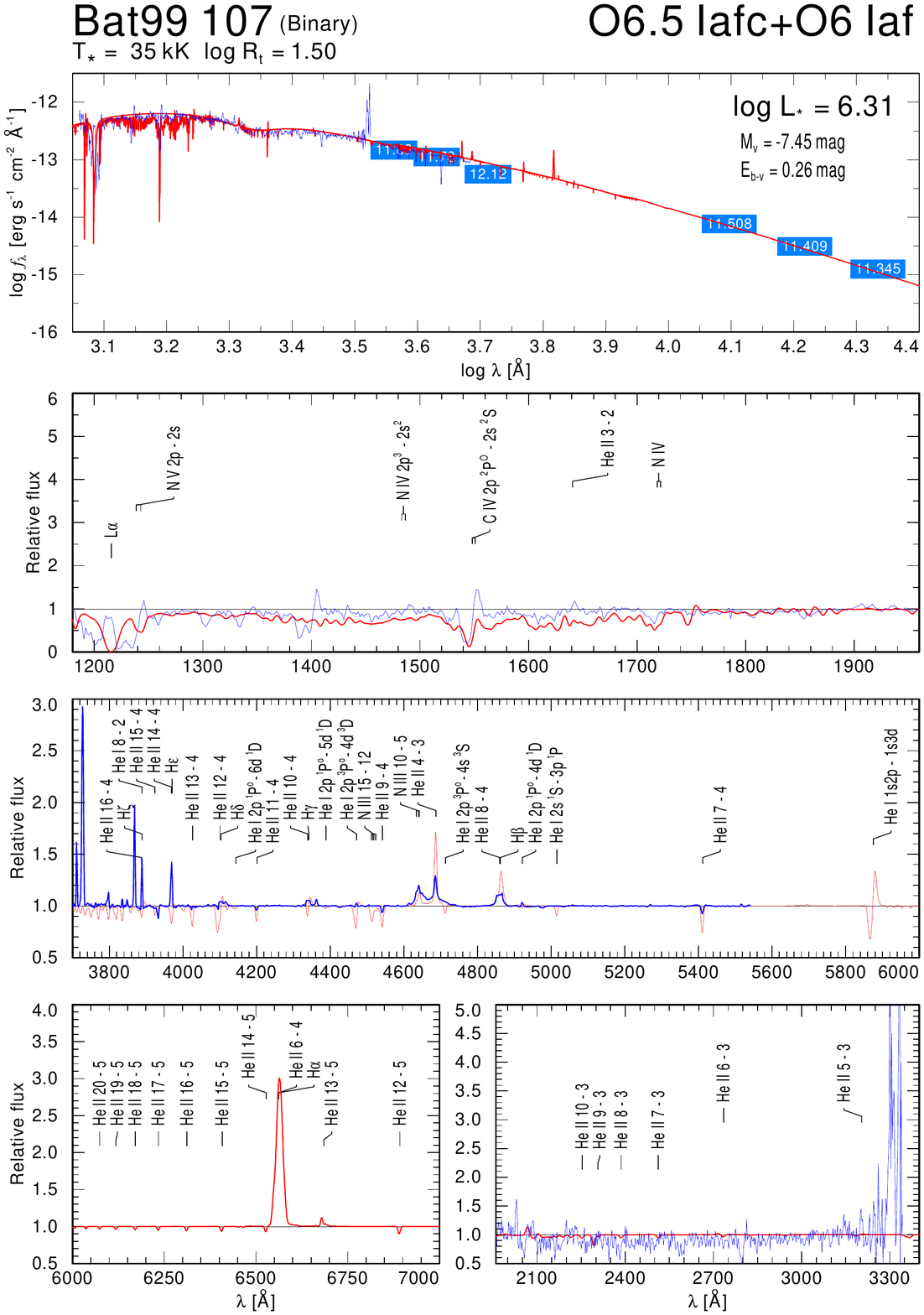}
  \qquad
  \includegraphics[width=0.46\hsize]{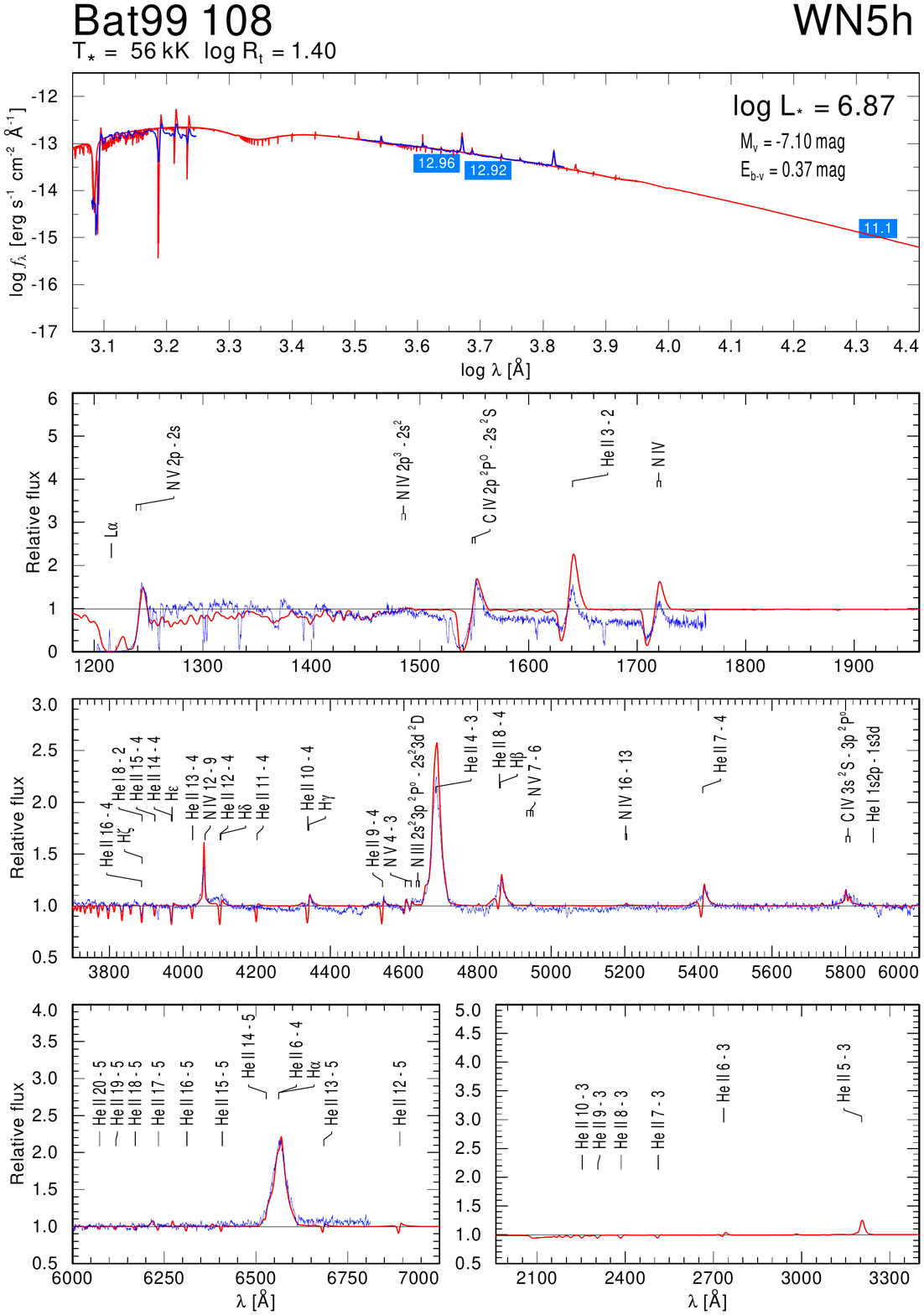}
  \vspace{-0.4cm}
  \caption{Spectral fit for BAT99\,107 and BAT99\,108}
  \label{fig:bat107}
  \label{fig:bat108}
\end{figure*}

\paragraph{BAT99\,108} is the most massive star in the core of R\,136 \citep{Crowther2010}. It is listed as WN5h star in the BAT99 catalog and has been analyzed by \citet{deKoter1997}, \citet{Crowther1998}, and \citet{Crowther2010}. In comparison to the most recent analysis by \citet{Crowther2010}, we obtained fairly similar stellar parameters. 
Note that our SED fit (Fig.\,\ref{fig:bat108}) matches the HST spectra (UV and optical) consistently with the $K_S$-band photometry \citep{Crowther2010}. We ignore the optical photometry (\citetalias{BAT99}, \citealt{Crowther1998}), which is inconsistent with the calibrated HST spectrum.
X-ray emission is associated with BAT99\,108 (cf. comment on BAT99\,106).

\paragraph{BAT99\,109} is another WN5h star \citepalias{BAT99} in the core of R\,136, previously analyzed by \citet{Crowther1998} and \citet{Crowther2010}.
We have UV and optical HST spectra for this object. According to \citet{deKoter1997}, the HST spectra of BAT99\,109 are contaminated by the flux of BAT99\,108 situated only $0.1\arcsec$ away. Since the optical HST spectrum is less affected by this contamination \citep{deKoter1997}, we primarily rely on this spectrum.  

This contamination is probably the reason that our SED fit cannot simultaneously reproduce the continuum slope of the UV and optical HST spectrum. However, the optical spectrum can be matched in conformity with ${K_S}$-band photometry \citep{Crowther2010}. Nevertheless, the luminosity of this star is subject to a large uncertainty, since the optical HST spectrum of BAT99\,109 is also contaminated to a certain extent. Despite these uncertainties, we obtained nearly the same stellar parameters as previously derived by \citet{Crowther2010}. With a luminosity of $\log\,(L/L_\odot) = 6.69$, it is one of the most luminous objects in our sample. We point out that a fit of the photometry (\citealt{Crowther1998},\citetalias{BAT99},\citealt{Crowther2010}) alone does not result in a lower luminosity.
Note that X-ray emission is associated with BAT99\,109 (cf. comment on BAT99\,106).

\paragraph{BAT99\,110} was classified as O2\,If* by \citet{Crowther2011}. The preceding studies by \citet{Heap1994} and \citet{deKoter1997} report nearly equal physical parameters. In contrast to these studies, our best fitting model has a stellar temperature of $T_* = 50\,\mathrm{kK}$ which is 7.5\,kK higher. Lower temperatures are excluded by our analysis, since, relative to the observed line strengths, our grid models with lower stellar temperatures overestimate the \ion{N}{iii} lines and underestimates the \ion{N}{iv} and \ion{N}{v} lines. This temperature discrepancy probably arises due to the inclusion of line blanketing in our models, which was not accounted for in the stellar atmosphere models at the time of the earlier studies. Moreover, our new study results in a luminosity which is factor of two higher, while the mass-loss rate is about 40\,\% lower compared to the results obtained by \citet{Heap1994} and \citet{deKoter1997}. We note that it is not possible to reproduce the UV and optical spectrum with the same reddening parameters. A satisfying fit of the UV spectra can only be achieved with a color excess of $E_{b-v}=0.1$ and a luminosity of $\log\,(L/L_\odot) = 5.33$. However, these values are considerably low compared to the results (see Table\,\ref{table:parameters}) derived from the photometry \citep{Crowther1998,BAT99} and the optical HST spectrum. 
Note that X-ray emission is associated with BAT99\,110 (cf. comment on BAT99\,106).

\paragraph{BAT99\,111} is a WN9ha star \citepalias{BAT99} in the center of R\,136. Although \citet{Schnurr2009} could not find radial velocity variations for this object, we treat it as a binary suspect because of the X-ray emission reported by \citet{Townsley2006}. We have two flux-calibrated HST spectra at hand, but we note that the model cannot perfectly reproduce the continuum shape of the UV and the optical spectra with the same reddening parameters. However, the luminosity derived from the UV spectrum is only 0.06\,dex lower than the value (Table\,\ref{table:parameters}) derived from the optical HST spectrum and optical photometry \citepalias{BAT99}. Due to a hydrogen mass-fraction of $X_{\element{H}} = 0.7$, this object appears not to be in an advanced evolution stage. Therefore, we disagree with the conclusion of \citet{Schnurr2009} that this star is more evolved than the other stars in the core of R\,136.

\paragraph{BAT99\,112} is another WN5h star in the core of R\,136 that is a candidate for a long-period binary system \citep{Schnurr2009}. An indirect argument in favor of the binary status is the hard X-ray emission detected by \citet{Townsley2006} and \citet{Guerrero2008I}. We expect that the potential companion does not contribute much to the bolometric luminosity, since no spectral lines of the potential companion can be recognized in the HST spectrum. Thus, we analyzed this star as a single star, despite its pending binary status. 

In the SED fit, the slope of the calibrated HST spectrum and the optical photometry \citepalias{BAT99} can be consistently reproduced with the same luminosity and color excess. In contrast, the ${K_S}$-band photometry \citep{Crowther2010} exhibits a clear excess, which might be caused by the potential companion or dust emission.
We note that the available HST spectrum does not cover the H$\alpha$ and H$\beta$ lines, so that the hydrogen abundance is derived using the higher members of the Balmer series alone. Since these lines are rather weak, the hydrogen abundance is subject to a relatively high uncertainty. 
We estimate a hydrogen mass-fraction of $X_{\element{H}} = 0.2$, which is 0.1\,dex lower than previously derived by \citet{Crowther2010}. The temperature obtained by these authors is slightly lower, while the mass-loss rate and the luminosity is more than 0.2\,dex higher compared to the results of the present paper. 

\paragraph{BAT99\,113} is a transition type O2\,If*/WN5 star \citep{Crowther2011,Evans2011} located close to the core of 30\,Doradus. We have two optical spectra at hand, an archival HST spectrum and a ground-based spectrum taken by \citet{Schnurr2008}. These spectra, however, clearly deviate from each other. For example, the equivalent width of the \ion{He}{ii}\,$\lambda\, 4686$ line differ by roughly a factor of two. Although the signal to noise ratio is lower in the HST spectrum, we primarily use this spectrum in our analysis because of the distinctive spatial resolution of the HST. The star was identified by \citet{Schnurr2008} as a binary system with a period of 4.7\,d. Since no indications of a companion were found in the spectra, we expect that the flux contribution of the companion is insignificant for the analysis of the WN star. Unfortunately, the HST spectra do not cover the H$\alpha$ and H$\beta$ lines. Consequently, the hydrogen abundance given in Table\,\ref{table:parameters} is derived from weak H\,$\gamma$ and H\,$\delta$ lines alone. 

\paragraph{BAT99\,114} is another transition type O2\,If*/WN5 star \citep{Crowther2011,Evans2011} in the vicinity of the 30\,Doradus core. We have two optical spectra at hand, an archival HST spectrum and a spectrum obtained by \citet{Schnurr2008}. Since the latter lacks a sufficient subtraction of the diffuse background, our analysis is mainly based on the HST spectrum. However, this spectrum covers only the wavelength range from 3300 to 4800\,{\AA}, where merely the H\,$\gamma$ and H\,$\delta$ lines can be found as indicators for the hydrogen abundance. 
Due to the X-ray emission detected by \citet{Guerrero2008I} we consider this object as a binary suspect, although \citet{Schnurr2008} could not find periodic radial velocity variations. 

\paragraph{BAT99\,116} was classified as WN5h:a by \citet{Schnurr2008}. These authors have reported radial velocity variations, but found no periodicity. \citet{Schnurr2009} noted that this object is likely a long periodic binary system, in agreement with the strong X-ray emission detected by \citet{Guerrero2008I,Guerrero2008II}. Thus, we consider this object as a binary suspect. Unfortunately, the only spectrum at hand \citep{Schnurr2008} lacks a subtraction of the diffuse background. Therefore, we are not able to establish a robust estimate of the hydrogen abundance. 

\paragraph{BAT99\,117} is a WN5ha star \citep{Foellmi2003b} located in the northern part of 30\,Doradus. The stellar temperature and luminosity derived in our analysis are moderately lower compared to the results obtained by \citetalias{HK2000}, whereas the mass-loss rate is 50\,\% higher. We also derive a 0.1\,dex higher hydrogen abundance. However, our analysis suffers from an insufficient subtraction of the diffuse background in the spectrum from \citet{Foellmi2003b}. This spectrum exhibits strong absorption lines in place of the \ion{O}{iii}-nebular emission lines at 4959\,{\AA} and 5007\,{\AA}, which might be caused by an overcorrection of the background. If this is true, the Balmer series will probably be impaired by the inadequate nebular subtraction as well. In this case, the hydrogen abundance listed in Table\,\ref{table:parameters} is a subject to high uncertainty. 

By comparing the HRD position to stellar evolution models performed by \citet{Meynet2005}, we derive an initial mass of roughly $M_{\mathrm{init}} = 120\,M_\odot$. Thus, this star belongs to the category of very massive stars.  \citet{Foellmi2003b} did not find periodic radial velocity variations.

\paragraph{BAT99\,118} is a WN6h star \citepalias{BAT99}, which is treated as a binary candidate in the radial velocity study by \citet{Schnurr2008}. An indirect argument in favor of the binary status is the strong and hard X-ray flux detected by \citet{Guerrero2008I}. New X-shooter observations performed by \citet{Sana2013} revealed it to be a SB2 binary with a mass ratio close to unity. Thus, the system consists of two similar WN stars, which were classified as WN5--6h + WN6--7h by \citet{Sana2013}. We analyzed this system as if it were a single star.

Our best fit is achieved at a stellar temperature of $T_* = 47\,\mathrm{kK}$, which is only marginally higher than the value recently derived by \citet{Doran2013}. The mass-loss rate obtained by these authors agrees well with the value presented in this work. In contrast, \citet{Crowther1998} obtained a mass-loss rate that is a factor of two lower, while the stellar temperature derived by these authors (on the basis of unblanketed model atmospheres) is 10\,kK lower. \citet{Crowther1998} and \citet{Doran2013} obtain luminosities that are lower compared to the value ($\log\,(L/L_\odot) = 6.66$) derived in this work. \citet{Crowther1998} obtained a value that is 0.32\,dex lower, whereas the analysis carried out by \citet{Doran2013} results in a luminosity that is 0.25\,dex lower. As opposed to this, the luminosity estimate by \citet{Sana2013} results in a luminosity that is 0.14\,dex higher compared to our new results. 

These differences need to be considered, if the luminosity is used to derive the current mass of the stellar content. On the basis of their high luminosity, \citet{Sana2013} derived a current mass between $80\,M_\odot$ and $205\,M_\odot$ for each component in BAT99\,118. 
For the initial masses, we obtained about $M_{\mathrm{init}} = 100\,M_\odot$ for each WN component by comparing the empirical HRD position of this object (see Figs.\,\ref{fig:hrd} and \ref{fig:hrd_geneva}) with the stellar evolution tracks by \citet{Meynet2005} and \citet{Yusof2013}. 

A description of the FUSE spectra (not considered in this work) can be found in \citet{Willis2004}. The authors derived a terminal velocity of $v_\infty = 1847\,\mathrm{km\,s^{-1}}$ which is about $250\,\mathrm{km\,s^{-1}}$ higher than the value used for the calculation of our grid models. 

\paragraph{BAT99\,119} is a WN6h star, which is listed as a single-line spectroscopic binary (SB1) in the BAT99 catalog with a period of 25.2\,d \citep{Moffat1989}. In contrast, \citet{Schnurr2008,Schnurr2009b} find a period of 158.8\,d, combining their radial velocity data with that of \citet{Moffat1989} and new polarimetric data. According to \citet{Schnurr2009b}, the companion is most likely an O-type star, although no obvious trace of the companion can be found in the spectrum. 
With the exception of a slightly smaller emission-line strength, the spectrum of BAT99\,119 resembles that of BAT99\,118, which is a SB2 of two similar WN stars \citep{Sana2013}. Considering that the resolving power of our optical spectra is only $R \approx 1000$, we also stress the possibility of a binary system encompassing two WN stars with a mass ratio close to unity.

Assuming an O-type companion, \citet{Schnurr2009b} were able to give constraints on the properties of both components in this binary system. They found the WN star to be the considerably more luminous component. Thus, we estimate the flux contribution of the companion to be negligible in the UV and optical spectral range, which is in accordance with the moderate infrared excess found in our analysis. 

Similar to BAT99\,118, we derive a stellar temperature of $T_* = 47\,\mathrm{kK}$, which is about 15\,kK higher than that obtained by \citet{Crowther1997} on the basis of unblanketed model atmospheres. In comparison to this former study, our substantially higher temperature entails a luminosity increase by a factor of roughly 2.5 ($\log\,(L/L_\odot) = 6.57$). The mass-loss rate derived here, on the other hand, is nearly identical to the value given by \citet{Crowther1997}, whereas \citet{Schnurr2009b} estimated a mass-loss rate on the basis of their polarimetric data, which is a factor of two higher. 

The latter authors derived a dynamical mass of $M_{\mathrm{dyn}} = 116 \pm 33\,M_\odot$ for the WN component.
The initial mass obtained from the HRD position (Figs.\,\ref{fig:hrd} and \ref{fig:hrd_geneva}) will be approximately $M_{\mathrm{init}} = 150\,M_\odot$, if the WN star contributes most to the overall flux of the binary system (WN + OB). 

\paragraph{BAT99\,120} is classified as WN9h star \citepalias{BAT99} and may be a dormant LBV, according to \citet{Crowther1995b}. We have two optical spectra at hand, an archival AAT spectrum (see Sect.\,\ref{subsect:data}) and a coadded spectrum observed by \citet{Foellmi2003b}. 
The stellar parameters presented in Table\,\ref{table:parameters} rely mainly on the latter spectrum, due to their high signal to noise ratio. This spectrum is best reproduced by a model with a stellar temperature of $T_* = 32\,\mathrm{kK}$, while a model with $T_* = 35\,\mathrm{kK}$ is more appropriate for the AAT spectrum. BAT99\,120 was previously analyzed by \citet{Pasquali1997} and \citet{Crowther1995b}. They obtained stellar temperatures of $T_* = 38.9\,\mathrm{kK}$ and  $T_* = 30\,\mathrm{kK}$, respectively. However, stellar temperatures higher than $T_* = 35\,\mathrm{kK}$ would spoil the fit of the \ion{He}{i} and \ion{He}{ii} lines in both optical spectra. 

In addition to the optical spectra, we used flux-calibrated UV spectra, which were obtained with the HST and the IUE satellite. Fitting the continua of these spectra and the available photometric data (2MASS and optical photometry from the \citetalias{BAT99} catalog), the luminosity is found to be $\log\,(L/L_\odot) = 5.58$, while the color excess amounts to $E_{b-v}=0.15$. In contrast, the luminosities presented by \citet{Crowther1995b} and \citet{Pasquali1997} are higher, while the derived color excess agrees with our study. The luminosity derived by \citet{Crowther1995b} is only slightly higher, whereas  \citet{Pasquali1997} derived a luminosity of a factor of 2.5 higher. This deviation can be attributed the higher temperature and thus higher bolometric correction derived by \citet{Pasquali1997}. In comparison to the former studies, the mass-loss rate is slightly lower in our new study. 

\begin{figure*}
  \centering
  \includegraphics[width=0.46\hsize]{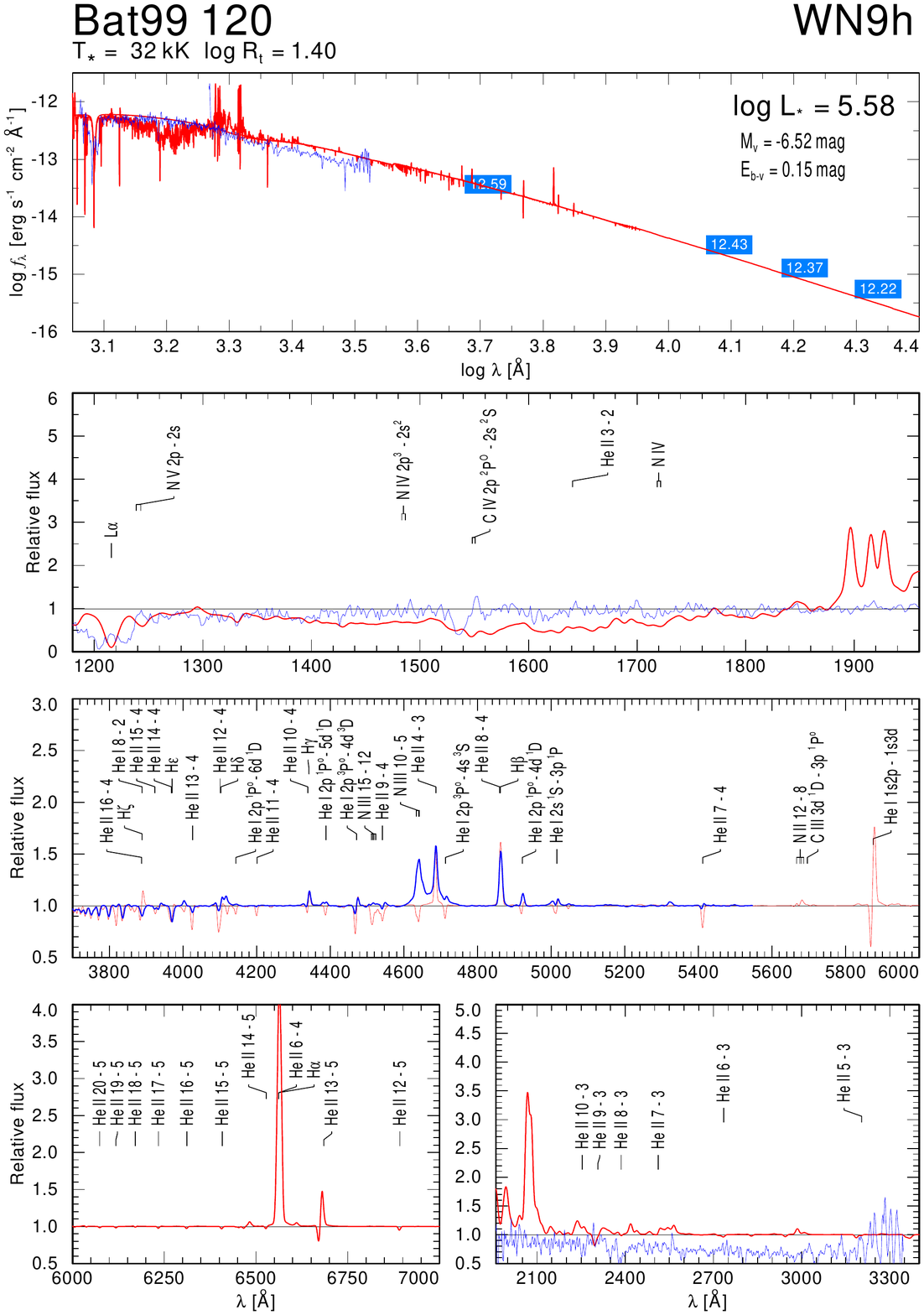}
  \qquad
  \includegraphics[width=0.46\hsize]{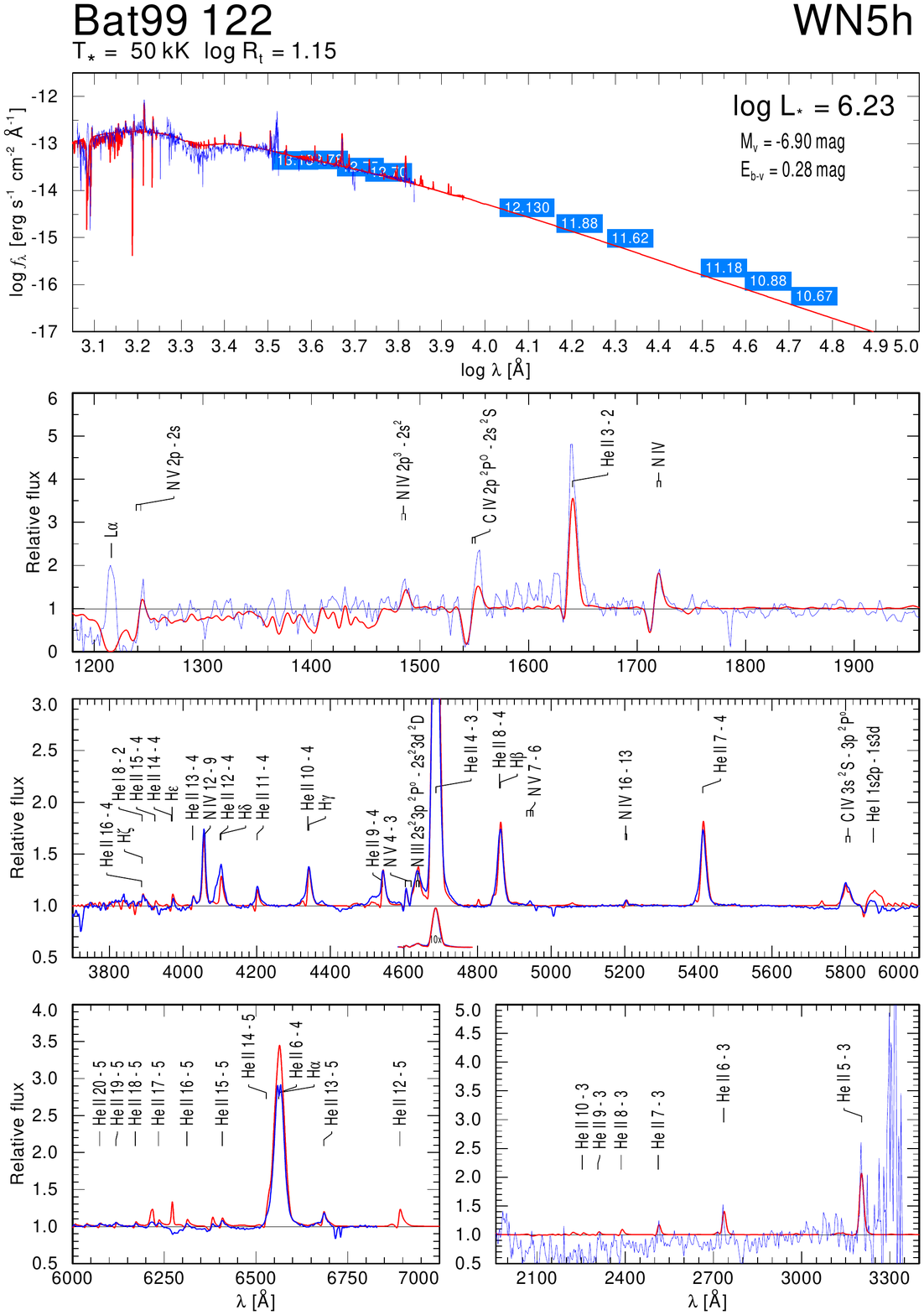}
  \vspace{-0.4cm}
  \caption{Spectral fit for BAT99\,120 and BAT99\,122}
  \label{fig:bat120}
  \label{fig:bat122}
\end{figure*}

\paragraph{BAT99\,122} had never been analyzed by means of model atmospheres before. The infrared excess reported by \citet{Hyland1978} can be seen in our fit of the SED (Fig.\,\ref{fig:bat122}) as well. The star was classified as WN5h by \citet{Evans2011}.

\paragraph{BAT99\,124} belongs to the WN4 subclass \citep{Foellmi2003b}. It is analyzed by means of model atmospheres for the first time in this work. The uncertainty of the obtained hydrogen abundance is large, since the available spectrum is strongly contaminated with nebular emission, which is evident by the strong \ion{O}{iii}\,$\lambda\lambda\, 4959, 5007$ nebular emission lines. According to \citet{MartinHernandez2005}, the shell structure of NGC\,2077 may be caused by the feedback of BAT99\,124. 

\paragraph{BAT99\,126} is listed in the BAT99 catalog as a WN3+O7 binary candidate. \citet{Foellmi2003b} found a period of 25.5\,d, but noted that more data are needed to verify this result. Therefore, we treat this object as a binary suspect, although it is likely a binary. The X-ray emission detected by \citet{Guerrero2008I} is further indirect evidence for the binary status. \citet{Foellmi2003b} reclassified the companion to O8 and the WN component to WN4b. The luminosity of this object is derived from photometry only, since no flux-calibrated spectra are available. 

\paragraph{BAT99\,128,} classified as WN3b \citep{Foellmi2003b}, is a typical WN3 star. An observational discrepancy exists for this star between the photometry obtained by \citet{Crowther2006b} and the flux-calibrated spectrum obtained by \citet{Torres-Dodgen1988}. The noisy spectrum measured by \citet{Torres-Dodgen1988} exhibits a higher flux, which results in a SED fit of only moderate quality and an unreliably low color excess of $E_{b-v}=0.01\,\mathrm{mag}$. The spectrophotometry from \citet{Crowther2006b} fits much better to the 2MASS and the IRAC photometry and results in a SED fit of higher quality and a more convincing color excess of $E_{b-v}=0.17\,\mathrm{mag}$. \citet{Foellmi2003b} found a radial velocity that is significantly below the mean $v_\mathrm{rad}$ of their sample, suggesting that this star might be a runaway. We note that BAT99\,128 falls into the regime of parameter degeneracy (cf.\ Sect. \ref{subsec:modelgrid}).

\paragraph{BAT99\,129} is an eclipsing binary with a WN3(h)a star \citepalias{BAT99} as primary component and an O5V companion \citep{Foellmi2006}. \citet{Foellmi2003b} find a radial velocity period of 2.76\,d, but no X-ray emission was detected by the {\em Rosat} satellite \citep{Guerrero2008II}. We estimate a hydrogen mass-fraction of $X_{\element{H}} = 0.2$, thus confirming the above classification. \citet{Foellmi2006} derived a luminosity of $\log\,(L/L_\odot) =  4.97$ for the WR component, which would make this the faintest WN star known in the LMC (cf.\ Fig.\,\ref{fig:hrd}). This result cannot be confirmed by our analysis. Assuming that the luminosity derived in our analysis is valid for the whole system, and applying the luminosity ratio of 0.3 derived by \citet{Foellmi2006}, we obtain a luminosity of $\log\,(L/L_\odot) =  5.68$ for the WN component. This luminosity is considerably higher than the value derived by \citet{Foellmi2006} and at the upper end of the luminosity range derived for the other presumably single WN3 stars in the LMC.

\paragraph{BAT99\,130} is the second WN11h star \citepalias{BAT99} in our sample. Our new analysis confirms the stellar parameters derived in the former study by \citet{Crowther1997} with the exception of a significantly lower hydrogen abundance.

\paragraph{BAT99\,131} was classified as WN4b \citep{Foellmi2003b} and had not been analyzed before. The available IUE spectra are not uniform and do not exhibit any prominent emission line. Therefore, we reject these spectra from our analysis, although the flux of the IUE long-wavelength spectrum is compatible to the available photometry. Thus, the luminosity obtained in our analysis is derived from visual narrowband, 2MASS and IRAC photometry alone. 

\paragraph{BAT99\,132} is a WN4b(h) star \citep{Foellmi2003b}, analyzed for the first time in this paper. The best fit of the spectra is achieved with a hydrogen-free model, although the presence of residual hydrogen in the stellar atmosphere of this object was reported by \citet{Foellmi2003b}.

\paragraph{BAT99\,133} is the third WN11h \citepalias{BAT99} in the LMC and one of the only three WN stars detected at $24\,\mu\mathrm{m}$ with the IRAC instrument aboard the Spitzer space telescope \citep{Bonanos2009}. According to \citet{Humphreys1994} and \citet{Weis2003}, this star is suspected to be an LBV in its quiescent phase. Likewise, \citet{Walborn1982} and \citet{Bonanos2009} have noted the spectroscopic similarities between BAT99\,133 and the LBV BAT99\,83 in its minimum. Contrary to this, our spectra of these two stars exhibit clear differences. For example, the \ion{He}{ii}\,$\lambda\, 4686$ line is absent in the spectrum of BAT99\,83, whereas a relatively small emission line is present in the spectrum of BAT99\,133. Further different features are the \ion{He}{i} lines, which are much more prominent in the spectrum of BAT99\,133. 

BAT99\,133 was previously analyzed by \citet{Crowther1997} and \citet{Pasquali1997}. The former obtained a stellar temperature of $T_* = 28.3\,\mathrm{kK}$, which is confirmed by our ana\-lysis. \citet{Pasquali1997}, on the other hand, derived a stellar temperature of roughly 8\,kK higher. However, our models clearly underpredict the observed \ion{He}{i}/\ion{He}{ii} ratio at this higher temperature. In principle, the same applies to the mass-loss rate and luminosity, where we can confirm the results obtained by \citet{Crowther1997}. Contrary to this, \citet{Pasquali1997} derived values for the mass-loss rate and the luminosity that are roughly twice as high. 

A study of the nebula associated with BAT99\,133 can be found in \citet{Pasquali1999} and \citet{Weis2003}.

\paragraph{BAT99\,134} is listed as WN4b star in the BAT99 catalog. In this first spectroscopic analysis with stellar atmosphere models, we derived physical parameters typical for the WN4 subclass. 

\citet{Dopita1994} discovered a ring nebula that surrounds BAT99\,134. As no \ion{He}{ii} nebular emission is detected by \citep{Naze2003b}, they obtain an upper limit for the number of \ion{He}{ii} ionizing photons delivered by the exciting star, which amounts to $< 3.2 \cdot 10^{45}$\,\ion{He}{ii} ionizing photons per second. This agrees with our final model, which does not produce a significant number of \ion{He}{ii} ionizing photons (see Table\,\ref{table:zansT}).

\section{Spectral fits}
\label{sec:specfits}

In this section, we present the spectral fits of all stars analyzed in this study. The individual plots encompass the fit of the spectral energy
distribution (top panel) to the photometric and flux-calibrated spectra as well as the fits to the normalized optical and UV spectra (lower panels), when available. The observations are plotted in blue, whereas the synthetic spectrum of the best-fitting model shown in red. 

Some of our stellar atmosphere models with stellar temperatures below $T_* = 32\,\mathrm{kK}$ exhibit spurious emission lines in the spectral range from about 1900\,{\AA} to 2100\,{\AA}. These emission features, which are not observed, originates from the third ionization stage of our generic model atom re\-presenting the iron-group elements. We note that the presence of these emission features is only a cosmetic issue and has no impact on the derived stellar parameters.

\clearpage

\begin{figure*}
  \centering
  \includegraphics[width=0.46\hsize]{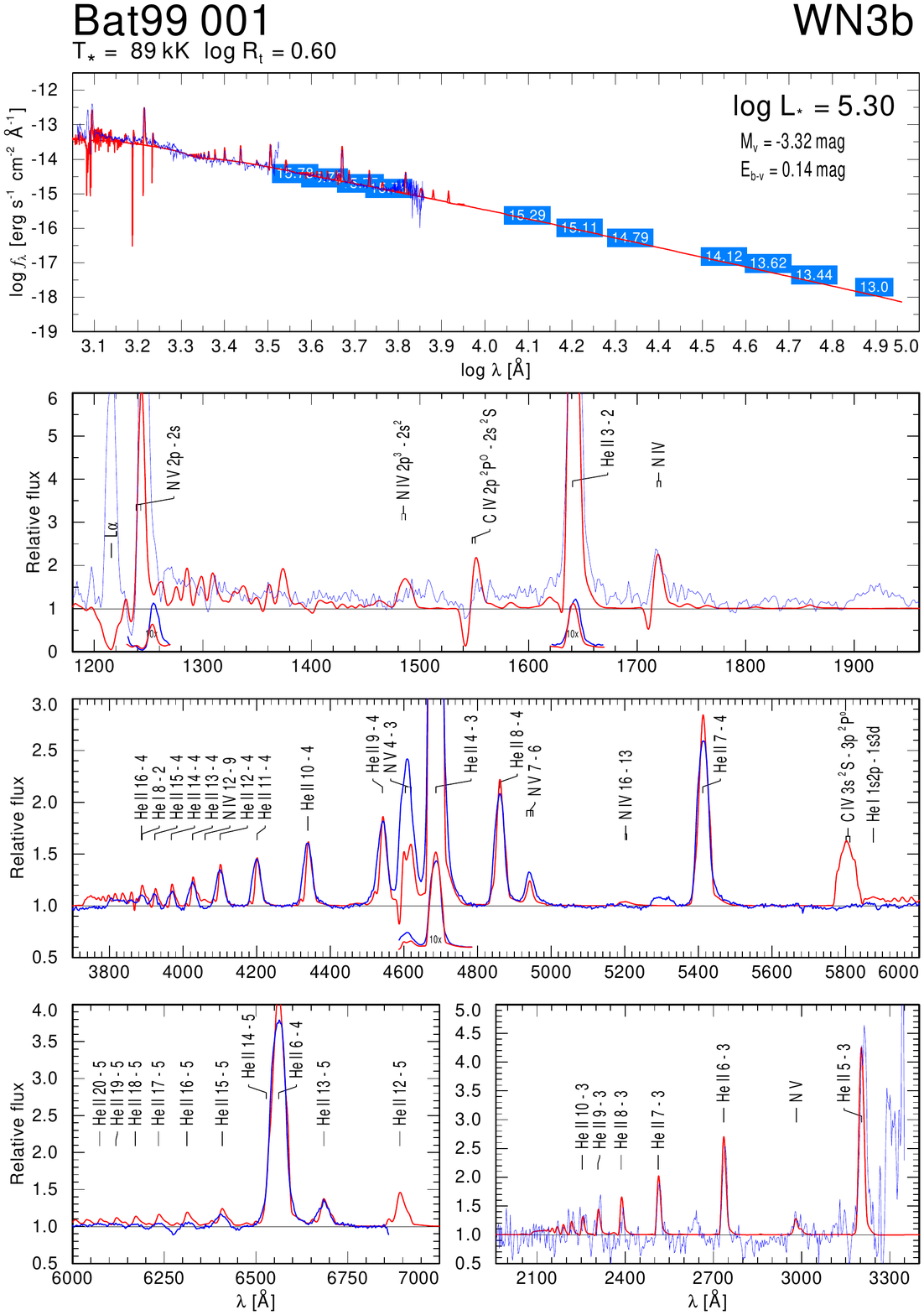}
  \qquad
  \includegraphics[width=0.46\hsize]{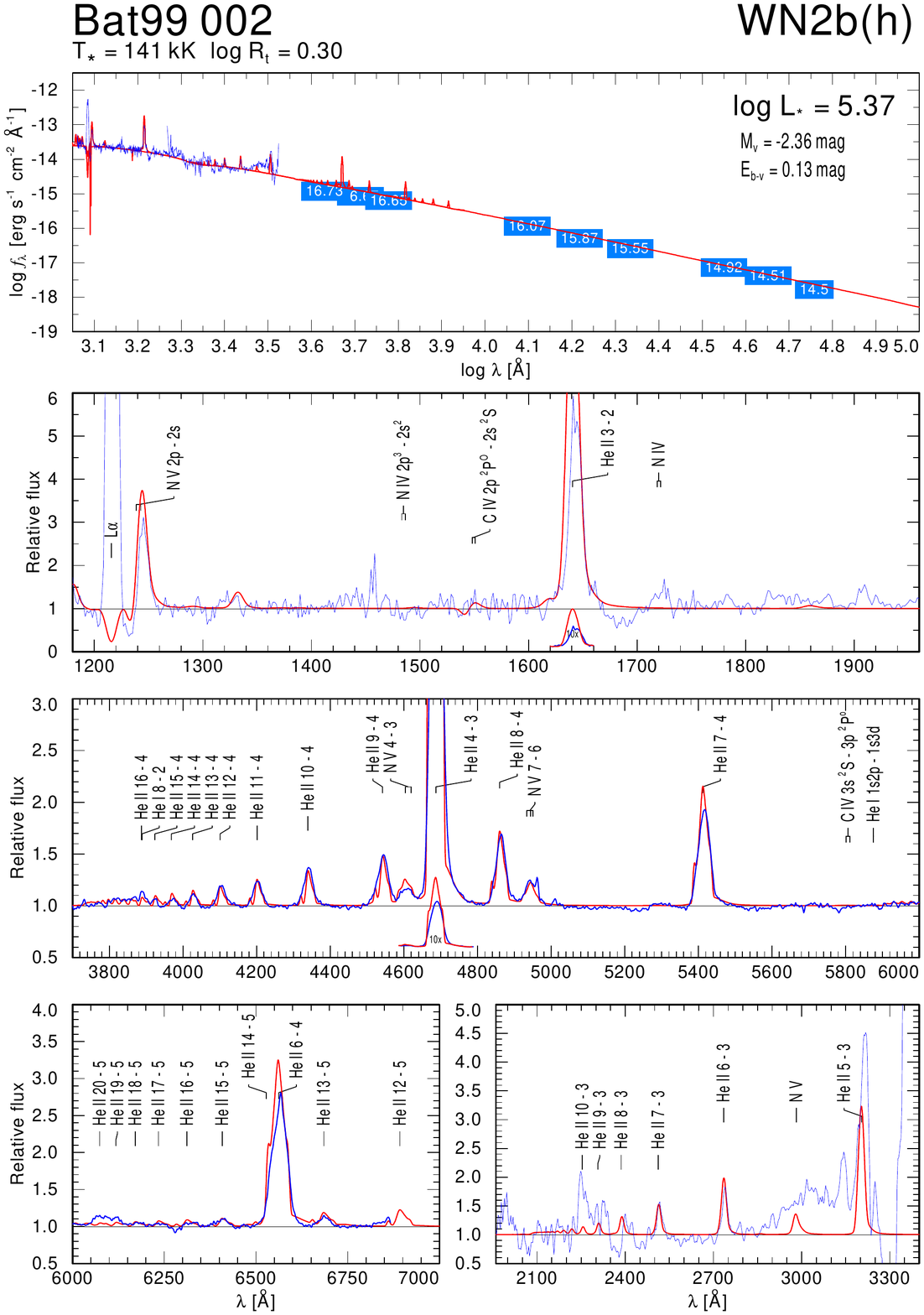}
  \vspace{-0.4cm}
  \caption{Spectral fit for BAT99\,001 and BAT99\,002}
  \label{fig:bat001}
  \label{fig:bat002}
\end{figure*}

\begin{figure*}
  \centering
  \includegraphics[width=0.46\hsize]{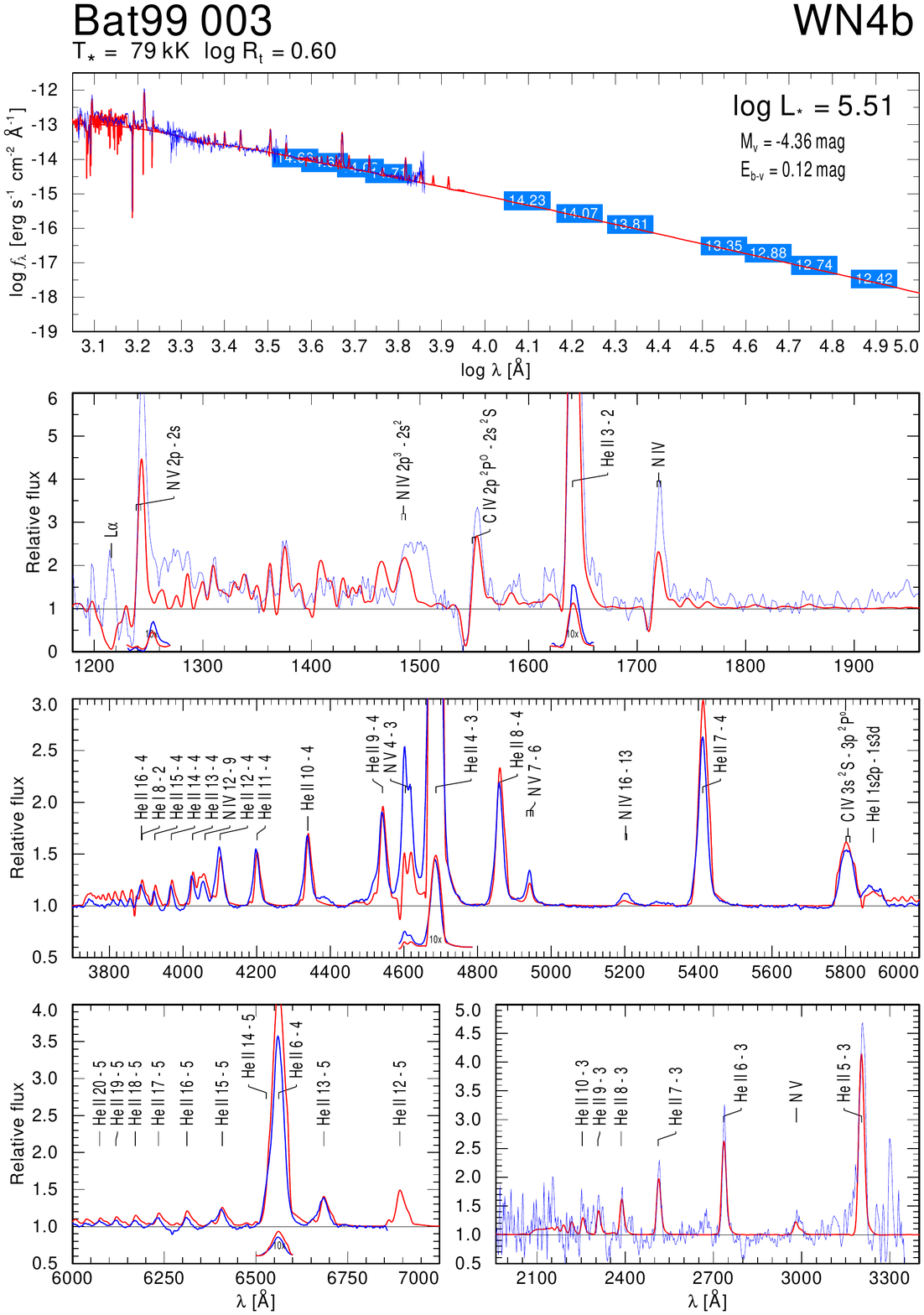}
  \qquad
  \includegraphics[width=0.46\hsize]{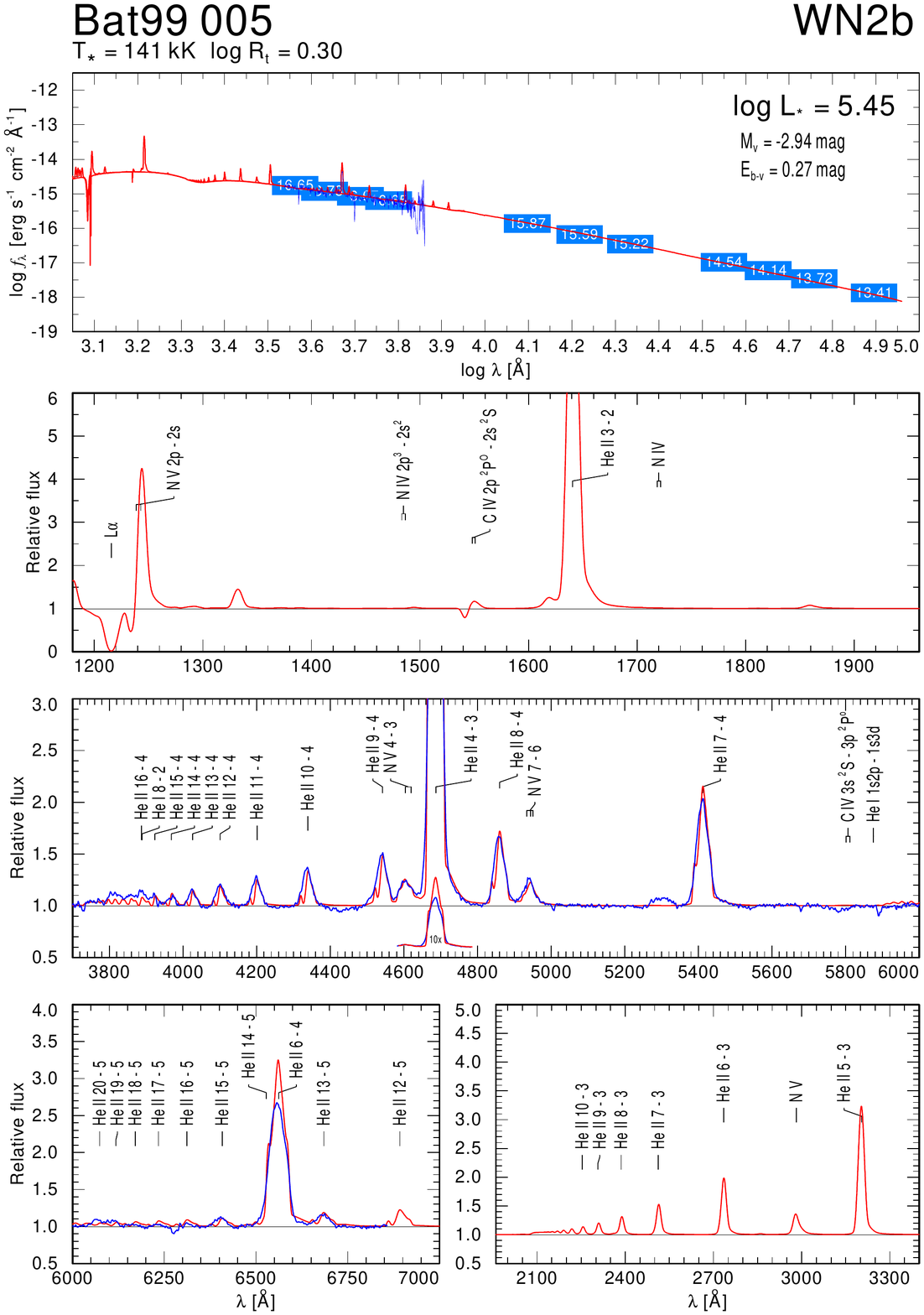}
  \vspace{-0.4cm}
  \caption{Spectral fit for BAT99\,003 and BAT99\,005}
  \label{fig:bat003}
  \label{fig:bat005}
\end{figure*}

\clearpage

\begin{figure*}
  \centering
  \includegraphics[width=0.46\hsize]{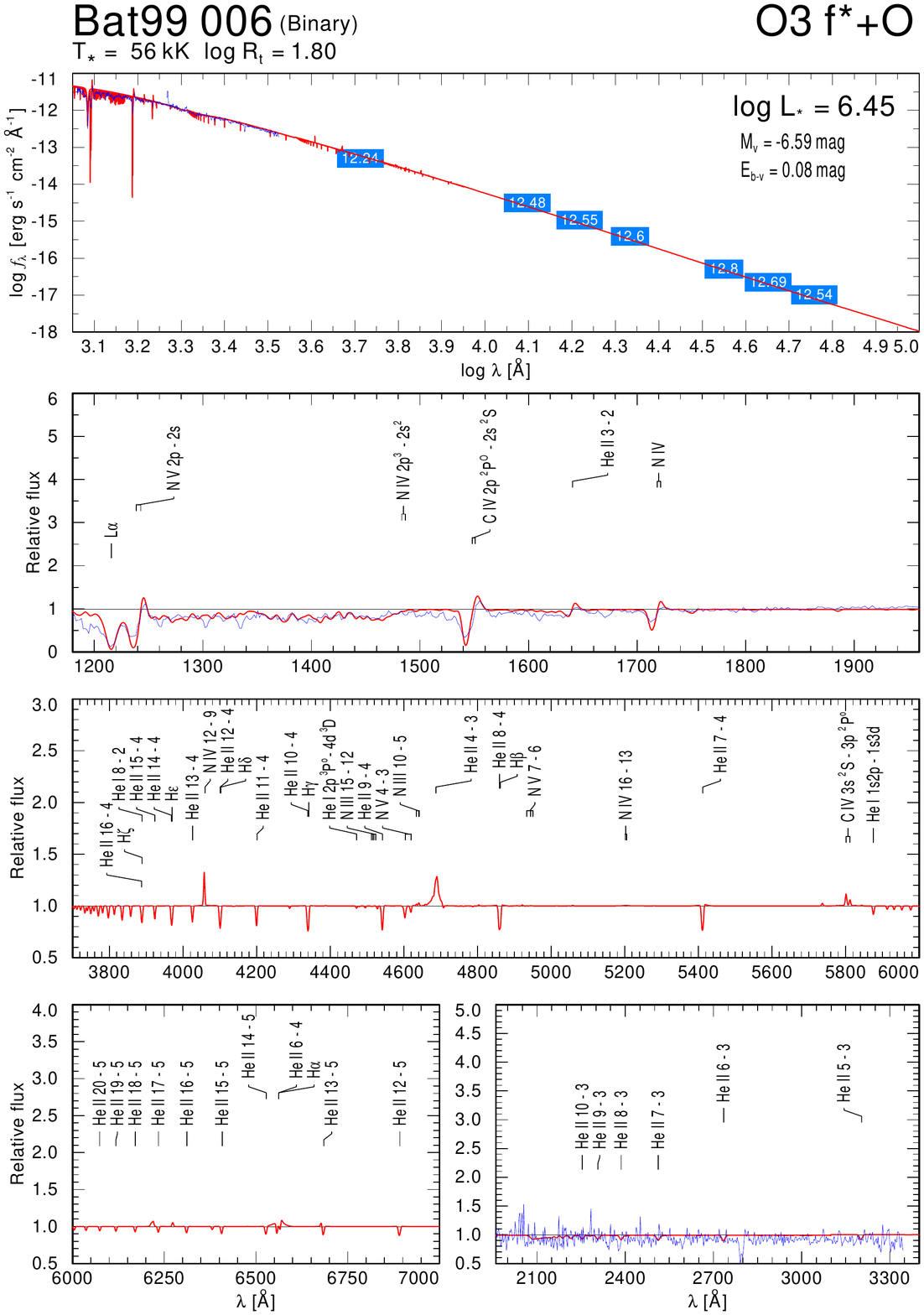}
  \qquad
  \includegraphics[width=0.46\hsize]{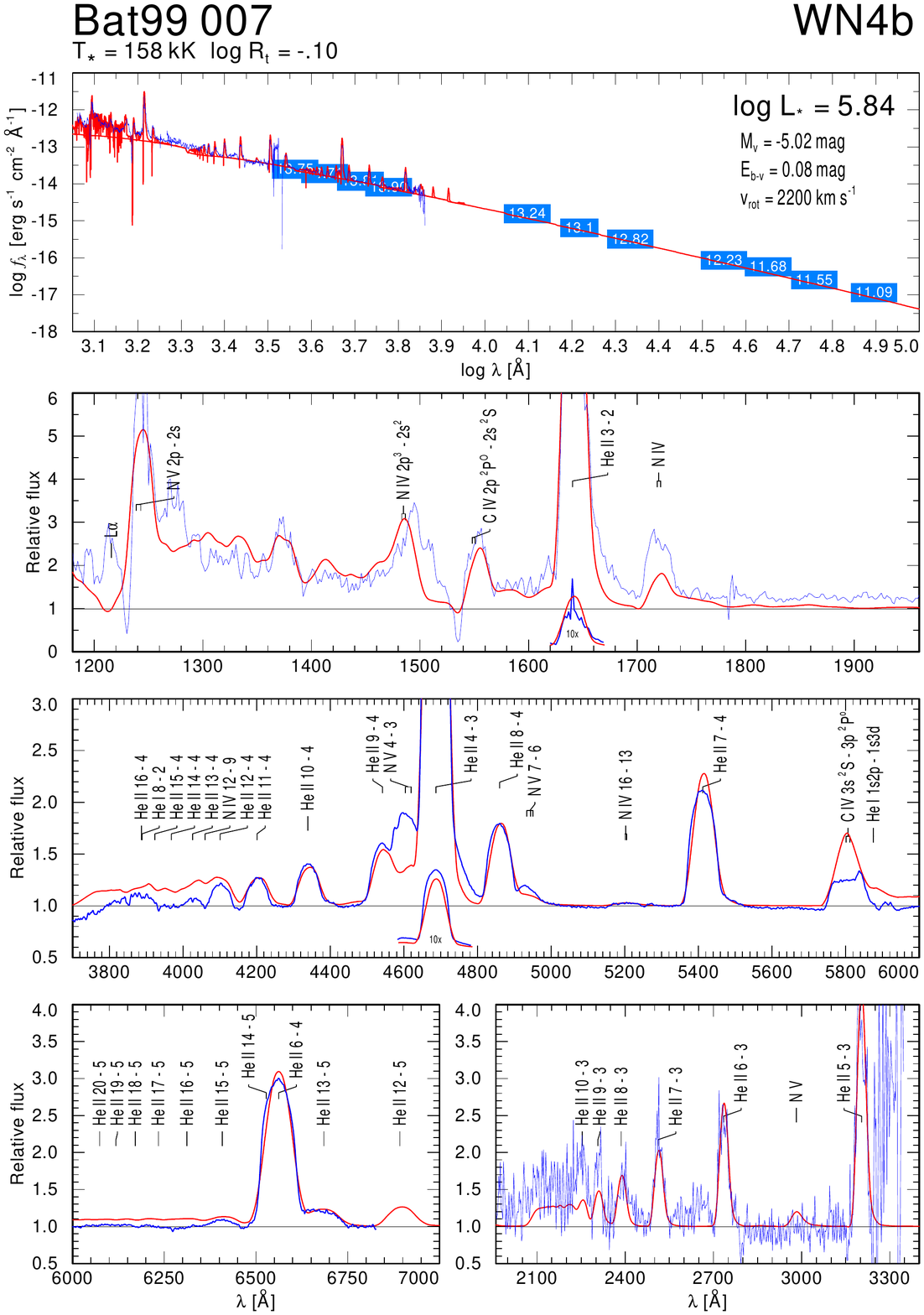}
  \vspace{-0.4cm}
  \caption{Spectral fit for BAT99\,006 and BAT99\,007}
  \label{fig:bat007}
  \label{fig:bat006}
\end{figure*}

\clearpage

\begin{figure*}
  \centering
  \includegraphics[width=0.46\hsize]{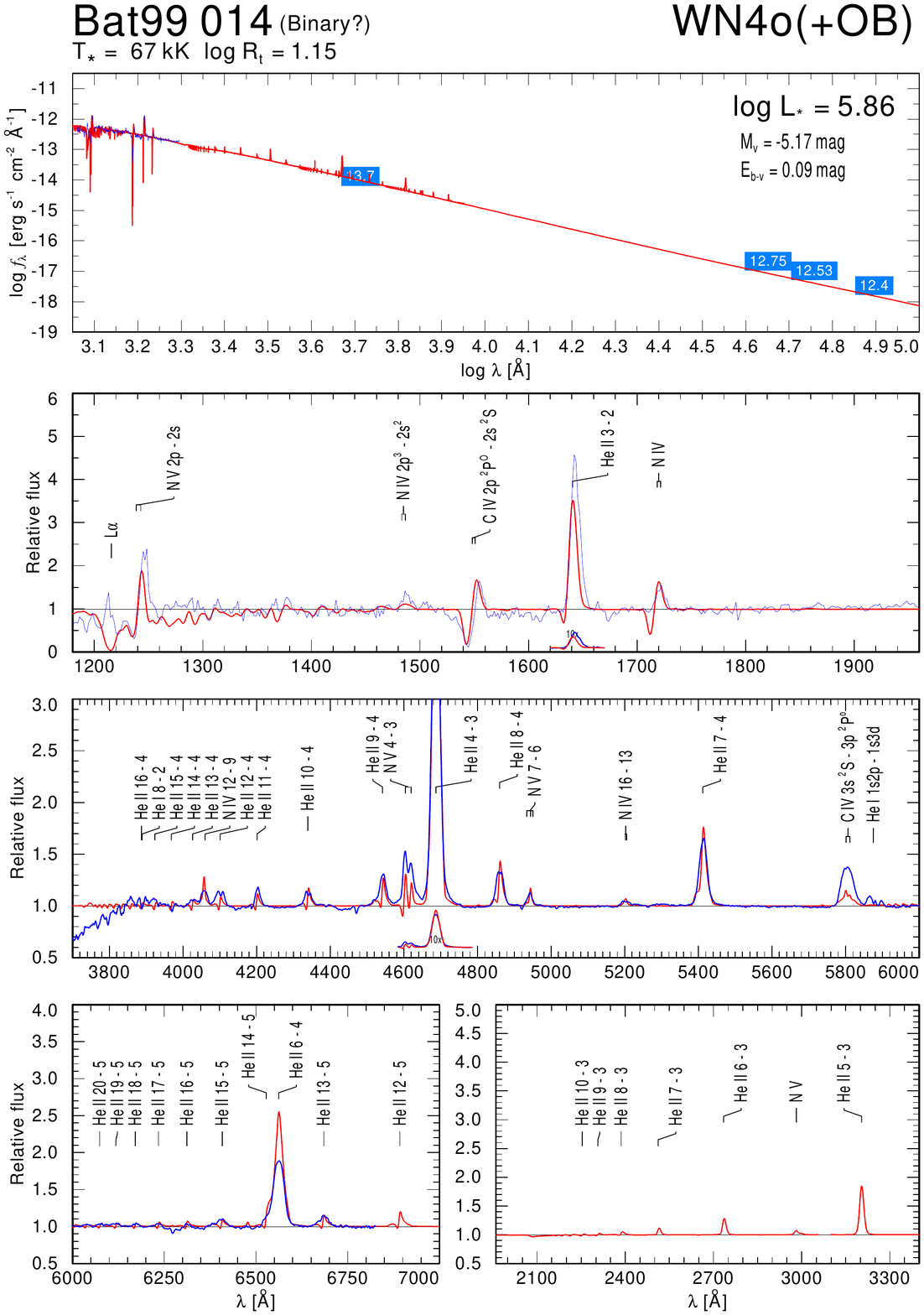}
  \qquad
  \includegraphics[width=0.46\hsize]{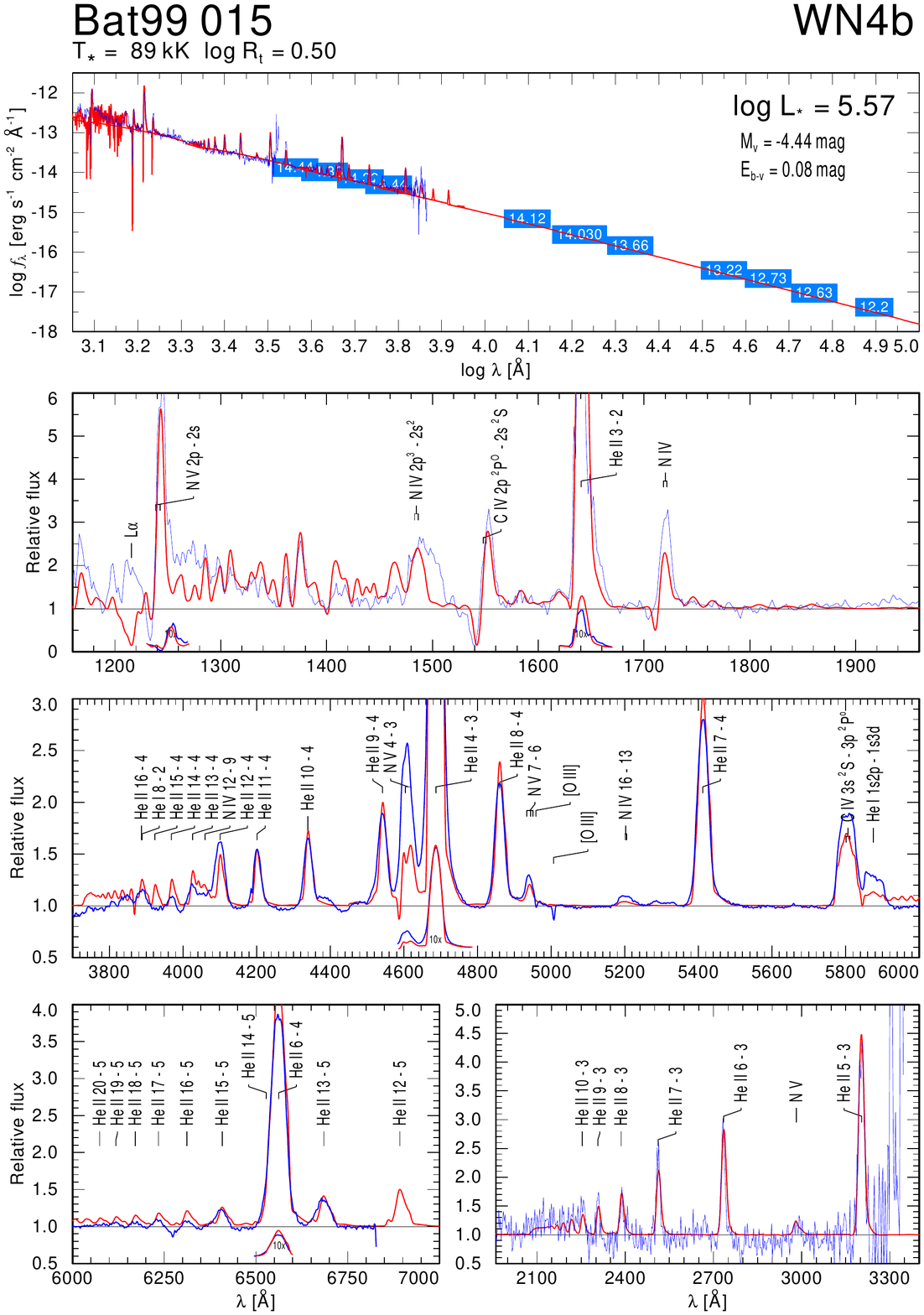}
  \vspace{-0.4cm}
  \caption{Spectral fit for BAT99\,014 and BAT99\,015}
  \label{fig:bat014}
  \label{fig:bat015}
\end{figure*}

\begin{figure*}
  \centering
  \includegraphics[width=0.46\hsize]{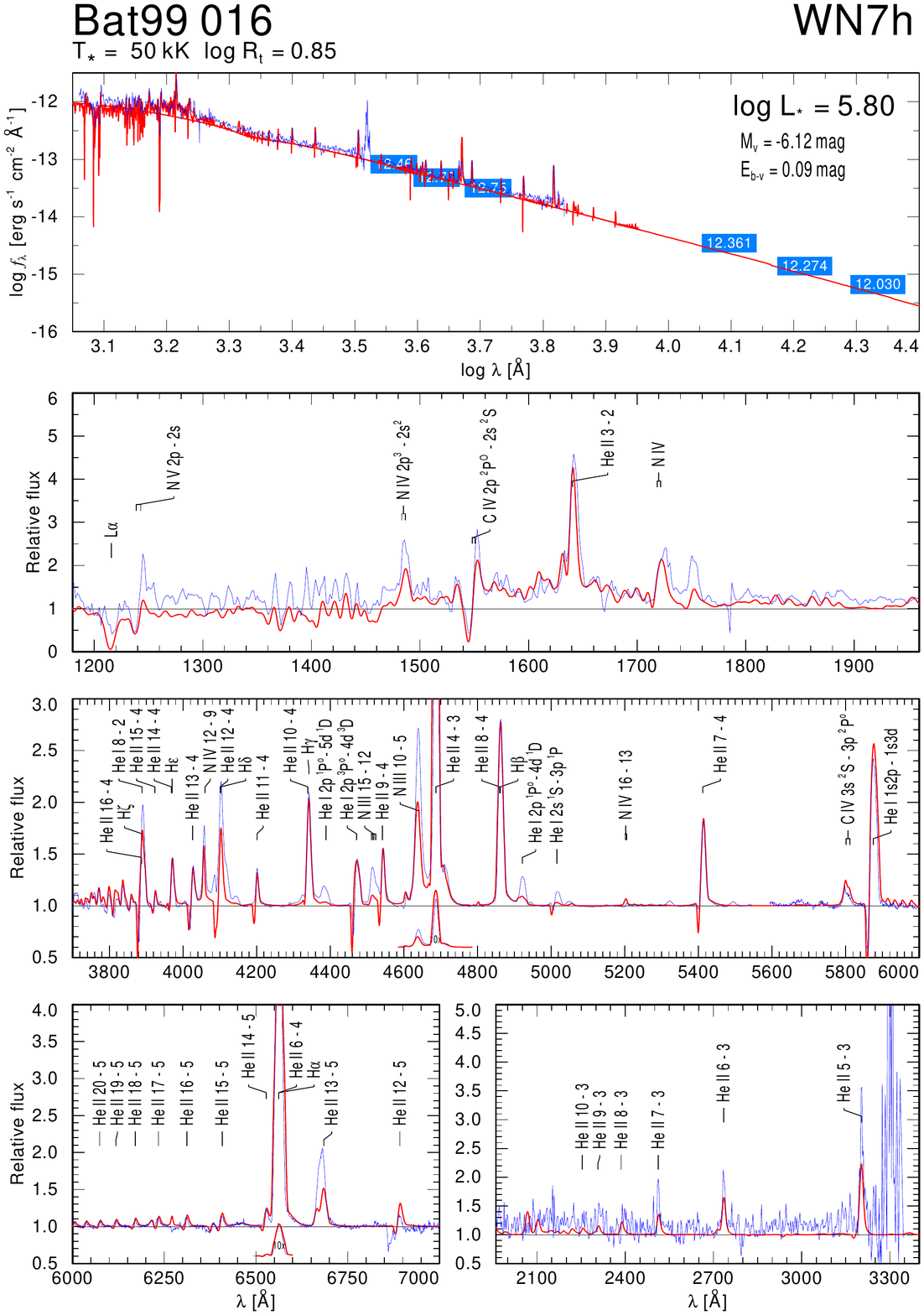}
  \qquad
  \includegraphics[width=0.46\hsize]{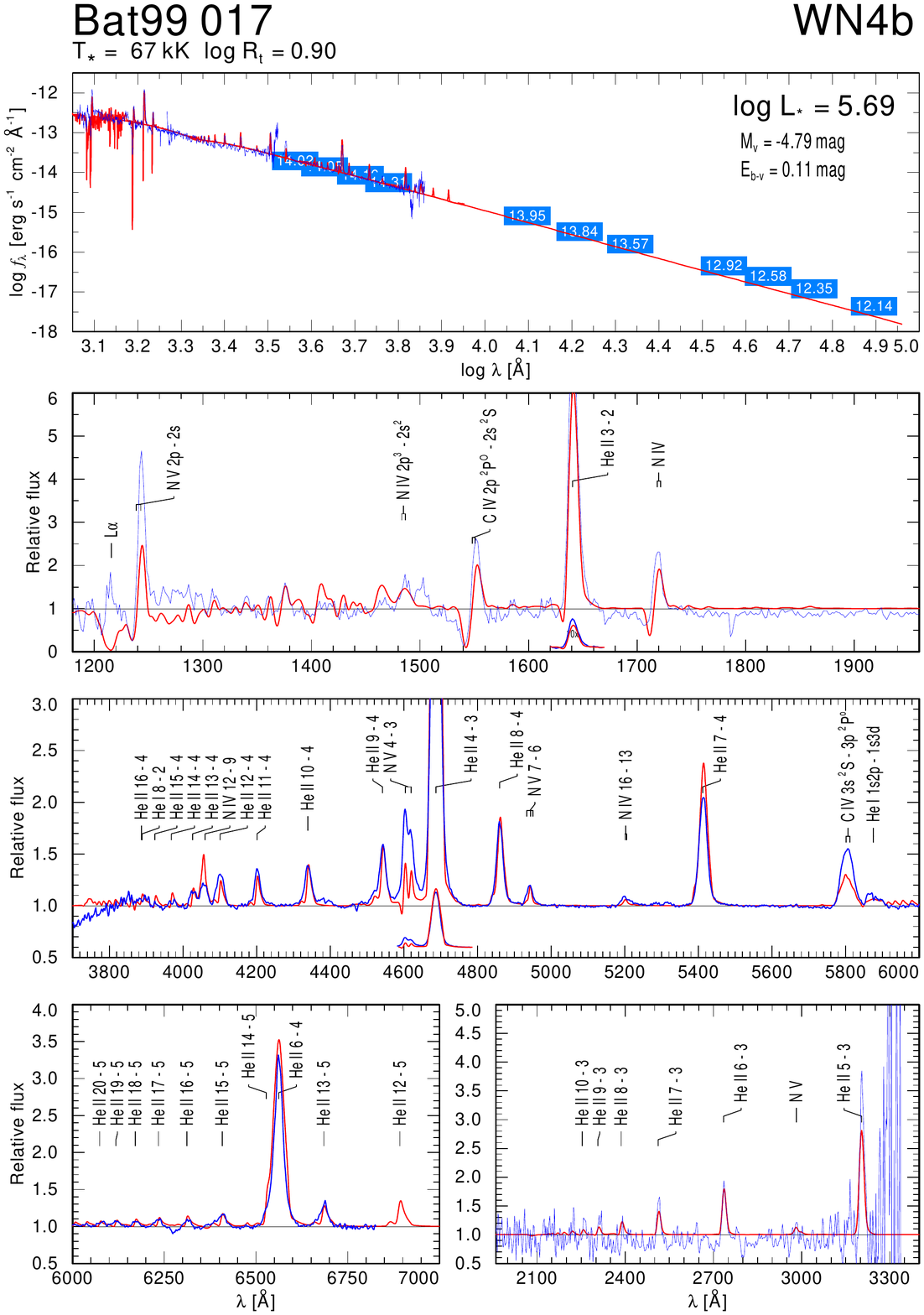}
  \vspace{-0.4cm}
  \caption{Spectral fit for BAT99\,016 and BAT99\,017}
  \label{fig:bat016}
  \label{fig:bat017}
\end{figure*}

\clearpage

\begin{figure*}
  \centering
  \includegraphics[width=0.46\hsize]{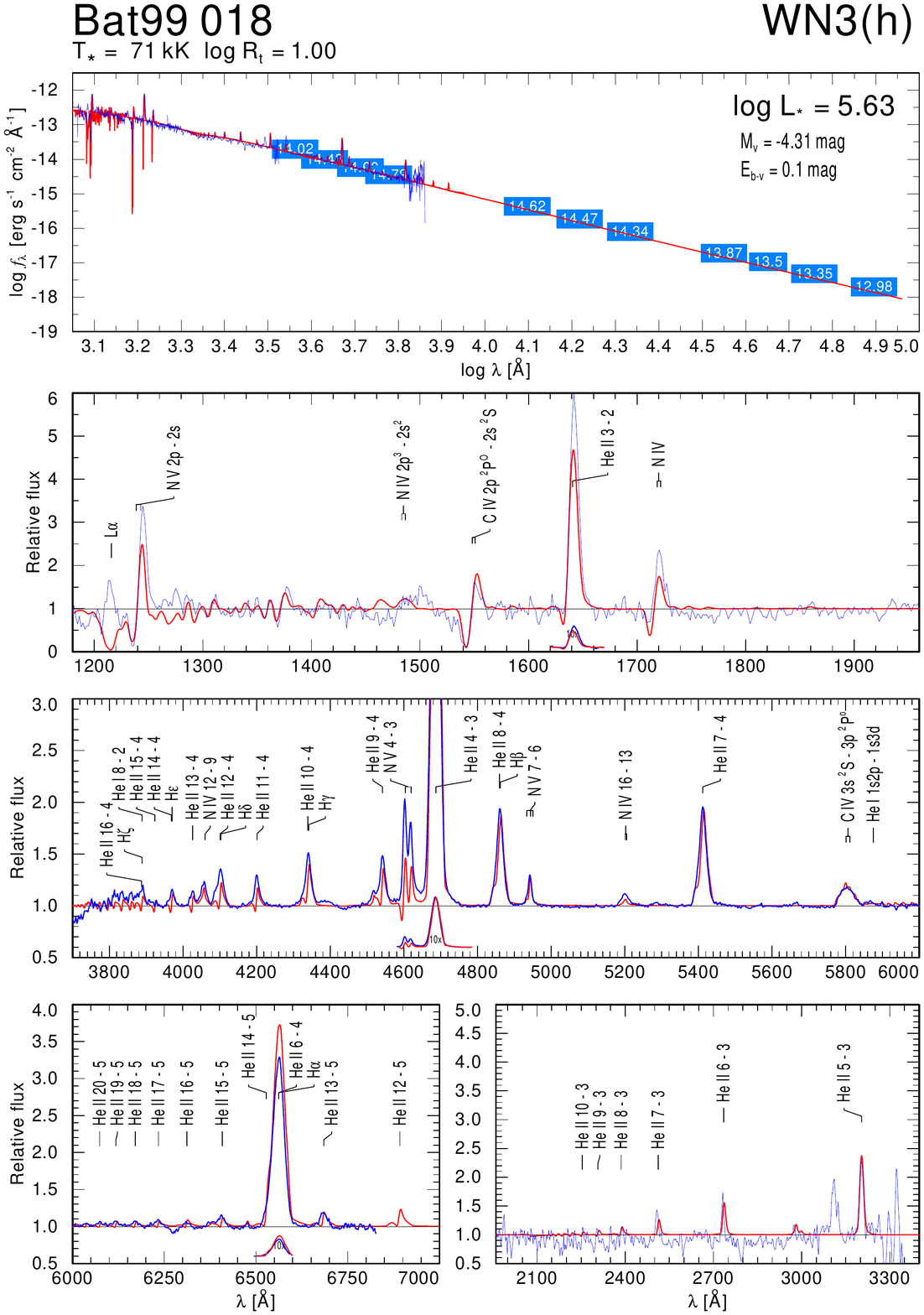}
  \qquad
  \includegraphics[width=0.46\hsize]{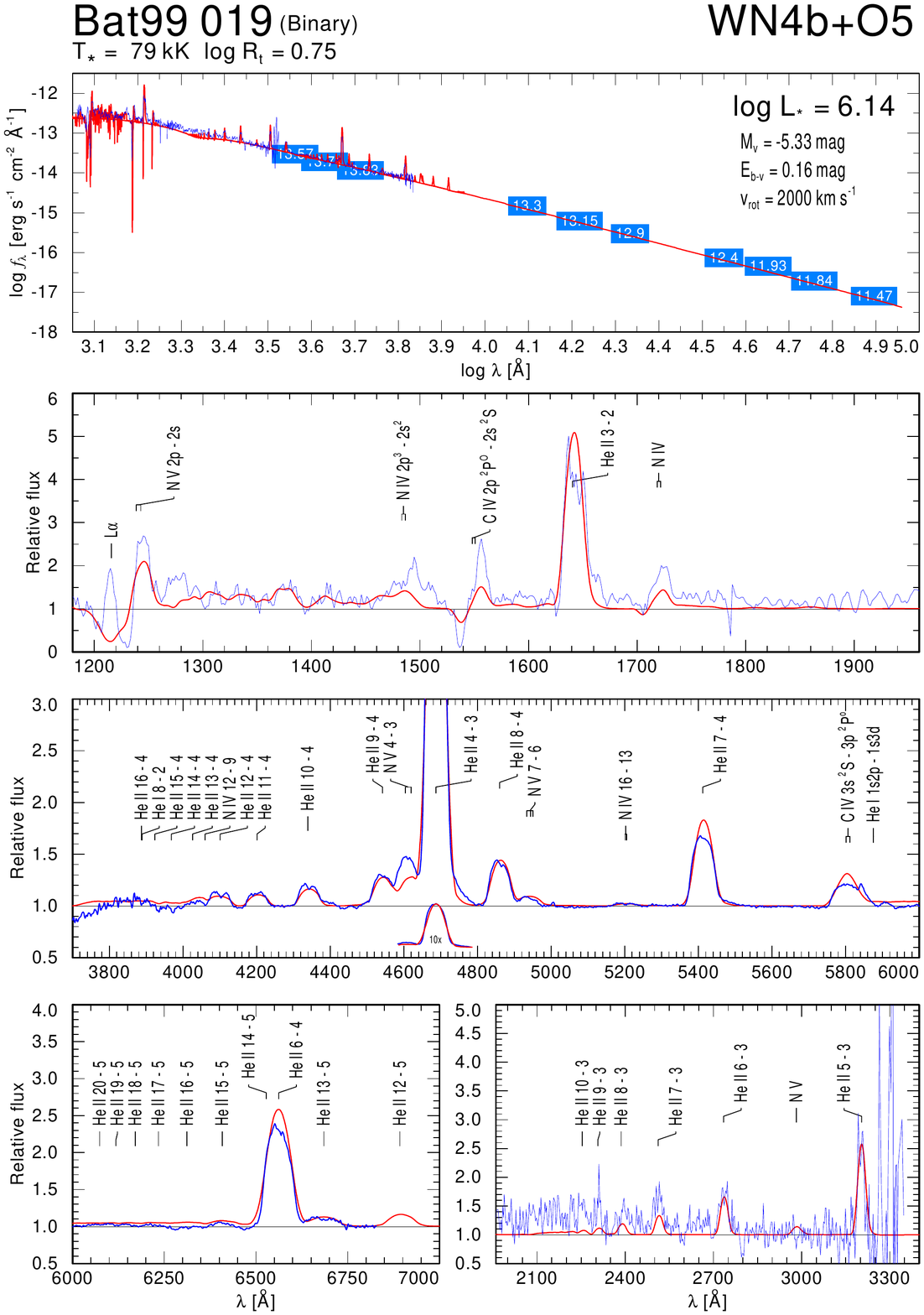}
  \vspace{-0.4cm}
  \caption{Spectral fit for BAT99\,018 and BAT99\,019}
  \label{fig:bat018}
  \label{fig:bat019}
\end{figure*}

\begin{figure*}
  \centering
  \includegraphics[width=0.46\hsize]{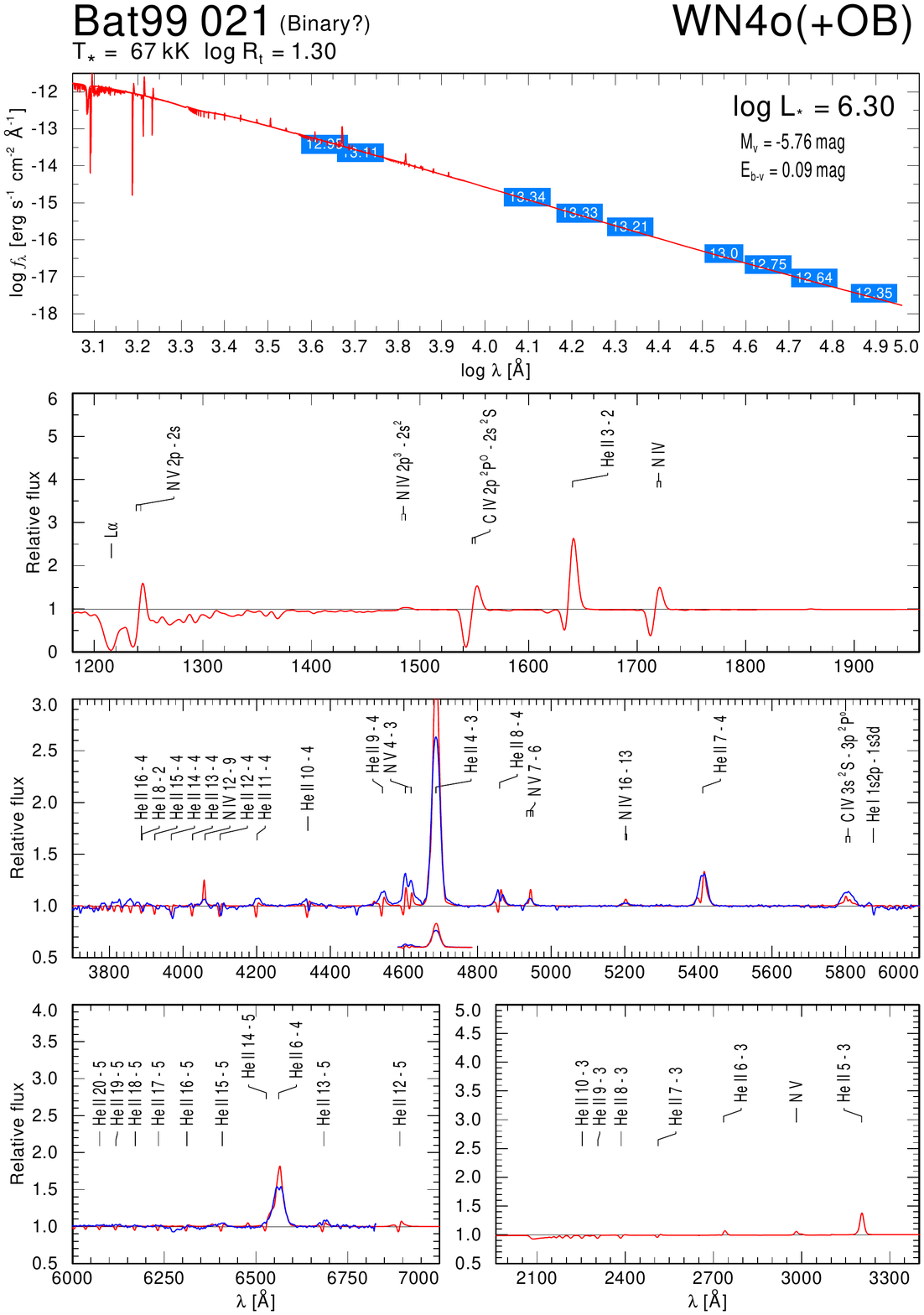}
  \qquad
  \includegraphics[width=0.46\hsize]{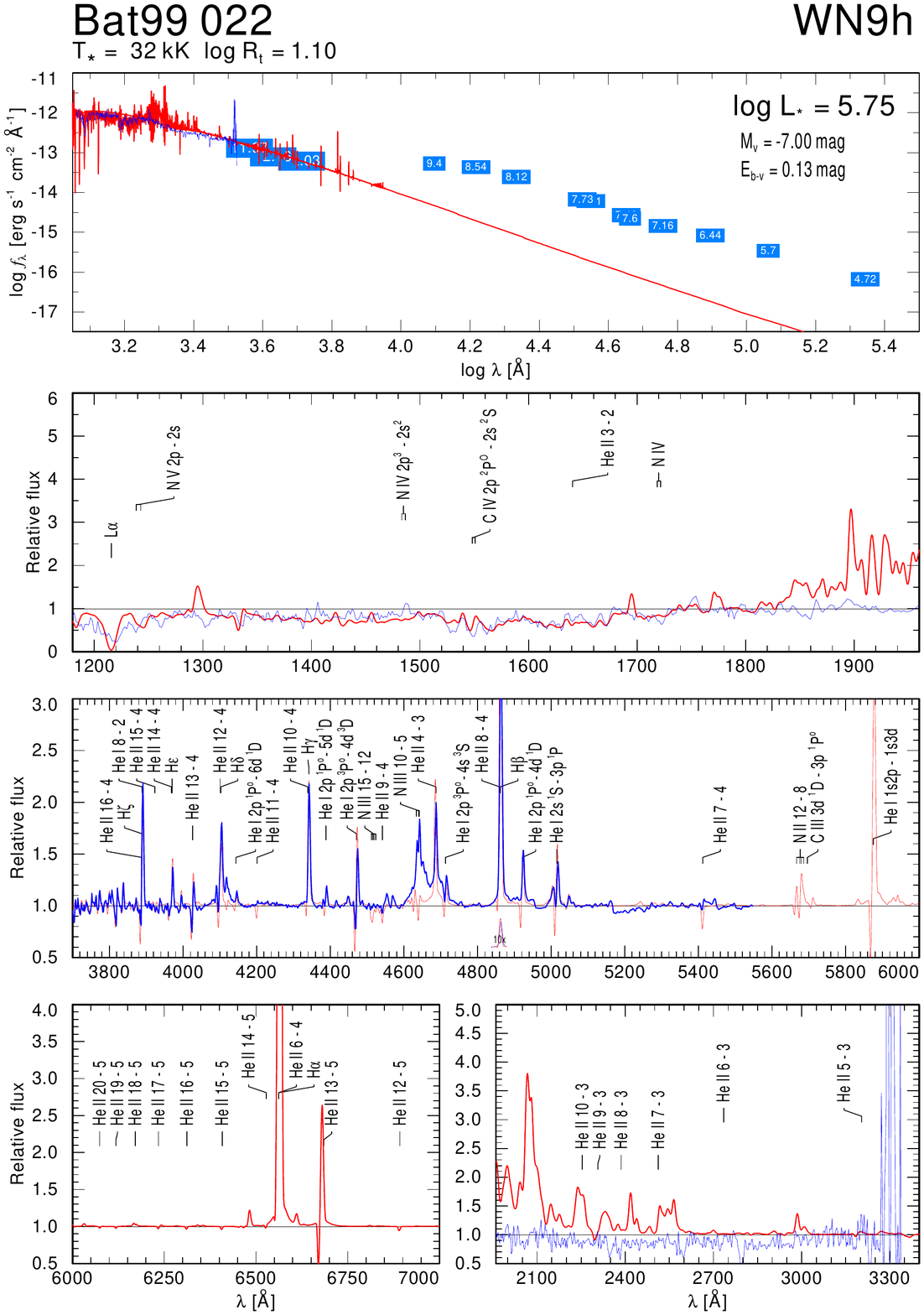}
  \vspace{-0.4cm}
  \caption{Spectral fit for BAT99\,021 and BAT99\,022}
  \label{fig:bat021}
  \label{fig:bat022}
\end{figure*}

\clearpage

\begin{figure*}
  \centering
  \includegraphics[width=0.46\hsize]{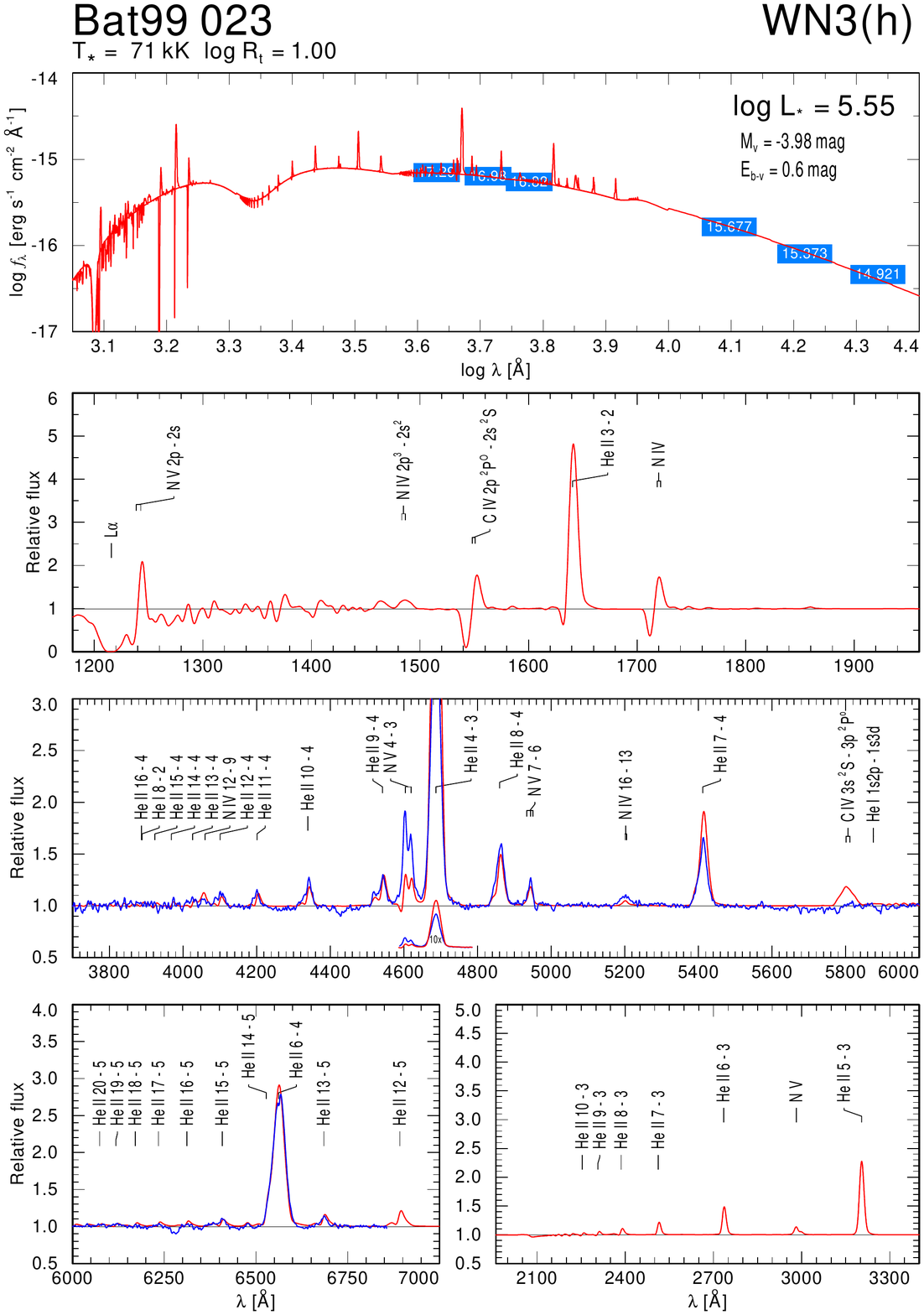}
  \qquad
  \includegraphics[width=0.46\hsize]{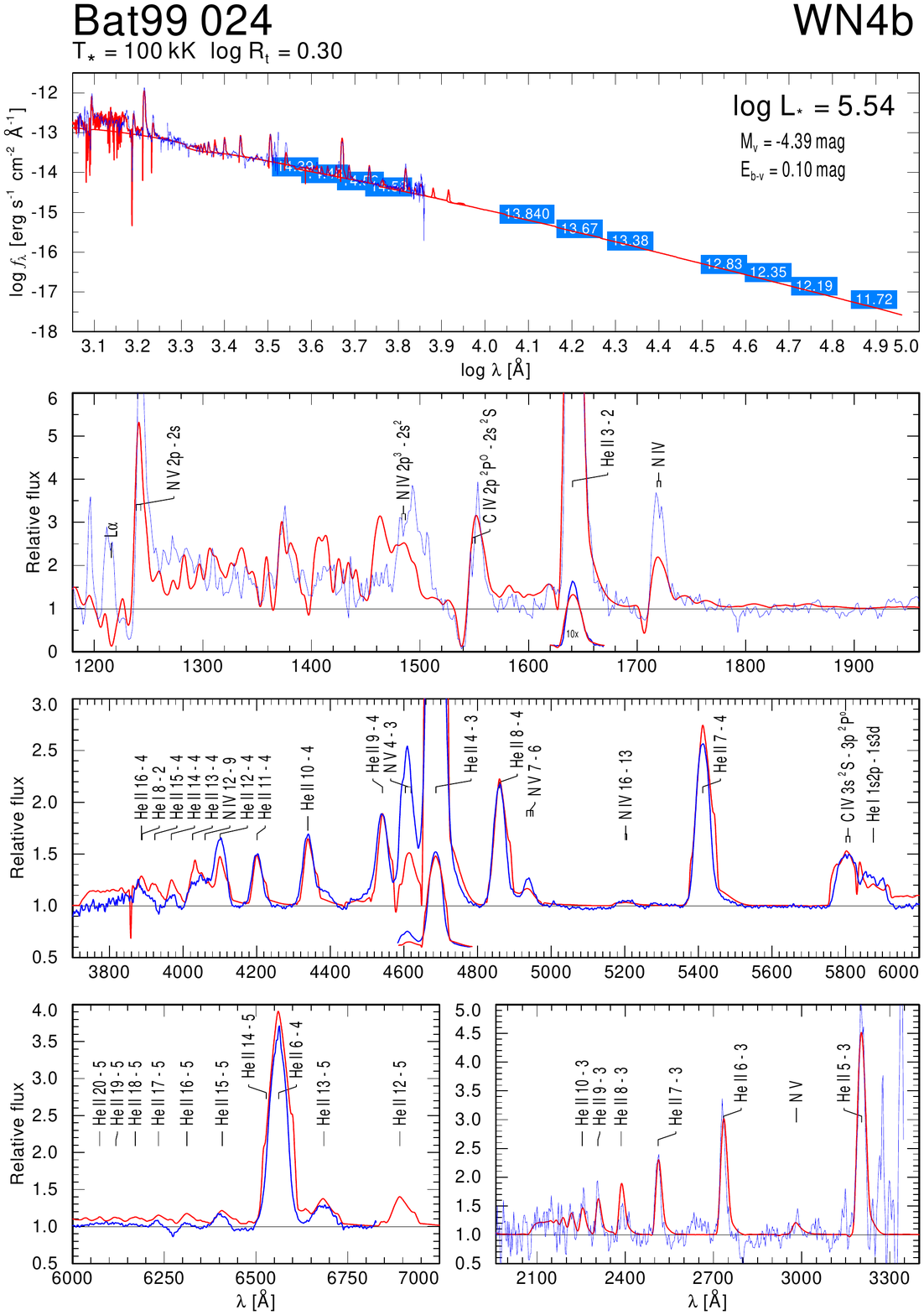}
  \vspace{-0.4cm}
  \caption{Spectral fit for BAT99\,023 and BAT99\,024}
  \label{fig:bat023}
  \label{fig:bat024}
\end{figure*}

\begin{figure*}
  \centering
  \includegraphics[width=0.46\hsize]{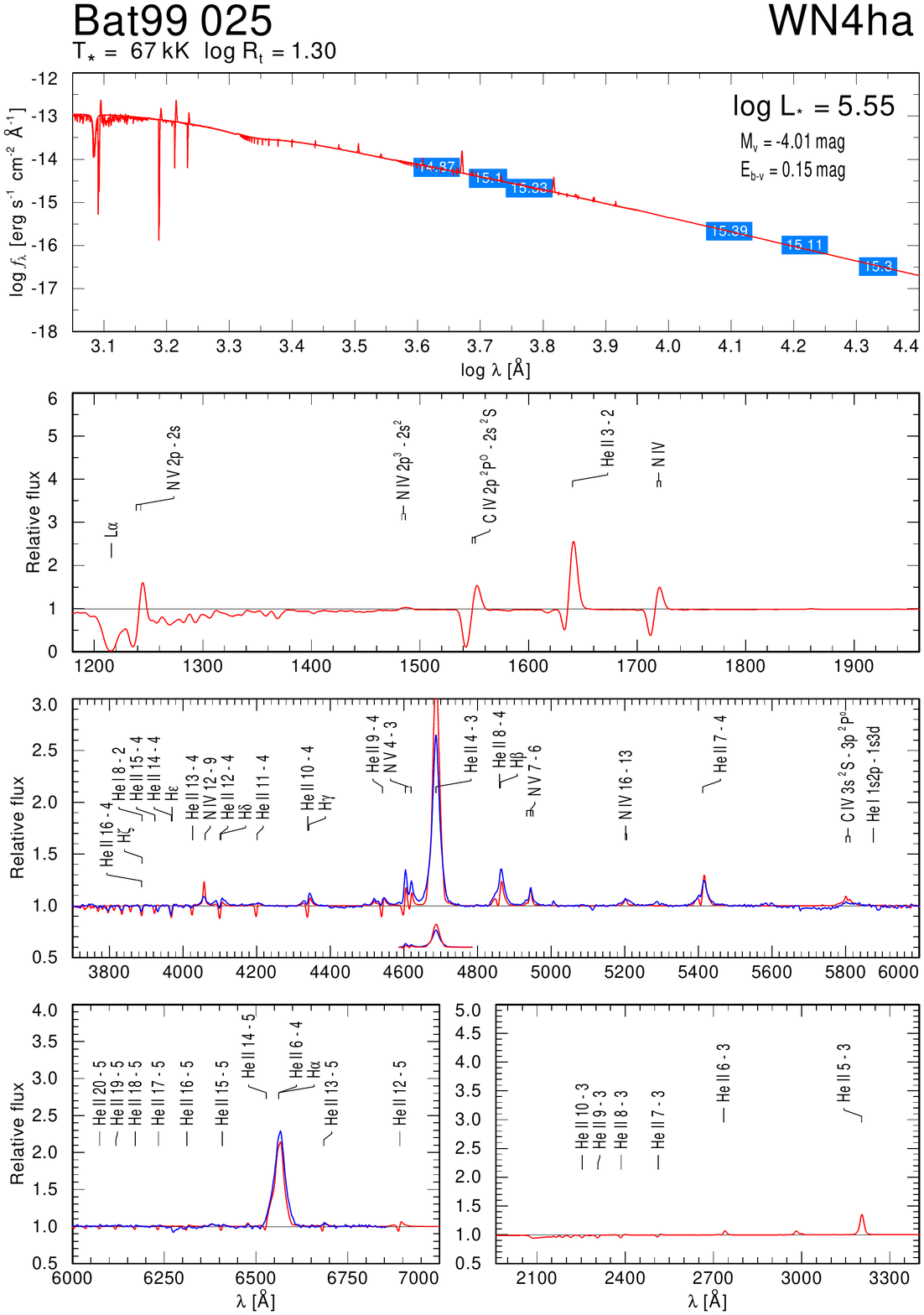}
  \qquad
  \includegraphics[width=0.46\hsize]{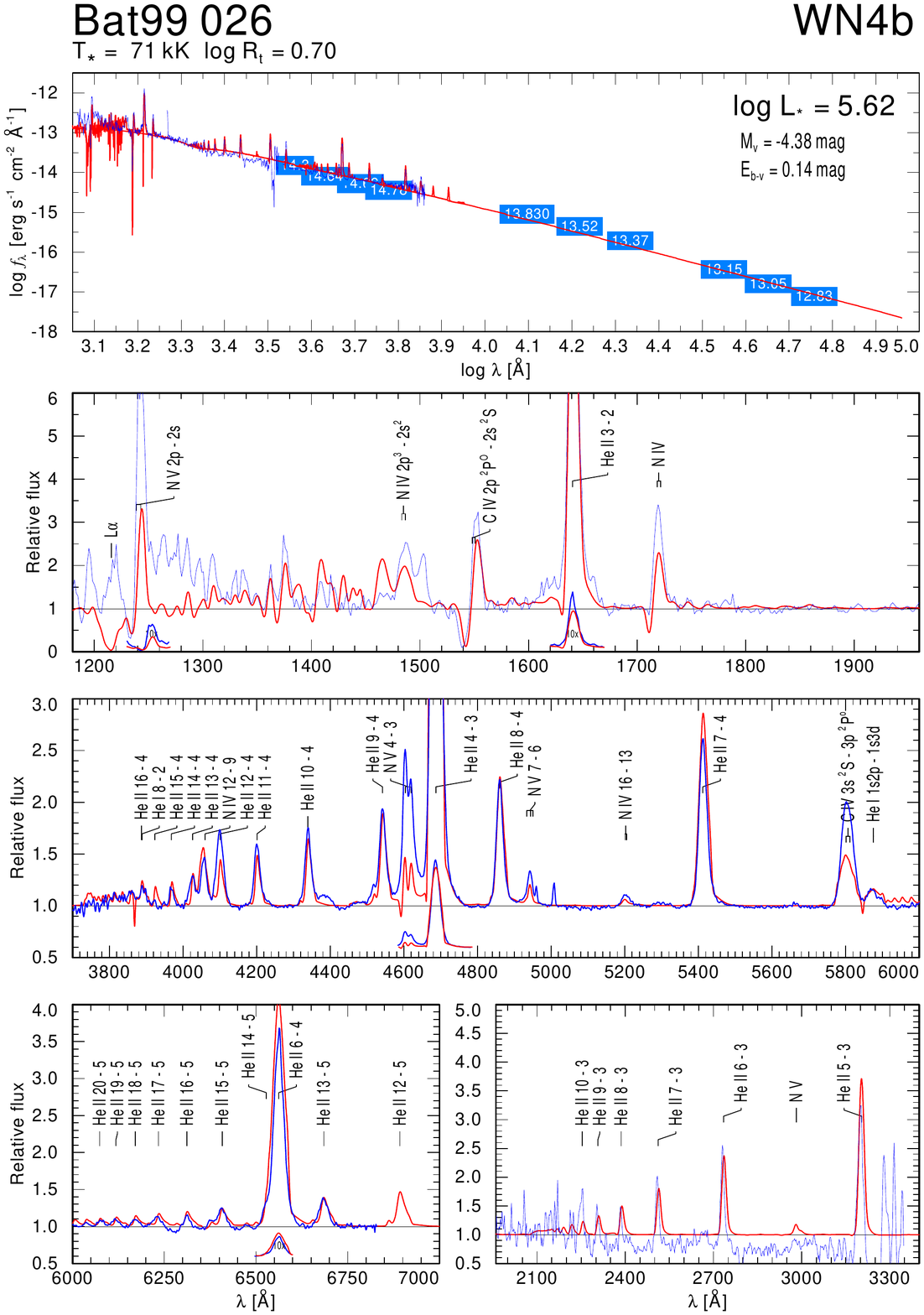}
  \vspace{-0.4cm}
  \caption{Spectral fit for BAT99\,025 and BAT99\,026}
  \label{fig:bat025}
  \label{fig:bat026}
\end{figure*}

\clearpage

\begin{figure*}
  \centering
  \includegraphics[width=0.46\hsize]{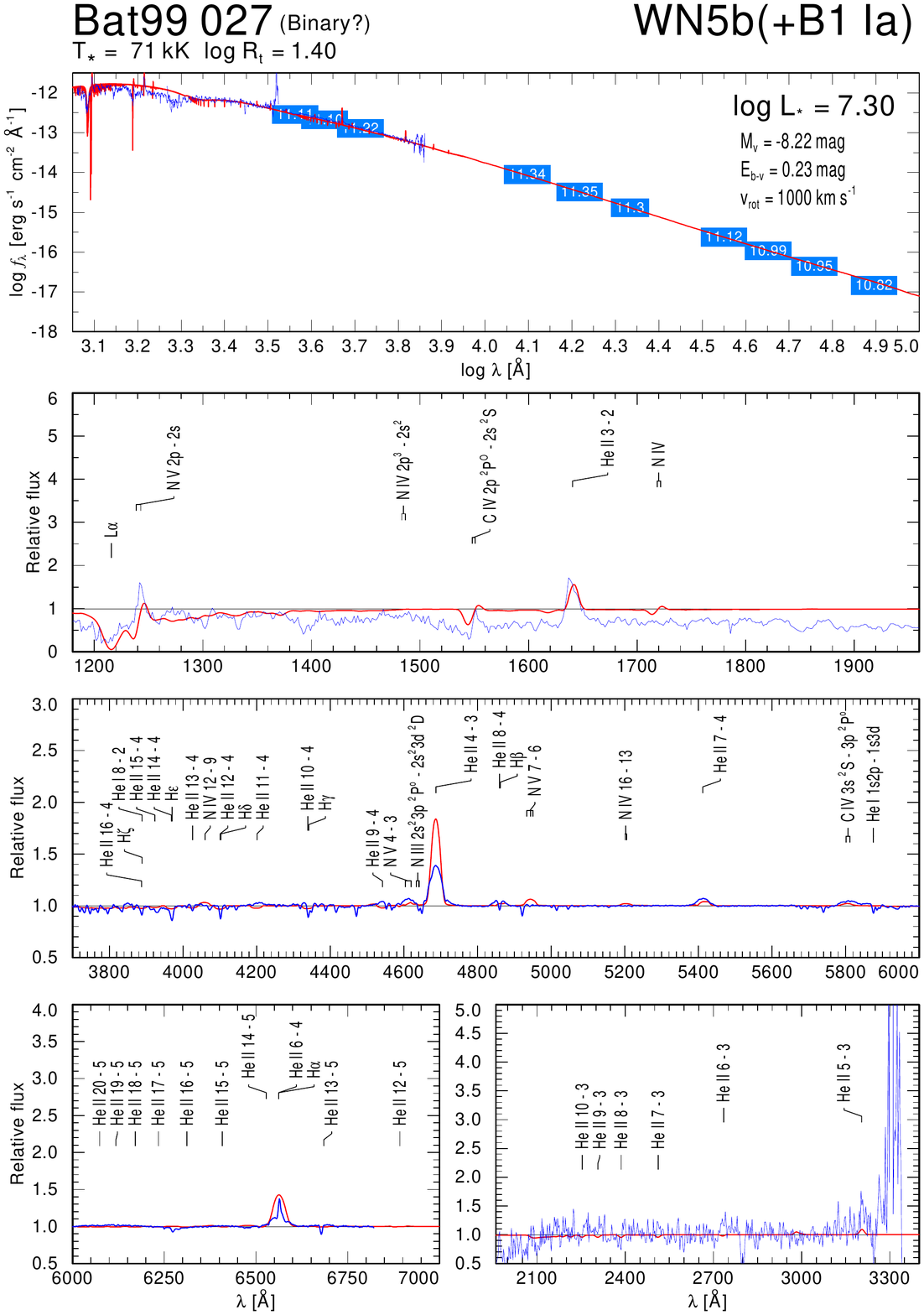}
  \qquad
  \includegraphics[width=0.46\hsize]{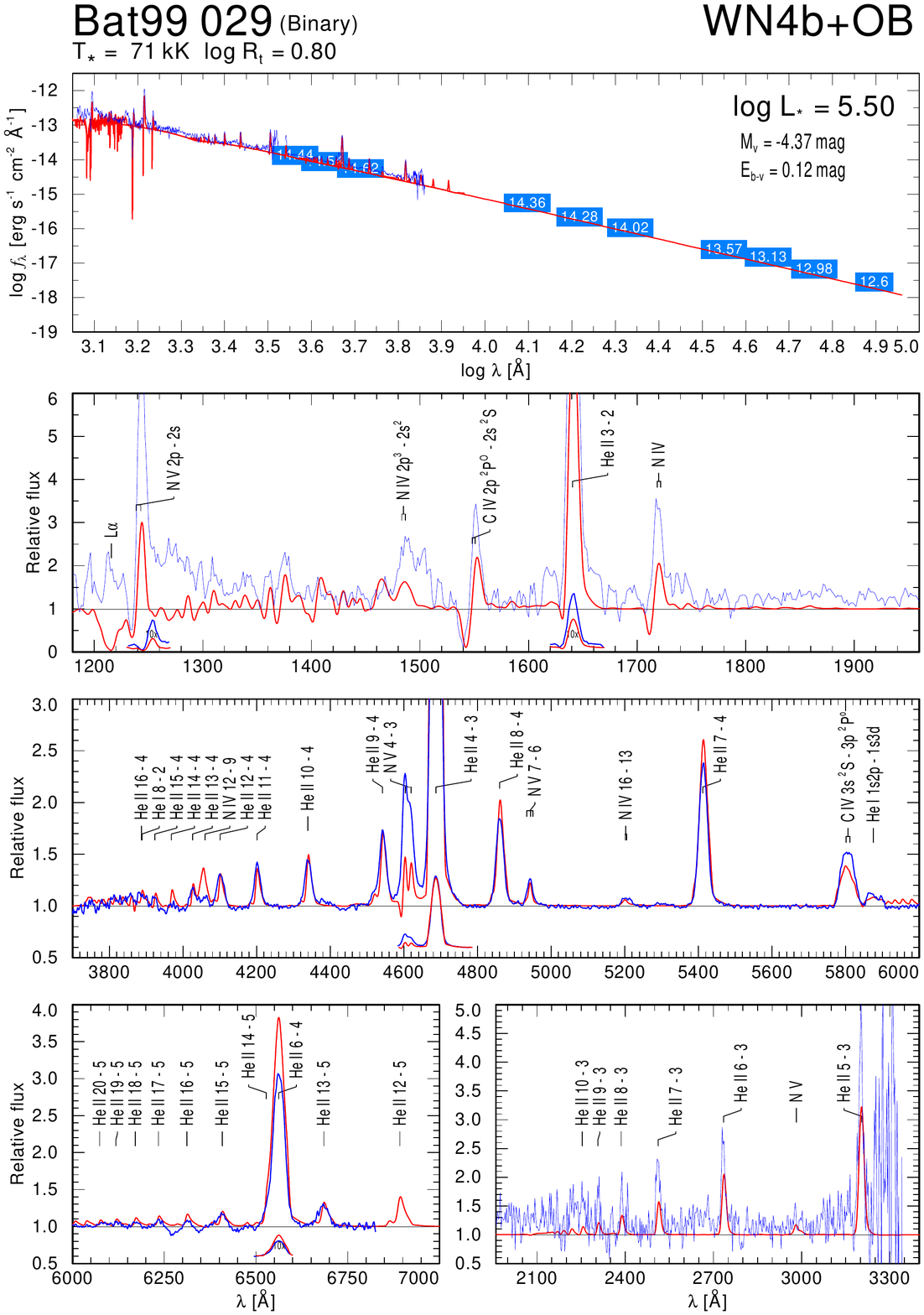}
  \vspace{-0.4cm}
  \caption{Spectral fit for BAT99\,027 and BAT99\,029}
  \label{fig:bat027}
  \label{fig:bat029}
\end{figure*}

\begin{figure*}
  \centering
  \includegraphics[width=0.46\hsize]{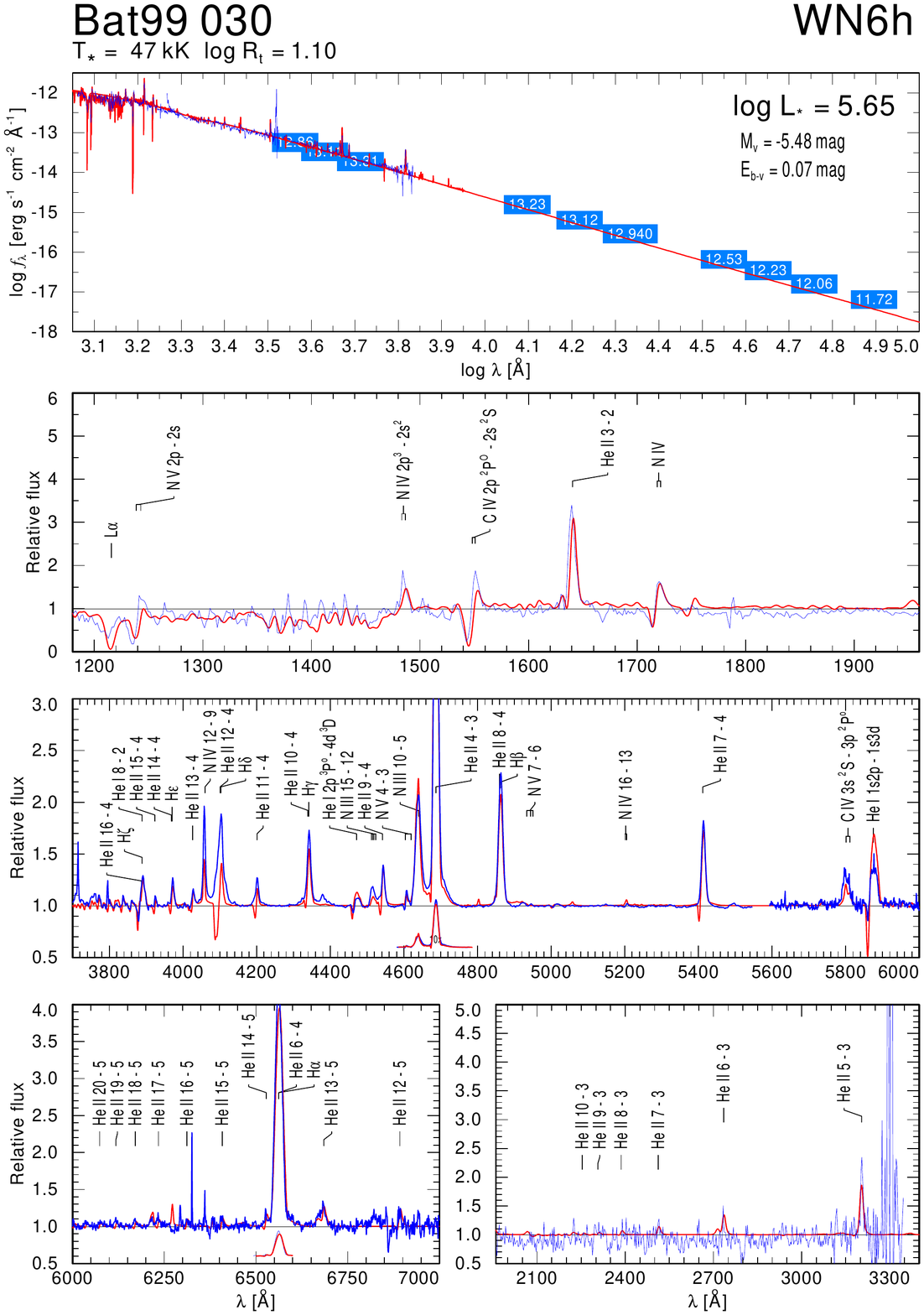}
  \qquad
  \includegraphics[width=0.46\hsize]{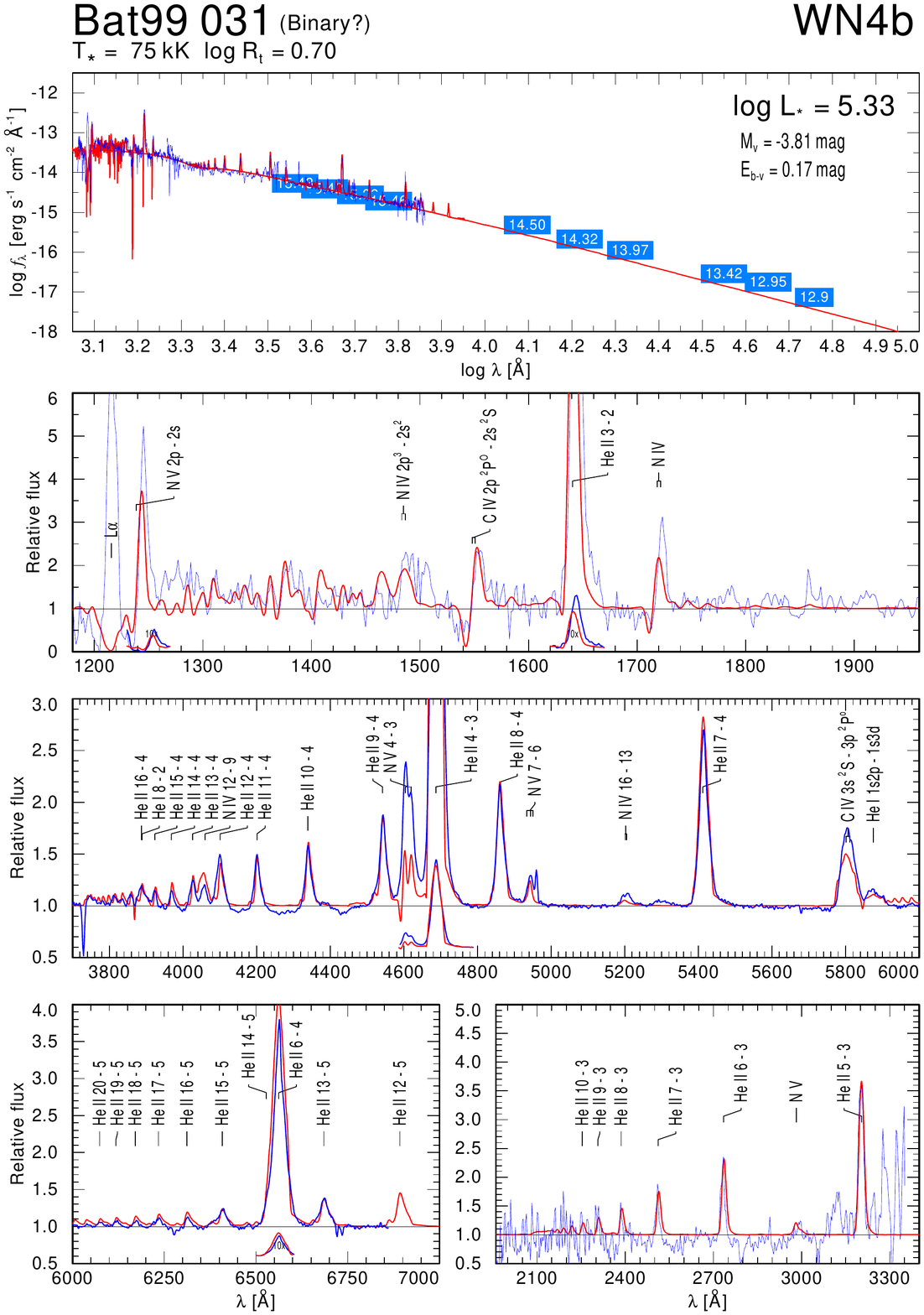}
  \vspace{-0.4cm}
  \caption{Spectral fit for BAT99\,030 and BAT99\,031}
  \label{fig:bat030}
  \label{fig:bat031}
\end{figure*}

\clearpage

\begin{figure*}
  \centering
  \includegraphics[width=0.46\hsize]{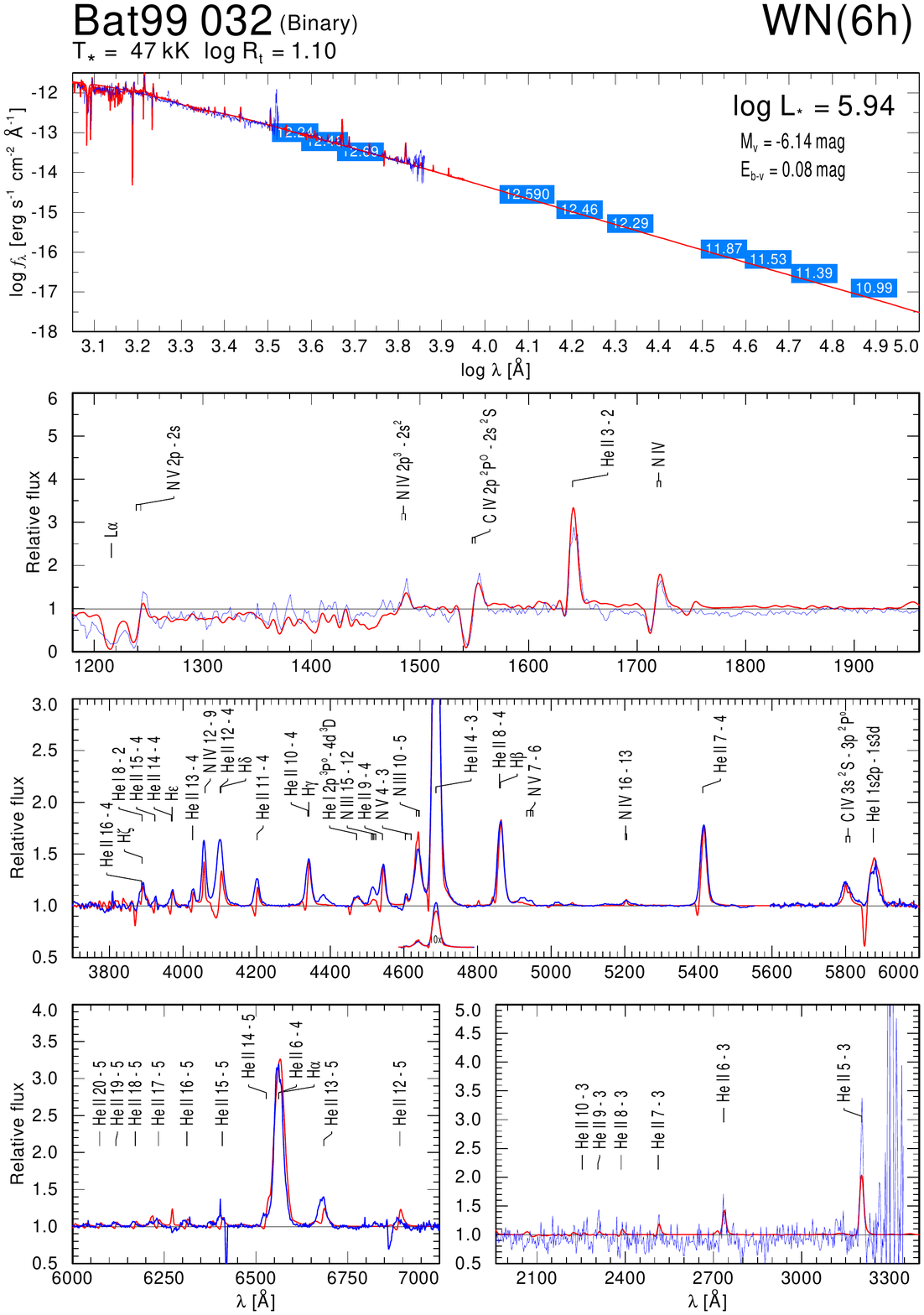}
  \qquad
  \includegraphics[width=0.46\hsize]{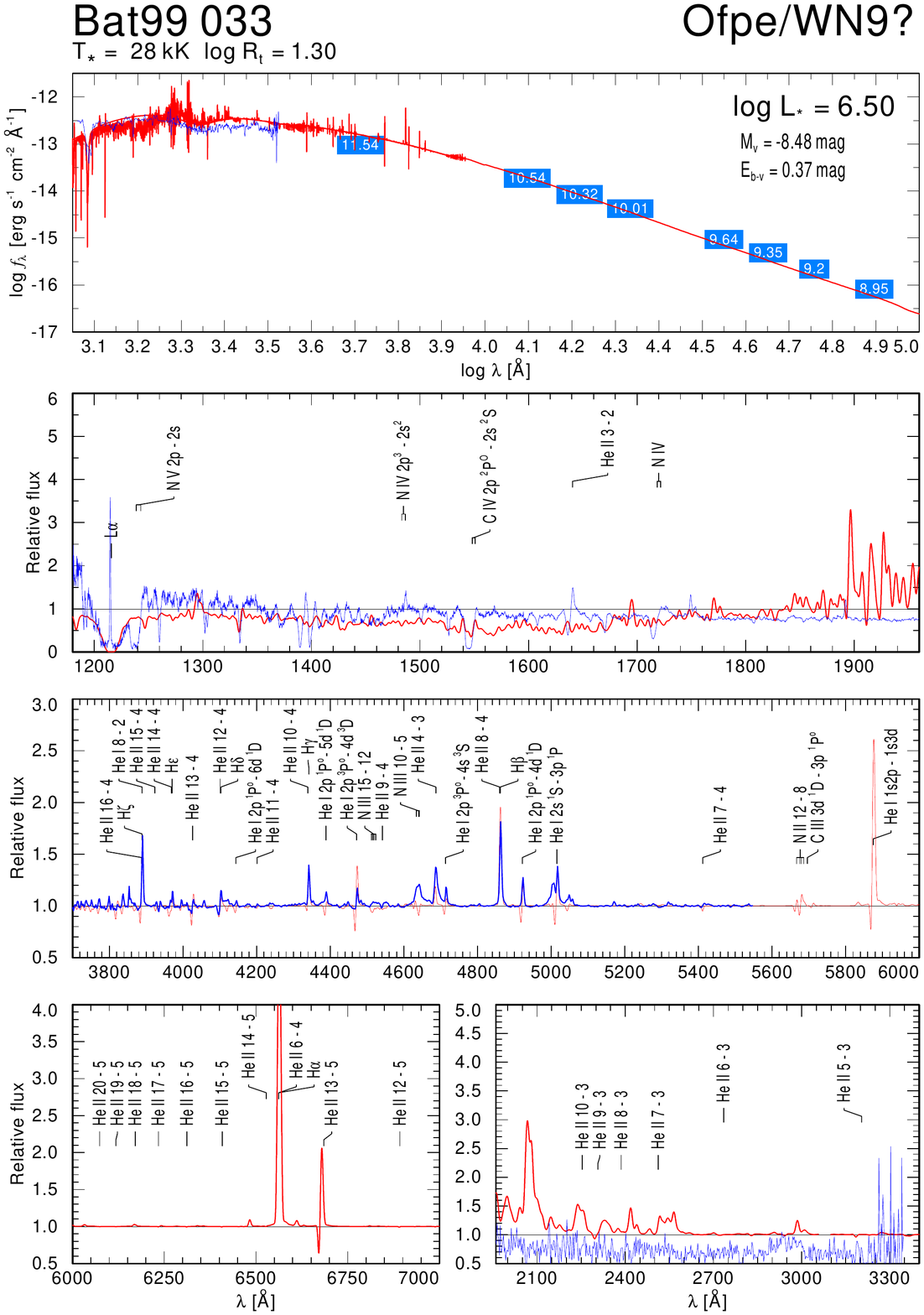}
  \vspace{-0.4cm}
  \caption{Spectral fit for BAT99\,032 and BAT99\,033}
  \label{fig:bat032}
  \label{fig:bat033}
\end{figure*}

\begin{figure*}
  \centering
  \includegraphics[width=0.46\hsize]{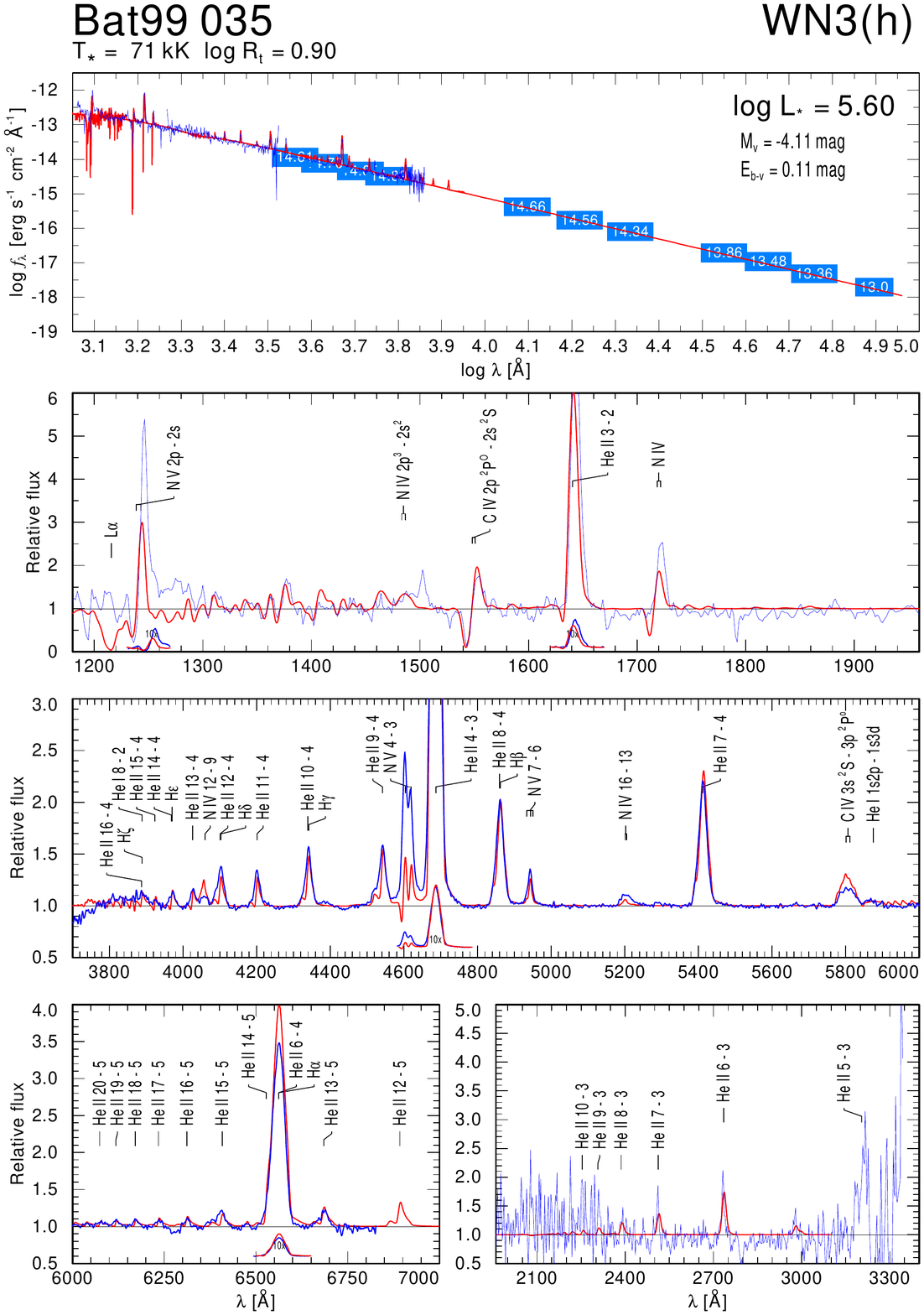}
  \qquad
  \includegraphics[width=0.46\hsize]{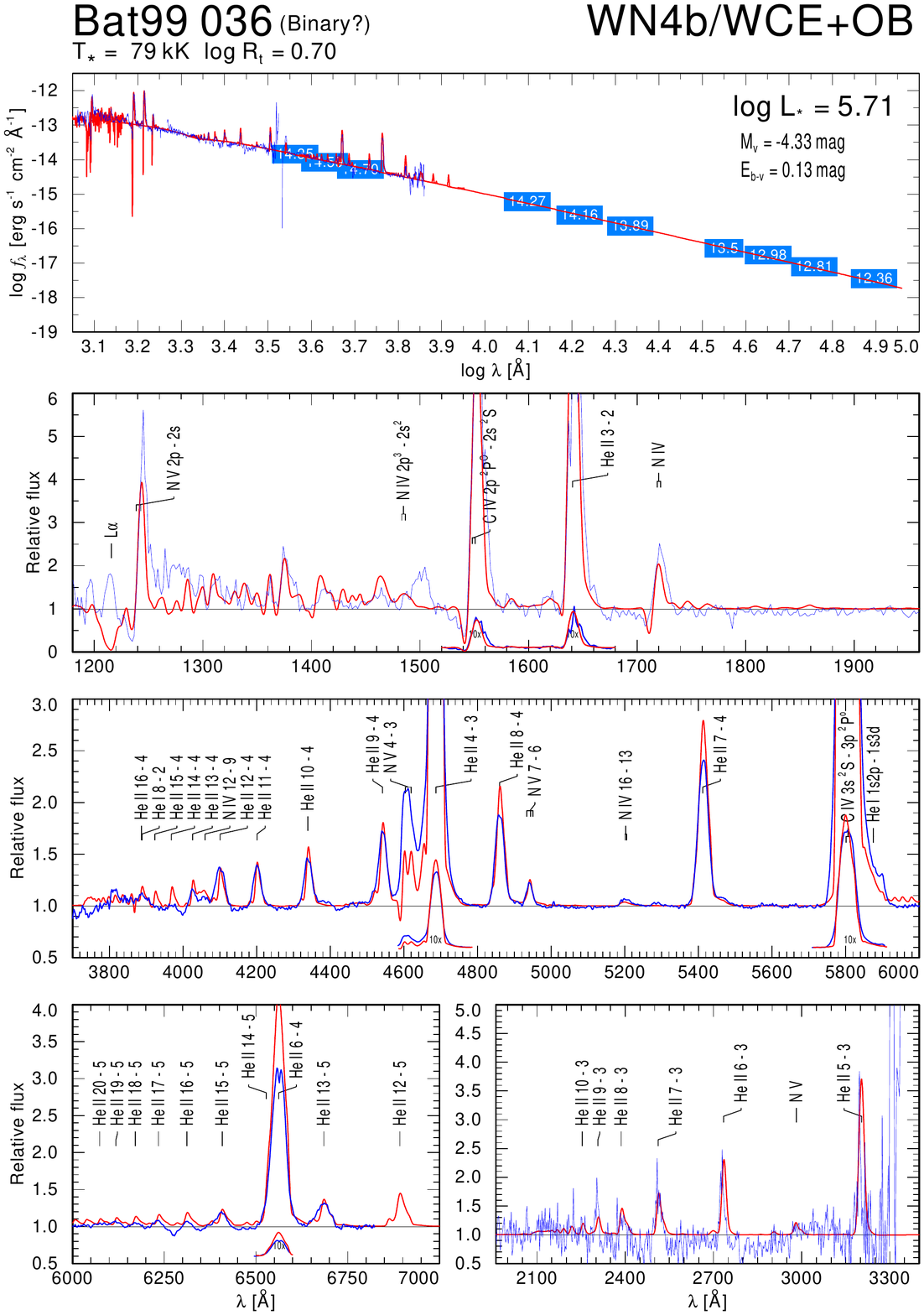}
  \vspace{-0.4cm}
  \caption{Spectral fit for BAT99\,035 and BAT99\,036}
  \label{fig:bat035}
  \label{fig:bat036}
\end{figure*}

\clearpage

\begin{figure*}
  \centering
  \includegraphics[width=0.46\hsize]{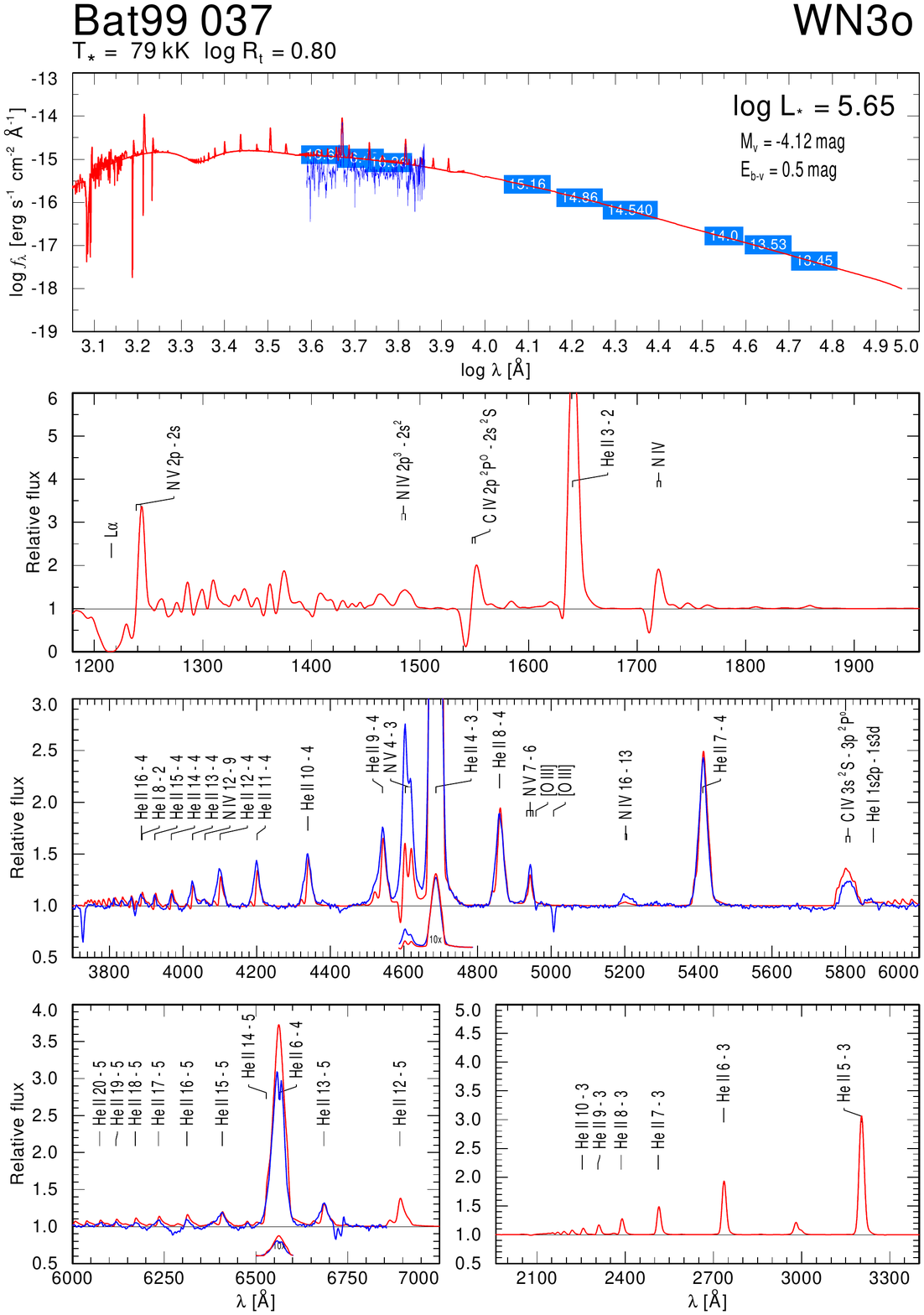}
  \qquad
  \includegraphics[width=0.46\hsize]{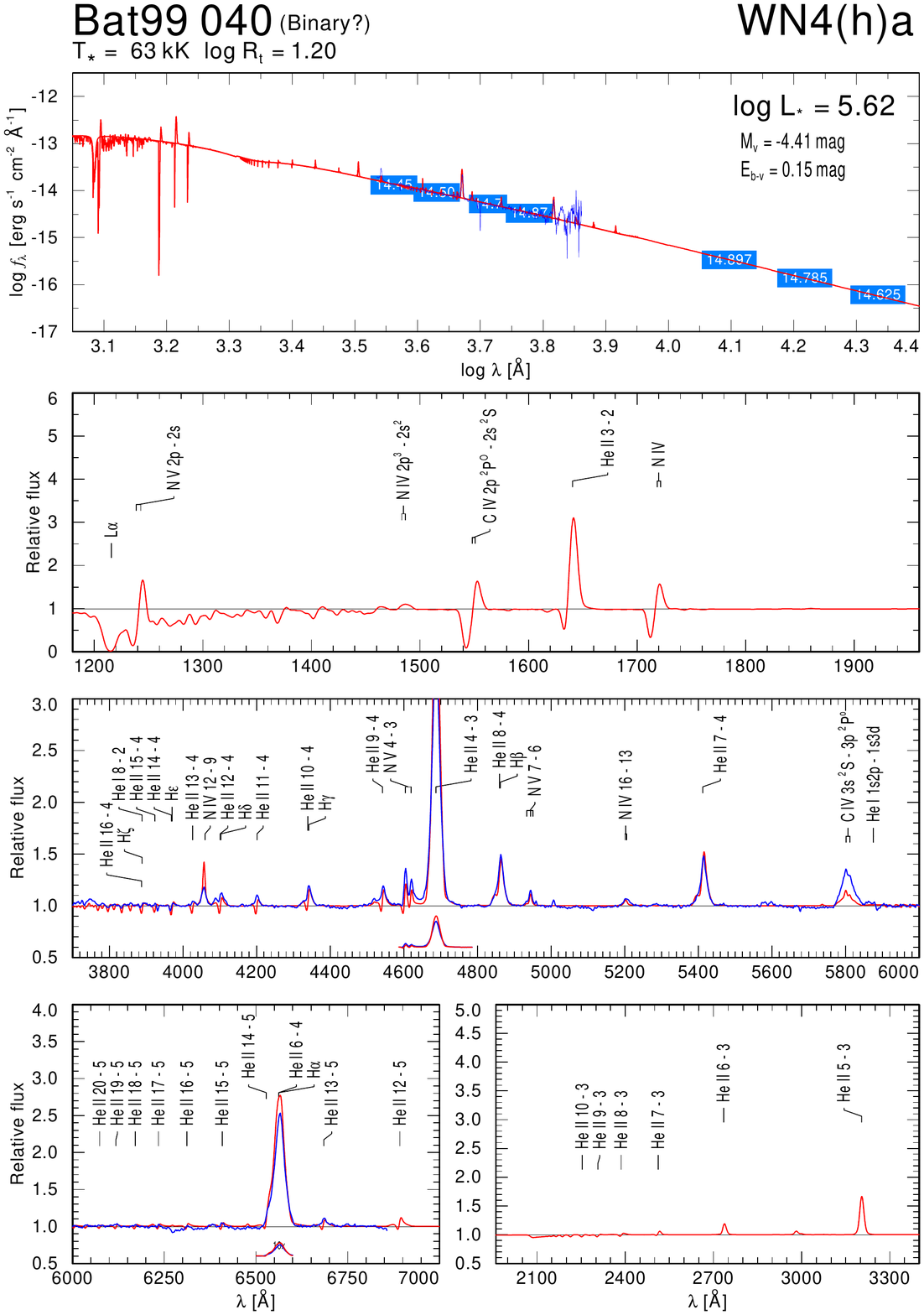}
  \vspace{-0.4cm}
  \caption{Spectral fit for BAT99\,037 and BAT99\,040}
  \label{fig:bat037}
  \label{fig:bat040}
\end{figure*}

\begin{figure*}
  \centering
  \includegraphics[width=0.46\hsize]{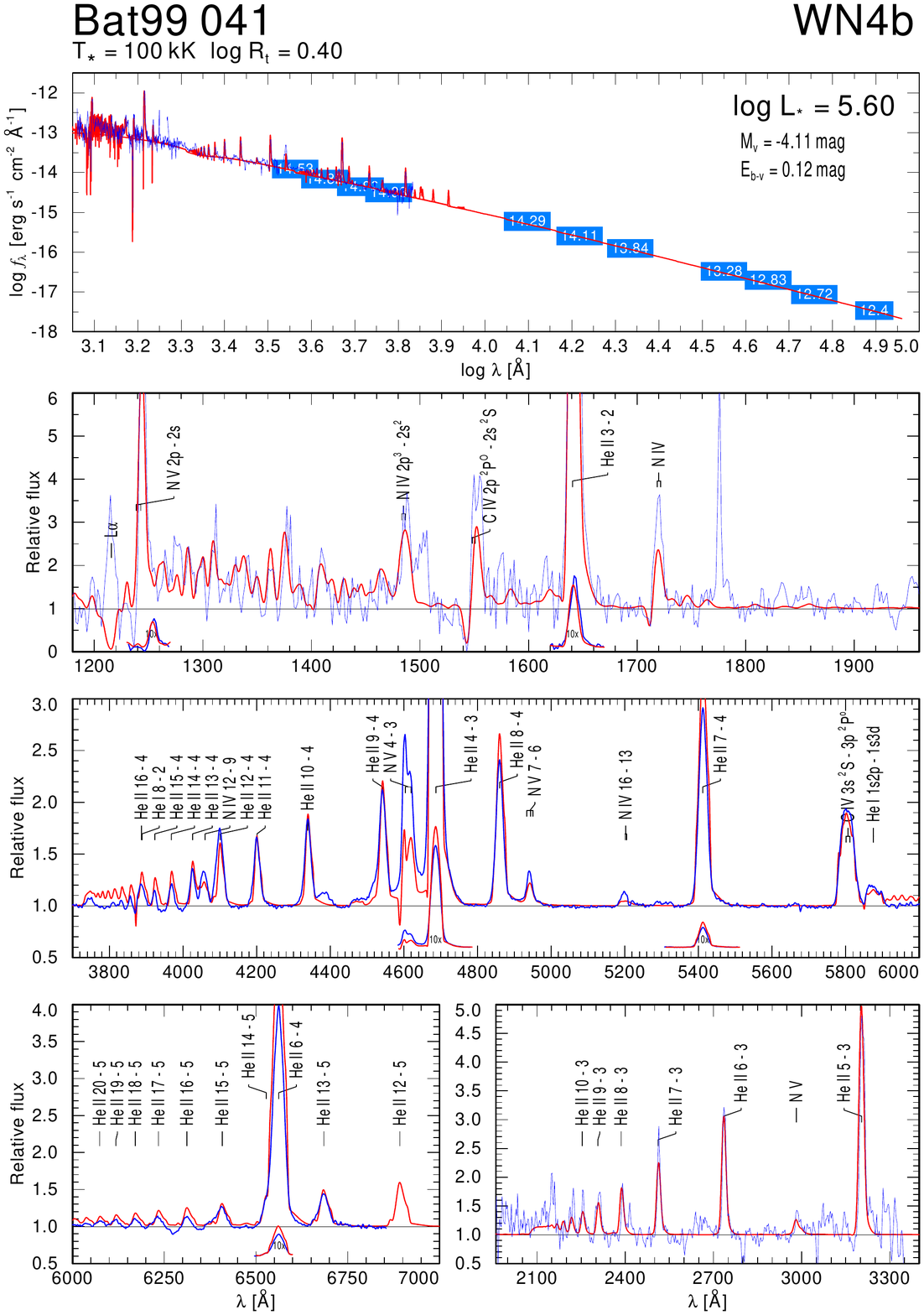}
  \qquad
  \includegraphics[width=0.46\hsize]{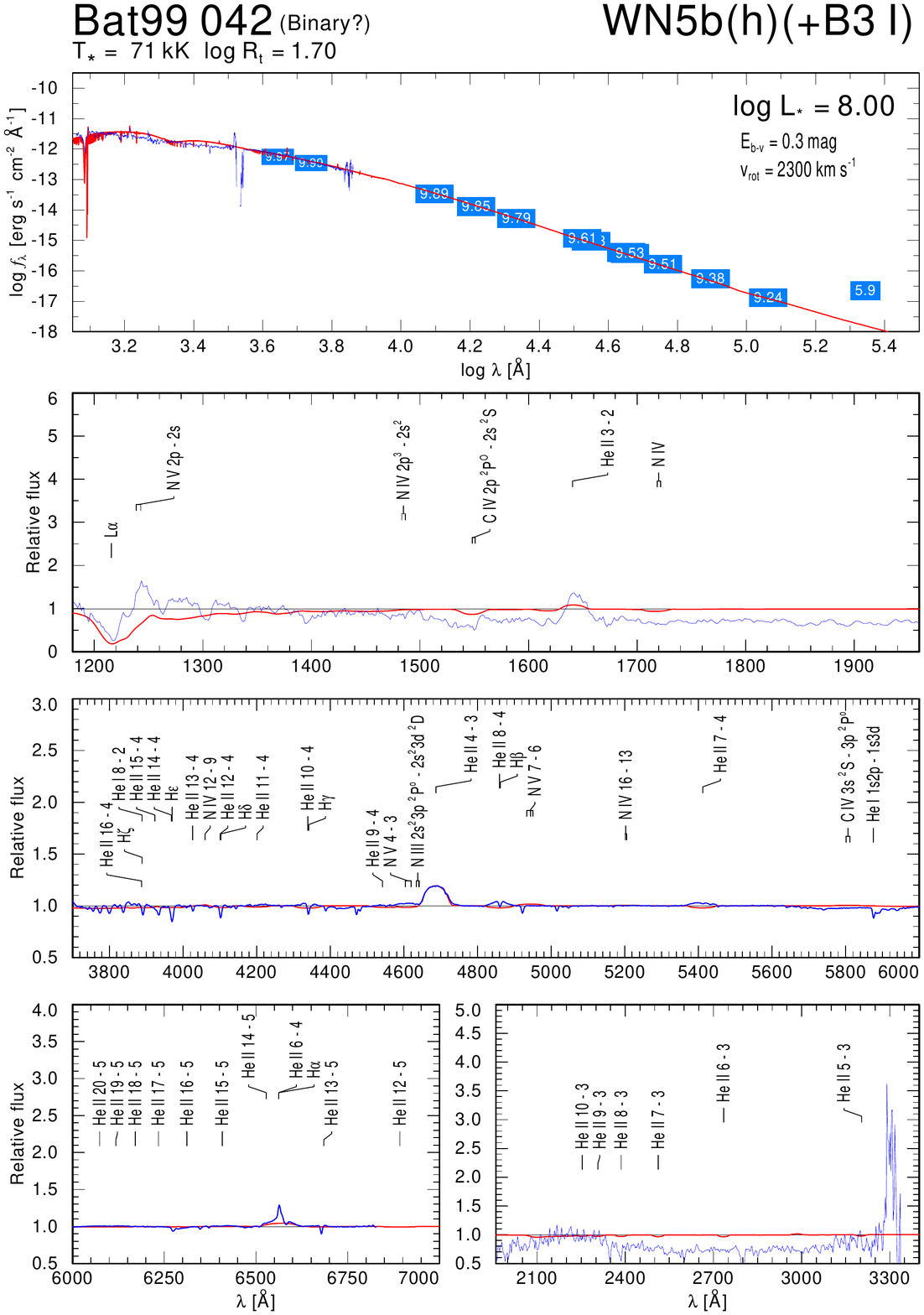}
  \vspace{-0.4cm}
  \caption{Spectral fit for BAT99\,041 and BAT99\,042}
  \label{fig:bat041}
  \label{fig:bat042}
\end{figure*}

\clearpage

\begin{figure*}
  \centering
  \includegraphics[width=0.46\hsize]{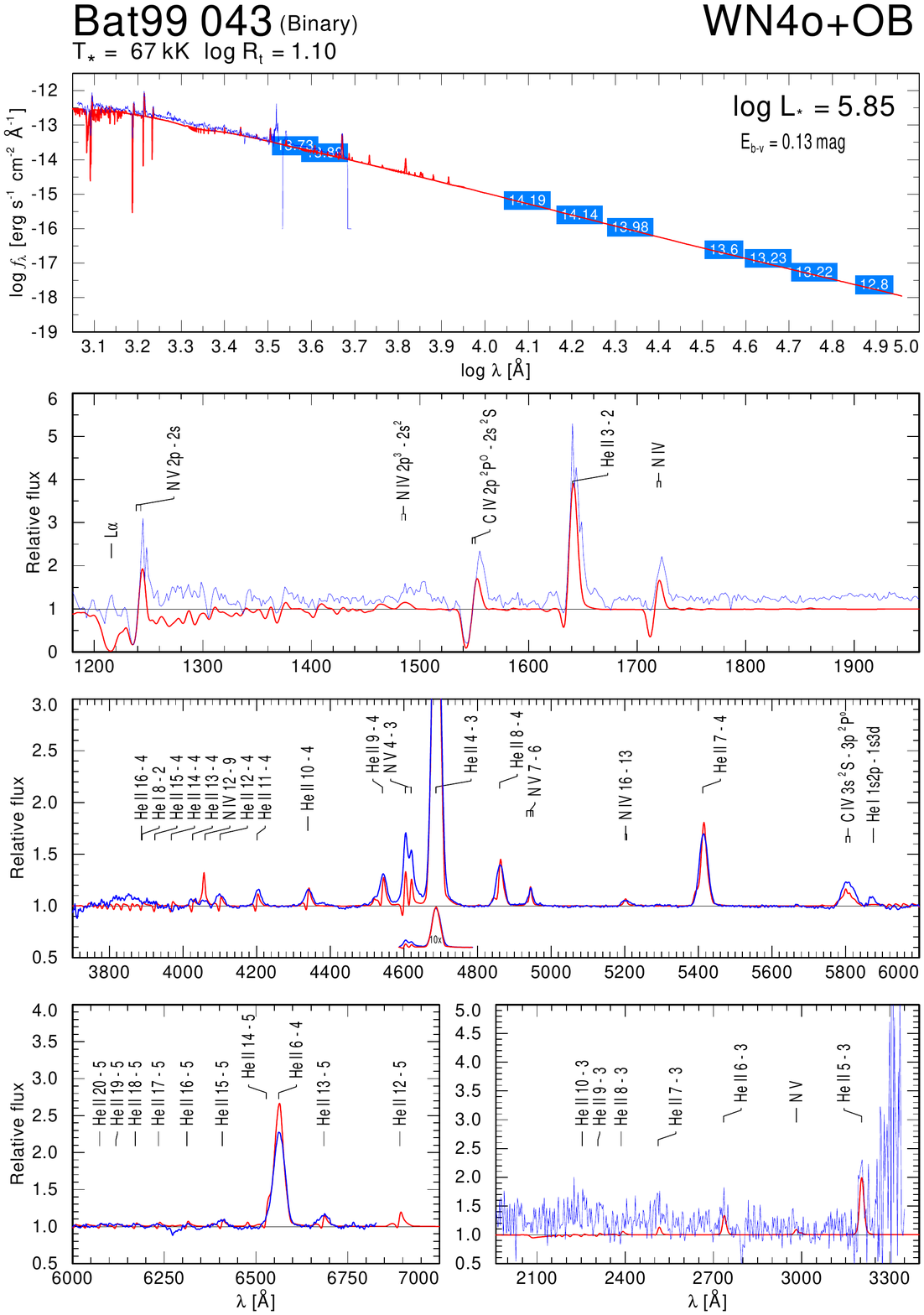}
  \qquad
  \includegraphics[width=0.46\hsize]{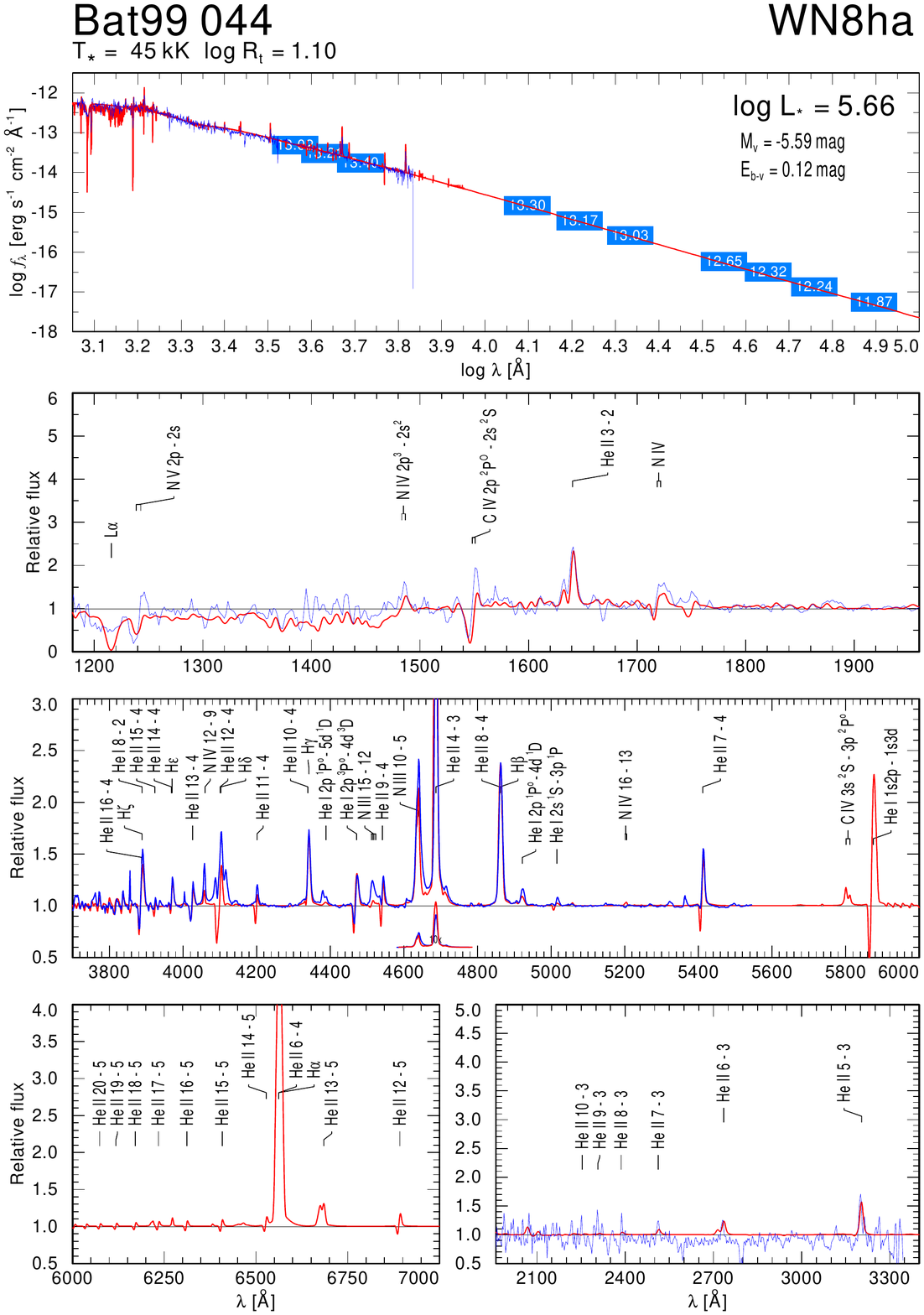}
  \vspace{-0.4cm}
  \caption{Spectral fit for BAT99\,043 and BAT99\,044}
  \label{fig:bat043}
  \label{fig:bat044}
\end{figure*}

\begin{figure*}
  \centering
  \includegraphics[width=0.46\hsize]{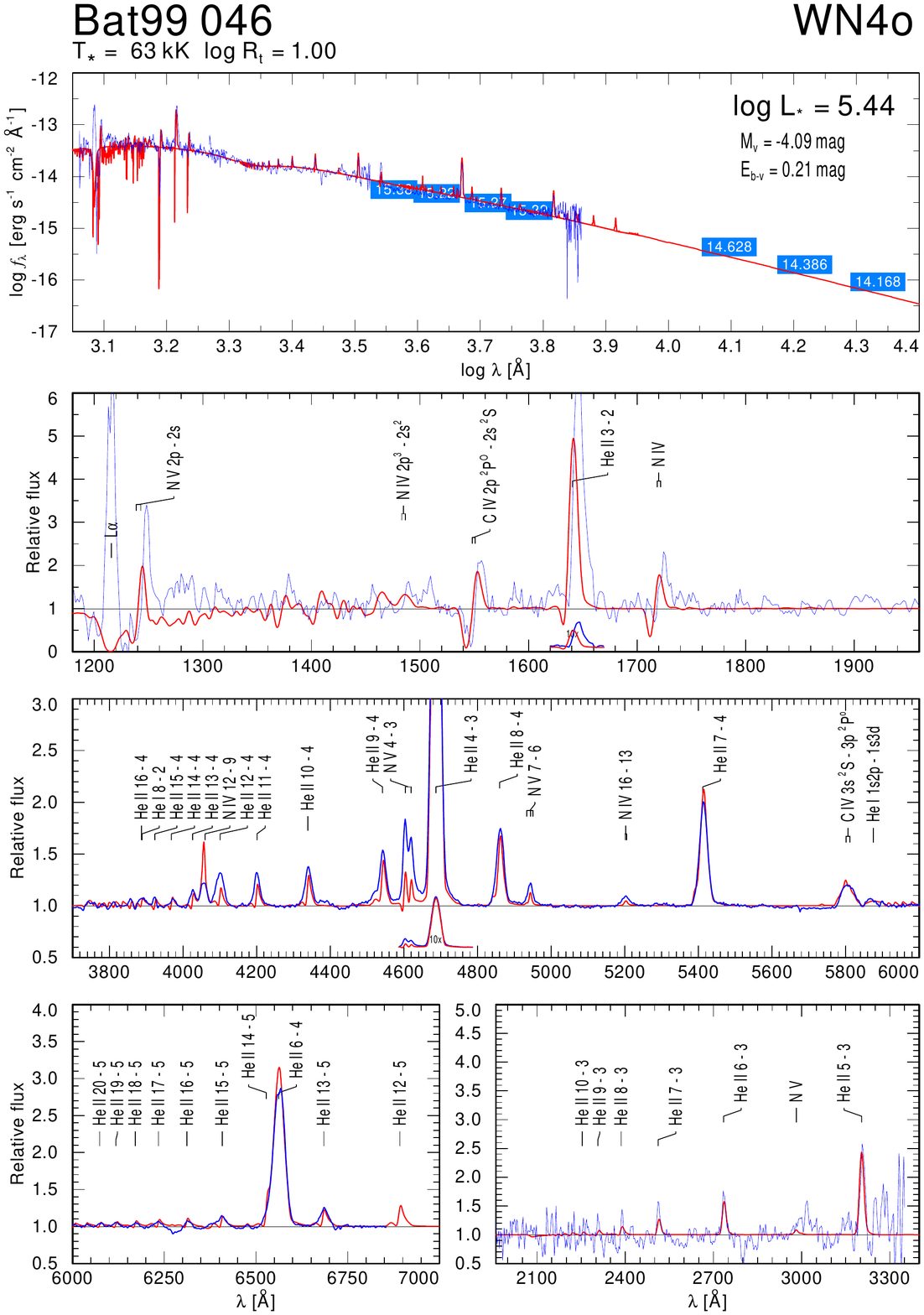}
  \qquad
  \includegraphics[width=0.46\hsize]{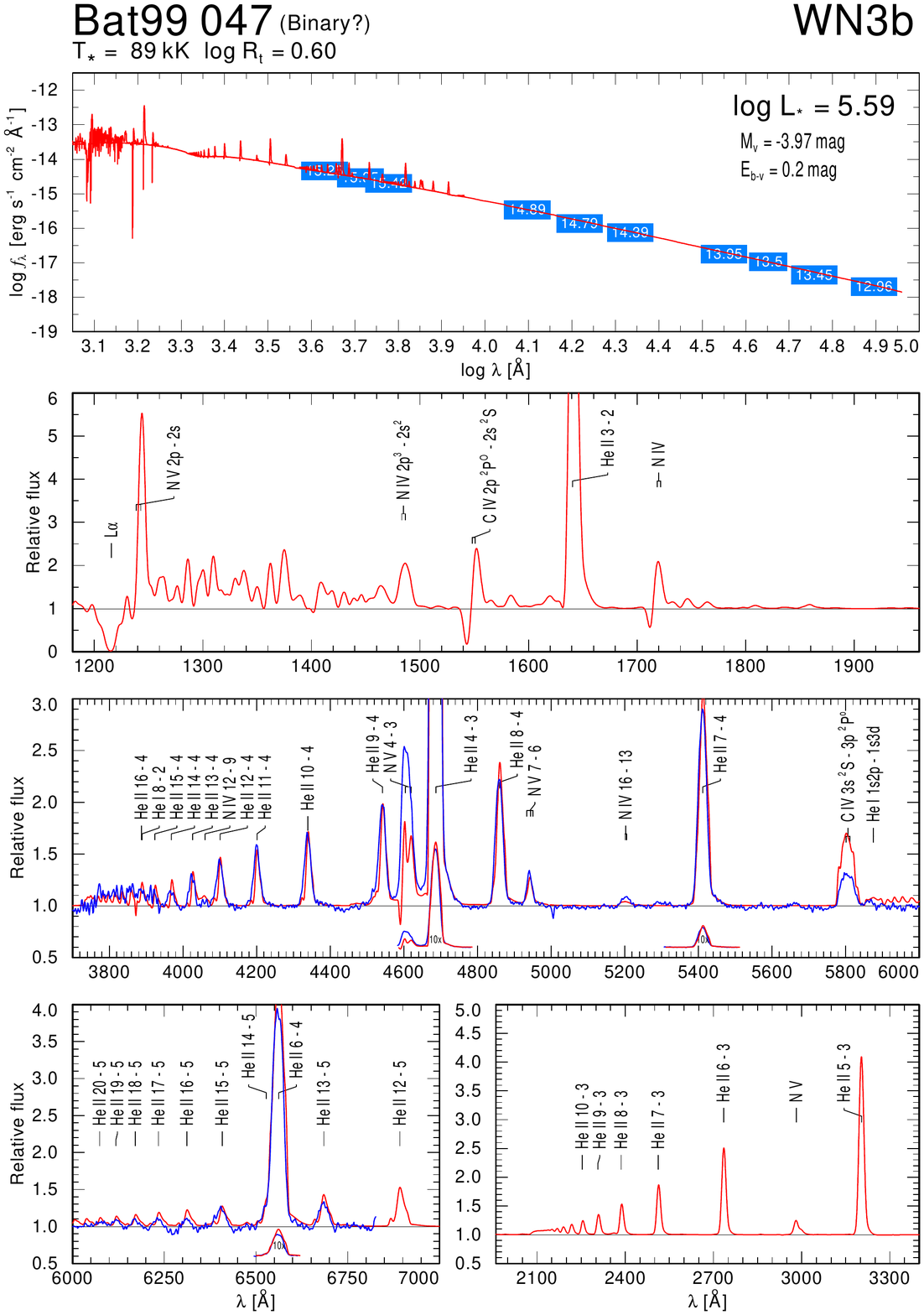}
  \vspace{-0.4cm}
  \caption{Spectral fit for BAT99\,046 and BAT99\,047}
  \label{fig:bat046}
  \label{fig:bat047}
\end{figure*}

\clearpage

\begin{figure*}
  \centering
  \includegraphics[width=0.46\hsize]{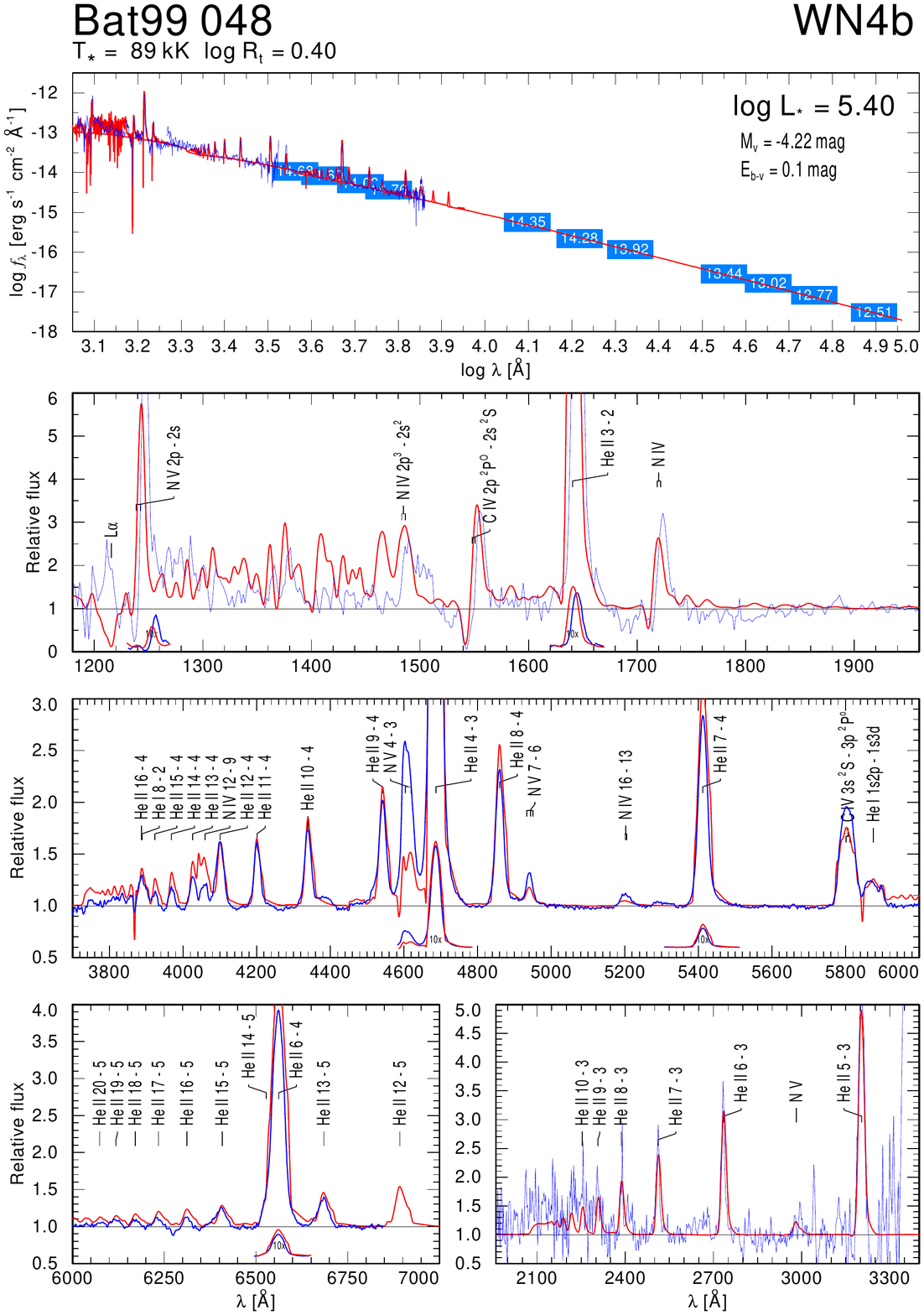}
  \qquad
  \includegraphics[width=0.46\hsize]{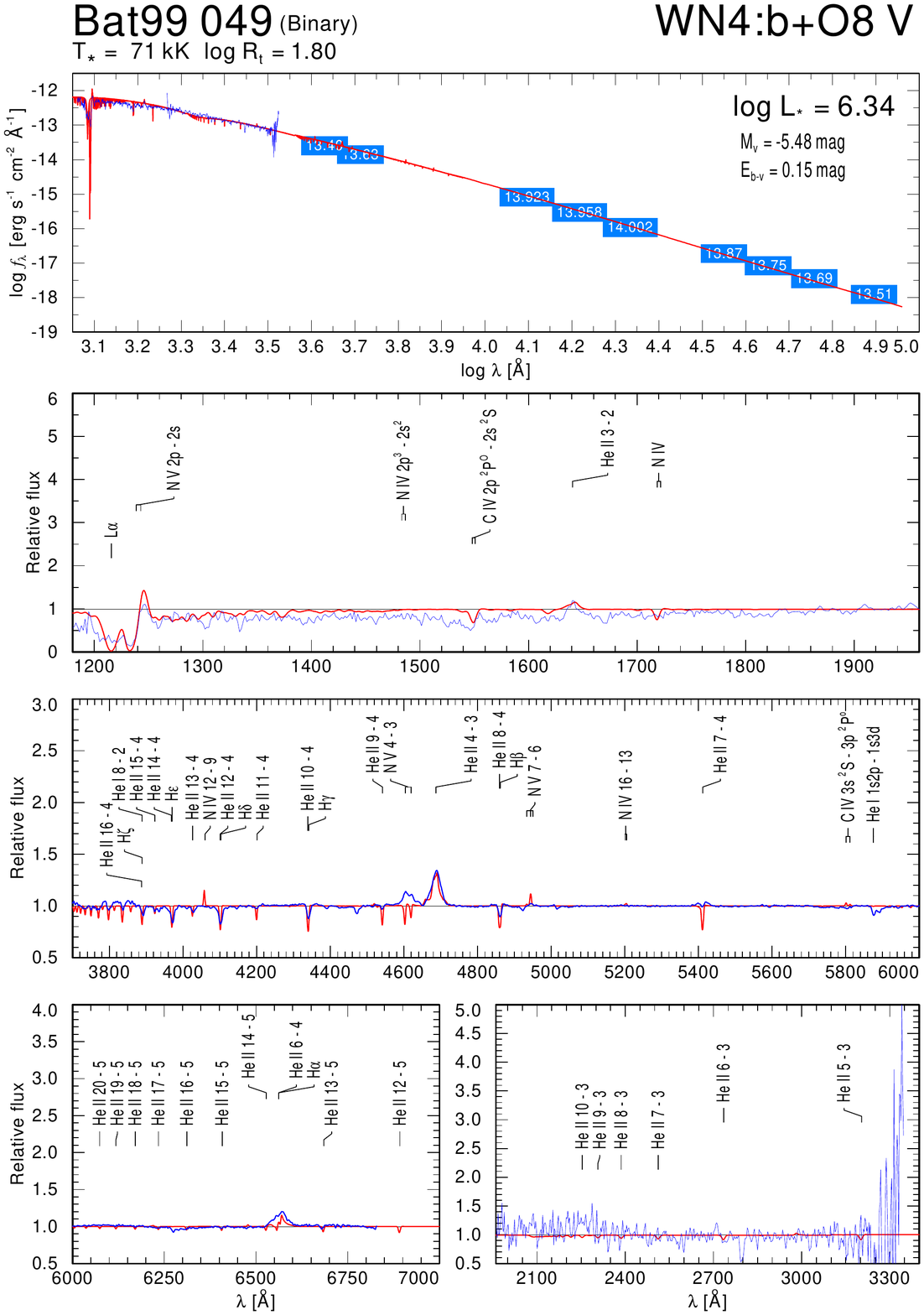}
  \vspace{-0.4cm}
  \caption{Spectral fit for BAT99\,048 and BAT99\,049}
  \label{fig:bat048}
  \label{fig:bat049}
\end{figure*}

\begin{figure*}
  \centering
  \includegraphics[width=0.46\hsize]{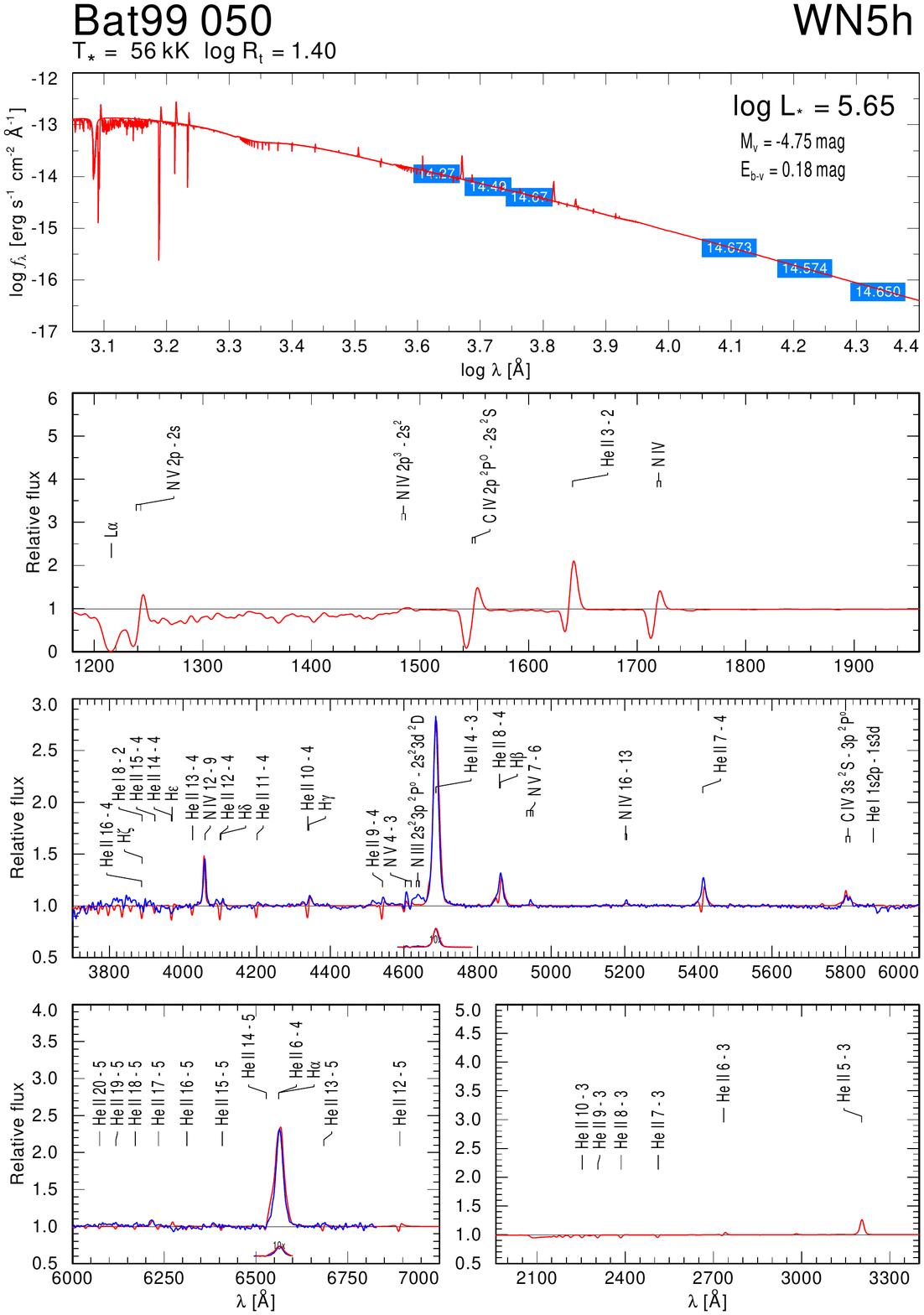}
  \qquad
  \includegraphics[width=0.46\hsize]{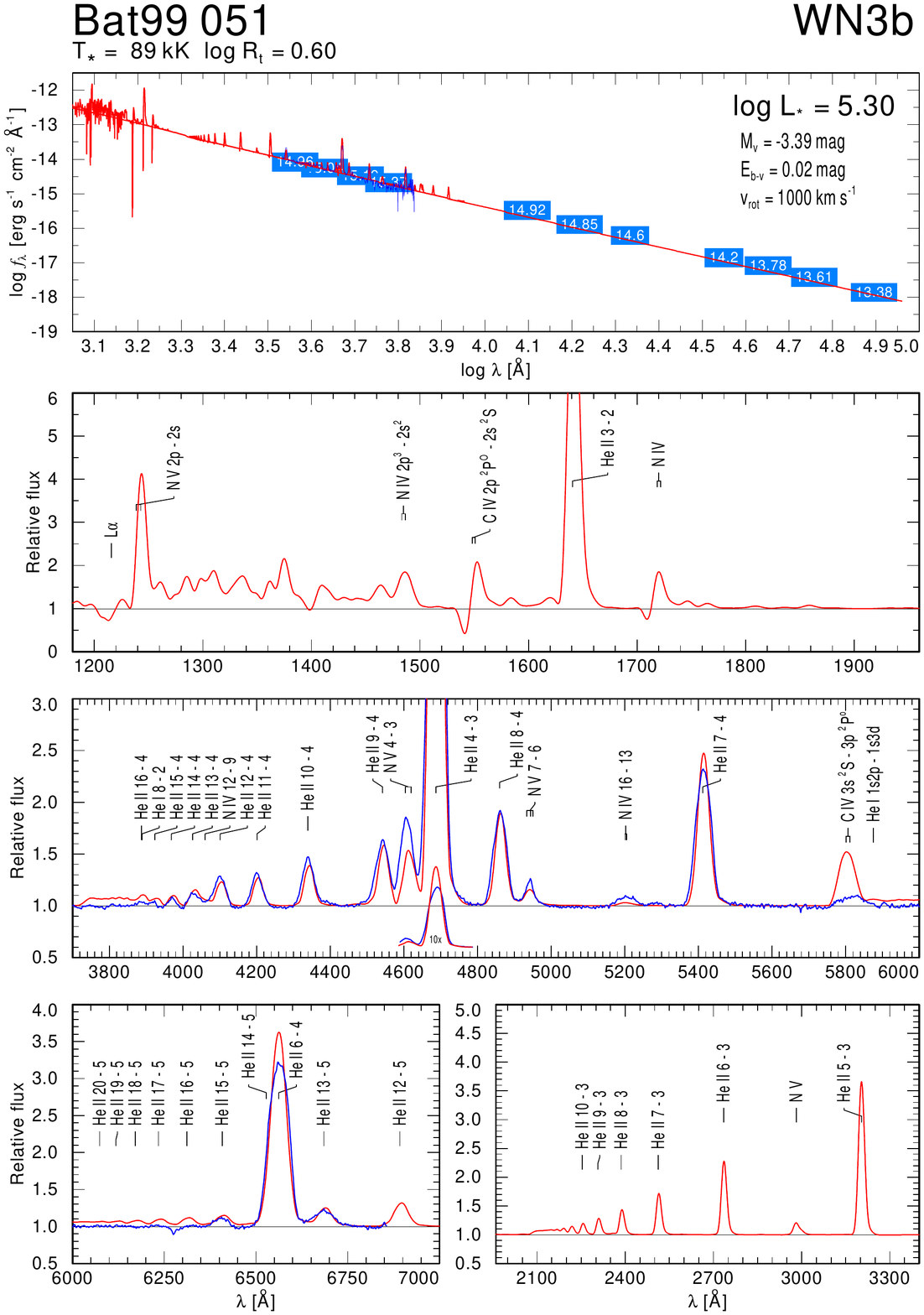}
  \vspace{-0.4cm}
  \caption{Spectral fit for BAT99\,050 and BAT99\,051}
  \label{fig:bat050}
  \label{fig:bat051}
\end{figure*}

\clearpage

\begin{figure*}
  \centering
  \includegraphics[width=0.46\hsize]{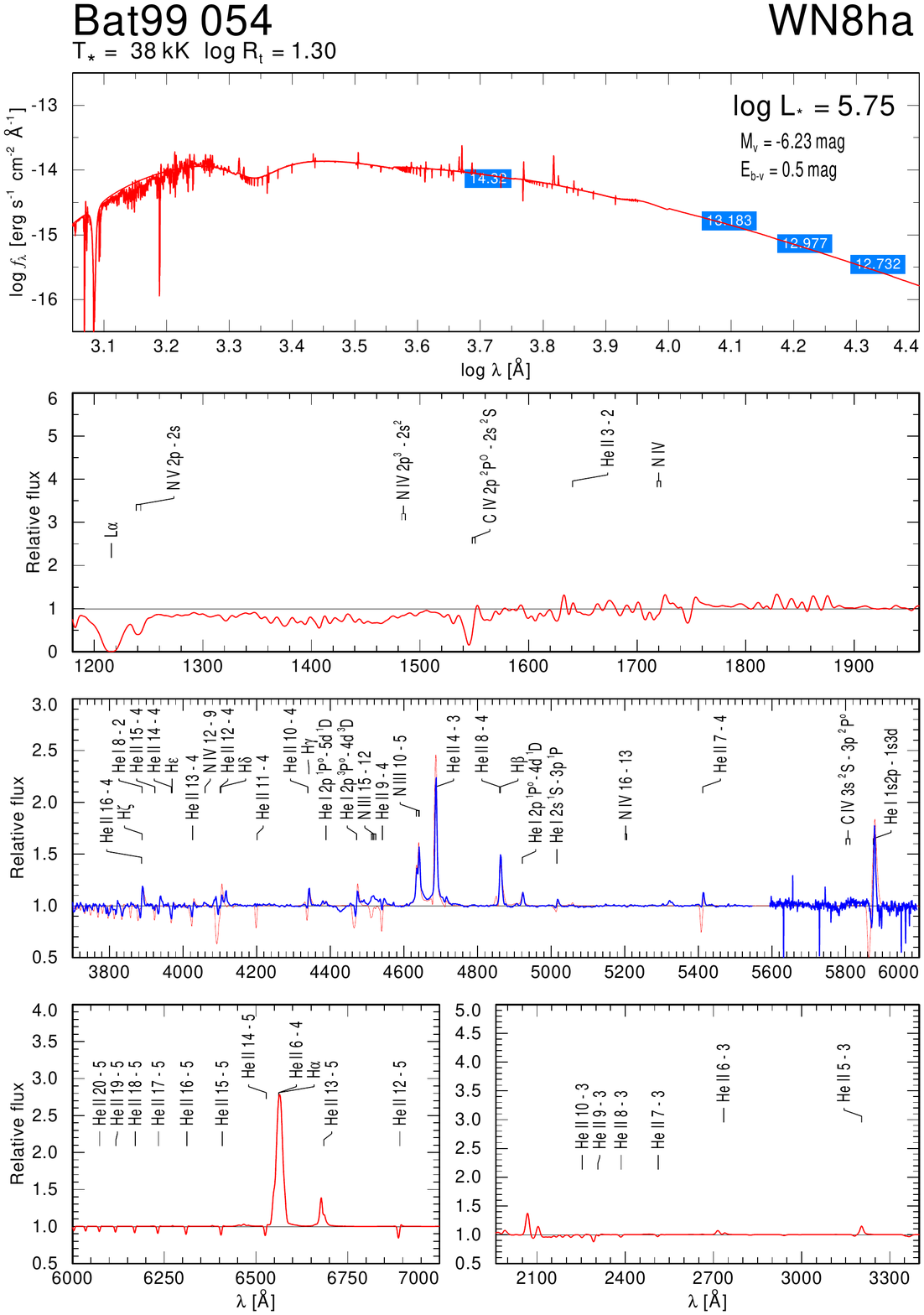}
  \qquad
  \includegraphics[width=0.46\hsize]{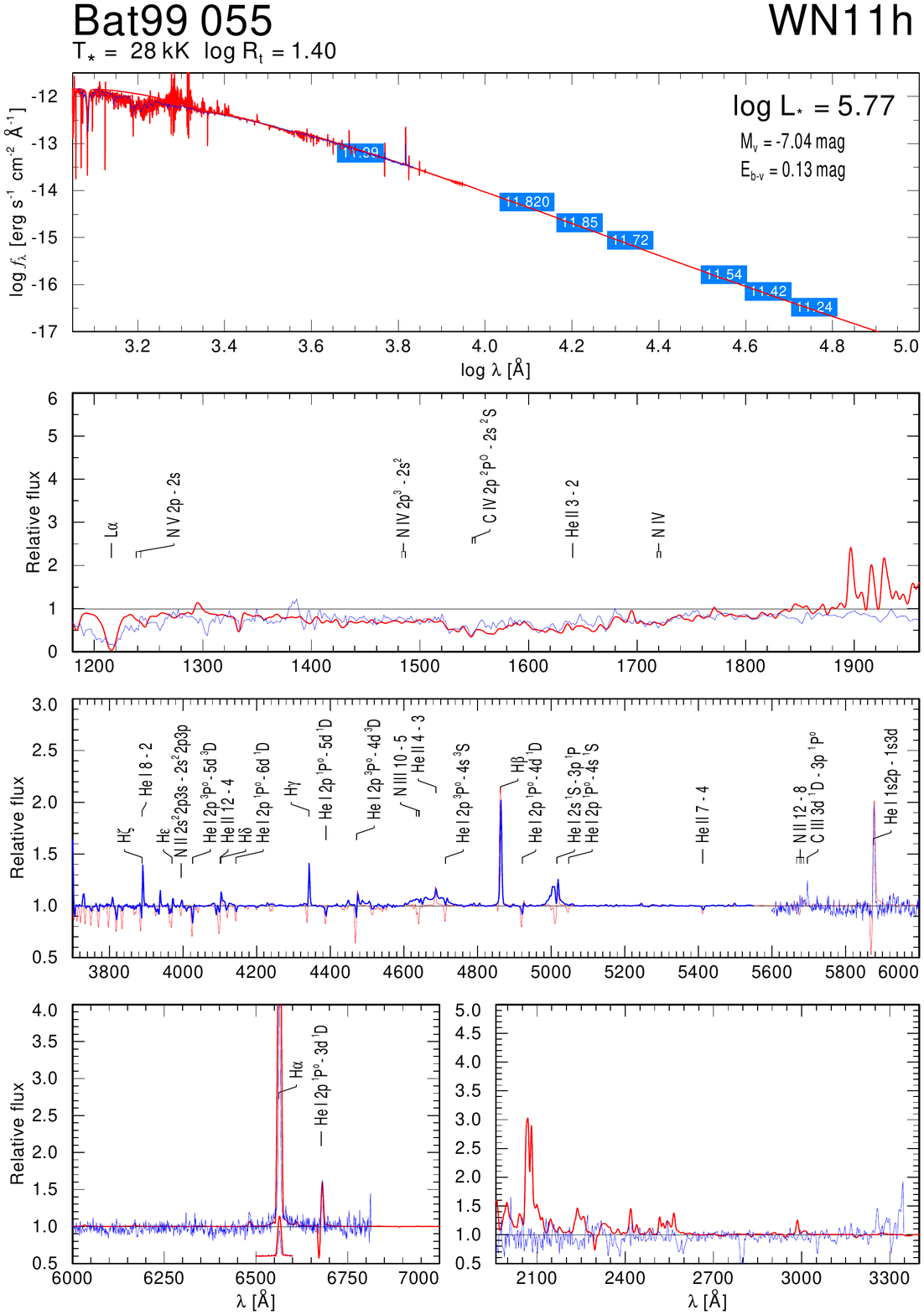}
  \vspace{-0.4cm}
  \caption{Spectral fit for BAT99\,054 and BAT99\,055}
  \label{fig:bat054}
  \label{fig:bat055}
\end{figure*}

\begin{figure*}
  \centering
  \includegraphics[width=0.46\hsize]{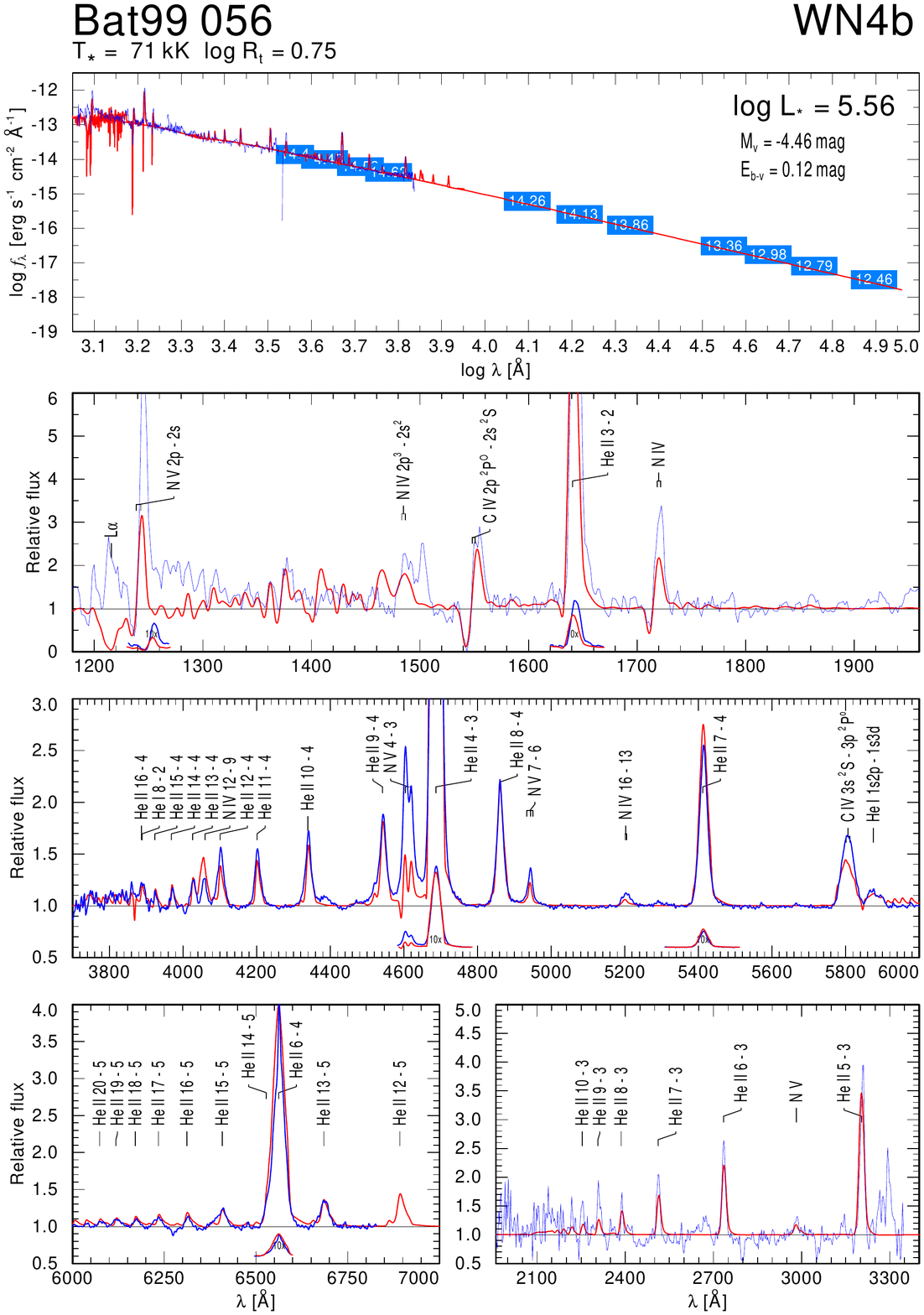}
  \qquad
  \includegraphics[width=0.46\hsize]{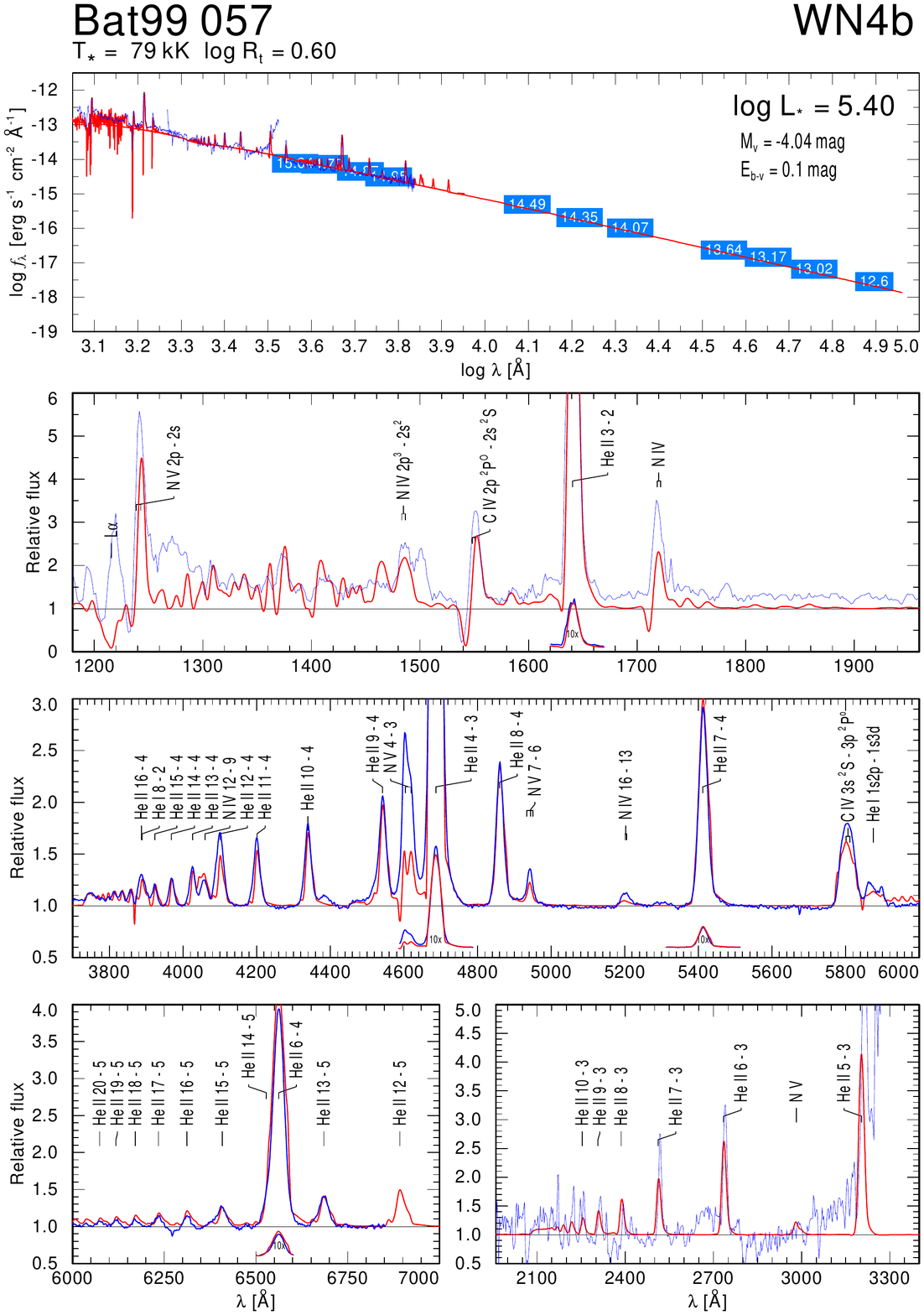}
  \vspace{-0.4cm}
  \caption{Spectral fit for BAT99\,056 and BAT99\,057}
  \label{fig:bat056}
  \label{fig:bat057}
\end{figure*}

\clearpage

\begin{figure*}
  \centering
  \includegraphics[width=0.46\hsize]{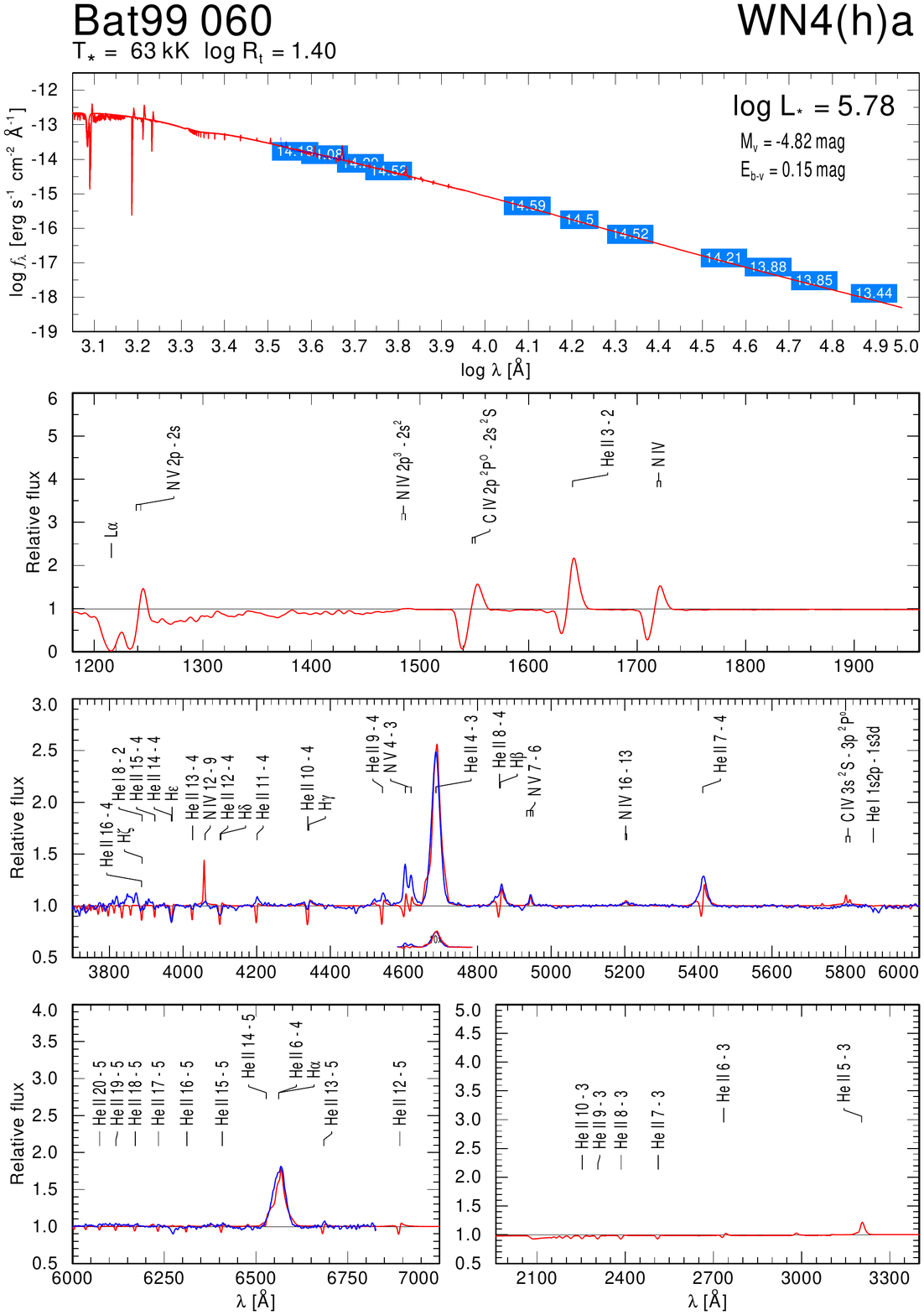}
  \qquad
  \includegraphics[width=0.46\hsize]{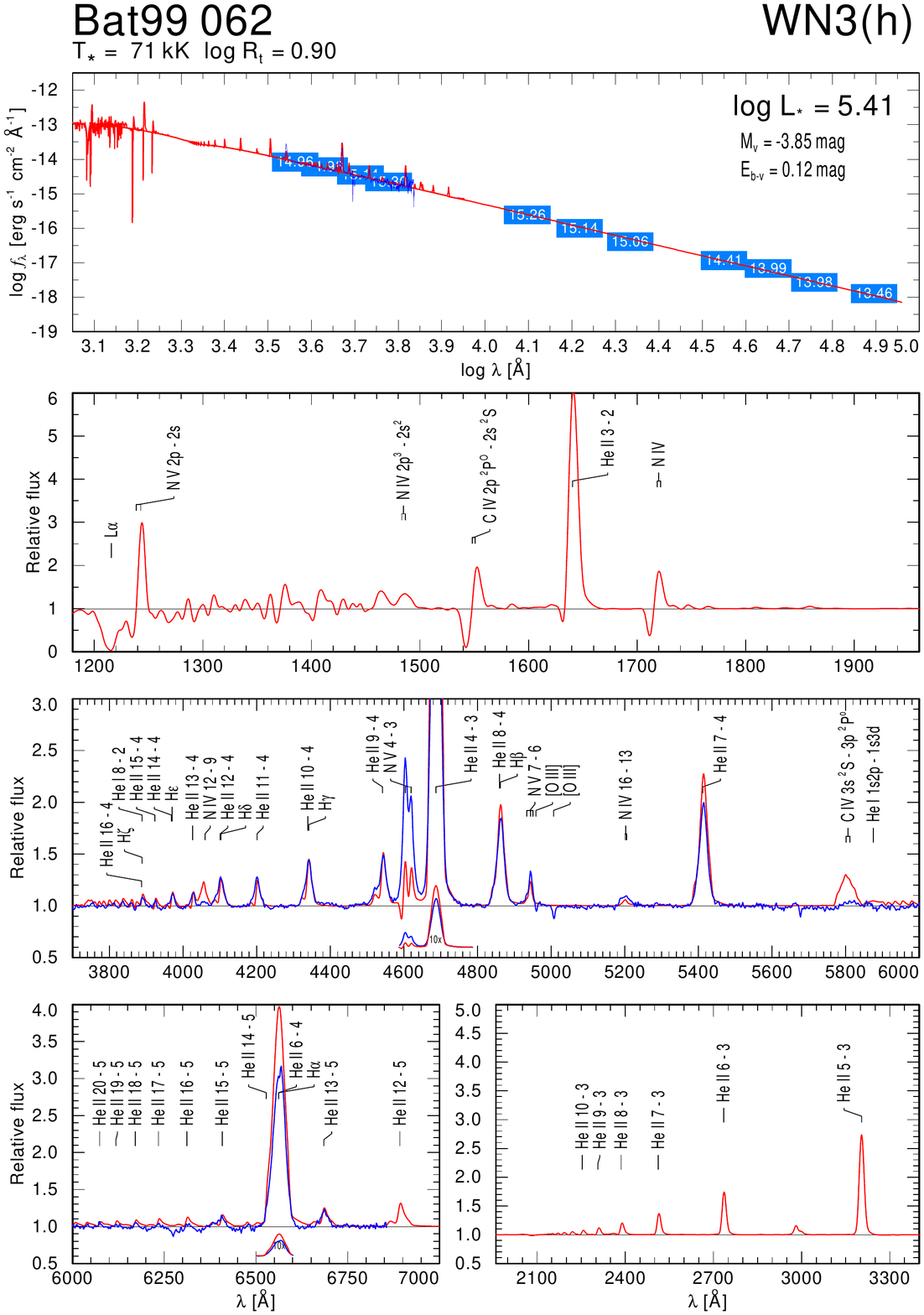}
  \vspace{-0.4cm}
  \caption{Spectral fit for BAT99\,060 and BAT99\,062}
  \label{fig:bat060}
  \label{fig:bat062}
\end{figure*}

\clearpage

\begin{figure*}
  \centering
  \includegraphics[width=0.46\hsize]{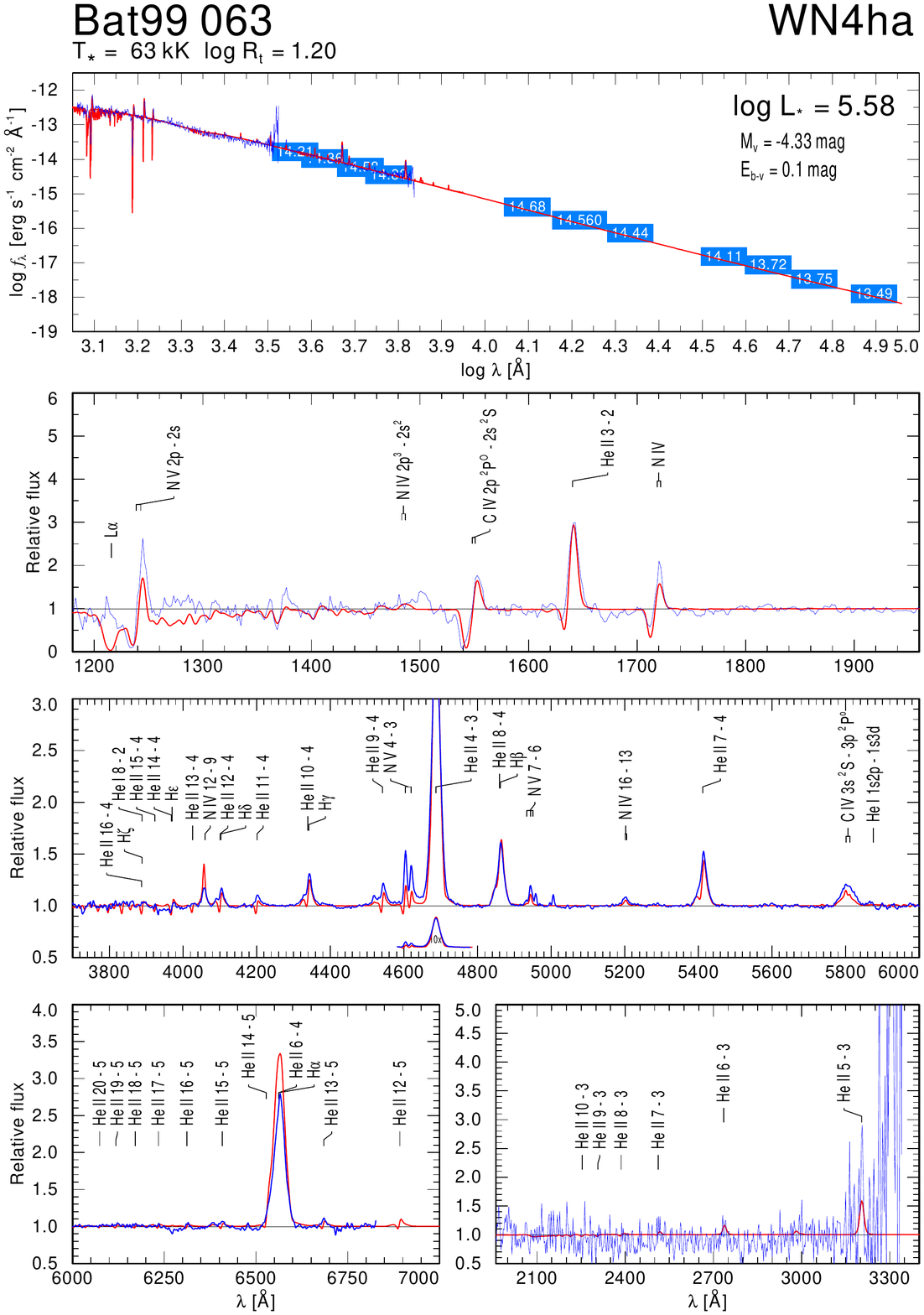}
  \qquad
  \includegraphics[width=0.46\hsize]{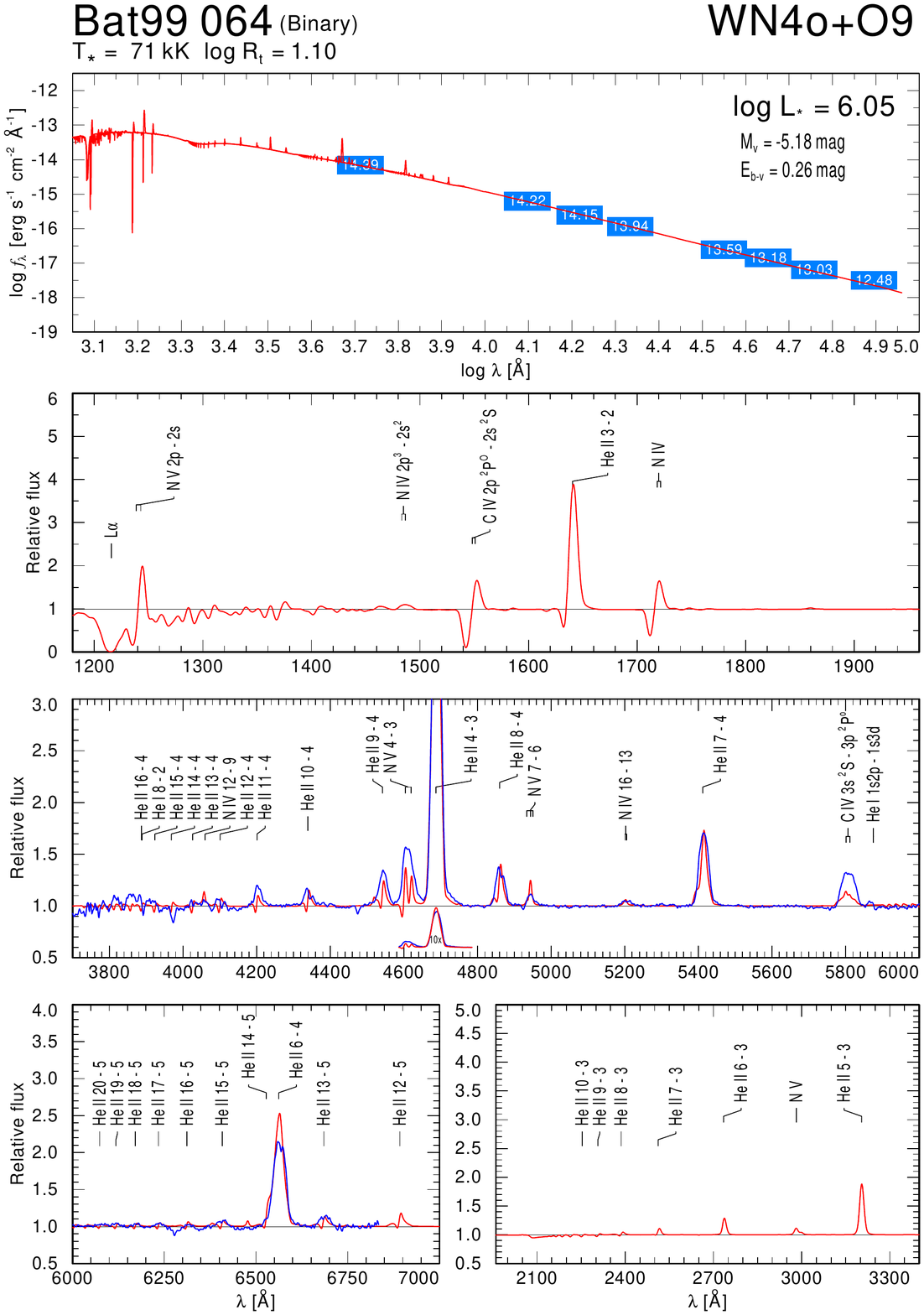}
  \vspace{-0.4cm}
  \caption{Spectral fit for BAT99\,063 and BAT99\,064}
  \label{fig:bat063}
  \label{fig:bat064}
\end{figure*}

\begin{figure*}
  \centering
  \includegraphics[width=0.46\hsize]{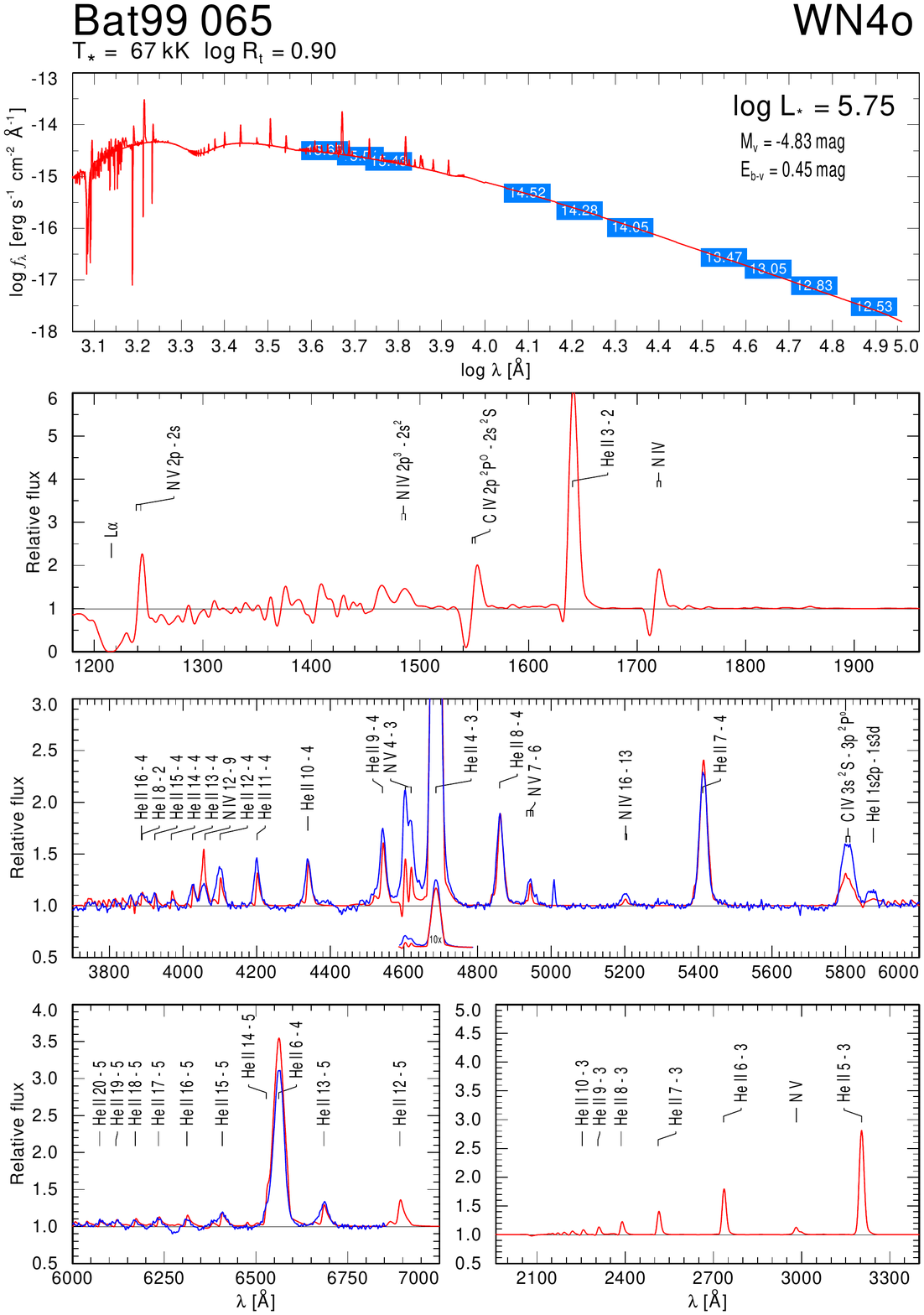}
  \qquad
  \includegraphics[width=0.46\hsize]{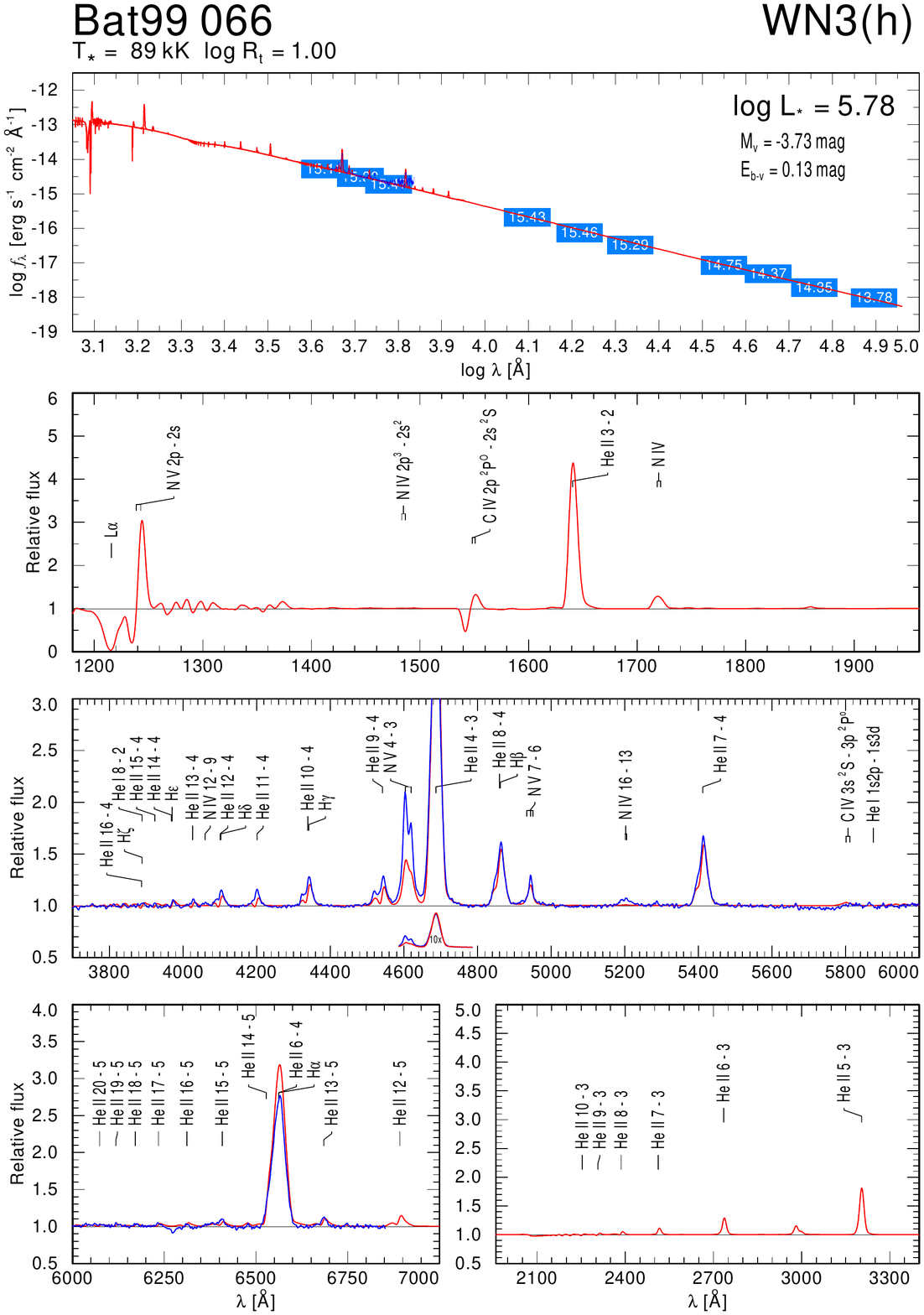}
  \vspace{-0.4cm}
  \caption{Spectral fit for BAT99\,065 and BAT99\,066}
  \label{fig:bat065}
  \label{fig:bat066}
\end{figure*}

\clearpage

\begin{figure*}
  \centering
  \includegraphics[width=0.46\hsize]{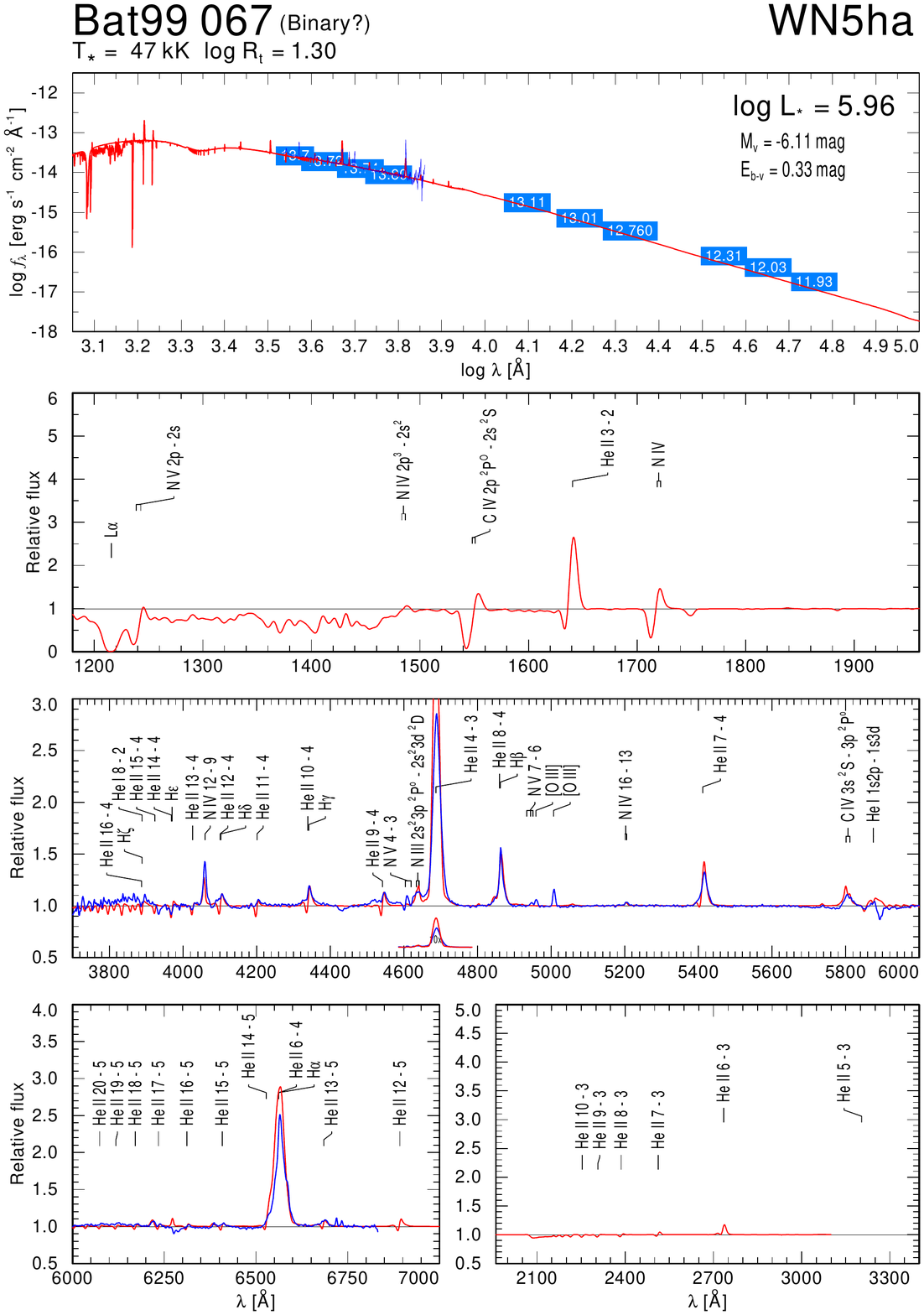}
  \qquad
  \includegraphics[width=0.46\hsize]{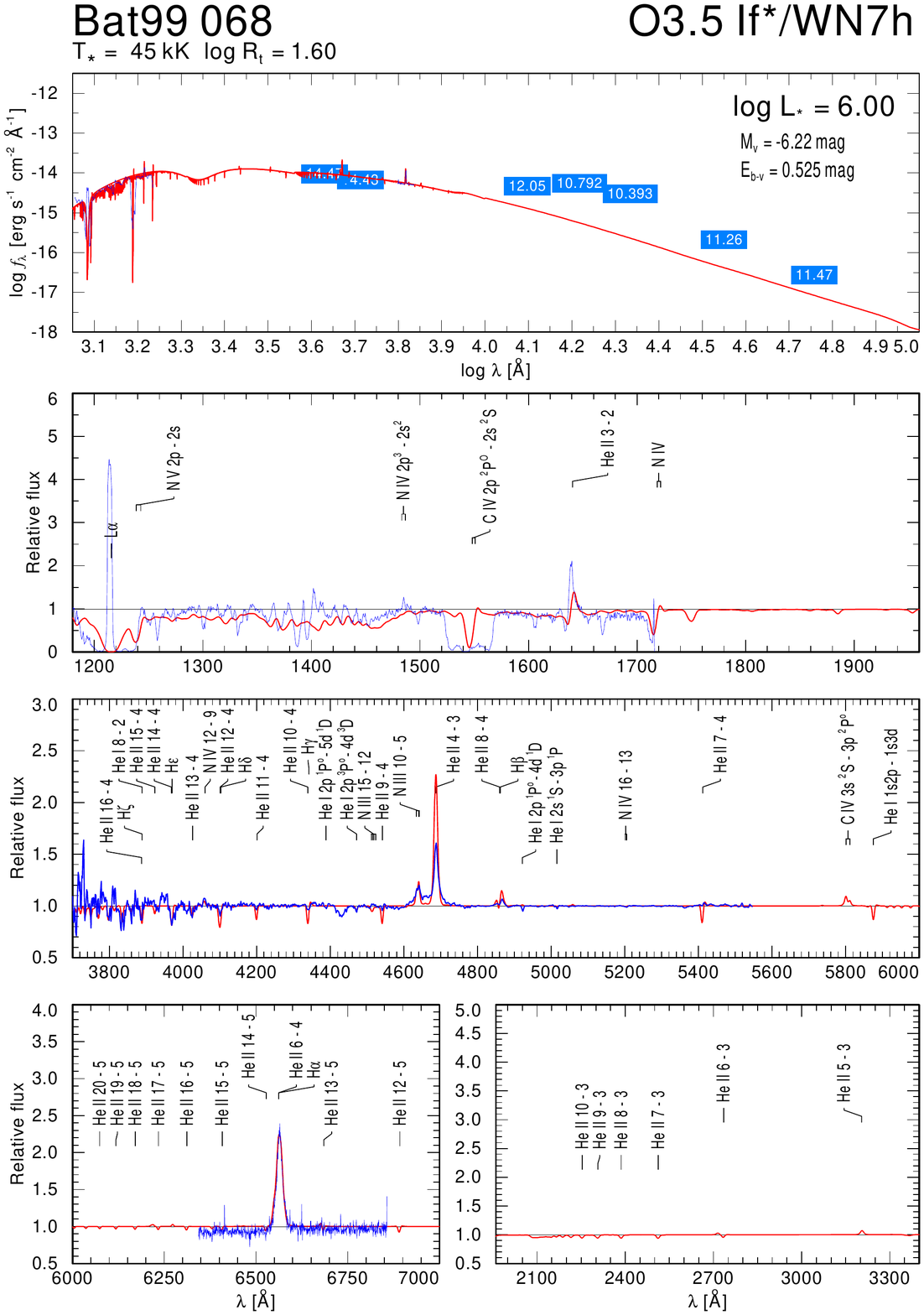}
  \vspace{-0.4cm}
  \caption{Spectral fit for BAT99\,067 and BAT99\,068}
  \label{fig:bat067}
  \label{fig:bat068}
\end{figure*}

\begin{figure*}
  \centering
  \includegraphics[width=0.46\hsize]{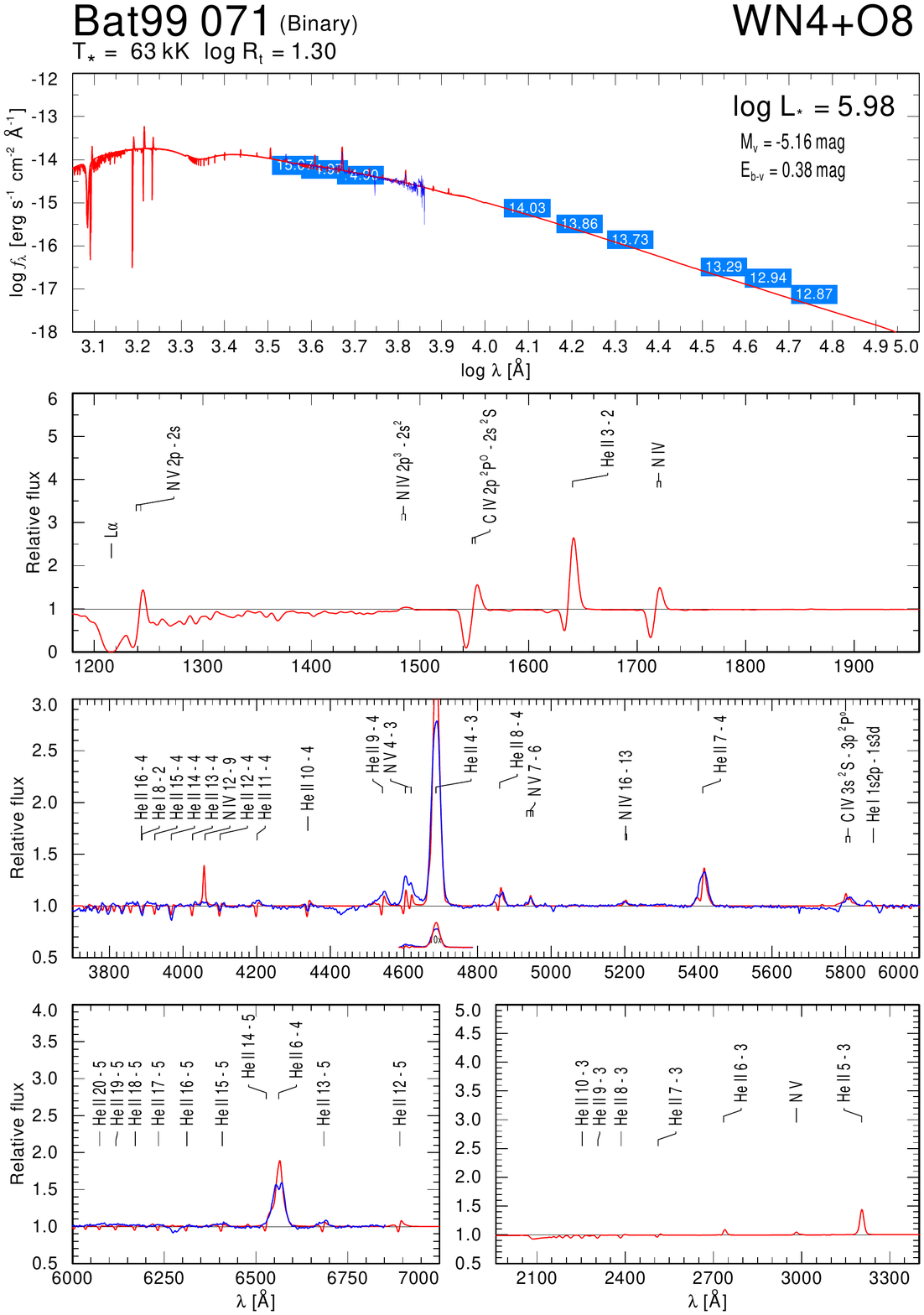}
  \qquad
  \includegraphics[width=0.46\hsize]{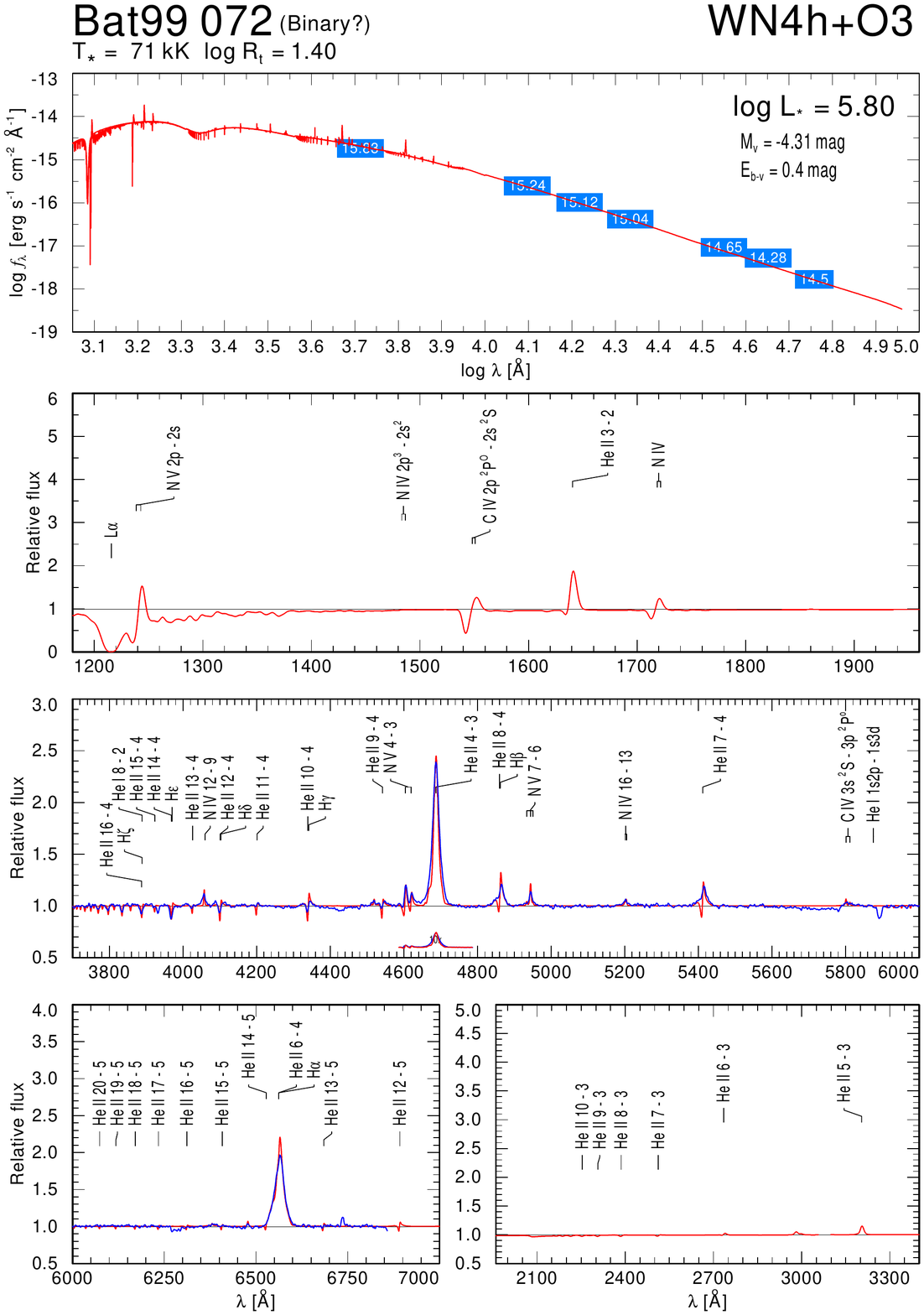}
  \vspace{-0.4cm}
  \caption{Spectral fit for BAT99\,071 and BAT99\,072}
  \label{fig:bat071}
  \label{fig:bat072}
\end{figure*}

\clearpage

\begin{figure*}
  \centering
  \includegraphics[width=0.46\hsize]{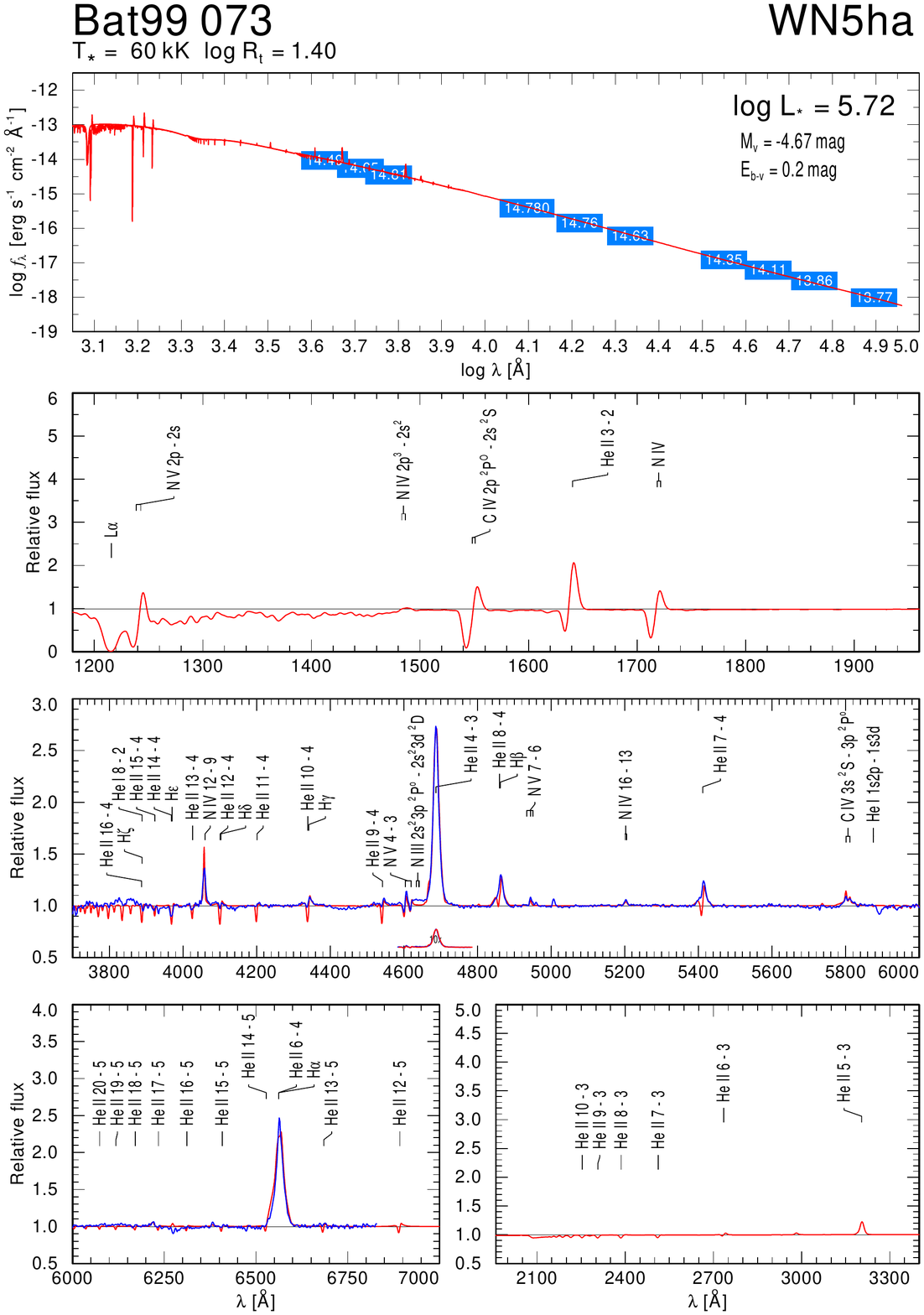}
  \qquad
  \includegraphics[width=0.46\hsize]{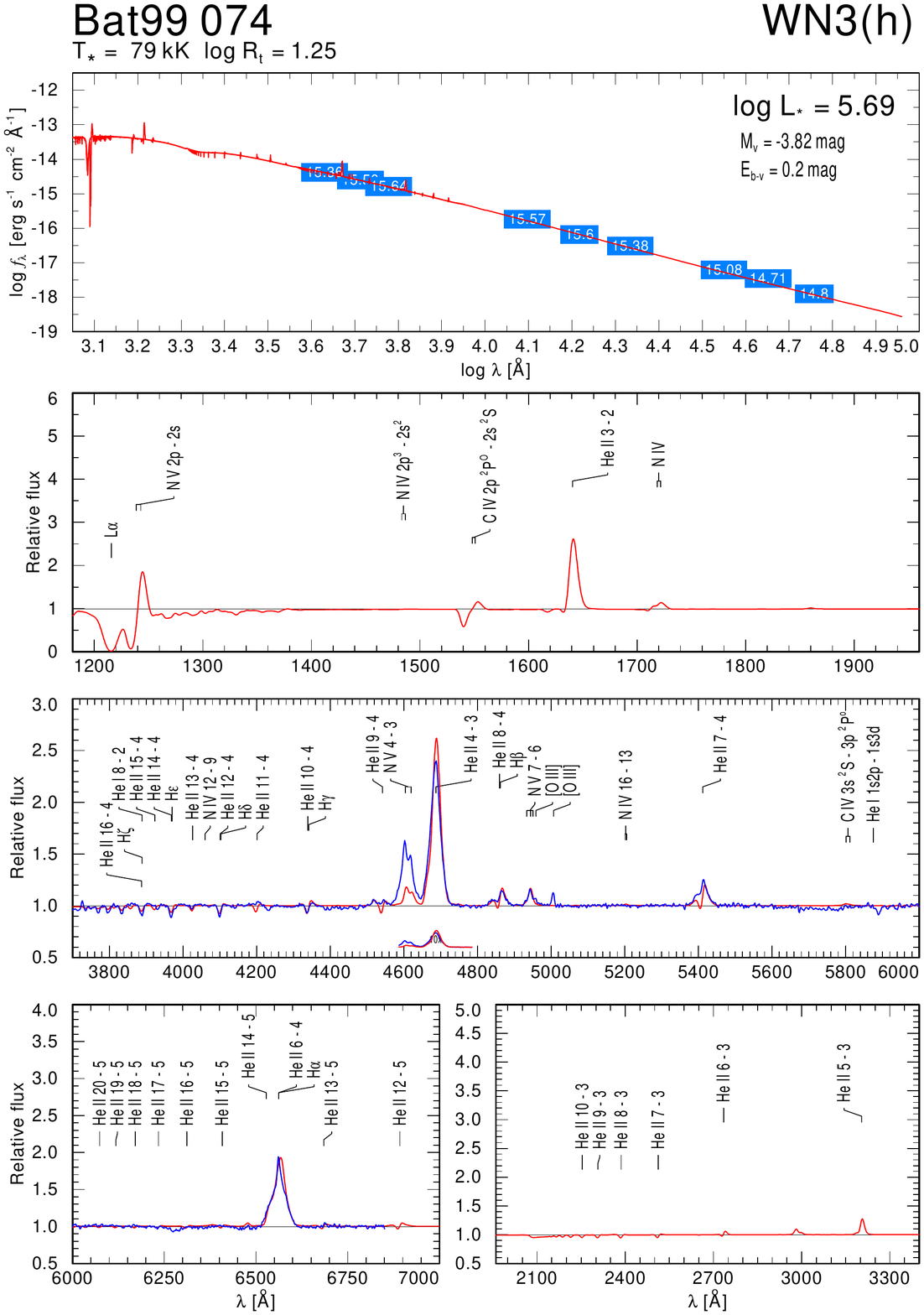}
  \vspace{-0.4cm}
  \caption{Spectral fit for BAT99\,073 and BAT99\,074}
  \label{fig:bat073}
  \label{fig:bat074}
\end{figure*}

\begin{figure*}
  \centering
  \includegraphics[width=0.46\hsize]{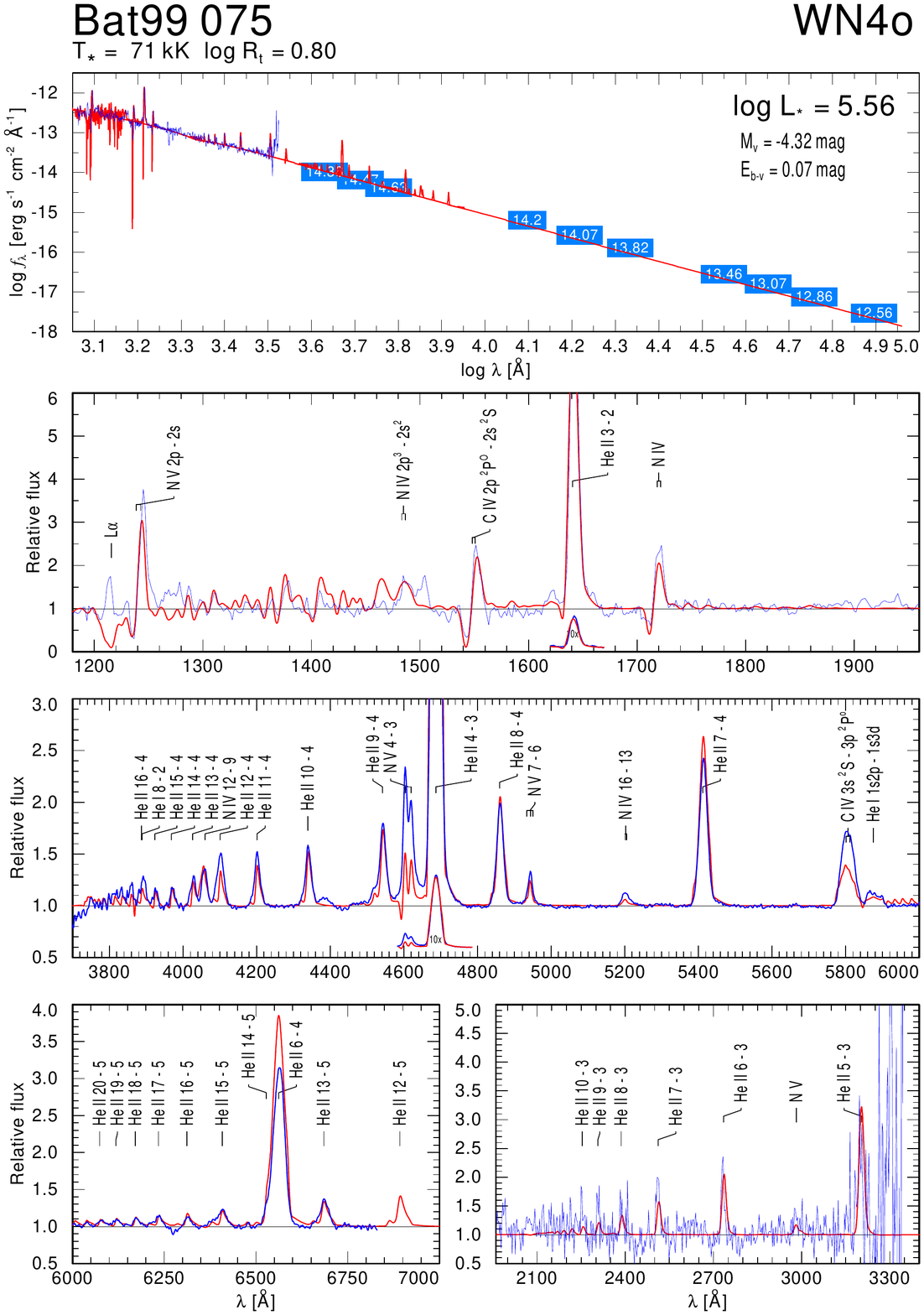}
  \qquad
  \includegraphics[width=0.46\hsize]{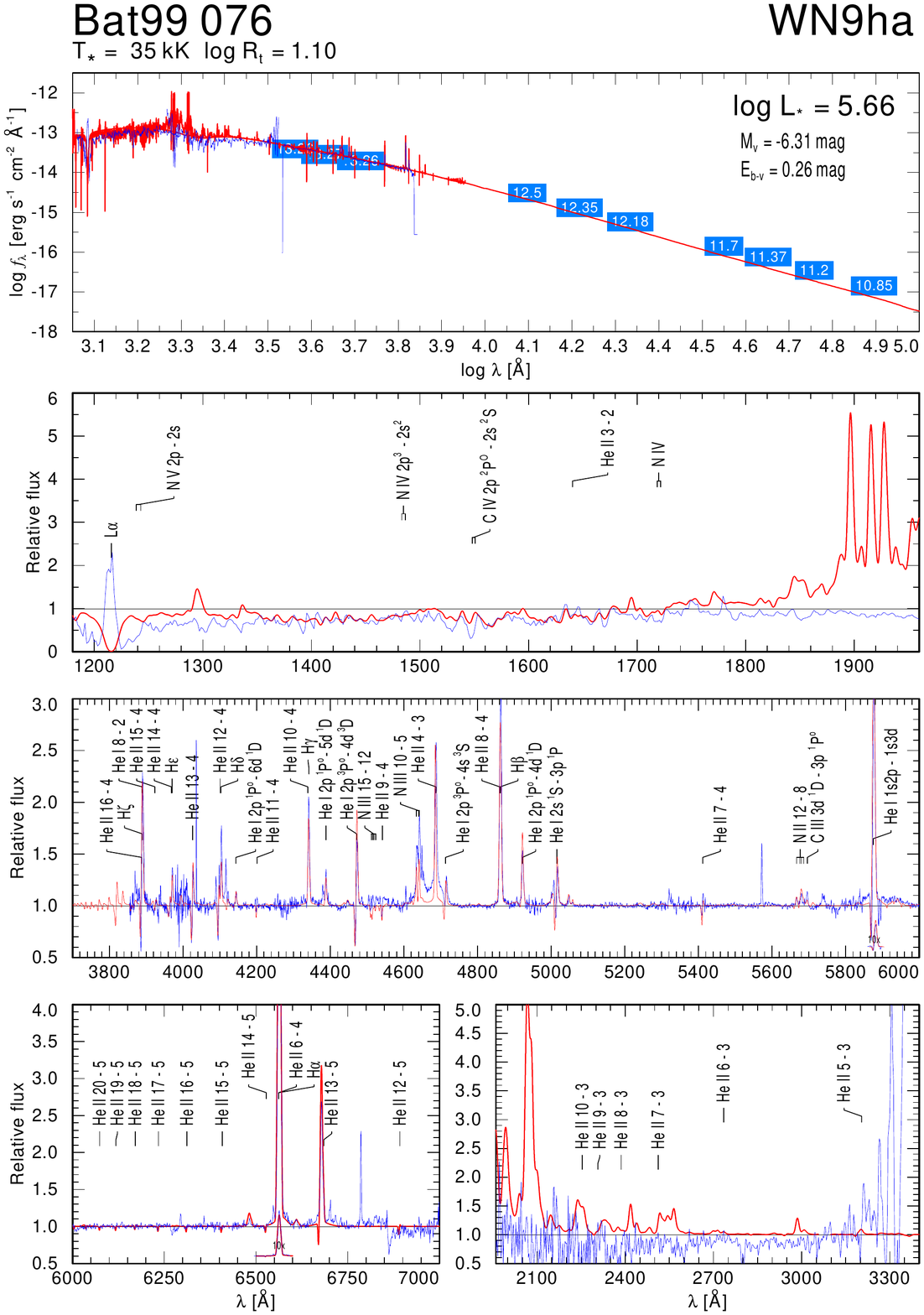}
  \vspace{-0.4cm}
  \caption{Spectral fit for BAT99\,075 and BAT99\,076}
  \label{fig:bat075}
  \label{fig:bat076}
\end{figure*}

\clearpage

\begin{figure*}
  \centering
  \includegraphics[width=0.46\hsize]{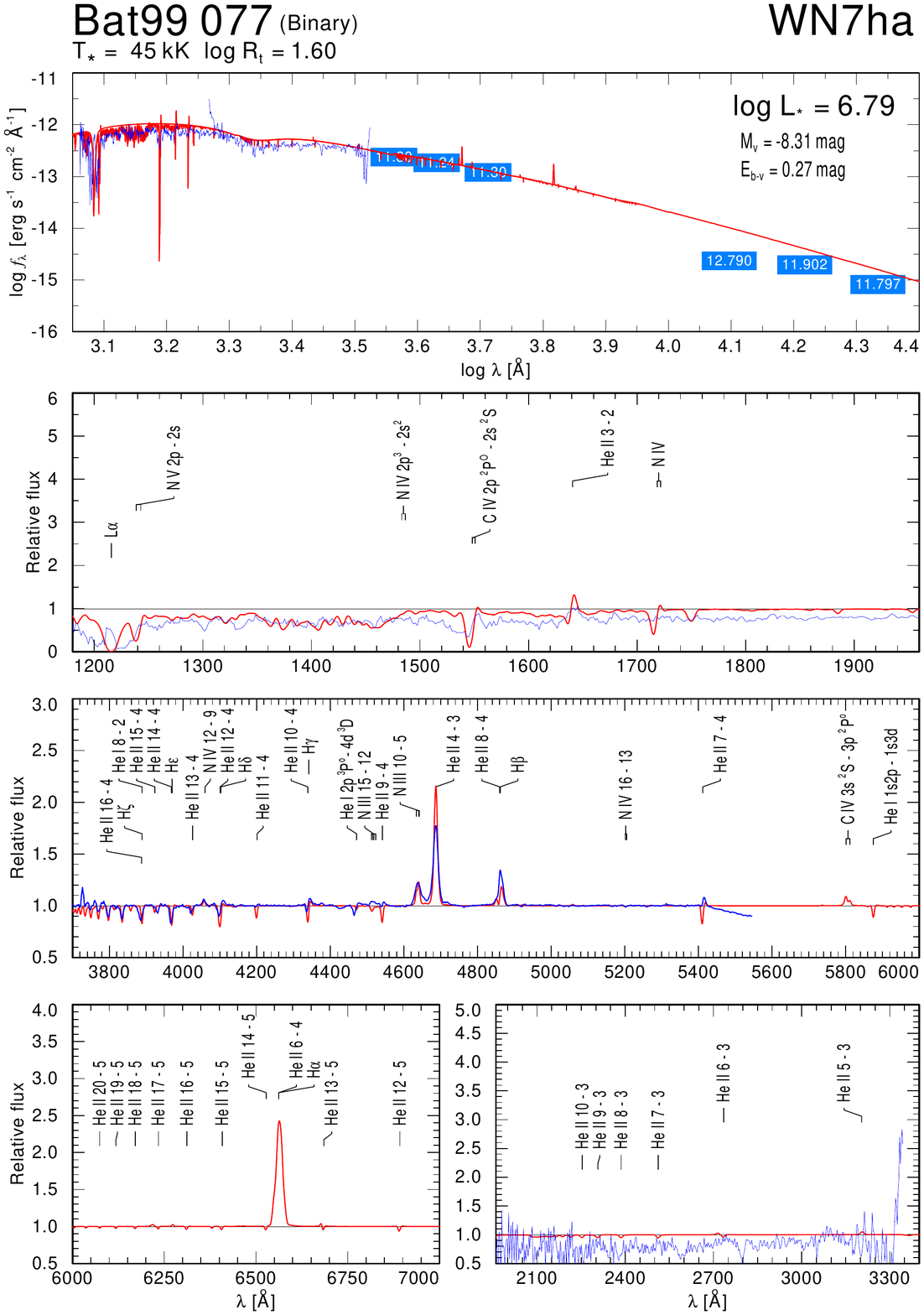}
  \qquad
  \includegraphics[width=0.46\hsize]{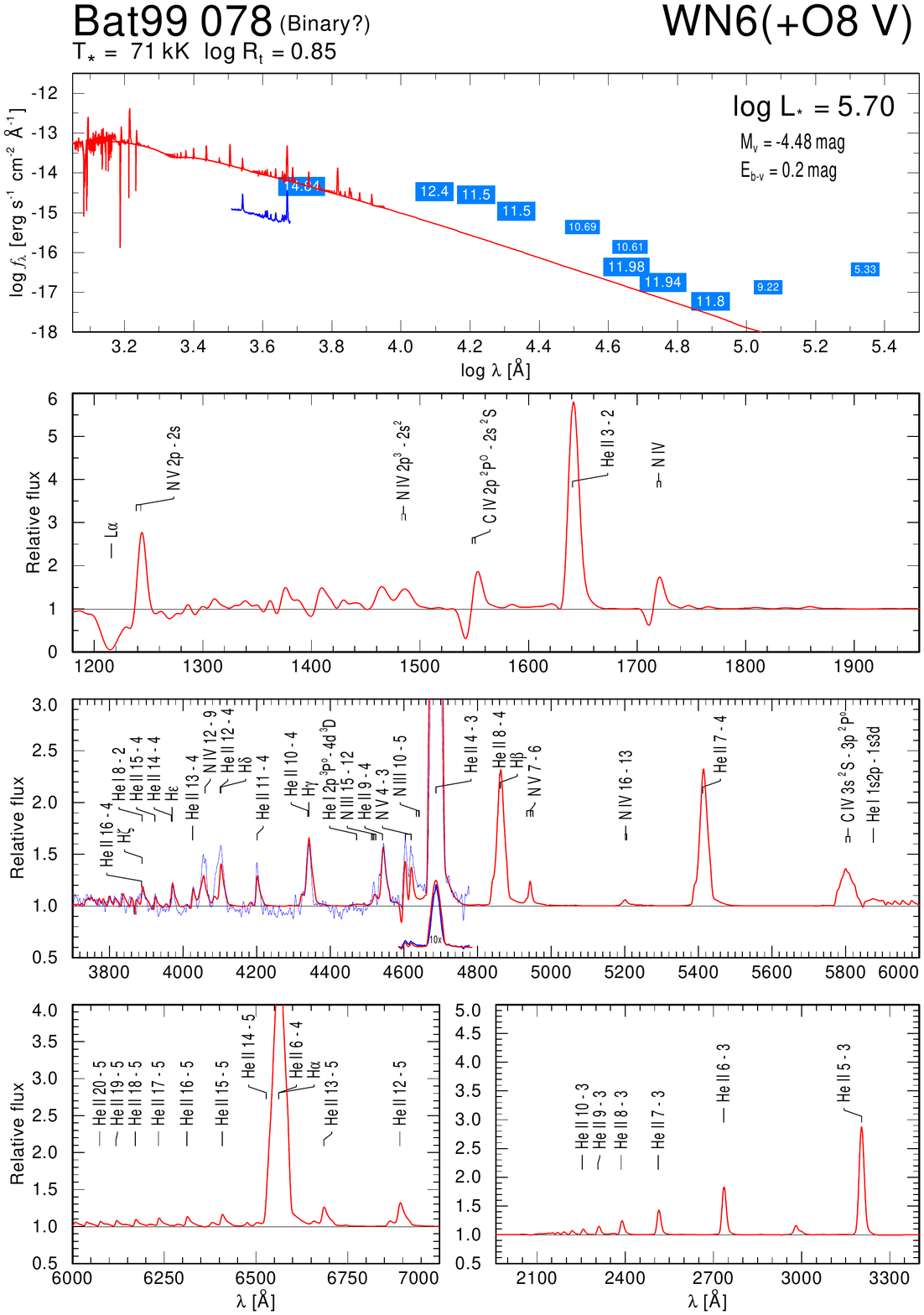}
  \vspace{-0.4cm}
  \caption{Spectral fit for BAT99\,077 and BAT99\,078}
  \label{fig:bat077}
  \label{fig:bat078}
\end{figure*}

\begin{figure*}
  \centering
  \includegraphics[width=0.46\hsize]{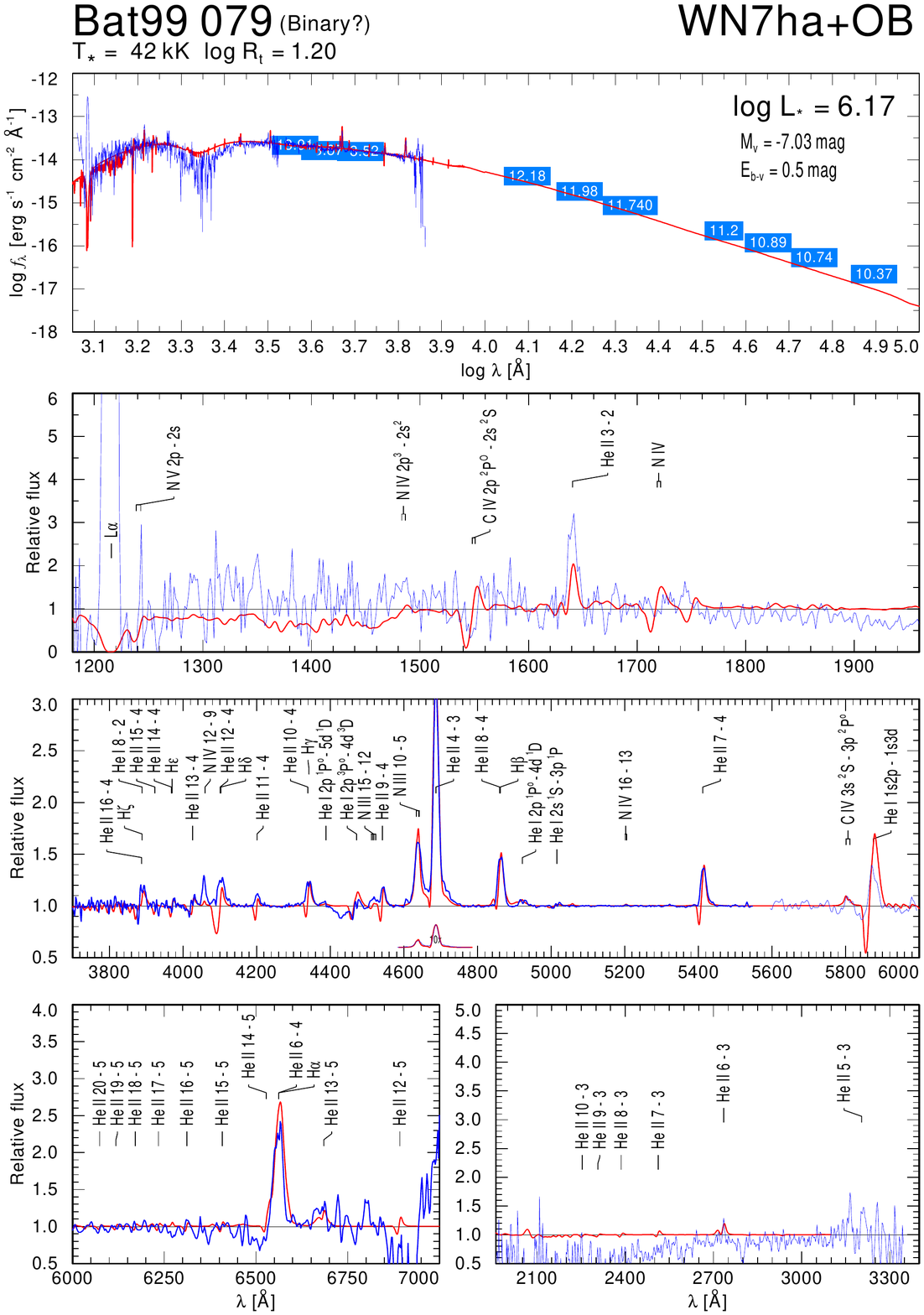}
  \qquad
  \includegraphics[width=0.46\hsize]{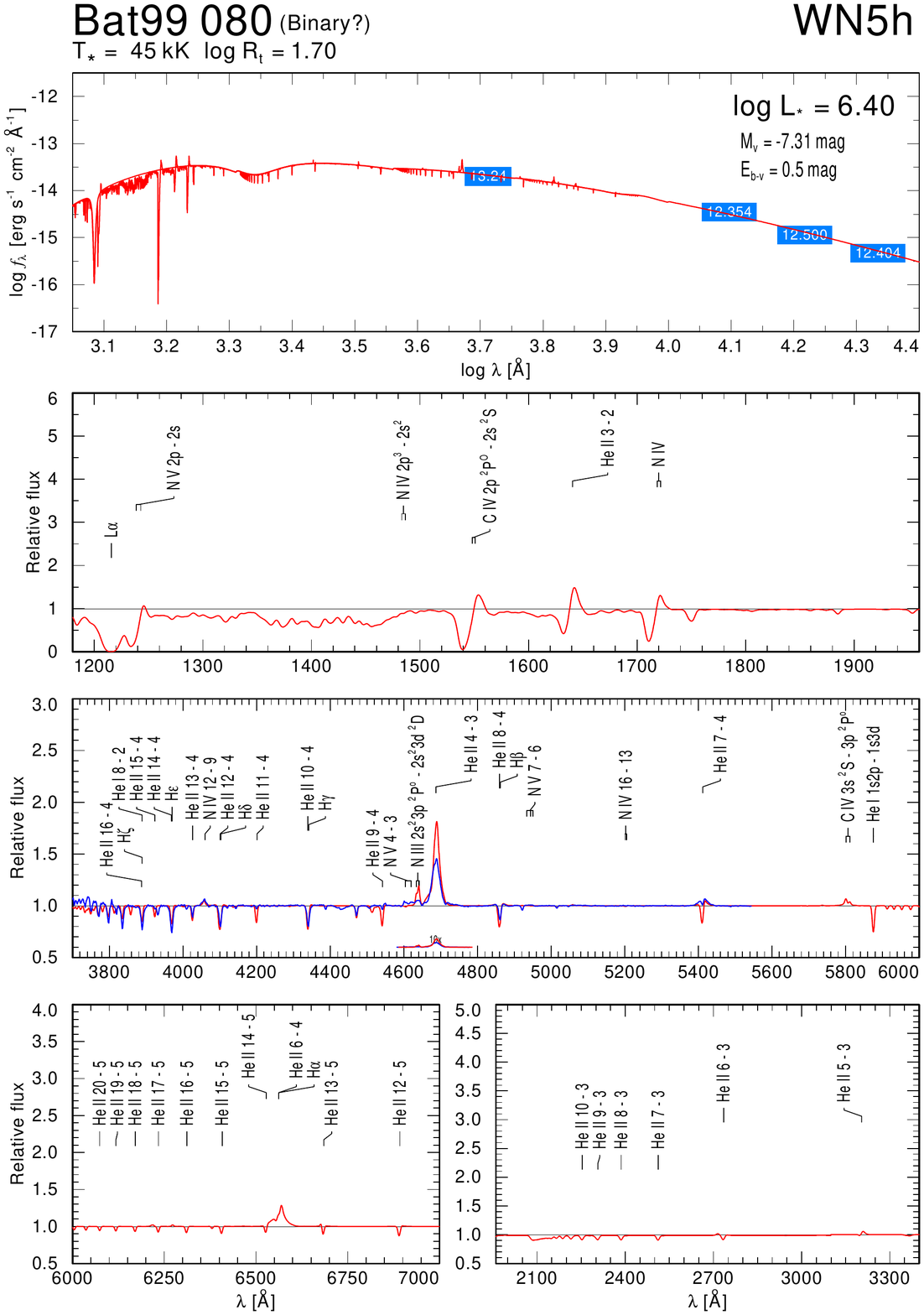}
  \vspace{-0.4cm}
  \caption{Spectral fit for BAT99\,079 and BAT99\,080}
  \label{fig:bat079}
  \label{fig:bat080}
\end{figure*}

\clearpage

\begin{figure*}
  \centering
  \includegraphics[width=0.46\hsize]{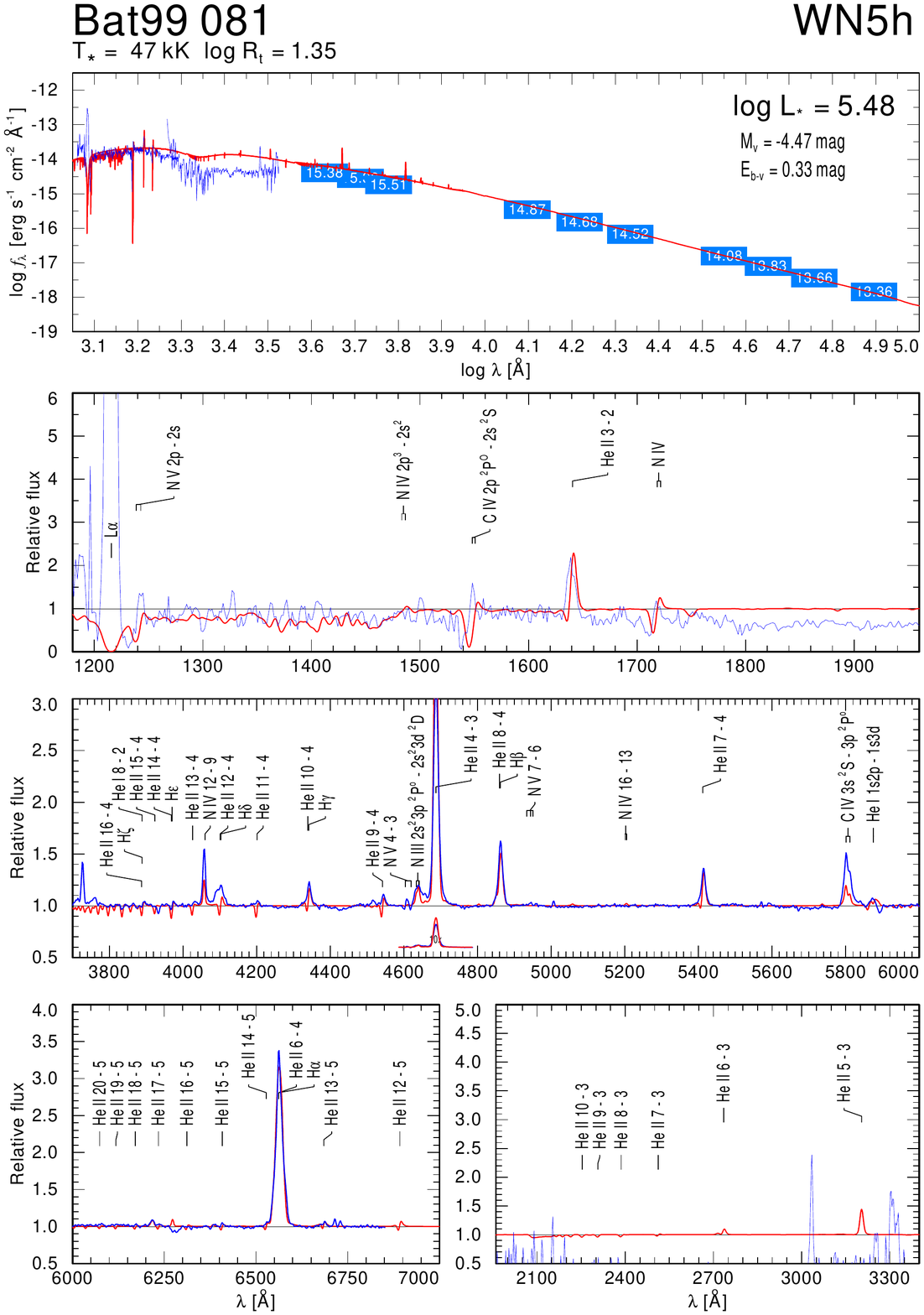}
  \qquad
  \includegraphics[width=0.46\hsize]{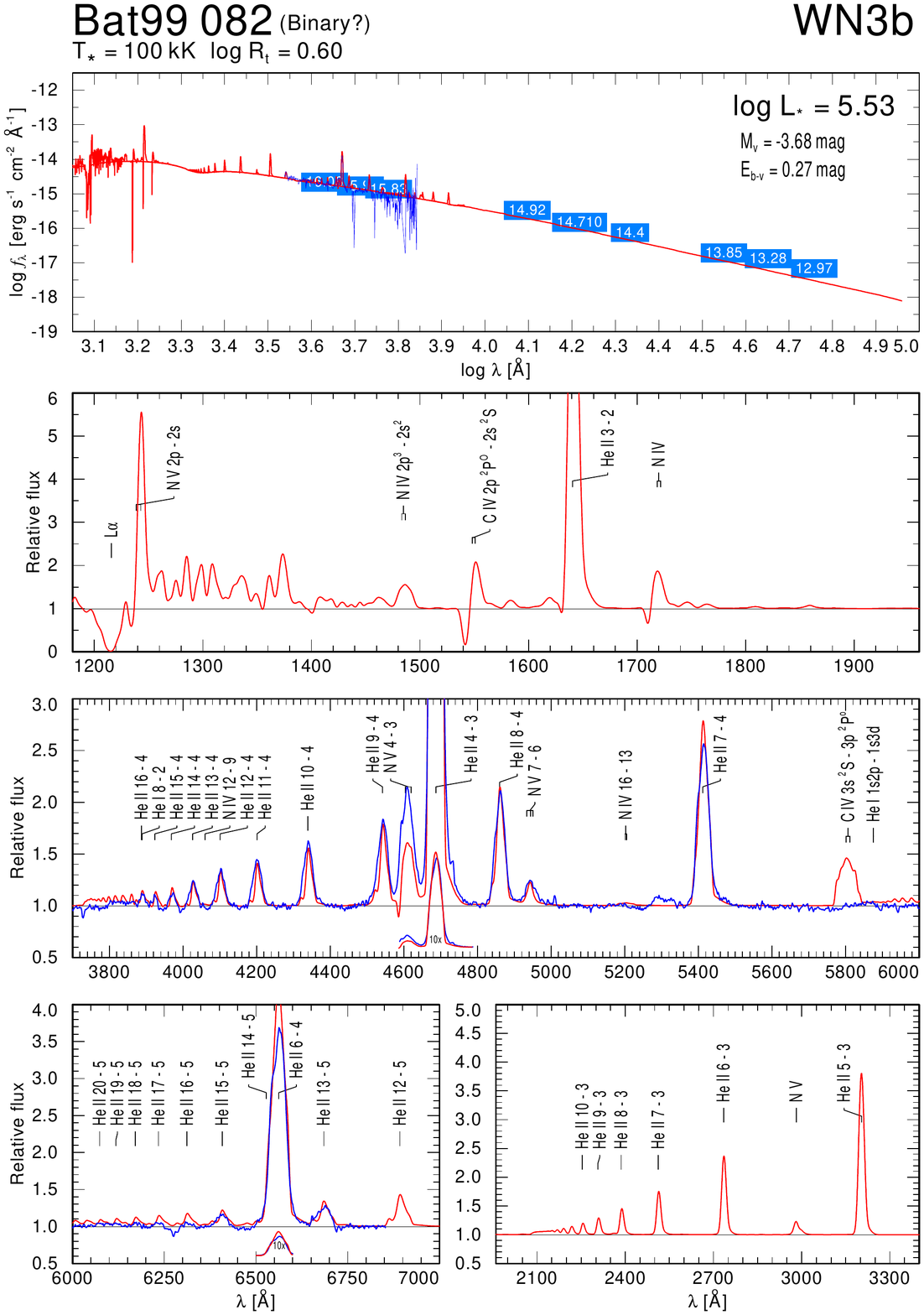}
  \vspace{-0.4cm}
  \caption{Spectral fit for BAT99\,081 and BAT99\,082}
  \label{fig:bat081}
  \label{fig:bat082}
\end{figure*}

\begin{figure*}
  \centering
  \includegraphics[width=0.46\hsize]{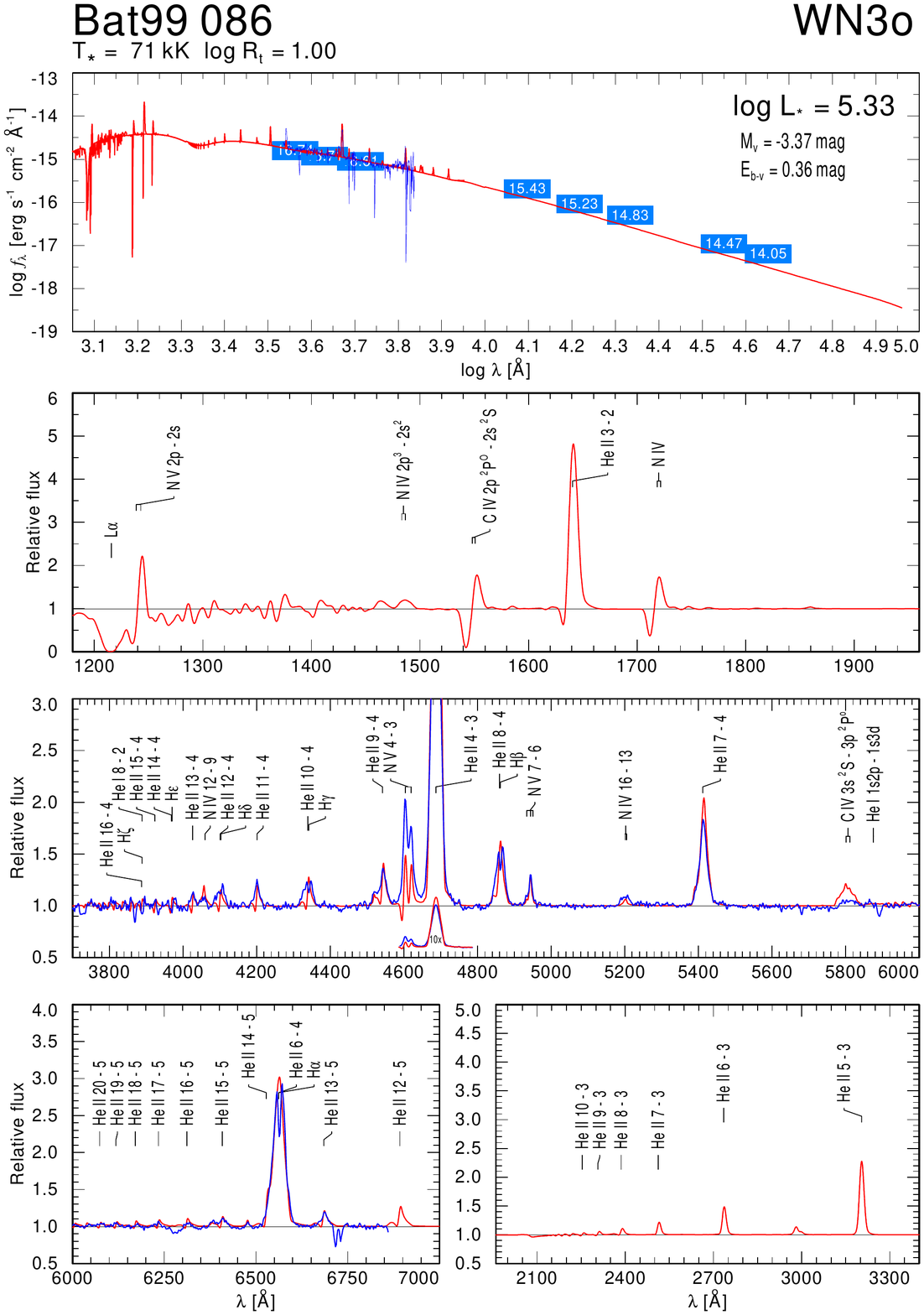}
  \qquad
  \includegraphics[width=0.46\hsize]{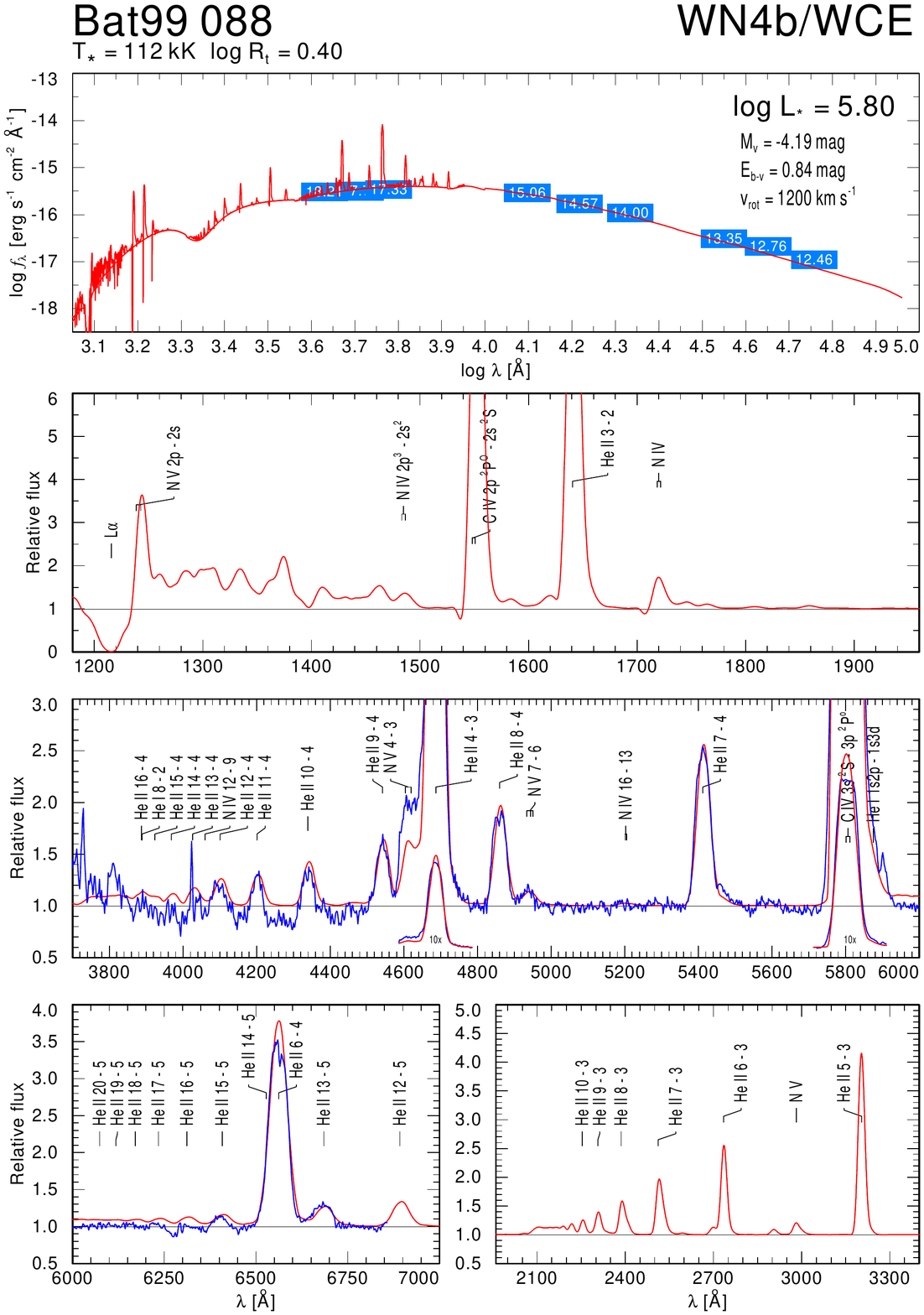}
  \vspace{-0.4cm}
  \caption{Spectral fit for BAT99\,086 and BAT99\,088}
  \label{fig:bat086}
  \label{fig:bat088}
\end{figure*}

\clearpage

\begin{figure*}
  \centering
  \includegraphics[width=0.46\hsize]{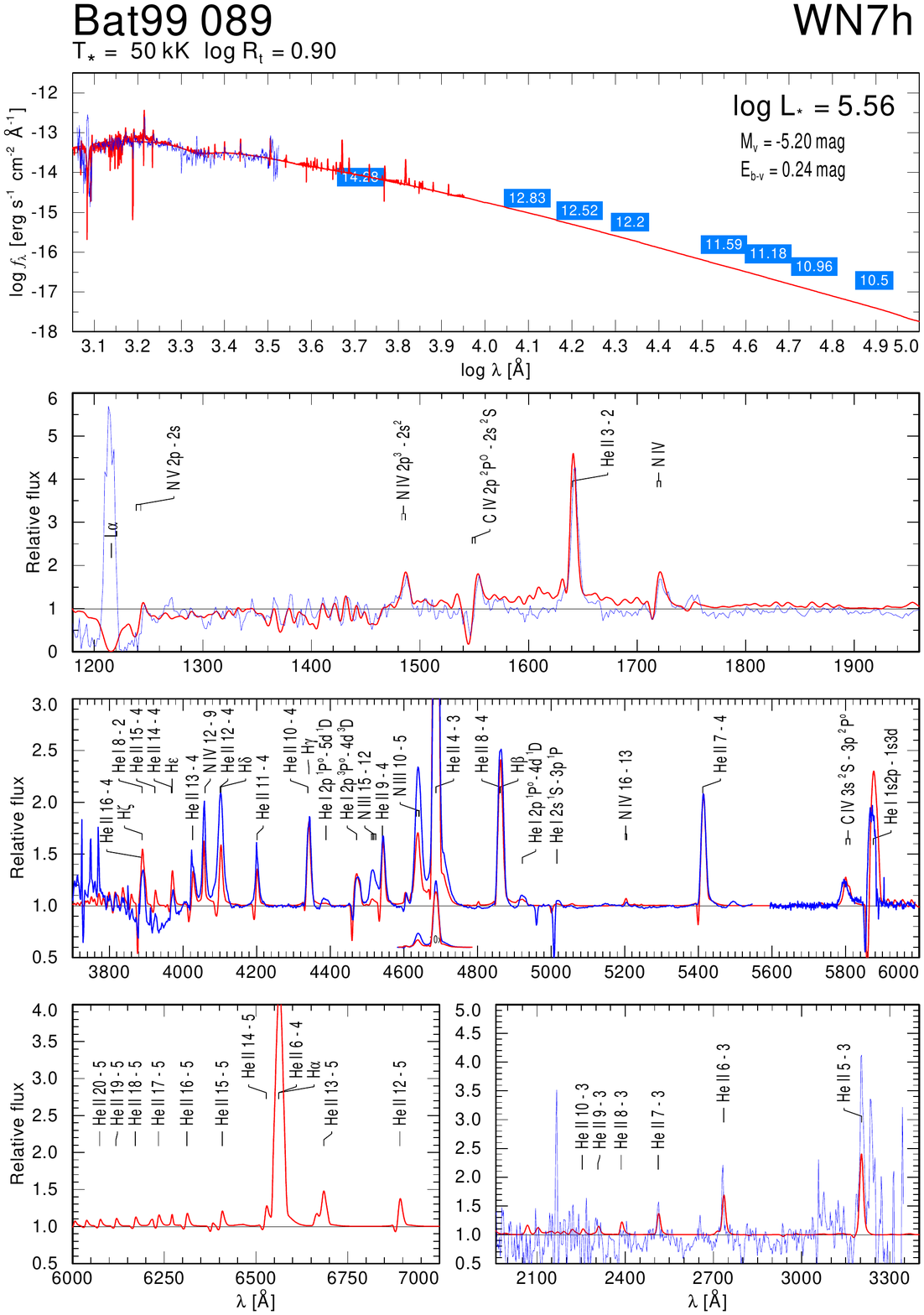}
  \qquad
  \includegraphics[width=0.46\hsize]{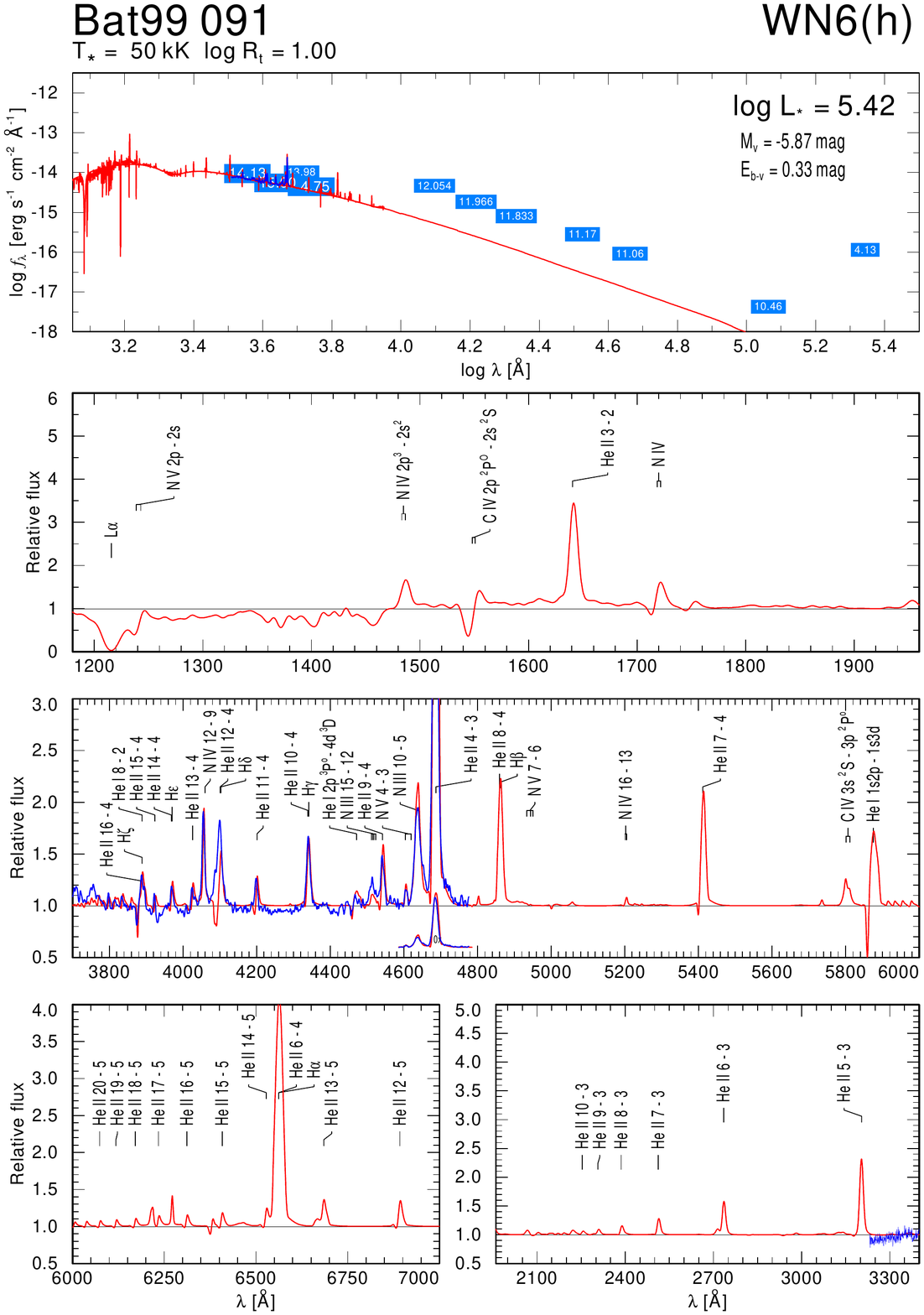}
  \vspace{-0.4cm}
  \caption{Spectral fit for BAT99\,089 and BAT99\,091}
  \label{fig:bat089}
  \label{fig:bat091}
\end{figure*}

\begin{figure*}
  \centering
  \includegraphics[width=0.46\hsize]{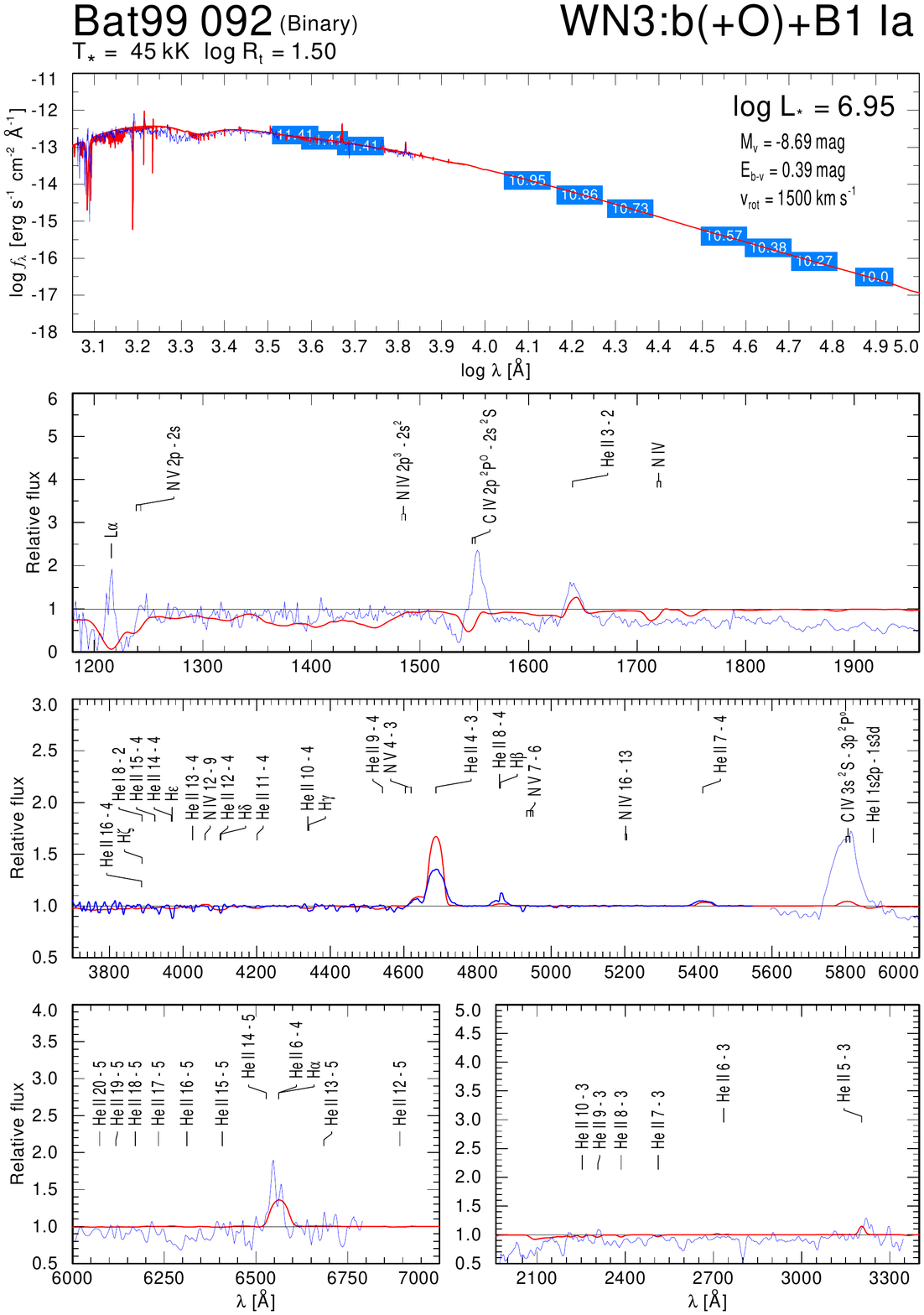}
  \qquad
  \includegraphics[width=0.46\hsize]{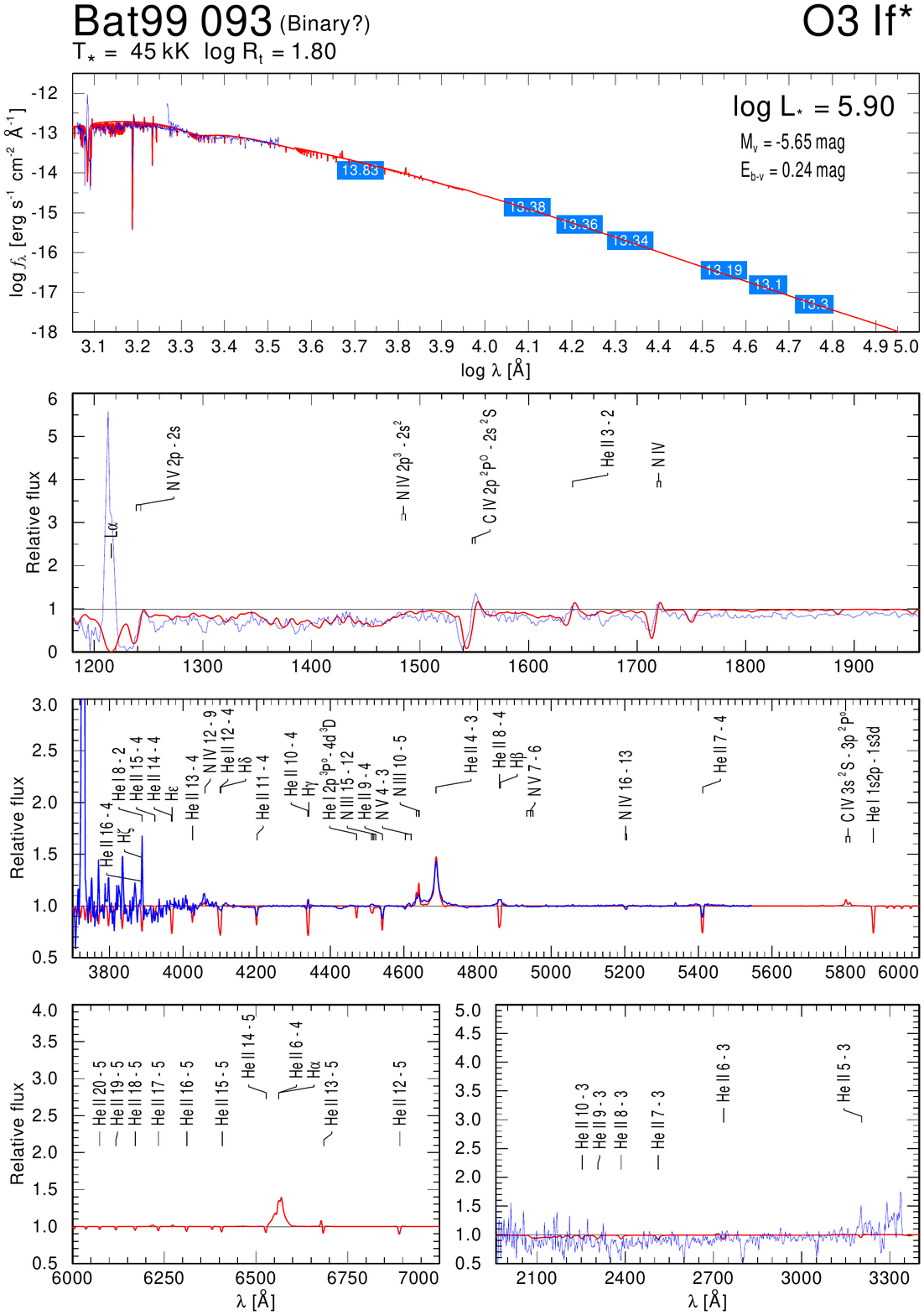}
  \vspace{-0.4cm}
  \caption{Spectral fit for BAT99\,092 and BAT99\,093}
  \label{fig:bat092}
  \label{fig:bat093}
\end{figure*}

\clearpage

\begin{figure*}
  \centering
  \includegraphics[width=0.46\hsize]{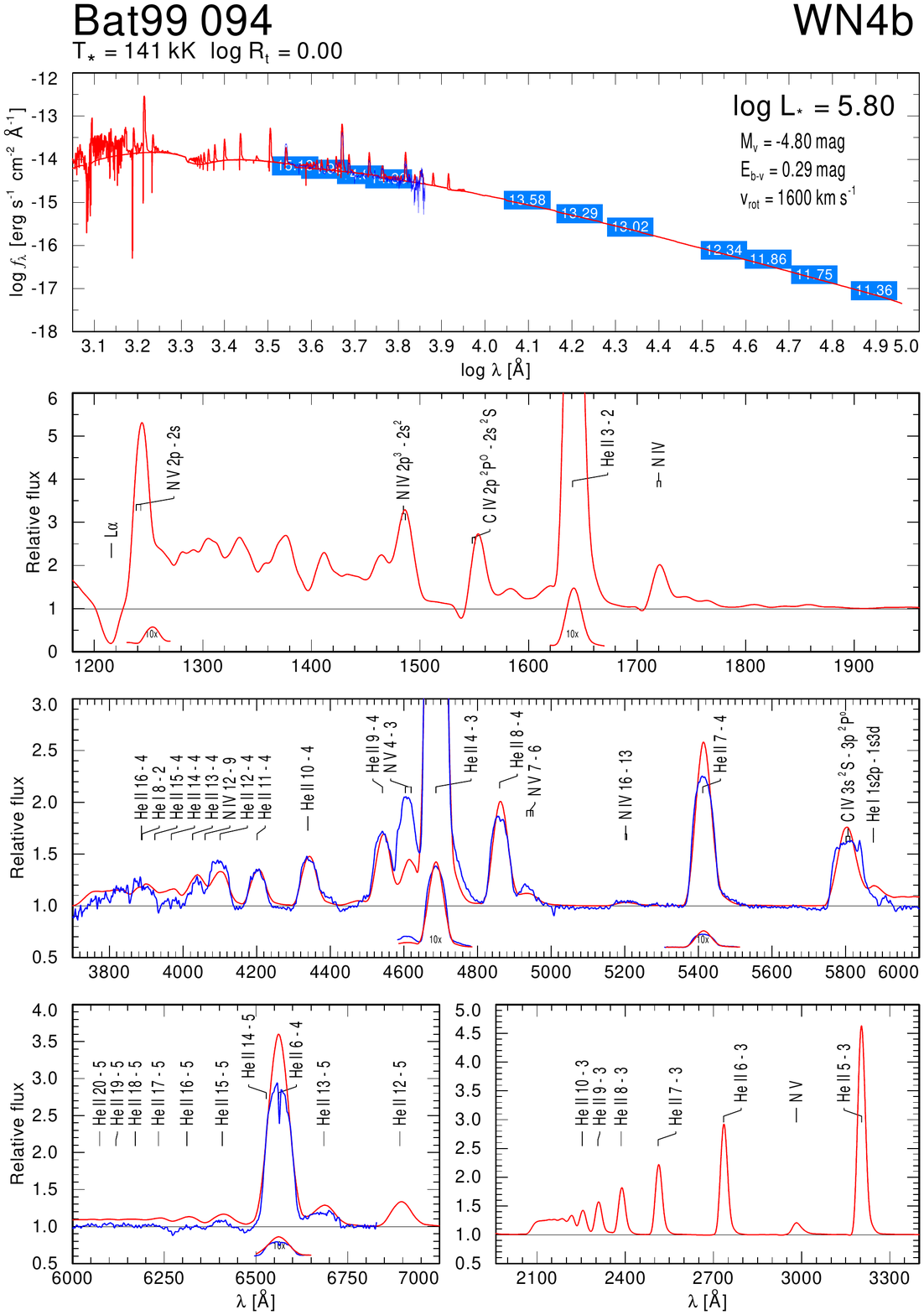}
  \qquad
  \includegraphics[width=0.46\hsize]{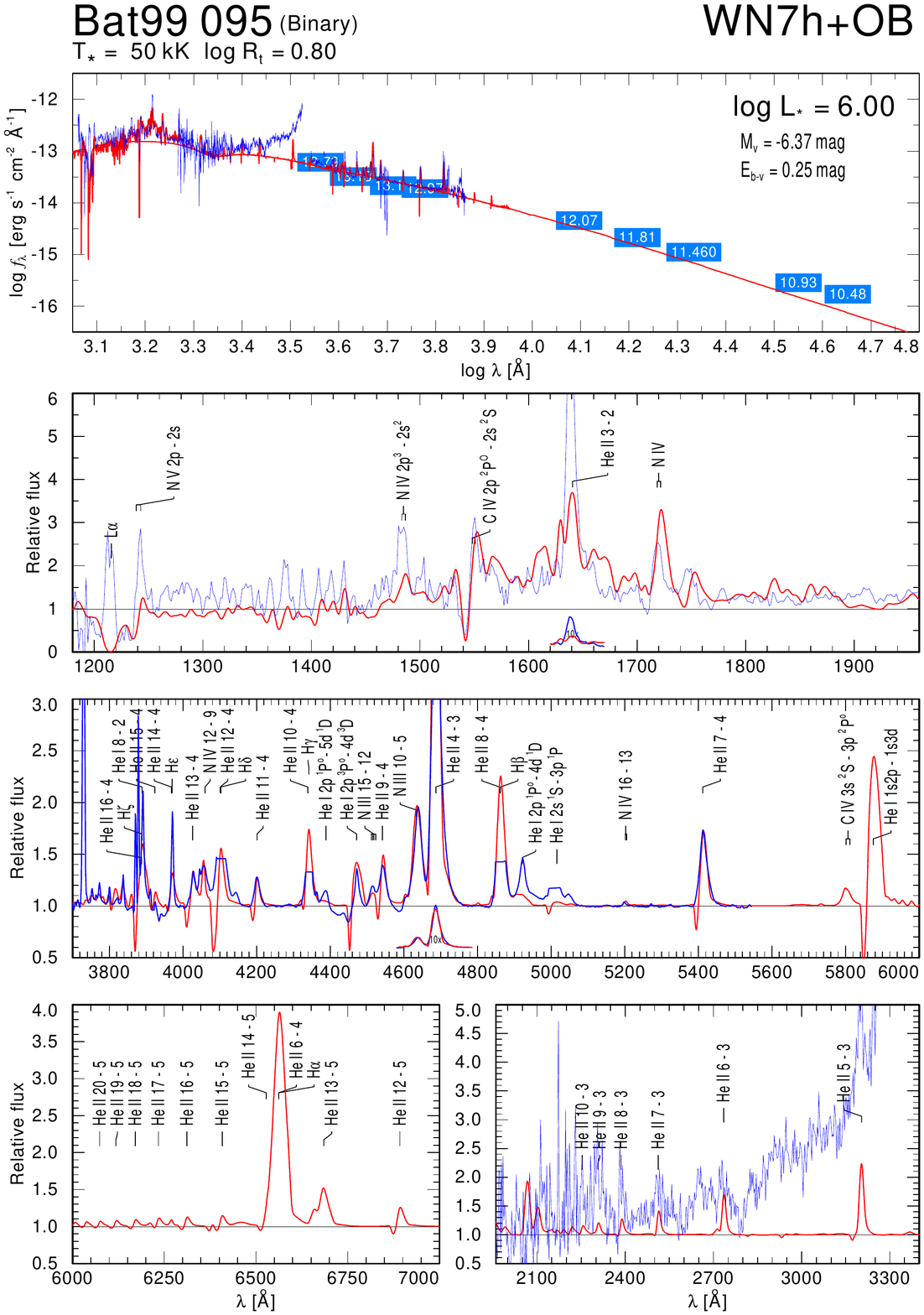}
  \vspace{-0.4cm}
  \caption{Spectral fit for BAT99\,094 and BAT99\,095}
  \label{fig:bat094}
  \label{fig:bat095}
\end{figure*}

\clearpage

\clearpage

\begin{figure*}
  \centering
  \includegraphics[width=0.46\hsize]{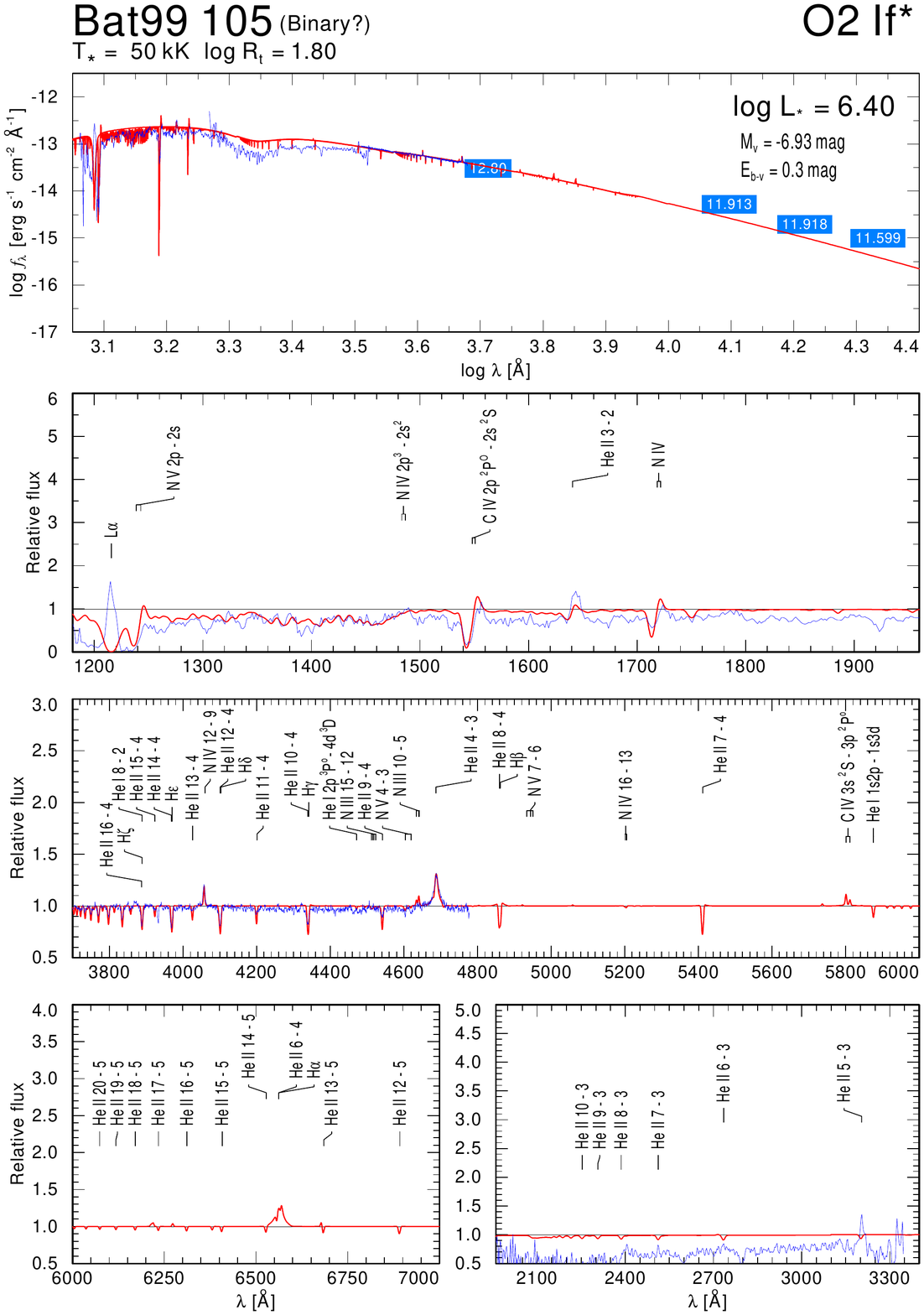}
  \qquad
  \includegraphics[width=0.46\hsize]{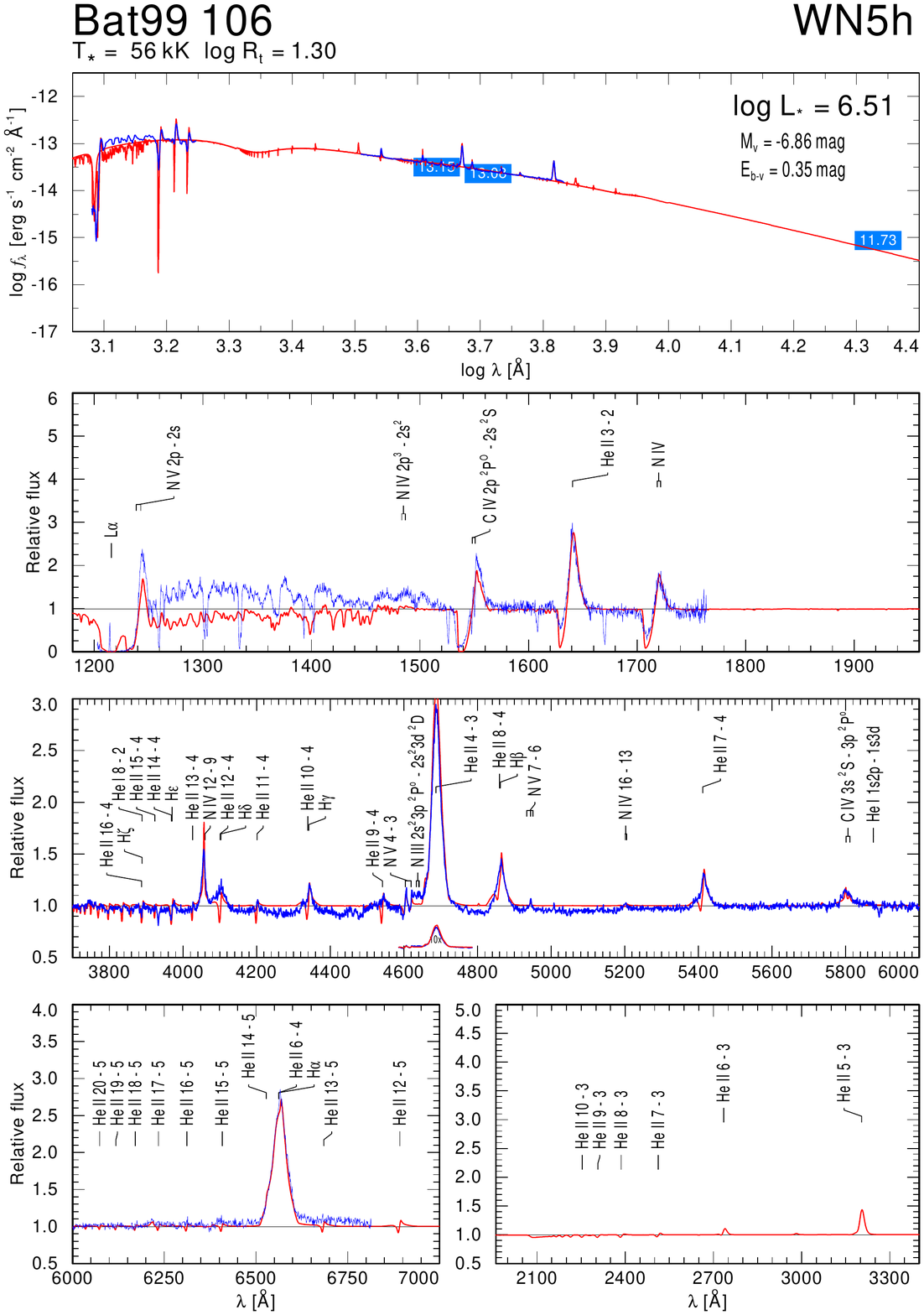}
  \vspace{-0.4cm}
  \caption{Spectral fit for BAT99\,105 and BAT99\,106}
  \label{fig:bat105}
  \label{fig:bat106}
\end{figure*}

\clearpage

\begin{figure*}
  \centering
  \includegraphics[width=0.46\hsize]{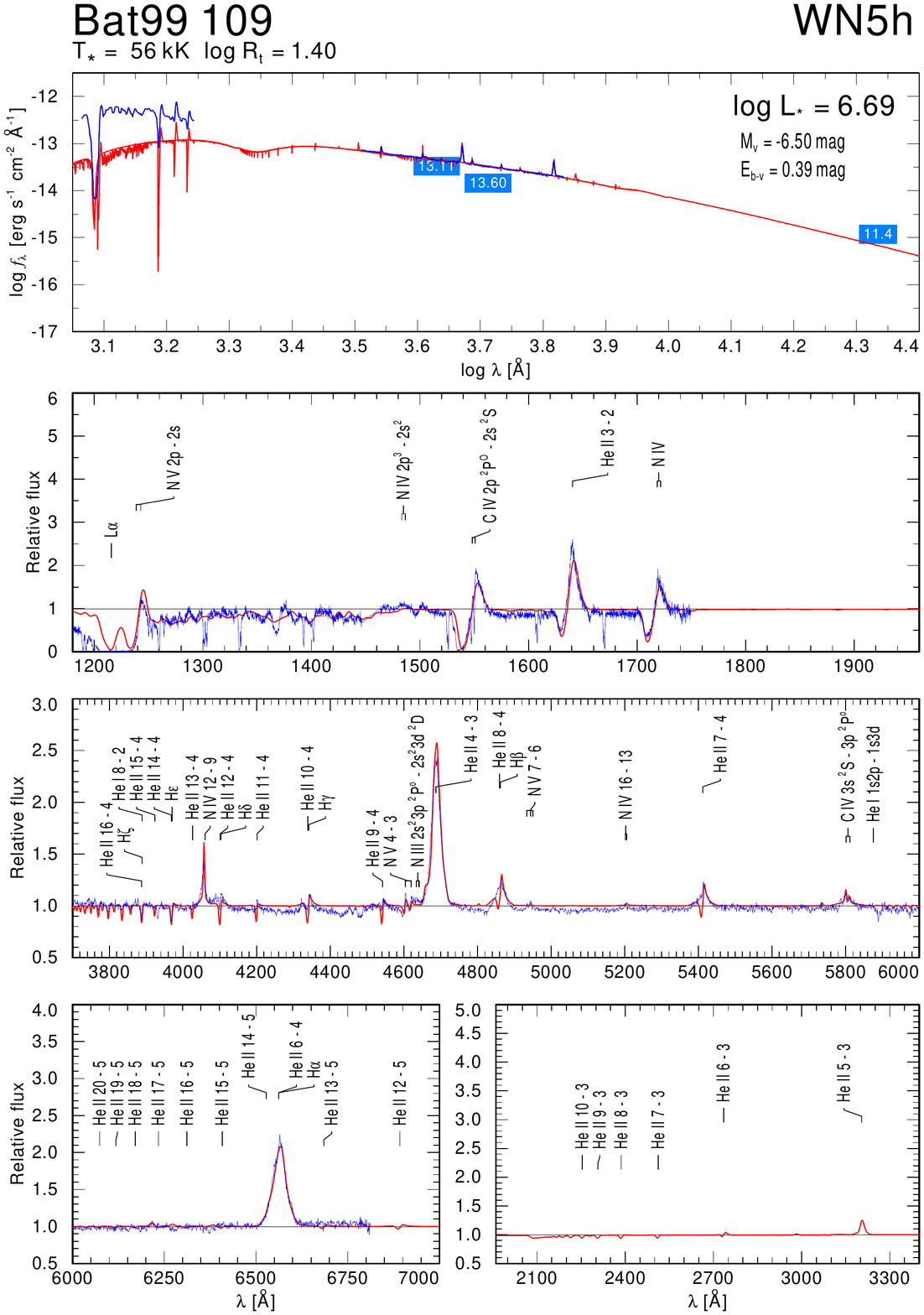}
  \qquad
  \includegraphics[width=0.46\hsize]{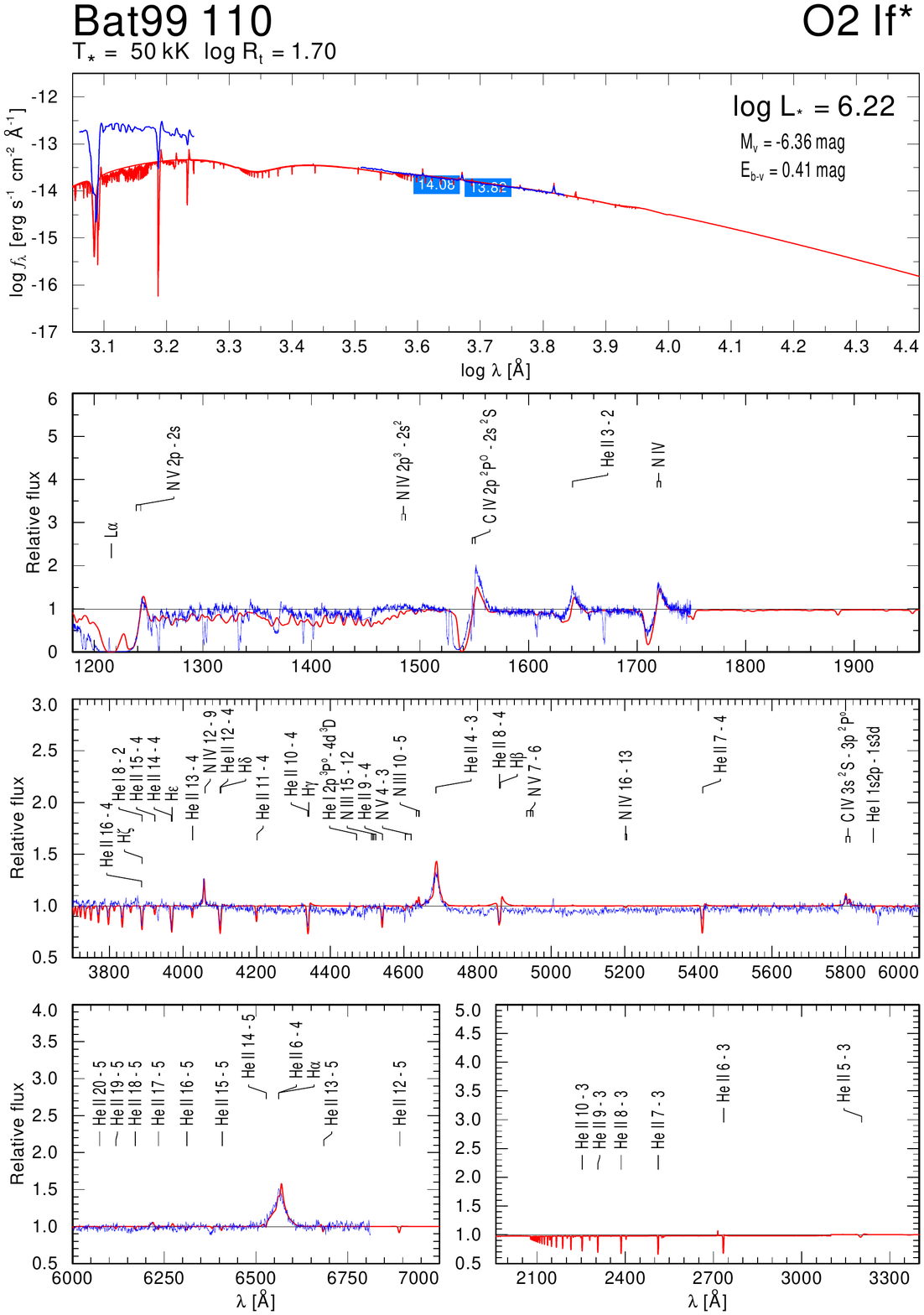}
  \vspace{-0.4cm}
  \caption{Spectral fit for BAT99\,109 and BAT99\,110}
  \label{fig:bat109}
  \label{fig:bat110}
\end{figure*}

\clearpage

\begin{figure*}
  \centering
  \includegraphics[width=0.46\hsize]{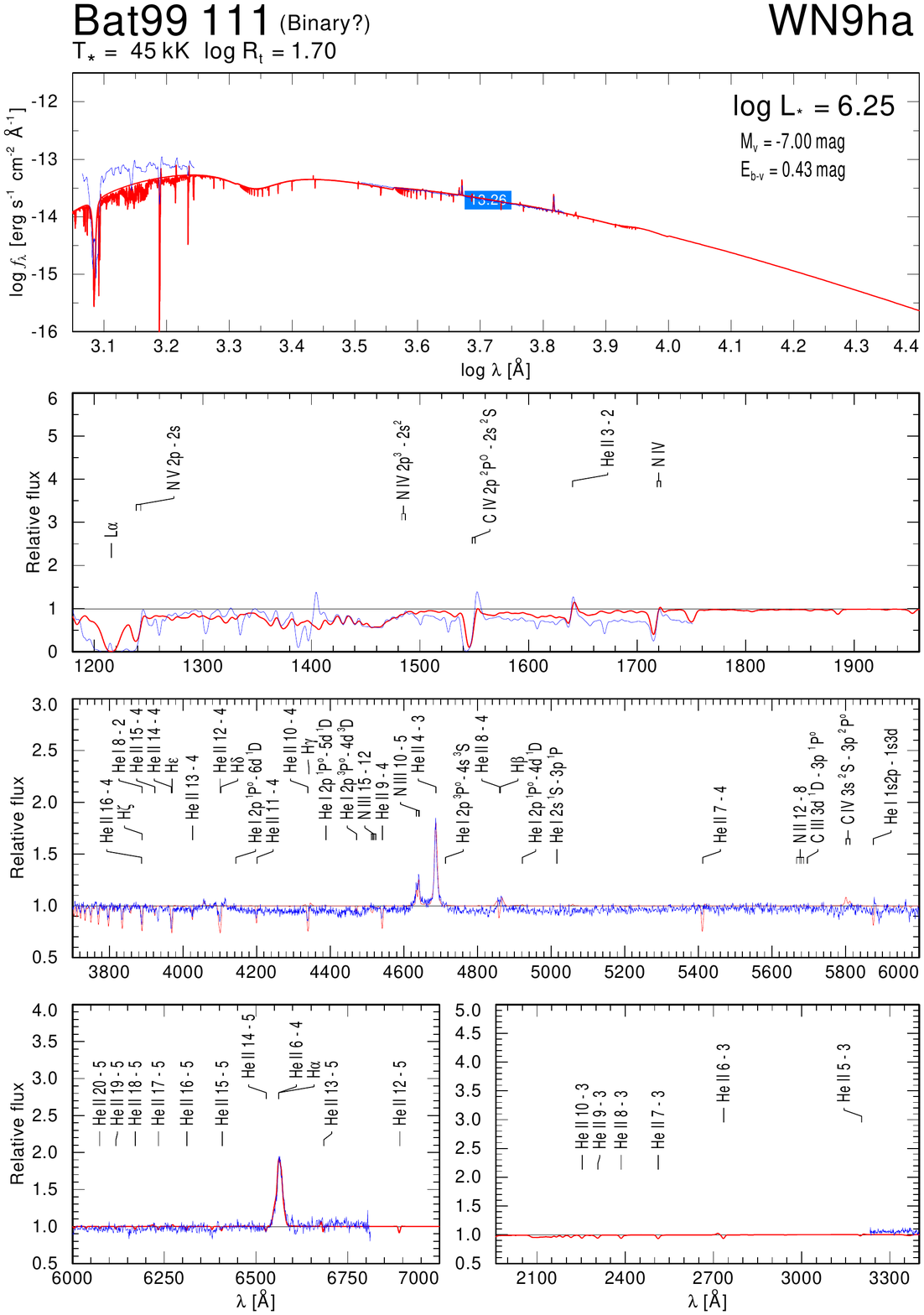}
  \qquad
  \includegraphics[width=0.46\hsize]{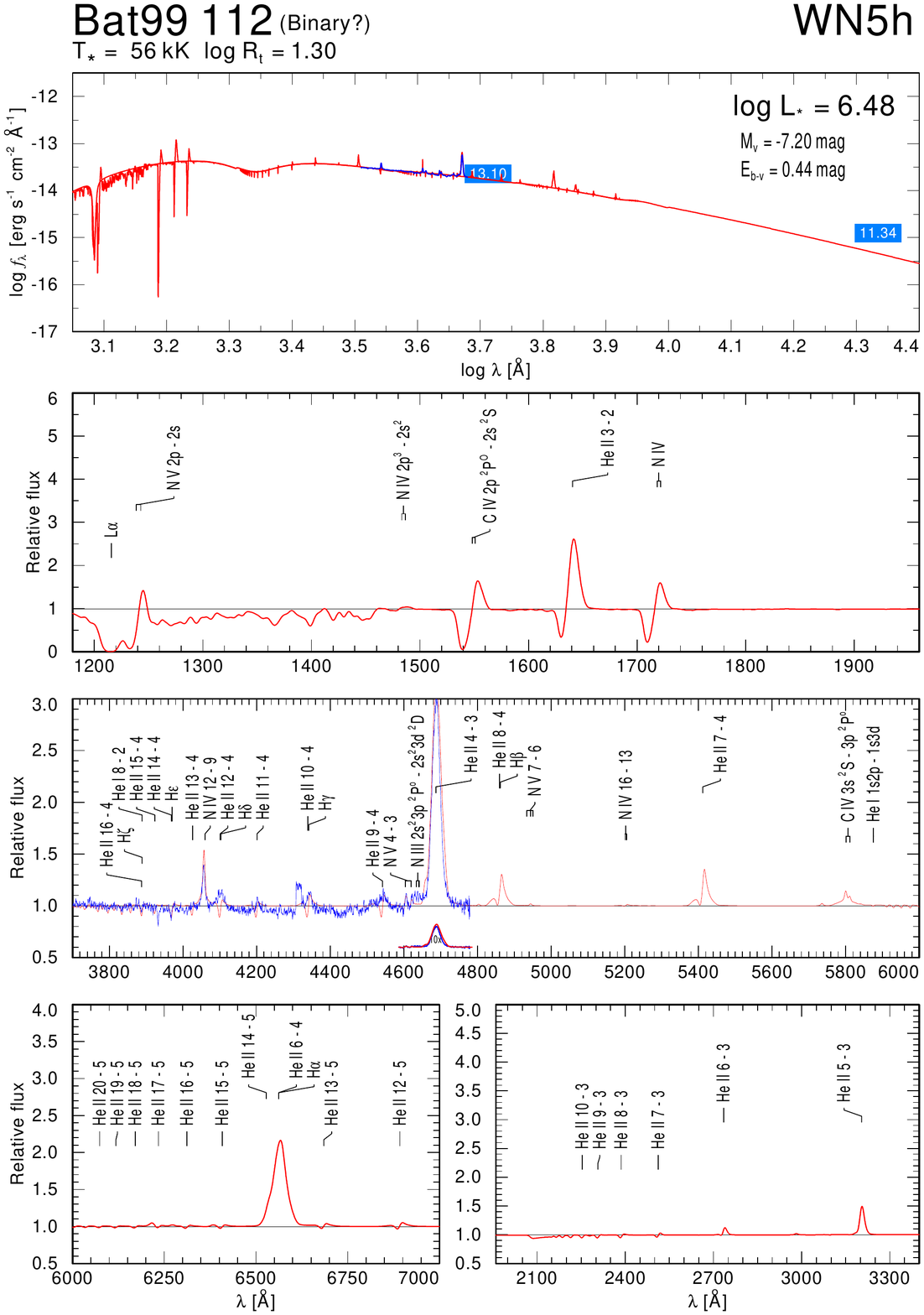}
  \vspace{-0.4cm}
  \caption{Spectral fit for BAT99\,111 and BAT99\,112}
  \label{fig:bat111}
  \label{fig:bat112}
\end{figure*}

\begin{figure*}
  \centering
  \includegraphics[width=0.46\hsize]{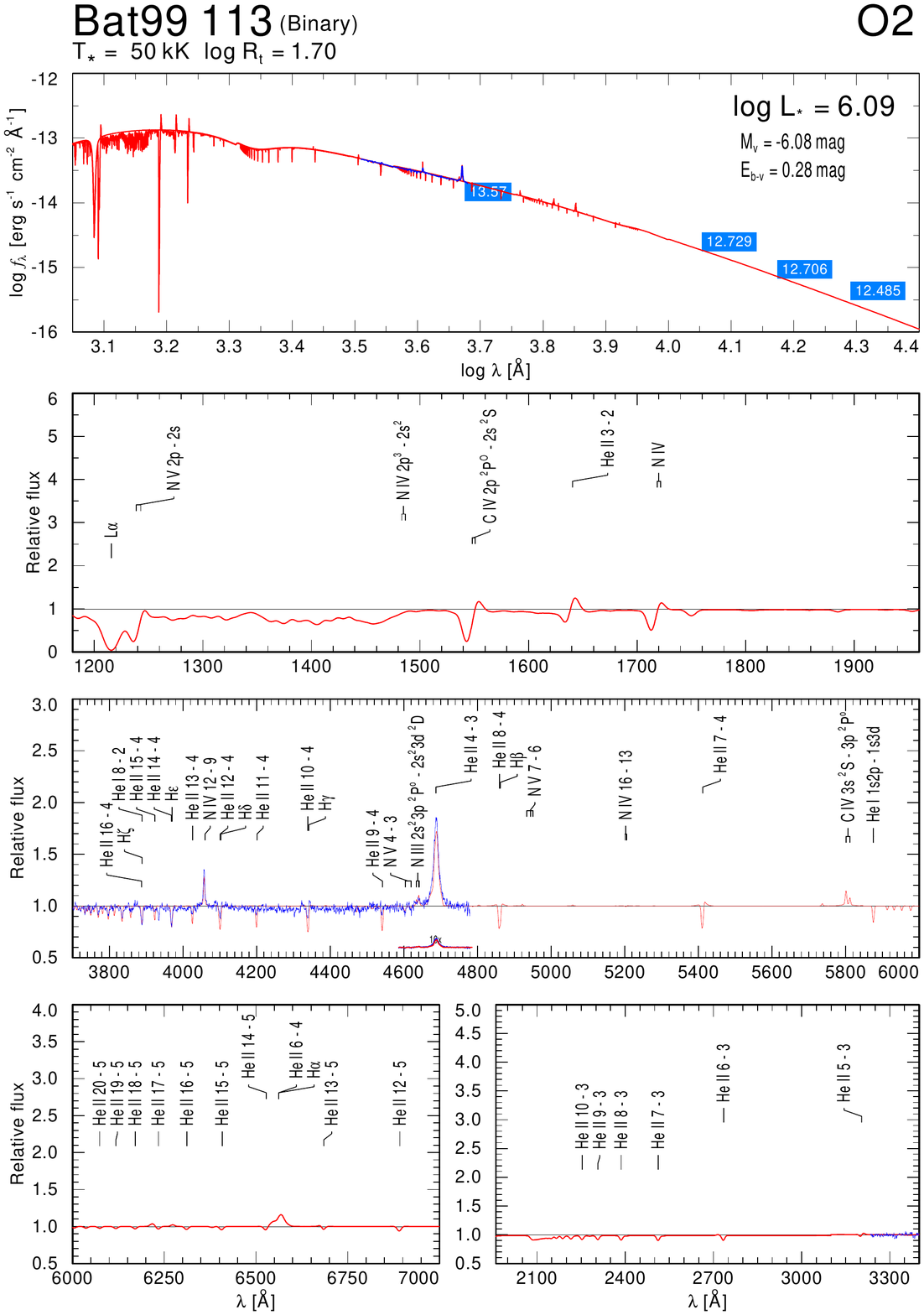}
  \qquad
  \includegraphics[width=0.46\hsize]{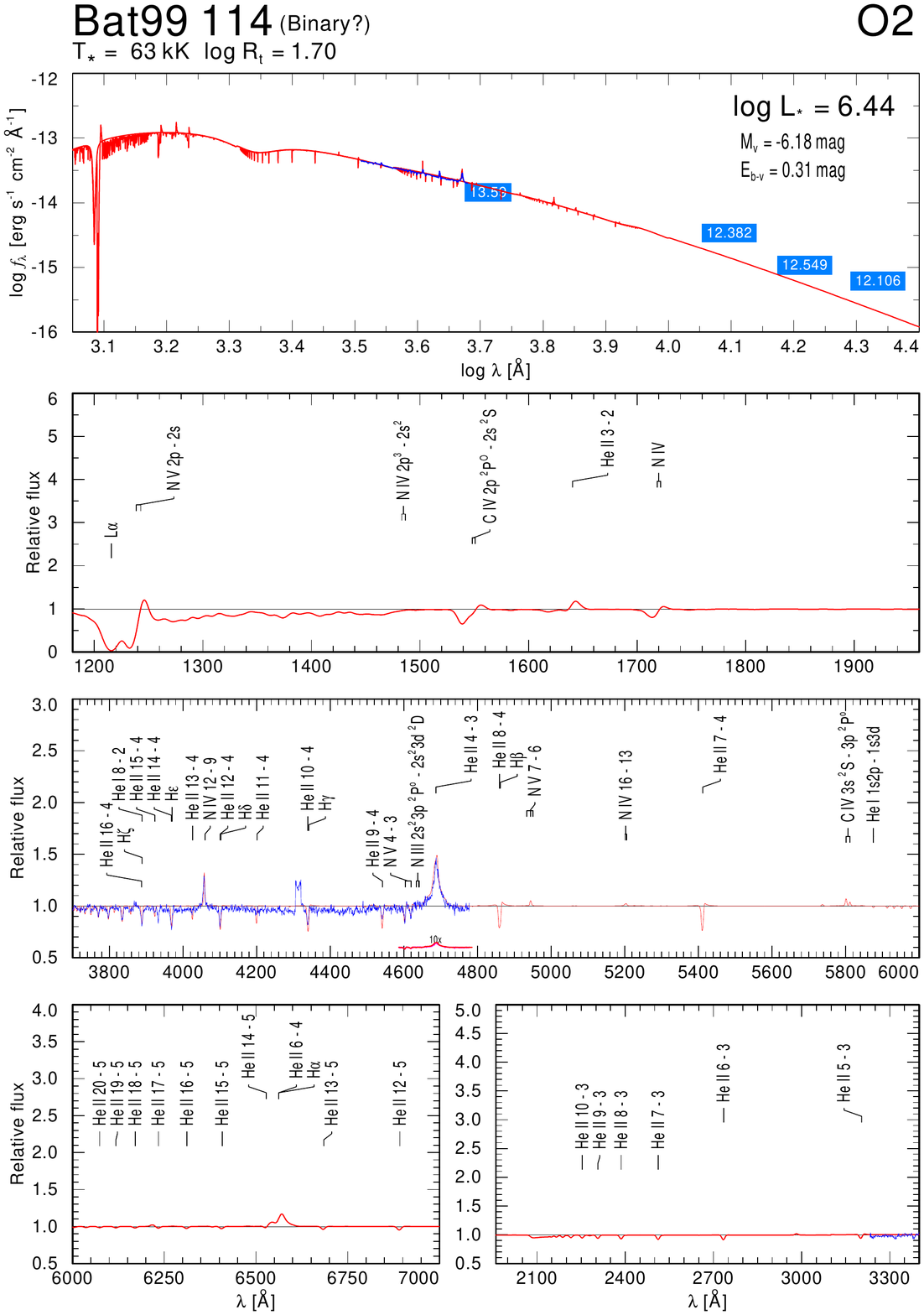}
  \vspace{-0.4cm}
  \caption{Spectral fit for BAT99\,113 and BAT99\,114}
  \label{fig:bat113}
  \label{fig:bat114}
\end{figure*}

\clearpage

\begin{figure*}
  \centering
  \includegraphics[width=0.46\hsize]{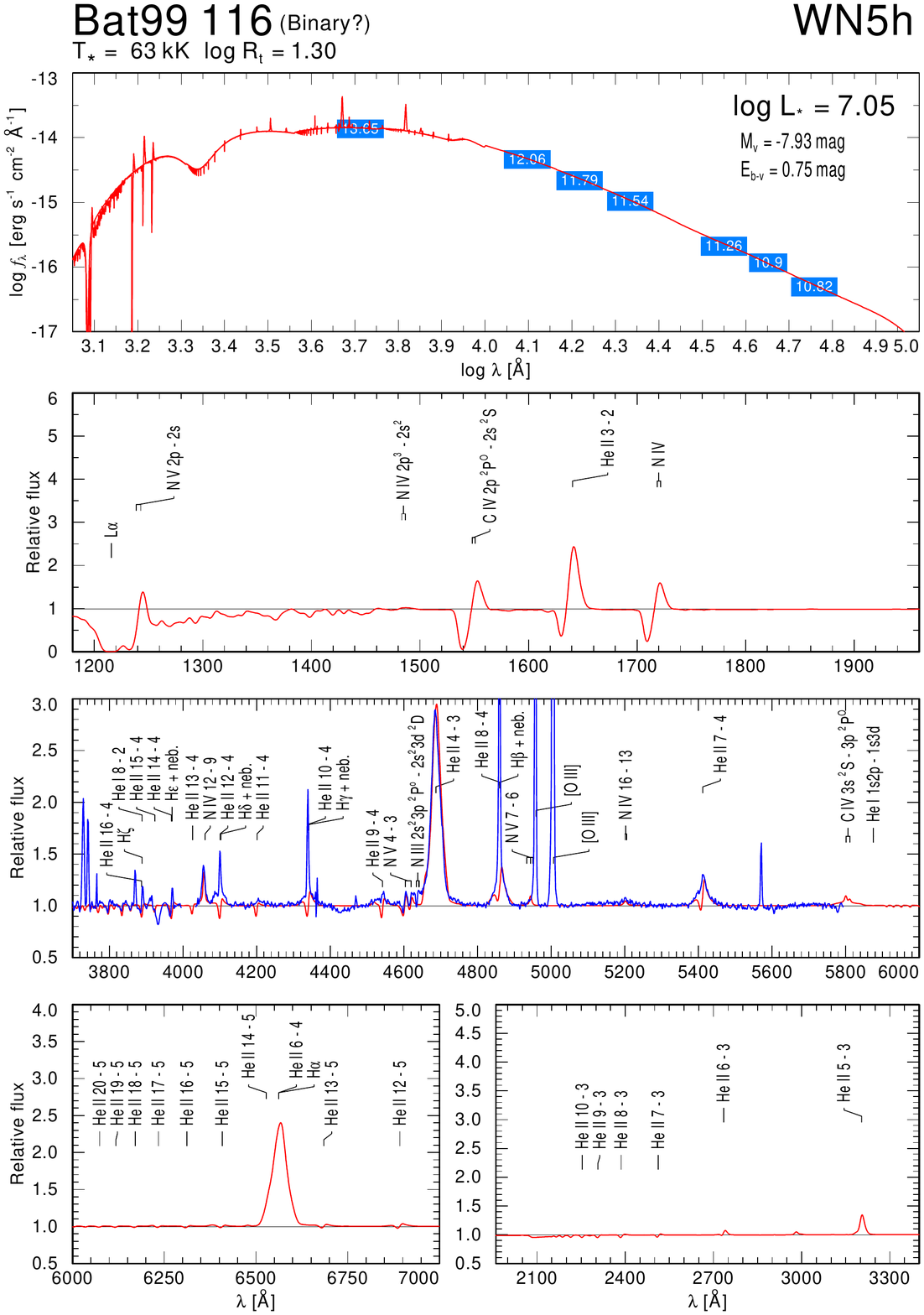}
  \qquad
  \includegraphics[width=0.46\hsize]{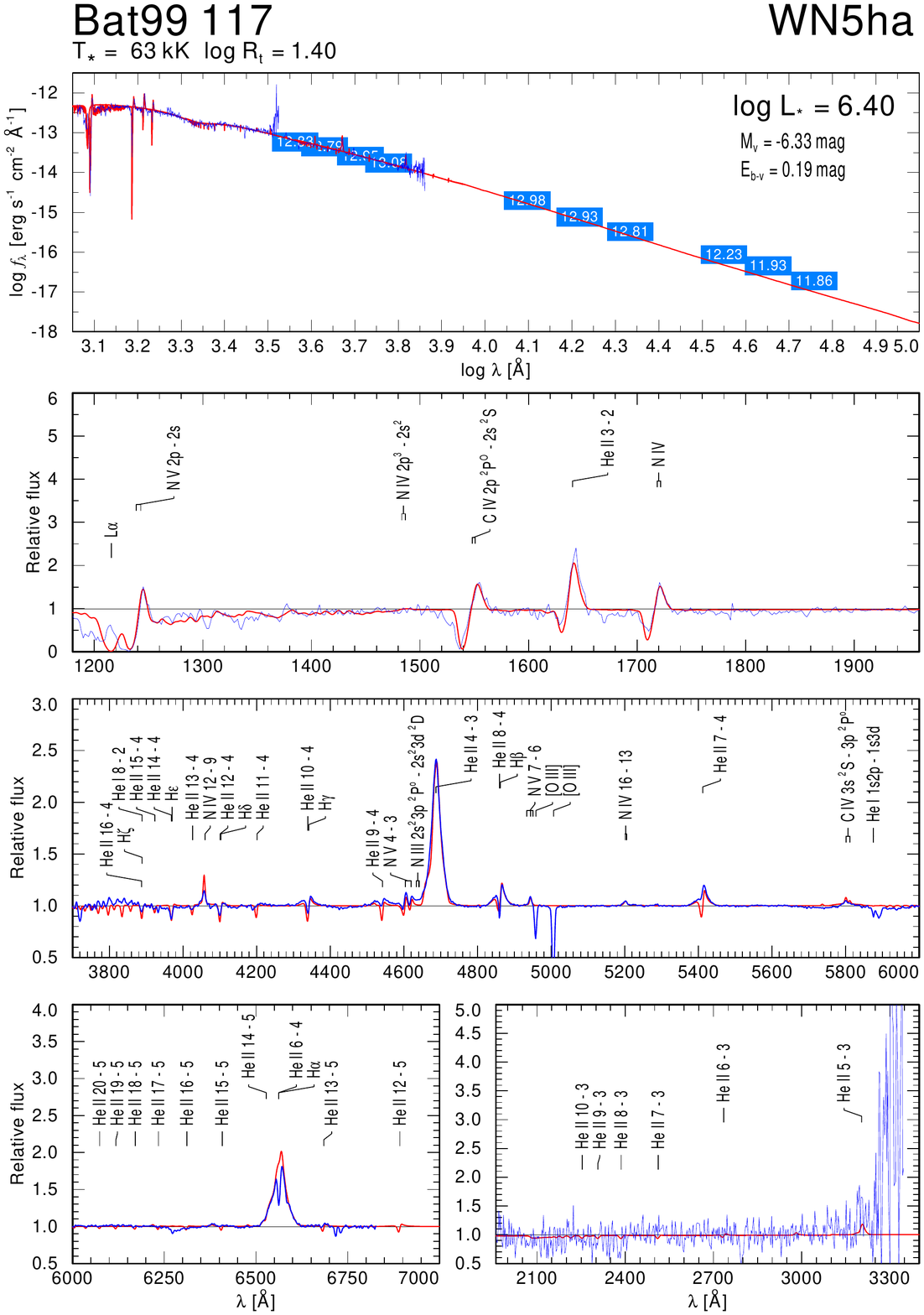}
  \vspace{-0.4cm}
  \caption{Spectral fit for BAT99\,116 and BAT99\,117}
  \label{fig:bat116}
  \label{fig:bat117}
\end{figure*}

\begin{figure*}
  \centering
  \includegraphics[width=0.46\hsize]{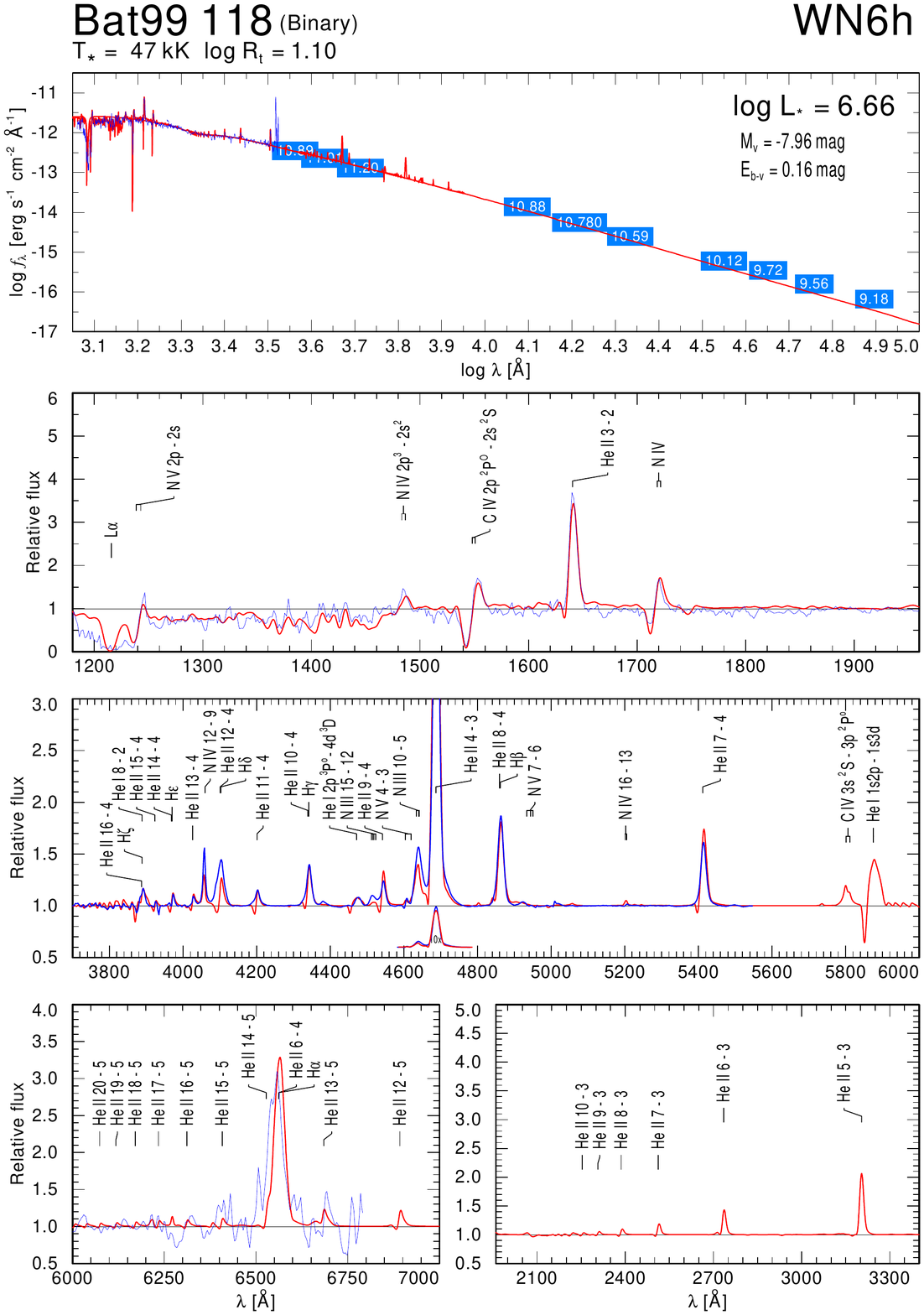}
  \qquad
  \includegraphics[width=0.46\hsize]{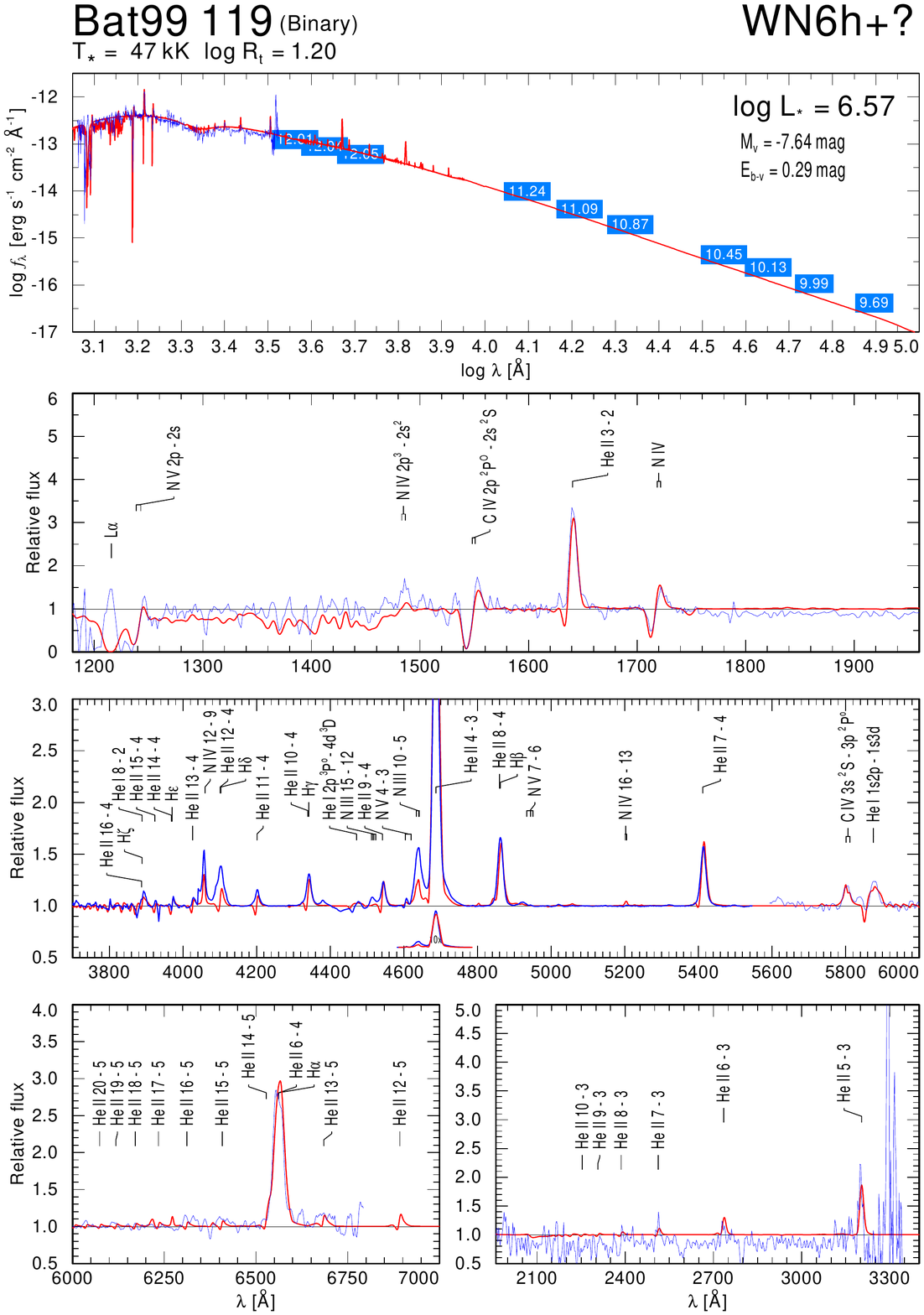}
  \vspace{-0.4cm}
  \caption{Spectral fit for BAT99\,118 and BAT99\,119}
  \label{fig:bat118}
  \label{fig:bat119}
\end{figure*}

\clearpage

\begin{figure*}
  \centering
  \includegraphics[width=0.46\hsize]{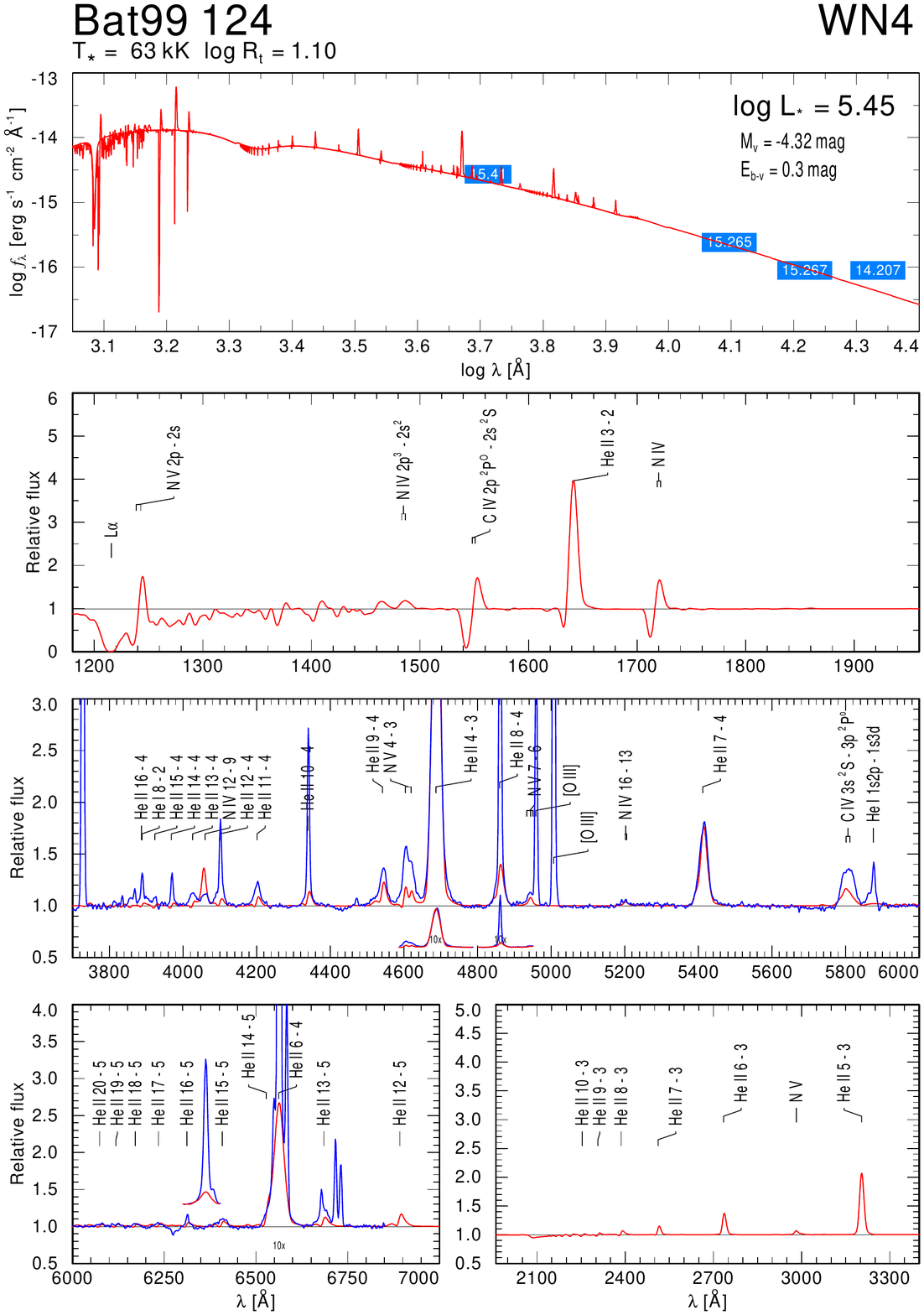}
  \qquad
  \includegraphics[width=0.46\hsize]{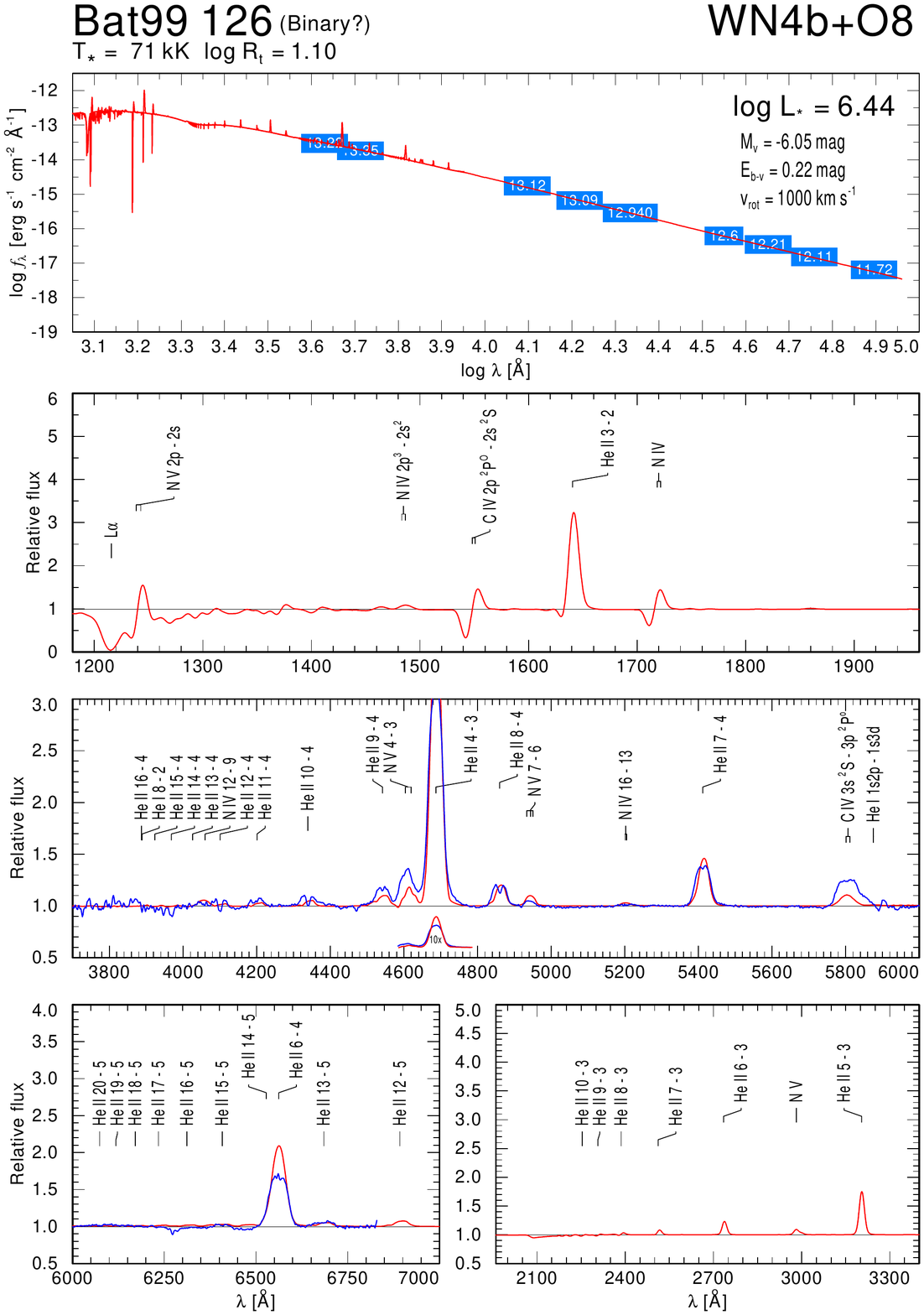}
  \vspace{-0.4cm}
  \caption{Spectral fit for BAT99\,124 and BAT99\,126}
  \label{fig:bat124}
  \label{fig:bat126}
\end{figure*}

\clearpage

\begin{figure*}
  \centering
  \includegraphics[width=0.46\hsize]{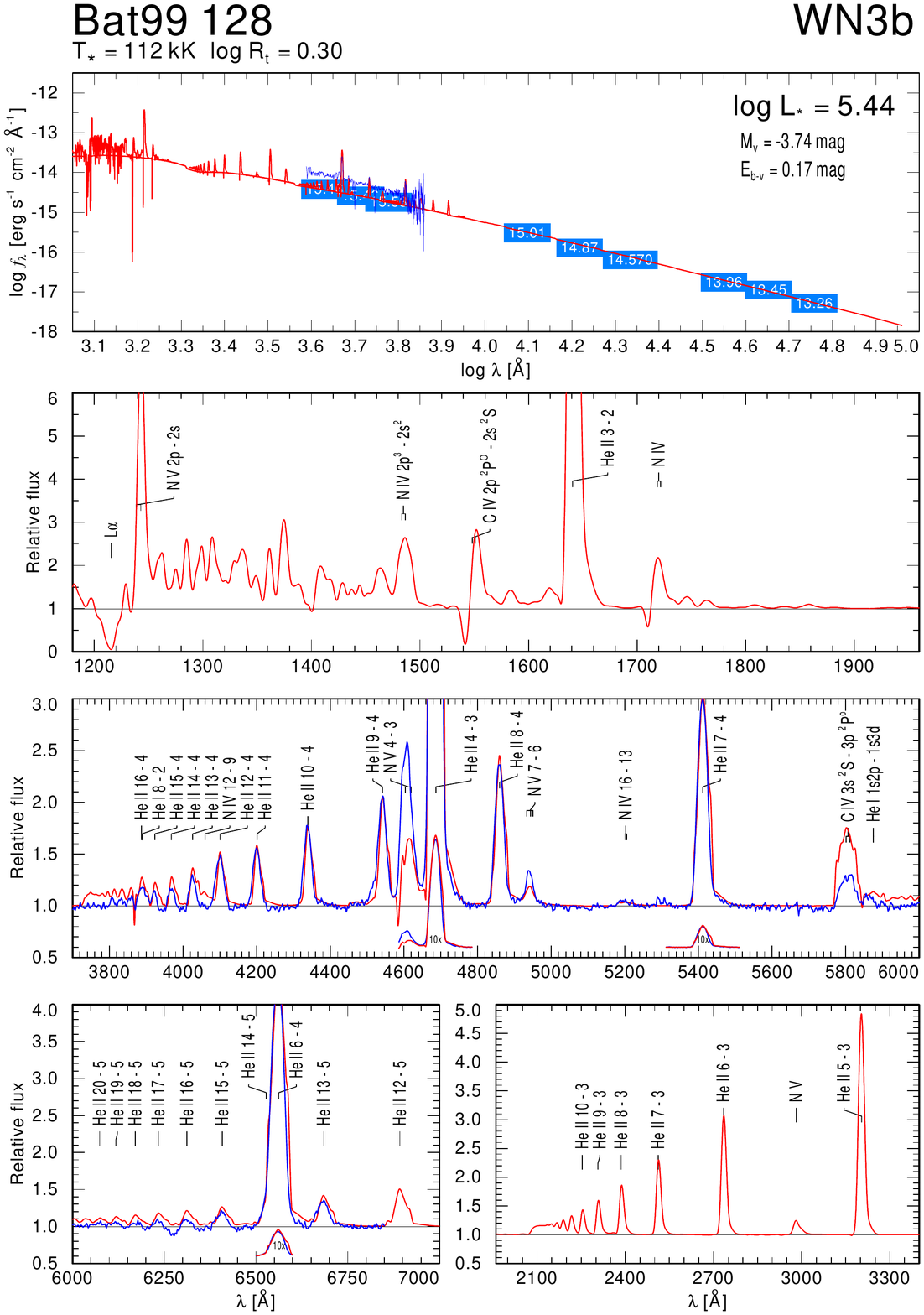}
  \qquad
  \includegraphics[width=0.46\hsize]{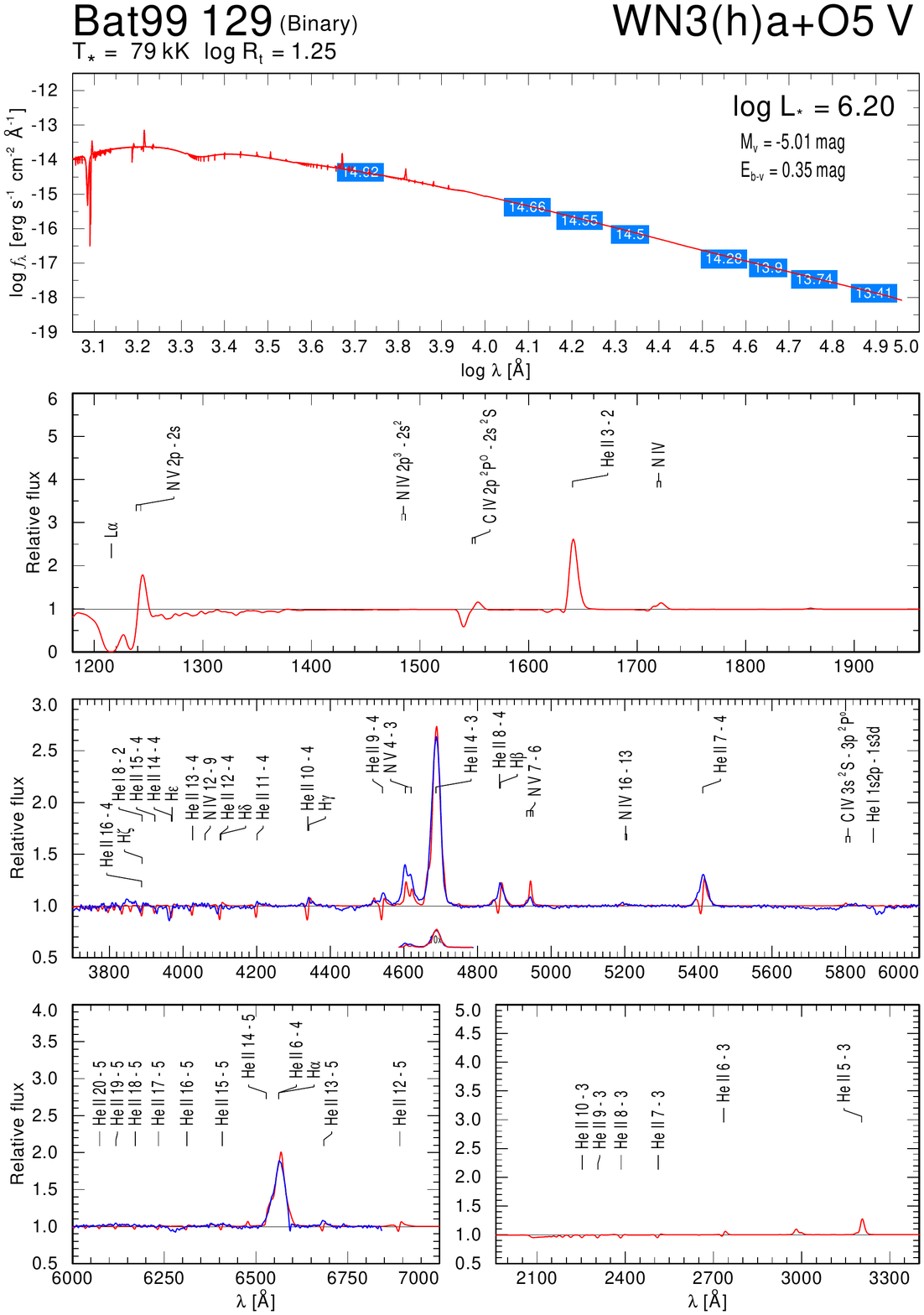}
  \vspace{-0.4cm}
  \caption{Spectral fit for BAT99\,128 and BAT99\,129}
  \label{fig:bat128}
  \label{fig:bat129}
\end{figure*}

\begin{figure*}
  \centering
  \includegraphics[width=0.46\hsize]{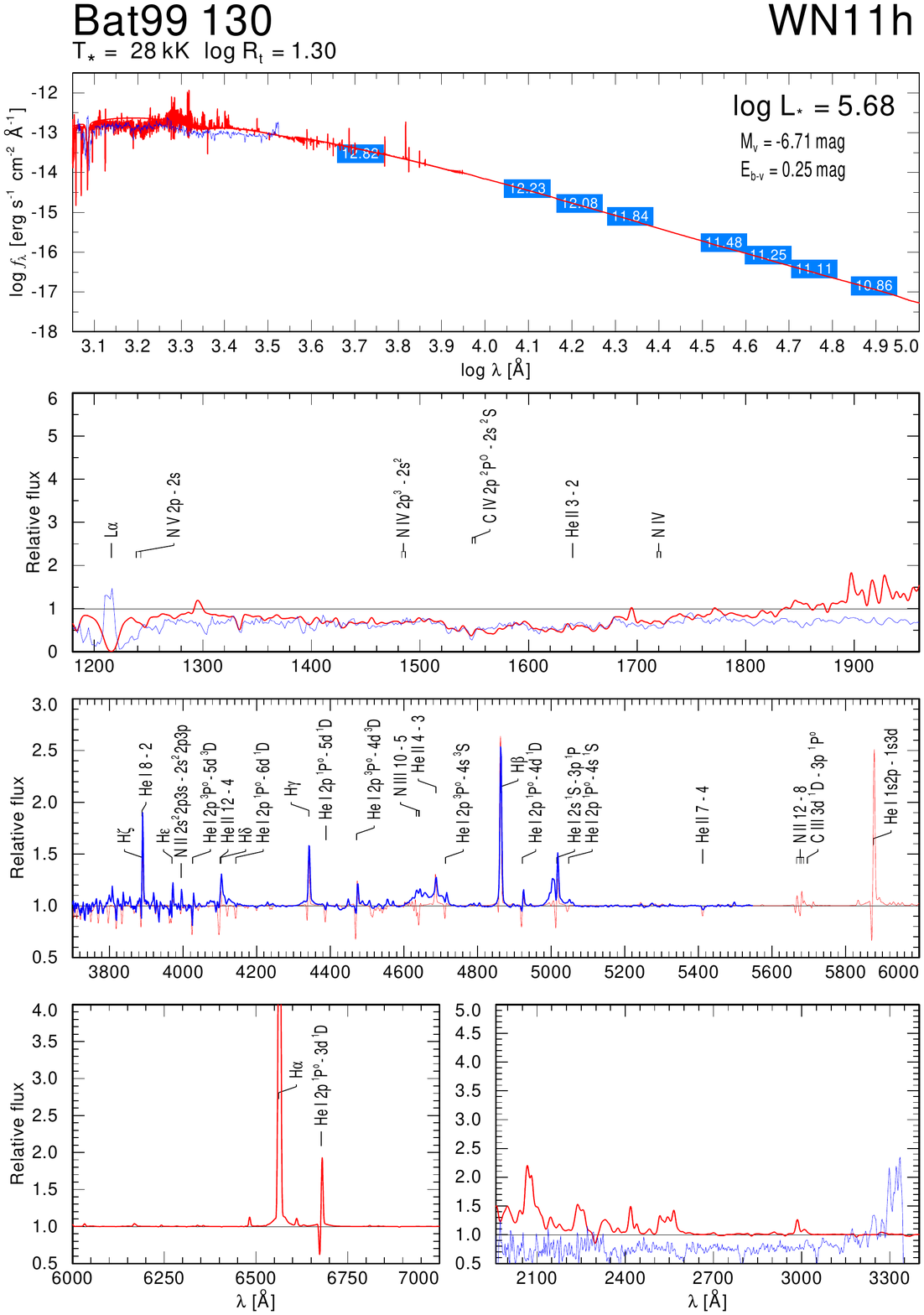}
  \qquad
  \includegraphics[width=0.46\hsize]{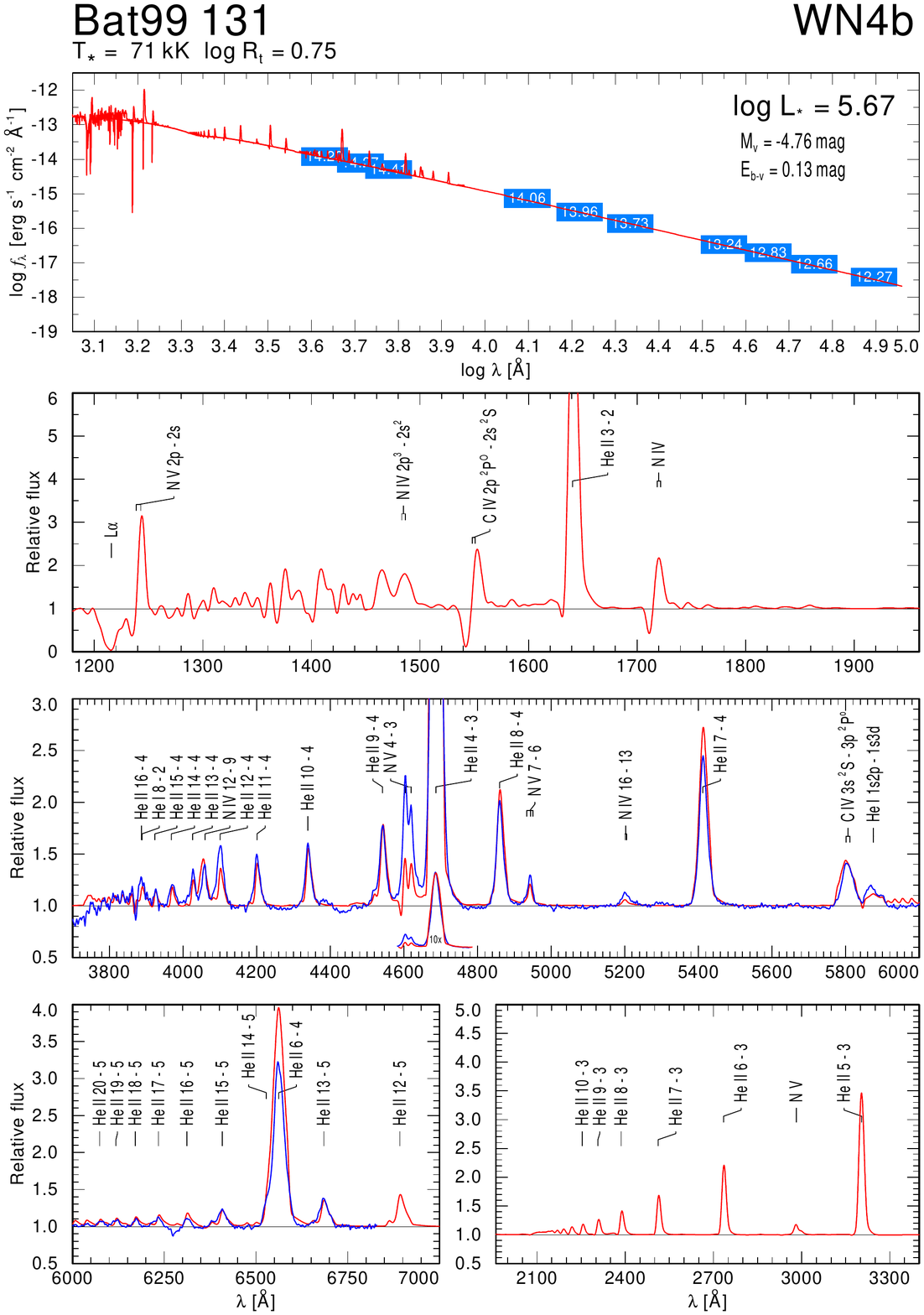}
  \vspace{-0.4cm}
  \caption{Spectral fit for BAT99\,130 and BAT99\,131}
  \label{fig:bat130}
  \label{fig:bat131}
\end{figure*}

\clearpage

\begin{figure*}
  \centering
  \includegraphics[width=0.46\hsize]{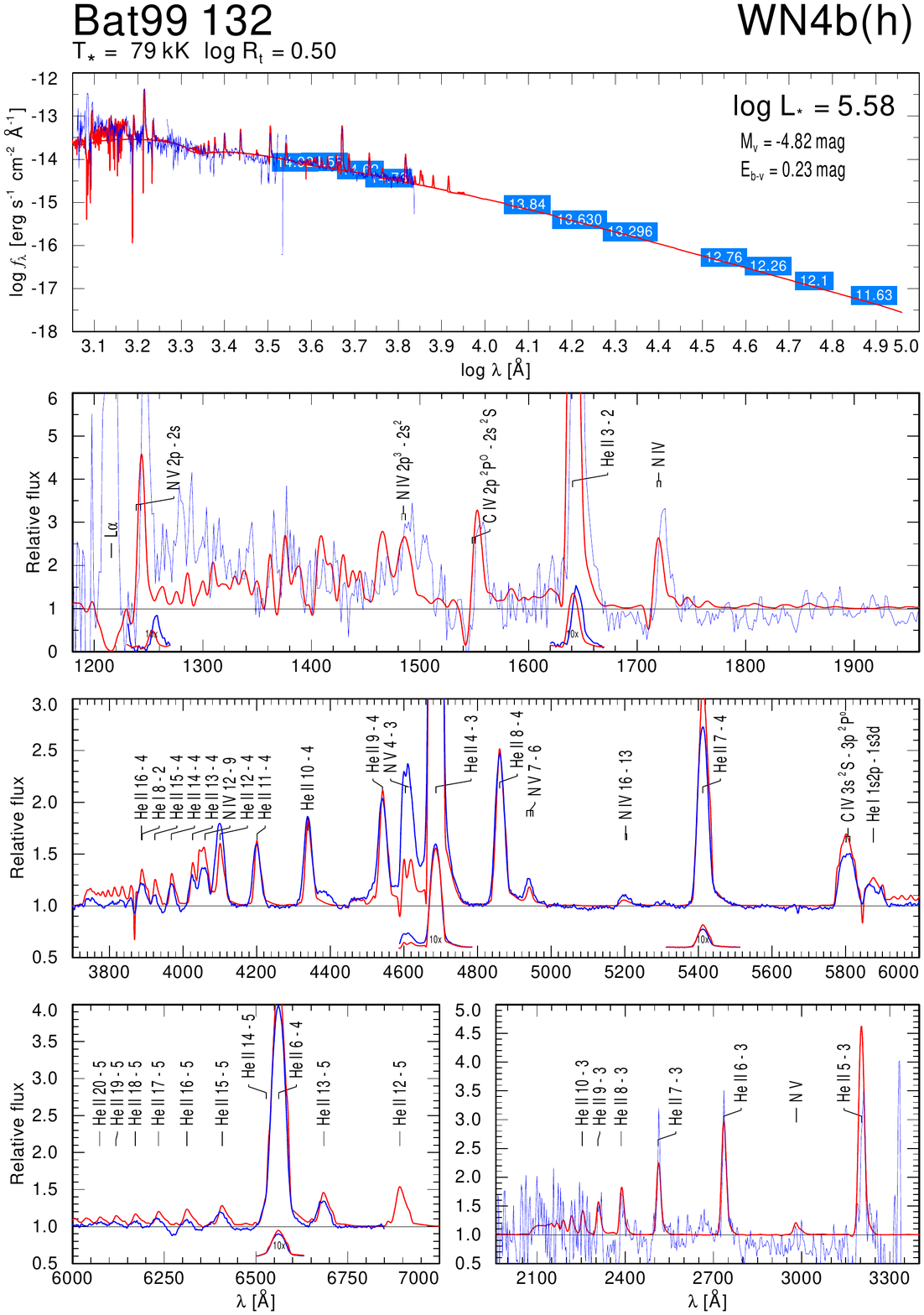}
  \qquad
  \includegraphics[width=0.46\hsize]{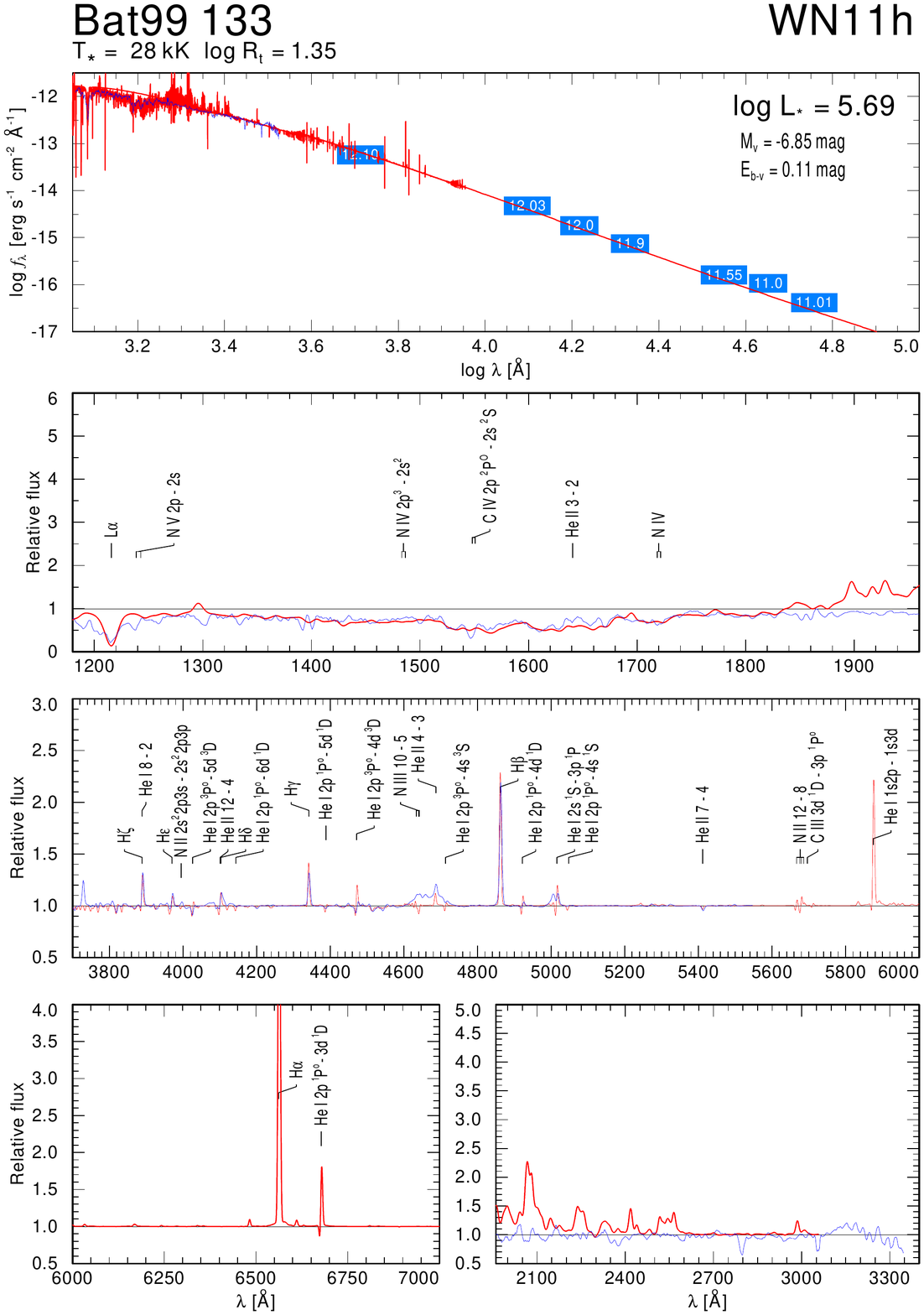}
  \vspace{-0.4cm}
  \caption{Spectral fit for BAT99\,132 and BAT99\,133}
  \label{fig:bat132}
  \label{fig:bat133}
\end{figure*}

\begin{figure*}
  \hspace{0.2cm}
  \includegraphics[width=0.46\hsize]{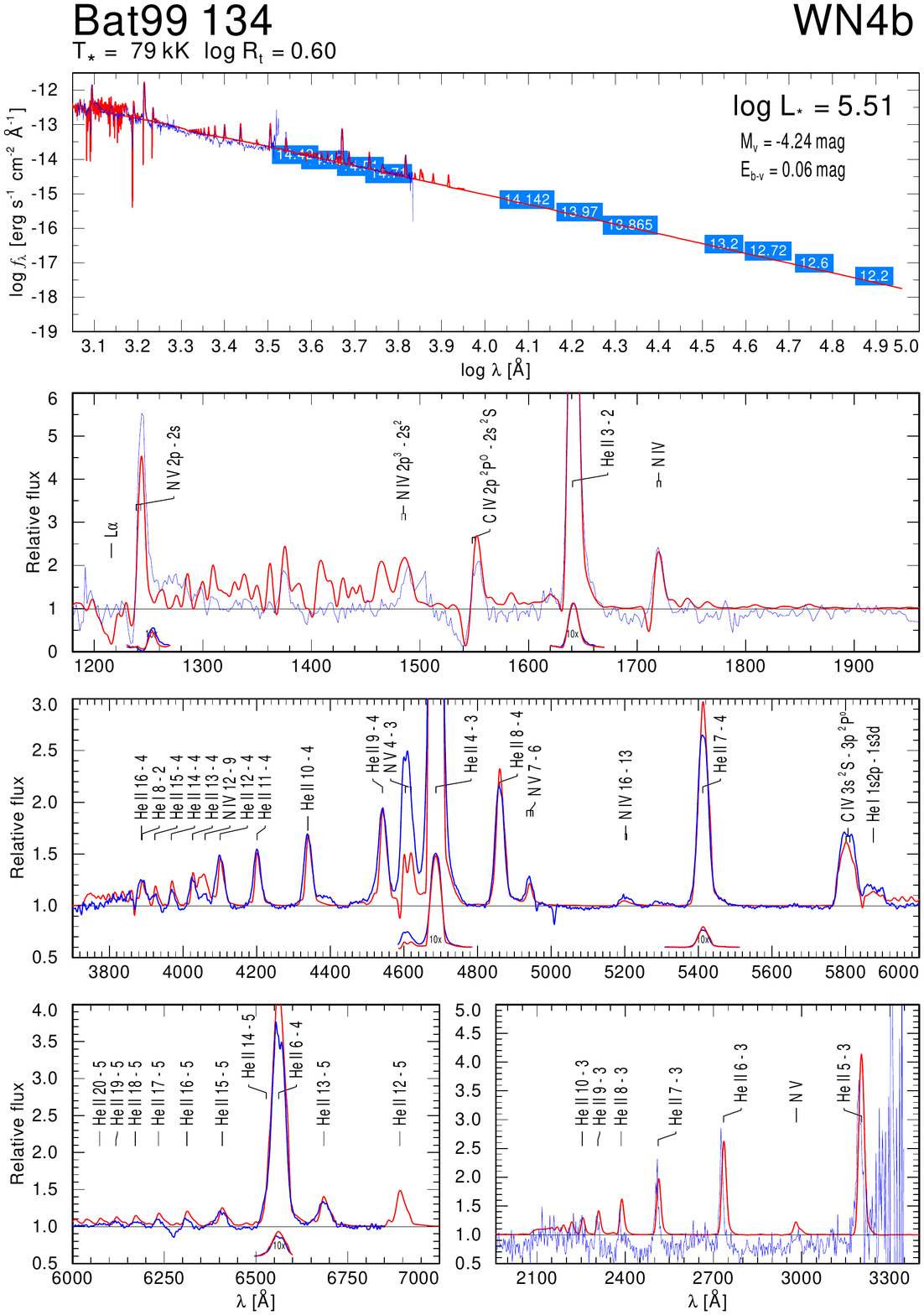}
  \vspace{-0.4cm}
  \caption{Spectral fit for BAT99\,134}
  \label{fig:bat134}
\end{figure*}

\end{appendix}

\end{document}